# Survey and Performance Evaluation of the Upcoming Next Generation WLAN Standard - IEEE 802.11ax

Qiao Qu[1], Bo Li[1], Member, IEEE, Mao Yang[*1], Member, IEEE, Zhongjiang Yan[1], Member, IEEE, Annan Yang[1], Jian Yu[2], Ming Gan[2], Yunbo Li[2], Xun Yang[2], Osama Aboul-Magd[2], Senior Member, IEEE, Edward Au[2], Senior Member, IEEE, Der-Jiunn Deng[3], Member, IEEE, and Kwang-Cheng Chen[4], Fellow, IEEE

[1]School of Electronics and Information, Northwestern Polytechnical University, Xi'an, China.

[2]Huawei Technologies Co., Ltd., China.

[3]the Department of Computer Science and Information Engineering, National Changhua University of Education, Changhua, Taiwan.

[4]the Department of Electrical Engineering, University of South Florida, Florida, USA.

**Abstract —With the ever-increasing demand for wireless traffic and quality of serives (QoS), wireless local area networks (WLANs) have developed into one of the most dominant wireless networks that fully influence human life. As the most widely used WLANs standard, Institute of Electrical and Electronics Engineers (IEEE) 802.11 will release the upcoming next generation WLANs standard amendment: IEEE 802.11ax. This article comprehensively surveys and analyzes the application scenarios, technical requirements, standardization process, key technologies, and performance evaluations of IEEE 802.11ax. Starting from the technical objectives and requirements of IEEE 802.11ax, this article pays special attention to high-dense deployment scenarios. After that, the key technologies of IEEE 802.11ax, including the physical layer (PHY) enhancements, multi-user (MU) medium access control (MU-MAC), spatial reuse (SR), and power efficiency are discussed in detail, covering both standardization technologies as well as the latest academic studies. Furthermore, performance requirements of IEEE 802.11ax are evaluated via a newly proposed systems and link-level**

integrated simulation platform (SLISP). Simulations results confirm that IEEE 802.11ax significantly improves the user experience in high-density deployment, while successfully achieves the average per user throughput requirement in project authorization request (PAR) by four times compared to the legacy IEEE 802.11. Finally, potential advancement beyond IEEE 802.11ax are discussed to complete this holistic study on the latest IEEE 802.11ax. To the best of our knowledge, this article is the first study to directly investigate and analyze the latest stable version of IEEE 802.11ax, and the first work to thoroughly and deeply evaluate the compliance of the performance requirements of IEEE 802.11ax.

# I. Introduction

Due to deep penetration of the mobile Internet and the continuous enrichment of wireless network services, demands for wireless traffic and quality of service (QoS) have dramatically increased in recent years [1]. Wireless local area networks (WLANs), together with the cellular networks, have emerged as the primary traffic bearing wireless networks due to their high speed, flexible deployment, and low cost. According to Cisco report, the wireless traffic in the world would sharply increase with a 47% Compound Annual Growth Rate (CAGR) from 2016 to 2021. Moreover, WLANs carried data traffic from 42% in 2015 to 49% in 2021 [2]. Therefore, to respond the rapid growth of traffic demands, researchers, enterprises, and standardization organizations are increasingly focusing on the key technologies and standardization process of the next generation WLANs. As the most widely used WLAN standard [3], Institute of Electrical and Electronics Engineers (IEEE) 802.11 is expected to release the next generation WLAN standard amendment: IEEE 802.11ax [4].

With the growing prosperity of smart terminals and the urgent demands of the Internet of Things (IoT) [5], high-dense deployment scenarios such as airports, stadiums, shopping malls, and enterprises develop into important scenarios of future wireless networks [6]. Future wireless networks need to deploy a large number of wireless access nodes such as base stations (BS) and access points (APs) in limited geographical areas to guarantee the required coverage and capacity; on the other hand, future wireless networks also need to support massive connectivities in a single cell, such as smart phones in a stadium and IoT devices in a smart home or enterprise network. Therefore, in Project Authorization Request (PAR) of IEEE 802.11ax, one of the most important documents of the standardization process, IEEE 802.11 highlights IEEE 802.11ax facing the requirements and objectives of such high-dense deployment scenarios [7].

From 1990s, IEEE started the standardization of IEEE 802.11 WLANs. IEEE 802.11 specifies the physical layer (PHY) and medium access control (MAC) layer of WLANs and has become the most widely used WLAN standard. To portrait the development of the IEEE 802.11 standard specification, we summarize the main amendments of IEEE 802.11 that have been or will be released in the future.

As shown in Fig. 1, IEEE 802.11 WLANs mainly work on three frequency bands: the sub-1 GHz band, the 2.4/5 GHz band, and the above 45 GHz band. In this figure, the left edge and right edge of each block indicate the establishing time and releasing time of the standard amendments respectively.

1) Sub-1 GHz band WLANs. Sub-1GHz band WLANs can support a long transmission distance; however, the transmission rate is limited; therefore, it is suitable for networking scenarios with large coverage and low transmission rate, such as IoT. Specifically, IEEE 802.11af constructs the network on television (TV) white band (white spaces spectrum) based on cognitive radio [8], while IEEE 802.11ah focuses on the networking consisting the tremendous IoT devices [9]}.

2) Above 45 GHz band WLANs. The millimeter wave is known by quite large capacity, directional transmission, and heavy path loss; therefore, above 45 GHz band WLANs are suitable for short distance and high-speed scenarios, such as high-definition video transmission at homes. Specifically, IEEE 802.11ad accomplishes the first standardization work in this band [10]. IEEE 802.11ay, as the successor of IEEE 802.11ad, introduces channel bonding (CB) and downlink (DL) multi-user (MU) multiple input multiple output (MU-MIMO) to further enhance the transmission rate [11]. Moreover, the IEEE 802.11aj standard amendment modifies IEEE 802.11ad to be compatible with the Chinese 60 GHz and 45 GHz bands.

3) 2.4/5 GHz band WLANs. Communications in 2.4/5G Hz band compromises both advantages, larger coverage and higher bandwidth, and attracts most interests in deployment due to global availability and suitability for CMOS implementation. Therefore, the 2.4/5 GHz band has become the most popular frequency band for WLAN standards and solutions. The upcoming next generation WLANs standard amendment IEEE 802.11ax is working on this band, and we therefore focus on the technology evolution of 2.4/5 GHz band WLANs. In the following of this article, without special instructions, the word ``WLANs" refers to WLANs in the 2.4/5 GHz band. By the way, the IEEE 802.11ba focuses on STA's power-saving and specifies PHY and MAC technologies for wake-up radio (WUR).

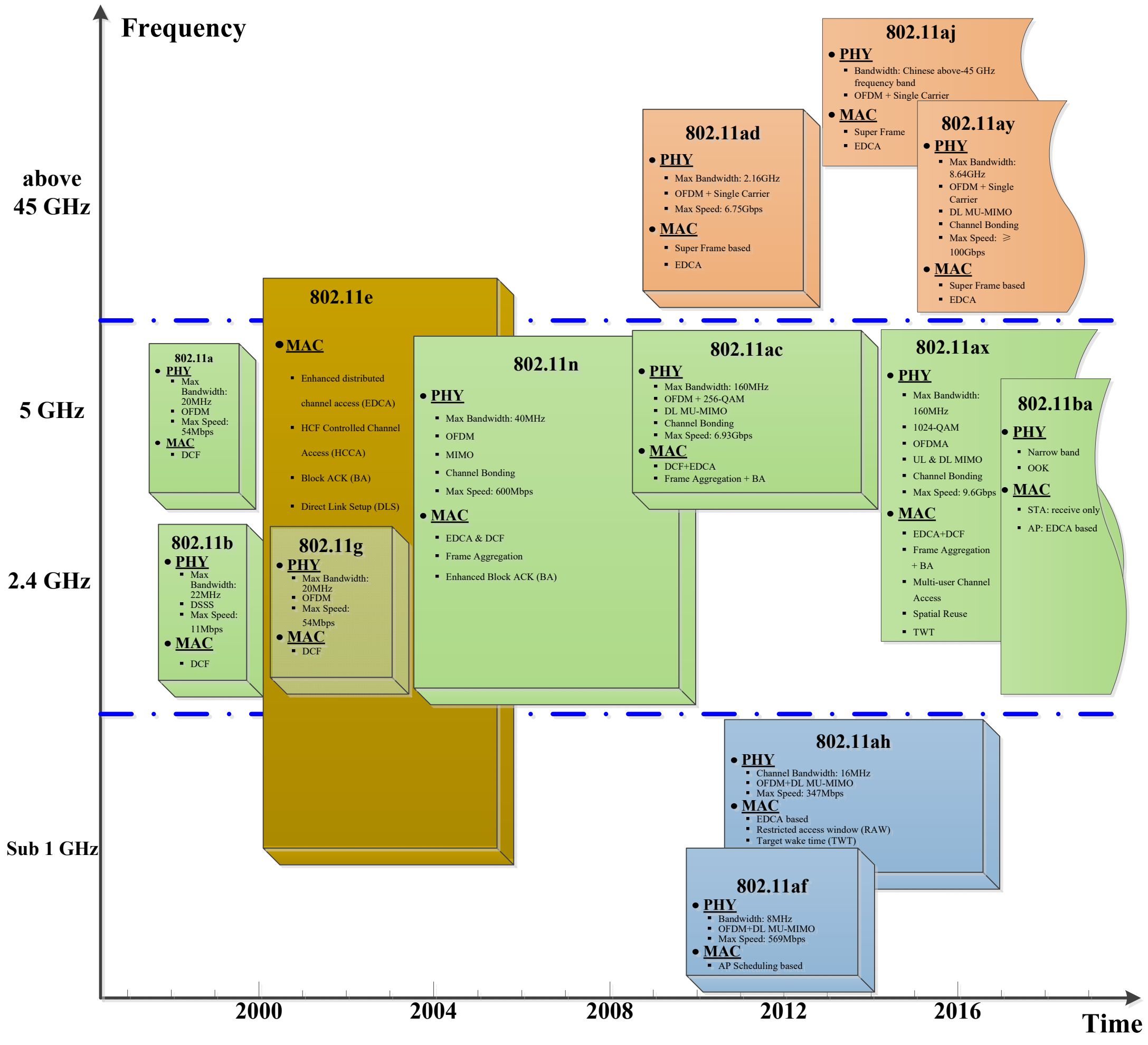

Fig. 1 Evolution of IEEE 802.11 WLANs.

The primary goal of the technology development and evolution of PHY is to improve the transmission rate. As shown in Fig. 1, IEEE 802.11a released in 1999 adopts orthogonal frequency multiplexing division (OFDM), and its maximum transmission rate reached 54 Mbps. During the same year, IEEE 802.11b adopts direct sequence spread spectrum (DSSS) with the maximum transmission rate of 11 Mbps. In 2003, IEEE 802.11g introduces OFDM to the 2.4 GHz band, so that the maximum transmission rate also reaches 54 Mbps. IEEE 802.11n, released in 2009, is supposed to be a milestone. It introduced CB and single user (SU) MIMO (SU-MIMO), thus the maximum transmission rate extends to 600 Mbps. Compared to the previous version, the maximum transmission rate of IEEE 802.11n increases more than 10-fold. In 2013, IEEE 802.11ac introduces 256 quadrature amplitude modulation (256-QAM) and CB enhancement technology, consequently, reaching a maximum transmission rate of 6.93 Gbps with DL MU-MIMO. Consequently, the PHY drives the transmission rate of IEEE 802.11 WLANs.

The primary goal of the development of the MAC layer is to improve the utilization efficiency of wireless resources shared by multiple users and to specify the corresponding protocol procedures. As shown in Fig. 1, IEEE 802.11 and amendments a/b/g used carrier sense multiple access with collision avoidance (CSMA/CA) based distributed coordination function (DCF). In subsequent amendments, the a major milestone to enhance the channel access of MAC layer is based on the IEEE 802.11e released in 2005 and IEEE 802.11n issued in 2009. Based on the characteristics in QoS, IEEE 802.11e specifies data traffic into four access categories (ACs). IEEE 802.11e utilizes hybrid coordination function (HCF) which includes both enhanced distribution channel access (EDCA) and controlled channel access (HCCA), transmission opportunity (TXOP), and block acknowledgement (BA) mechanisms. Thus, the following IEEE 802.11n supports DCF, EDCA, and HCCA. These not only improves the QoS guarantee, but also improves the access efficiency through the TXOP mechanism. Moreover, the frame aggregation (FA) mechanism improves the access efficiency. IEEE 802.11ac follows the MAC technology of IEEE 802.11n. Therefore, the MAC layer evolution promotes the continuous improvement of access efficiency of WLANs.

With the release of IEEE 802.11ac, the maximum transmission rate as well as the access efficiency of WLANs has been significantly improved. However, the current WLANs standard still faces challenges to meet the high-dense deployment scenarios of the future wireless network. Therefore, in 2013, after the release of IEEE 802.11ac, IEEE 802.11 immediately launched an early study of the next generation WLANs standard amendment IEEE 802.11ax: high efficiency WLAN (HEW). It is worth emphasizing that, in contrast to previous versions of IEEE 802.11 standard amendments that mainly focused on the improvements of the transmission rate, IEEE 802.11ax focuses more on the network performance and user experience of the high-dense deployment scenarios, such as improvements of single user average throughput and area throughput [7]. Therefore, according to Fig. 1, IEEE 802.11ax significantly enhances MAC layer technologies by introducing multi-user MAC (MU-MAC), spatial reuse (SR), and target wakeup time (TWT), which improves the access efficiency and the user experiencein in high-dense deployment scenarios. On the other hand, IEEE 802.11ax also achieves a maximum transmission rate up to 9.6 Gbps by introducing 1024-QAM, orthogonal frequency division multiple access (OFDMA), uplink (UL) MU-MIMO, enhanced CB, *and etc*. Currently, IEEE 802.11ax has released Draft 2.0. With regard to the fact that IEEE 802.11ax has introduced many new features in both PHY and MAC layers, we expect that IEEE 802.11ax will become a new milestone in the evolution of the IEEE 802.11 standard. Therefore, during recent years, both industrial and academia focus strongly on the standarization process and key technologies of IEEE 802.11ax [12]-[16].

The analysis mentioned above mainly focuses on some of the key amendments of IEEE 802.11, which make it easy to understand the backbone of IEEE 802.11 standard. It is worth noting that from the initial IEEE 802.11 standard released in 1997 to December 2016, the IEEE task group (TG) has worked on more than 20 amendments [17][18] that covered all aspects of wireless networking. To easily understand the development of the IEEE 802.11 standard, Tab. 1 summarizes all released versions of IEEE 802.11 and their amendments. In addition, eight amendments are being standardized [19], as shown in Tab. 2.

Tab. 1 The released standard amendments of of IEEE 802.11

| **Standard Name** | **Publish Year** | **Brief description** |
|---|---|---|
| IEEE 802.11-1997 [17] | 1997 | The original IEEE 802.11 standard lays the foundation of the IEEE 802.11 standard, and introduces the basic concept of basic service set (BSS). It defines the networking mode of infrared and |
| | | 2.4 GHz unlicensed bands. Frequency-hopping spread spectrum (FHSS) and direct sequence spread spectrum (DSSS) are adopted in 2.4 GHz. The workflow and the whole process framework of DCF and point coordination function (PCF) are defined in the MAC layer. The IEEE 802.11-1997 standard was clarified in 1999, and IEEE 802.11-1999 was generated. Therefore, both IEEE 802.11-1997 and IEEE 802.11-1999 refer to the most primitive IEEE 802.11 standard. |
| IEEE 802.11a | 1999 | For 5 GHz oriented unlicensed band networking. PHY uses OFDM technology. The channel width is 20 MHz. Modulation methods include binary phase-shift keying (BPSK), quadrature phase-shift keying (QPSK), 16-QAM, and 64-QAM. Eight modulation and coding schemes (MCS) are supported. |
| IEEE 802.11b | 1999 | For 2.4 GHz oriented unlicensed band networking. The physical layer adopts DSSS and complementary code keying (CCK) technology. The channel width is 20 MHz. IEEE 802.11b launched a corrigendum in 2003. |
| IEEE 802.11d | 2001 | For adding additional regulatory domains. The MAC layer defines the country information element, and adds the national information elements to the management frames. |
| IEEE 802.11g | 2003 | For 2.4 GHz oriented unlicensed band networking. OFDM technology is introduced in 2.4 GHz band, which enhances data rate on the basis of IEEE 802.11b. The MAC layer introduces a protection mechanism compatible to IEEE 802.11b. |
| IEEE 802.11h | 2003 | Dynamic frequency selection (DFS) and transmit power control (TPC) are introduced, and are oriented to IEEE 802.11a standard revision working at a 5 GHz unlicensed band. The MAC layer defines the DFS process to avoid using the same channel with the radar system, and ensures balanced use of the channel. The power control process is defined to reduce the interference to the satellite service. |
| IEEE 802.11i | 2004 | For the security enhancement of WLANs, a new security measure is introduced. MAC layer defines temporal key layer integrity protocol (TKIP) and CCM [counter mode (CTR) with cipher block chaining (CBC) with message authentication code (MAC)] protocol (CCMP), and introduces security association concept and the definition of security association management process. |
| IEEE 802.11j | 2004 | Revised standards for the Japanese market to support the 4.9 to 5 GHz band communication, using the physical layer of IEEE 802.11a. |
| IEEE 802.11e | 2005 | For QoS guarantee of multiple service types in WLANs. The MAC layer divides service into four ACs according to QoS requirement, introduces EDCA and HCCA to guarantees QoS, introduced TXOP mechanism and BA mechanism, and |
| | | introduces the need for an AP to participate in the establishment of direct link setup (DLS). |
| IEEE 802.11-2007 [20] | 2007 | & 2007 & IEEE 802.11 standard released in 2007 integrates the standard amendments of IEEE 802.11a, IEEE 802.11b, IEEE 802.11d, IEEE 802.11g, IEEE 802.11h, IEEE 802.11i, IEEE 802.11j, and IEEE 802.11e. |
| IEEE 802.11k | 2008 | For wireless resource measurement mechanism. When a station (STA) needs to associate with an AP, it does not only consider the signal strength as the judgment basis, while also considers the load intensity of each AP as important factors. |
| IEEE 802.11r | 2008 | The MAC layer redefines the security key negotiation protocol, and enhances the efficiency of handoff between multiple APs. |
| IEEE 802.11y | 2008 | Revised standards for the U.S. market to support the 3650 to 3700 MHz band communication, using the physical layer of IEEE 802.11a. |
| IEEE 802.11w | 2009 | For management frame protection oriented. The MAC layer enhances the protection of security in the management frames, enables the check to the integrity and source of the management frame, and adds protection for replay attacks. |
| IEEE 802.11n | 2009 | Work in the 2.4/5 GHz unlicensed band with the goal to improve the throughput of WLAN. The physical layer introduces MIMO technology (up to four space streams), and introduces CB technology to support the use of 40 MHz channel. MAC layer introduces a channel access method based on primary channel and secondary channel, introduces frame aggregation technology, and enhances the BA mechanism. |
| IEEE 802.11p | 2010 | For wireless access vehicular environments (WAVE) in 5.9 GHz band intelligent transportation systems (ITS). IEEE 802.11a physical layer is adopted, and the channel bandwidth is 10 MHz. The MAC layer does not need to establish the BSS and waits for association and authentication, and the node is synchronized via the broadcast management frame. |
| IEEE 802.11z | 2010 | For the establishment of direct link between users. The MAC layer defines the process of establishing a direct link between two STA (without participation of AP) and a tunneled direct-link setup (TDLS) based mechanism. |
| IEEE 802.11v | 2011 | For wireless network management. To support the simple network management protocol (SNMP) in WLANs and match with AP’s more complex trend, MAC layer defines the mechanism of network information interaction between STAs, and enhances management information base (MIB). |
| IEEE 802.11u | 2011 | For the interaction between WLANs and external network. It enables WLANs to search for more external networks, defines the interconnection function between different wireless networks, and |
| | | defines the QoS mapping between WLANs and external network. |
| IEEE 802.11s | 2011 | Wireless devices can build wireless mesh networks in a connected manner. The MAC layer adds the mesh coordination function (MCF) access protocol for the mesh network, and defines the architecture of the mesh network and the default routing protocol hybrid wireless mesh protocol (HWMP). |
| IEEE 802.11-2012 [21] | 2012 | The IEEE standard published in 2012, and integrats IEEE 802.11k, IEEE 802.11r, IEEE 802.11y, IEEE 802.11w, IEEE 802.11n, IEEE 802.11p, IEEE 802.11z, IEEE 802.11v, IEEE 802.11u, and IEEE 802.11s standard amendments. |
| IEEE 802.11ae | 2012 | Strengthenes the management frame differentiated transmission priority. The MAC layer introduces QoS management frame (QMF), divides the management frame into different priority, and allowes nodes to access and transmit management frames based on their priority classification in QMF. |
| IEEE 802.11aa | 2012 | To improve the robustness of audio and video stream transmission, MAC divides service into six queues according to stream classification layer by service (SCS), adds groupcast with retries service (GCR), and enhanced OBSS management and interaction to guarantee the QoS of multimedia services. |
| IEEE 802.11ad | 2012 | For the short distance communication network, work in 60 GHz unlicensed band. The physical layer supports directional antenna data transmission, and is called directional multi-gigabit (DMG), which supports OFDM, single carrier, and low power single carrier technology of data transmission. The channel bandwidth is 2.16 GHz. The MAC layer adds association beam forming training (A-BFT) access, announcement transmission interval (ATI) access, contention-based accessperiod (CBAP) access, and service period (SP) access. |
| IEEE 802.11ac | 2013 | An enhanced amendment of IEEE 802.11n for unlicensed bands below 6 GHz (excluding 2.4 GHz) with the goal to further improve WLAN throughput. The physical layer introduces 256-QAM modulation, enhances MIMO technology, and supports most eight space streams, supports DL MU-MIMO transmission, and enhances CB technology to support the use of 80 MHz and 160 MHz channel. Several modifications have been made in the MAC layer to support the above features, such as DL MU-MIMO frame, MAC frame structure enhancement, and enhanced CB access process. |
| IEEE 802.11af | 2014 | Allow White Space Devices (WSD) to share the TV white band via cognitive radio, works in the 54-790 MHz band. The PHY defines the OFDM technology for the 6/7/8 MHz channel and the corresponding CB technology to support MIMO and DL MU-MIMO technology. The MAC layer determines the currently |
| | | available channel information by querying the geolocation data base (GDB). |
| IEEE 802.11-2016 [18] | 2016 | IEEE 802.11 standard released in 2016, the new integration of IEEE 802.11ae, IEEE 802.11aa, IEEE 802.11ad, IEEE 802.11ac, and IEEE 802.11af standard amendments. |
| IEEE 802.11ah | 2016 | The long distance communication network application for outdoor scenarios, supporting the IoT scenarios and works in a frequency band below 1 GHz. The physical layer defines the OFDM transmission for the 1 MHz channel to 16 MHz channel using CB technology, supporting MIMO and DL MU-MIMO technology. The MAC layer introduced the concept of Relay AP, TWT mechanism, energy saving mechanism based on Traffic Indication Map (TIM)STA, and non-TIM STA, and sectorized BSS technology. |
| IEEE 802.11ai | 2016 | On the premise of ensuring security, STA can quickly establish fast initial link setup (FILS). The MAC layer optimizes network discovery, security settings, network protocol address allocation, and FILS capability indication procedures to reduce the time to build initial links. |

Tab. 2 The amendments being developed of IEEE 802.11

| **Standard Name** | **Brief description** |
|---|---|
| IEEE 802.11aj | The standard amendment of 60 GHz and 45 GHz bands for the Chinese market is the specific standard revision of IEEE 802.11ad for the Chinese market. |
| IEEE 802.11ak | Standard amendment to enhance connectivity between WLAN and other networks. The MAC layer can create transit links through general link (GLK) STA in a network, and conforms to the IEEE 802.1Q. |
| IEEE 802.11aq | AP can provide services to external or non-AP service to STA, i.e. AP acts as a proxy server. MAC layer adds the pre-association discovery (PAD) function, so that STA can discover the service provided by AP. |
| IEEE 802.11ax | Work in 2.4 GHz or 5 GHz unlicensed bands, and the goal is to achieve efficient WLANs in high-dense deployment scenarios. The PHY introduces OFDMA, UL MU-MIMO, 1024-QAM and dual carrier modulation (DCM) technologies, and further enhances the CB technology. The access mechanism in the MAC layer has been significantly improved, including MU-MAC and SR. |
| IEEE 802.11ay | For the above 45 GHz band WLAN, IEEE 802.11ay is the enhanced version of IEEE 802.11ad, and its goal is to achieve higher throughput (at least achieving MAC throughput to 20 Gbps). The physical layer introduces CB technology to increase the channel width to four times of IEEE 802.11ad, introduces MIMO technology, and supports MU-MIMO. The MAC layer introduces the channel access technology based on CB, enhances the BA mechanism, and introduces SR technology. |
| IEEE 802.11az | To meet the requirements of users to use high-performance location services anytime and anywhere, IEEE 802.11 needs to expand its own positioning capabilities to adapt to new scenarios (such as dense deployment scenarios). Fine timing measurement (FTM) is introduced into the physical layer, and different positioning methods and workflow are developed for 2.4/5 GHz and 60 GHz, respectively. |
| IEEE 802.11ba | In some application scenarios (such as IoT), on the one hand, it is required to extend the battery lifetime of the device, while on the other hand, low delay is guaranteed in some application scenarios. The concept of wake-up radio (WUR) is proposed to allow the primary connectivity radio as much as possible to remain sleeping while maintaining another low power transceiver in the working state. The ultimate goal is to reduce the energy consumption of the device to one watt. |

Currently, there are several looking into key technologies of IEEE 802.11ax from different perspectives [12][13][16][22]-[25]. Based on Draft 0.4, Afaqui et al. [13] analyze the scenarios and demands of IEEE 802.11ax, and then investigate several possible key technologies, especially overlapping BSS (OBSS) related technologies. Khorov et al. [22] analyze several potential technical characteristics of IEEE 802.11ax during the earlier stage. Gong et al. [25] illustrate CB, DL MU-MIMO, and 256-QAM that are adopted in IEEE 802.11ac, and then introduce the possible features of IEEE 802.11ax via comparison with IEEE 802.11ac. Cheng et al. [24] discuss the application scenarios, channel access, SR, and several other issues based on multiple IEEE standard proposals. Bellalta et al. [23] survey several new standard amendments including IEEE 802.11ax, IEEE 802.11ah, and IEEE 802.11af, and analyze the support for the multimedia scenarios in IEEE 802.11ax. Omar et al. [26] and Bellalta [27] mainly focus on results of academic studies, and list a series of potential technologies for IEEE 802.11ax. Deng et al. [12][16] analyze the problems of legacy IEEE 802.11, and discuss the QoS requirements and the technical challenges faced by IEEE 802.11ax, and then propose a protocol framework for MU access. Our previous study [13] analyzes and investigates the MU-MAC protocol based on OFDMA, and proposes an efficient OFDMA based multi-user MAC protocol framework.

These illustrate high quality analysis on the performance requirements and potential technologies for IEEE 802.11ax. However, the IEEE 802.11ax Draft 2.0 was just released at the end of 2017, it is necessary to analyze and summarize the key technologies based on the latest stable version of IEEE 802.11ax (Draft 2.0) in a holistic manner. While, on the other hand, these studies mainly focused on academic results and less on the standardization, industry perspectives, and concerns. Therefore, it is necessary to combine the academic and industrial viewpoints to study the key technologies and standardizations of 802.11ax to achieve a more objective discussion. To clearly analyze and summarize IEEE 802.11ax, one possible way is to directly introduce the specifications. However, to provide a more complete picture of the key technologies to the readers, we in this article try to introduce both the key technologies specified in standard and the recent studies of these key technologies in the academia. Finally, we notice that few existing studies show objective performance results of IEEE 802.11ax evaluated by the designed specifically network simulation platform. In response to these challenges and via integration of industrial and academic points of view, this article surveys the key technologies and latest standardization progress of IEEE 802.11ax, providing a comprehensive and objective analysis, and evaluating the performance through an integrated simulation platform.

The contributions of this article can be summarized as follows:

- To the best of our knowledge, this is the first work to directly investigate and analyze the latest stable version (Draft 2.0) of IEEE 802.11ax. Focusing on the standardization process of IEEE 802.11ax, this article surveys the application scenarios, objectives, and requirements as well as key technologies including detailed analyses and discussions of PHY enhancements, MU-MAC, SR, and power efficiency. This article covers the standardization as well as academic results. Therefore, it better matches the latest standardization progress of IEEE 802.11ax.
- To the best of our knowledge, this is the first study to thoroughly and deeply evaluate the performance towards the performance requirements of IEEE 802.11ax. Based on the simulation results presented in this article, researchers may obtain a more profound understanding of the achievable performance of IEEE 802.11ax. We believe that such an understanding is useful for future standardizations. It has been highlight that the simulation platform named system & link level integrated simulation platform (SLISP) used in this article is the first integrated simulation platform for IEEE 802.11ax in the world by integrating link level simulation and system level simulation, which meets the requirements of the standard document [28]-[30]. Therefore, the performance results given in this article are objective and convincing.

The remainder of this article is organized as follows: To quickly obtain a general understanding of IEEE 802.11ax, Sec. II presents a general overview of IEEE 802.11ax, including application scenarios, technical requirements, key technologies, and the standardization process. Sec. III - Sec. VI present the key technologies of IEEE 802.11ax in detail. Specifically, Sec. III introduces the PHY enhancement technologies; Sec. IV focuses on MU-MAC; Sec. V describes and analyzes the spatial reuse technology; Sec. VI introduces some other enhancement techniques. The proposed integrated simulation platform and performance evaluation of IEEE 802.11ax are shown in Sec. VII. In Sec. VIII we conclude this article and show some perspectives beyond 802.11ax WLANs.

# II. A Quick Glance at IEEE 802.11ax

## A. Background and Requirement Analysis

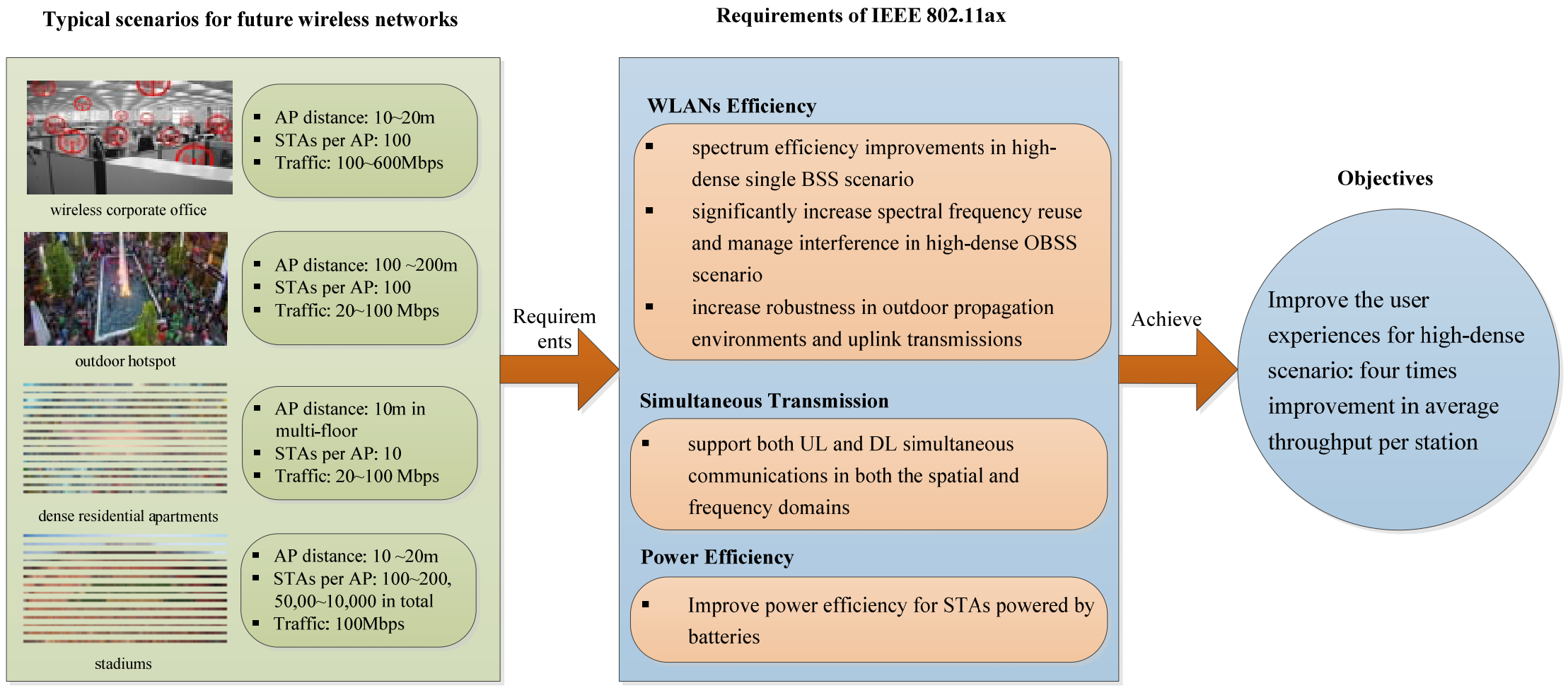

Fig. 2 Scenarios and requirements for IEEE 802.11ax.

Future wireless networks are required to meet high-dense deployment scenarios, including both indoor and outdoor scenarios. Different interpretations exist on the definition of high-dense deployment wireless networks. For example, one definition is that the density of the access infrastructures, such as BS and AP, is larger than the density of active users [31]. Another such definition is that the density of the access infrastructures is above 1000/km$^2$ [31]. Specific to the high-dense deployment of WLANs, it requires a large number of APs deployed in a limited area [15][32]-[35]; meanwhile, massive connectivity needs to be served in each cell [1]. Fig. 2 shows four typical scenarios that IEEE 802.11ax needs to support.

1) Enterprise office scenario

The wireless office has become an important trend for many enterprises to reduce the deployment costs, while enhancing working efficiency [36]. For example, employees can consult literature, send files, download data, and take video conferences via wireless networks. In an enterprise, since the work stations among employees can be very close, it is probably necessary to deploy one AP every 10-20 meters, and each AP often supports dozens or even hundreds of STAs. In addition to ordinary business requirements, video conferencing services generally require the transmission rate of hundreds of Mbps [37].

2) Outdoor large hotspots

A smart city requires a large number of hotspots deployed in the city, particularly in large shopping malls and other places that attract a large number of people. This often requires to deploy an AP every 100-200 meters, and each needs to cover dozens or even hundreds of STAs. In addition, it is necessary to support user demands with different QoS due to ever-diverse user demands.

3) Densely residential apartments

In 2011, more than 25% of households in the world had WiFi connections, according to the Strategy Analytics, and this ratio was even 61% in the United States [38][39]. Therefore, it is not difficult to imagine that a large number of WLANs deploy in densely populated residential apartments. First of all, each household on each floor, e.g., deploys at least one AP. With the continuous growth of the intelligent home furnishing and IoT, AP deployed in every family needs to support about 10 STAs and furthermore, and the QoS requirements may be quite different among these STAs.

4) Stadium scenario

Stadium holding sports games, big concerts, and large-scale social activities need to support wireless network services. This extremely challenges the performance of WLANs. It probably requires deploying one AP every 10-20 m. Moreover, each AP needs to support 100-200 STAs, and the number of potential users in the entire stadium may be 50,000-100,000. The WLANs in stadium should provide video stream and video-call services for users [40][41].

Evidently, these typical scenarios pose a great challenge to WLANs, and the performance of traditional WLANs in high-dense deployment scenarios will rapidly deteriorate and can hardly meet the demands. Therefore, IEEE immediately launched the next generation of WLANs standard amendment: IEEE 802.11ax after the release of IEEE 802.11ac. This means that IEEE 802.11ax should directly face high-dense deployment scenarios.

PAR of IEEE 802.11ax points out that IEEE 802.11ax specifies both PHY and MAC layer technologies. The overall objective of IEEE 802.11ax is to significantly improve the user experience in high-dense deployment scenarios. Specifically, IEEE 802.11ax needs to achieve at least a four-fold improvement in the average throughput per STA in specific high-dense deployment scenarios. In addition, the 5% percentile of per STA throughput, packet delay, and packet error ratio (PER) also need to be guaranteed. To achieve this objective, PAR asks IEEE 802.11ax to meet the following technical requirements:

- Enhance WLAN efficiency

First of all, the efficiency of traditional IEEE 802.11 in high-dense scenarios dramatically reduces [13]. Therefore, in the high-dense scenarios, IEEE 802.11ax is required to achieve a more efficient use of spectrum resources in scenarios with a high density of STAs per BSS.

Secondly, in OBSS scenarios, interference severely affects the throughput of the entire network and unfortunately, and the traditional IEEE 802.11 lacks effective means of spectrum reuse and interference management. Therefore, IEEE 802.11ax needs to significantly improve the spectrum reuse and interference management capability in OBSS scenarios with a high density of both STAs and BSSs.

Thirdly, the traditional IEEE 802.11 mainly focused on indoor networking. The support for the outdoor networking is very limited since the transmission distance in the outdoor scenario is farther and the channel environment is much more complicated. However, since the maximum transmit power of STA is often lower than AP the traditional IEEE 802.11 is unlikely to satisfy the robustness of UL transmission. Therefore, IEEE 802.11ax is required to improve both the robustness of outdoor scenario and of UL transmission.

- Support parallel transmission in frequency domain and spatial domain

The traditional IEEE 802.11 only supports SU transmission in the frequency domain. Even for the space domain, the traditional IEEE 802.11ac just enables DL parallel transmission, i.e., DL MU-MIMO, while the UL MU parallel transmission is not suported. This severely inhibits the efficiency of the MAC layer. Therefore, IEEE 802.11ax shall enable both UL and DL MU parallel access capabilities in the frequency domain, and introduce the UL MU parallel access and transmission capabilities in the space domain.

- Improve power efficiency

Batteries often power STAs; therefore, power efficiency is always an important requirement for IEEE 802.11. Furthermore, smart phones became increasingly powerful during recent years; thus, the requirements of power efficiency became more urgent. Therefore, IEEE 802.11ax needs to further improve power efficiency. More importantly, as IEEE 802.11ax introduces several new features in PHY and MAC, especially MU access, power efficiency enhancements that efficiently match these new features should be considered.

- Key Technologies for 802.11ax

In response to the high-dense deployment scenarios and the objective and technical requirements, IEEE 802.11ax proposes a series of key technologies, as shown in Tab. 3. It is worth noting that the structure of the following sections is consistent with Tab. 3.

Tab. 3 Key technologies and requirements analysis for IEEE 802.11ax

| Key technologies | | Enhance WLANs efficiency | | | Parallel transmission in frequency domain and spatial domain | Power efficiency | Section |
|---|---|---|---|---|---|---|---|
| | | Specteum efficiency improvements in singe BSS | Spectrum reuse and interference management in OBSS scenatios | Robustness of outdoor scenario and UL transmission | | | |
| PHY Enhancements | New Modulation and Coding | √ | | √ | | | Sec. III.B Sec. |
| | | | | | | | III.C |
| | Subcarrier division mechanism | √ | | | √ | | Sec. III.D |
| | New multiple access technology (OFDMA, MU-MIMO) | √ | | √ | √ | | Sec. III.E |
| | Enhanced CB | √ | | | | | Sec. III.F |
| | New PPDU | √ | | √ | √ | | Sec. III.G |
| MU-MAC | UL MU MAC | √ | | √ | √ | | Sec. IV.C |
| | DL MU MAC | √ | | √ | √ | | Sec. IV.D |
| | Cascaded MU-MAC | √ | | √ | √ | | Sec. IV.E |
| Spatial Reuse | BSS Color | | √ | | | | Sec. IV.C |
| | Two NAVs | | √ | | | | Sec. V.D |
| | OBSS_PD | | √ | | | | Sec. V.E |
| | SRP | | √ | | | | Sec. V.F |
| Other Technologies | TWT | | | | | √ | Sec. VI.A |
| | Power Efficiency Enhancements | | | | | √ | Sec. VI.B |

1) PHY enhancements

IEEE 802.11ax adopts new modulations and coding strategies. Firstly, by introducing 1024-QAM, the maximum transmission rate could be further improved. Theoretically, the maximum transmission rate of IEEE 802.11ax is 9.6 Gbps. Secondly, DCM enhances the robustness of transmissions in both outdoor scenario and UL transmission. Finally, low-density parity check (LDPC), as well as binary convolutional encoding (BCC) are chosen as a mandatary coding technique in IEEE 802.11ax.

A new subcarrier division mechanism is adopted for IEEE 802.11ax, which is more fine-grained compared to traditional IEEE 802.11. In IEEE 802.11ax, the 20 MHz band is divided into 256 subcarriers, which is four times that of the legacy IEEE 802.11. This new subcarrier division mechanism leads to more precise and efficient scheduling of OFDMA resources, and further improved spectrum efficiency.

IEEE 802.11ax enhances the multiple access technology based on OFDMA and UL MU-MIMO (DL MU-MIMO has been used in IEEE 802.11ac) to guarantee the MU parallel transmission in both frequency domain and spatial domain. This provides a solid foundation for improving the efficiency of WLANs.

The enhanced CB in IEEE 802.11ax enables APs and/or STAs to transmit frames on non-continuous channels, which improves channel utilization and fully uses the larger bandwidth. To support these new PHY technologies, physical layer convergence protocol (PLCP) protocol data unit (PPDU) is also enhanced accordingly.

Sec. III shows the details for PHY enhancements.

2) MU-MAC enhancements

The most important enhancement for the MAC layer of IEEE 802.11ax is the enhancement of MU-MAC. MU-MAC is a type of high efficiency multiple access technologies, which enables multiple users to obey certain access rules to transmit UL data concurrently (known as UL MU-MAC), or to transmit DL data concurrently (known as DL MU-MAC) through given network access resources (i.e. space domain, time domain, and frequency domain resources). MU-MAC covers OFDMA based and MU-MIMO based multiple access processes introduced by IEEE 802.11ax. Thus, in the absence of confusion, this article uses the term MU-MAC to refer to all the multiple user access proposed in 802.11ax.

Firstly, IEEE 802.11ax first introduces UL MU-MAC, including UL OFDMA and UL MU-MIMO. Secondly, for DL MU-MAC, IEEE 802.11ac only supports DL MU-MIMO, but IEEE 802.11ax introduces DL OFDMA to further enhance DL parallel access.

Moreover, IEEE 802.11ax proposes a cascaded MU MAC, allowing the DL MU transmission and the UL MU transmission to occur alternately in a TXOP time duration, which further improves the MAC efficiency.

In summary, MU-MAC in IEEE 802.11ax makes the spectrum efficiency more efficient, and makes the MAC layer overhead much lower.

Sec. IV shows the details for MU-MAC enhancements.

3) SR technology

SR has been suggested to be an optional technology in IEEE 802.11ax to enhance the spectrum reuse capability and interference management ability in high-dense OBSS scenarios. By introducing the BSS color mechanism, the node can easily distinguish whether the received frame originated from the intra-BSS or inter-BSS, where intra-BSS represents the STA’s own BSS and inter-BSS indicates other BSS. BSS color is the technical premise for the following three mechanisms of SR.

IEEE 802.11ax enhances the virtual carrier sensing mechanism by requiring the node to maintain two NAVs counters: intra-BSS NAV counter and basic NAV counter.

Not only is two NAVs based virtual carrier sensing mechanism compatible with MU-MAC, but it also avoids the TXOP-ending chaos problem in high-dense deployment scenarios since all the existing NAVs would be ended by any CF-End frame.

OBSS_PD based SR mechanism introduced in IEEE 802.11ax allows one node to use a higher clear channel assessment (CCA) threshold for physical carrier sensing if the received data packet originated from inter-BSS, which improves the probability of parallel transmission.

Moreover, IEEE 802.11ax also introduces the spatial reuse parameter (SRP) mechanism, which allows STAs to perform spatial reuse in shorter time granularity, i.e., intra-PPDU granularity. Similarly as OBSS_PD, SRP also enhances the probability of parallel transmission.

Sec. V shows the details for the SR technology.

4) Other MAC layer enhancements

Another important breakthrough of IEEE 802.11ax is the service reservation mechanism. Specifically, based on the AP's global vision as well as its powerful control and management capabilities, IEEE 802.11ax adopts TWT, thus enhancing the scheduling ability and QoS guarantee of WLANs by allocating different service times for different STAs. The TWT mechanism achieves service reservation in the time dimension, reduces the collision, and enhances the QoS guarantee of WLANs. It is worth noting that the original purpose of introducing TWT mechanism is to enhance power-saving. But, from the angle of technical generality, TWT mechanism introduces service reservation. Therefore, in this article, we will comprehensively describe service reservation based on TWT mechanism.

Moreover, to increase power efficiency, IEEE 802.11ax introduces a TWT based power save mechanism and intra-PPDU based power save mechanism. These mechanisms are better adapted to the MU-MAC introduced in IEEE 802.11ax, and improve the power efficiency for many scenarios.

Sec. VI shows the details for other MAC layer enhancements.

### B. Standardization Process

To simplify understanding the standardization process of the IEEE 802.11, this subsection uses the standardization process of the IEEE 802.11ax as an example. As shown in Fig. 3, to improve the efficiency of WLANs, after the release of IEEE 802.11ac in March 2013, IEEE 802.11 immediately sets up a study group (SG) named as HEW [42]. Via full discussion, a series of features have been confirmed as the main technologies for next generation WLANs. Then, HEW completed two important documents: PAR [7] and criteria for standards development (CSD) [43]. PAR, as one of the most important documents in the standardization process, specifies the scope and technical requirements, while CSD specifies the technical criteria standardization process.

In March 2014, the IEEE 802 executive committee approved the PAR and the an 802.11 Task Group (802.11ax) was formed to develop the technical specification. Dr. Osama Aboul-Magd from HUAWEI was elected as chairman [44]. In the development process of TGax, for the first time, MU and SR were regarded as the most important research directions. To form the final IEEE 802.11ax amendment, TGax decided to form specification framework document (SFD) [45] first. In general, SFD is used to stipulate the outline of the standard amendment and clarifies some technologies the draft should contain.

At the beginning of 2016, the development of SFD ended, and the first draft of IEEE 802.11ax was released in March 2016. In November 2016, IEEE 802.11ax released Draft 1.0 [46]. In October 2017, IEEE 802.11ax released Draft 2.0 [4], which was referenced in this article. In addition, the final version of IEEE 802.11ax will be released in 2019. Fig. 3 shows the timeline of the IEEE 802.11ax standardization process. To facilitate a better understanding of the basic process of standardization, we explain the main role of each important document in TGax in Tab. 4 [7][29][30][43][45][47]-[49].

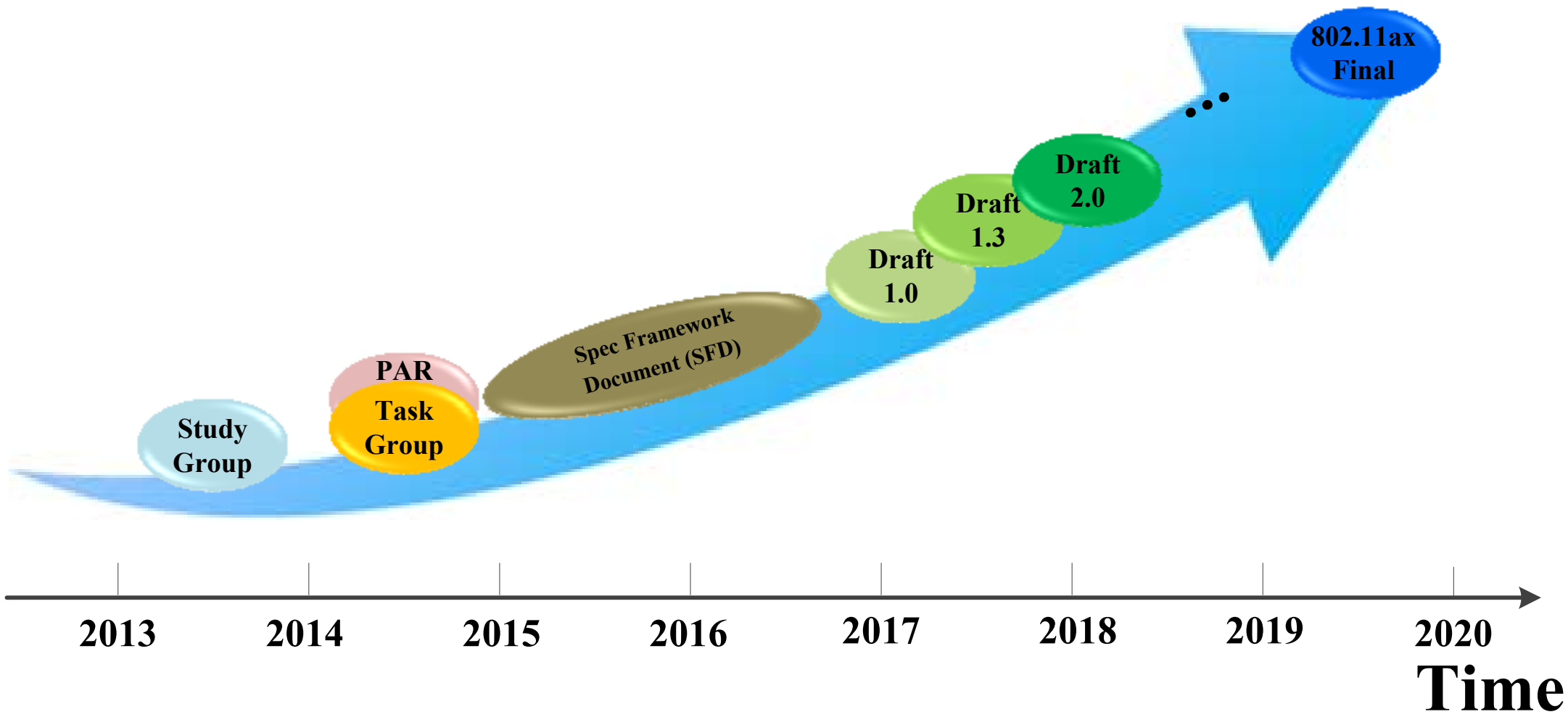

Fig. 3 Standarization process of 802.11ax.

Tab. 4 Important documents for the standardization of IEEE 802.11ax

| Doc. NO. | Doc. Title | Content |
|---|---|---|
| 11-14-0165 | Project Authorization Request (PAR) [7] | It defines the goals that IEEE 802.11ax needs to achieve and the technical requirements that need to be met. |
| 11-14-0169 | Criteria for Standard Developement (CSD) [43] | This is the refinement of PAR, and defines the technical standards in the development process of IEEE 802.11ax standards, including the discussion and requirements for compatible coexistence, market potential, and technical feasibility. |
| 11-14-0571 | TGax Evaluation Methodology [29] | The evaluation criteria and evaluation methods of IEEE 802.11ax standard are defined, and the necessity of evaluating the performance of IEEE 802.11ax through link level and system level integration simulation platform is defined. |
| 11-14-0882 | TGax Channel Models [30] | The channel model for simulation verification of IEEE 802.11ax is defined. |
| 11-14-0938 | TGax Selection | The development process of IEEE 802.11ax Task |
| | Procedure [47] | Group is defined. |
| 11-14-0980 | TGax Simulation Scenarios [28] | The simulation scene, business model, and simulation parameters of IEEE 802.11ax standard are defined. |
| 11-14-1009 | TGax Functional Requirements [49] | The functional requirements of IEEE 802.11ax are defined, such as the specific requirements for system performance, spectrum efficiency, working frequency band, and compatible coexistence, which indicate the direction of various solutions. |
| 11-15-0132 | Specification Framework Document (SFD) [45] | Defines the outline of developing IEEE 802.11ax and guides the compilation of IEEE 802.11ax draft, which specifies some technical solutions needed to be included in the draft. |

# III. PHY Enhancements

## A. PHY Enhancements Overview

The development of PHY technology is always an essential part of the evolution of wireless networks. IEEE 802.11ax introduces several PHY enhancement technologies, which enable the IEEE 802.11ax to achieve a higher transmission rate of up to 9.6 Gbps, as shown in Fig. 4.

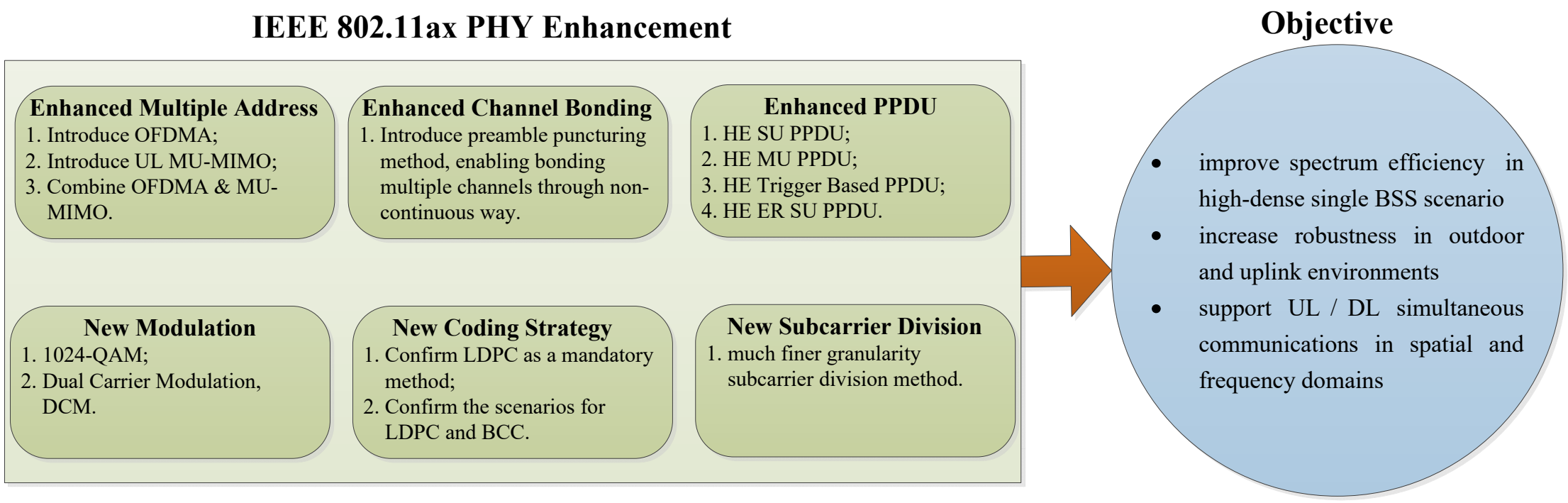

Fig. 4 Overview of PHY enhancements.

- New modulation technology

Since IEEE 802.11ac uses 256-QAM as its highest order modulation, IEEE 802.11ax introduces 1024-QAM to further improve the peak rate.

Then, IEEE 802.11ax introduces the DCM, enabling the same information to be modulated in a pair of subcarriers to enhance the signal-to-noise ratio (SNR) and the robustness of data transmission.

Finally, IEEE 802.11ax adopts a much finer granularity of subcarrier interval and a longer (four times) symbol length. Longer symbol length and longer protection interval can improve the robustness of data packet transmission in outdoor scenarios with complicated channel characteristic such as multipath fading.

This part is discussed in detail in Sec. III.B.

- New code strategy

LDPC is an optional coding technology in IEEE 802.11ac; however, it achieves bigger gain than BCC with the long codeword length. Therefore, both LDPC with BCC are considered as the mandatory technologies in IEEE 802.11ax under different cases. Moreover, IEEE 802.11ax clearly specifies the application scenario of LDPC and BCC.

This part is discussed in detail in Sec. III.C.

- New subcarrier division mechanism

For legacy IEEE 802.11, a channel with 20 MHz bandwidth is divided into 64 subcarrier spaces and consequently, the bandwidth of each subcarrier is 312.5 KHz. The number of subcarriers in IEEE 802.11ax is four times that of the legacy IEEE 802.11. In this case, the 20 MHz band is divided into 256 subcarriers, and the bandwidth of each subcarrier is reduced to 78.125 KHz. This finer granularity mechanism helps IEEE 802.11ax to obtain more precise and efficient scheduling for OFDMA resources, further improving spectrum efficiency.

This part is discussed in detail in Sec. III.D.

- Enhanced multiple access technology

IEEE 802.11ax introduces OFDMA, which divides the channel(s) into several resource units (RUs), each of the RU consisting of multiple subcarriers. Then, each STA is supported in one RU for UL or DL transmission. Enabling parallel transmission, OFDMA reduces the overhead and collision, and further enhances spectrum efficiency.

DL MU-MIMO is adopted by IEEE 802.11ac, 802.11ax further introduces UL MU-MIMO to ensure symmetrical high throughput for both DL and UL. IEEE 802.11ax allows up to eight STAs to transmit simultaneously through MU-MIMO. Moreover, MU-MIMO and OFDMA are allowed to work together; i.e., multiple STAs can parralelly send or receive frames in the same RU through MU-MIMO, which further increases transmission efficiency.

This part is discussed in detail in Sec. III.E.

- Enhanced CB Technology

For channel bonding, IEEE 802.11ac supports several bandwidth modes: 20/40/80/160(80+80) MHz. However, except for the 80+80 MHz mode, all other modes require the bonded channels to be continuous. Thus, larger channel bandwidth cannot be bonded when multiple idle channels are separated by the busy channel. Therefore, the preamble puncturing mechanism is introduced in IEEE 802.11ax, allowing the channel to be bonded in a non-continuous way. This increases the available bandwidth, while improving the transmission rate.

This part is discussed in detail in Sec. III.F.

- Enhanced PPDU

To support different technologies and scenarios, IEEE 802.11ax defines four different efficient PPDU formats: the high efficiency (HE) SU PPDU, the HE MU PPDU, the HE trigger based (TB) PPDU, and the HE extended range (ER) SU PPDU.

This part is discussed in detail in Sec. III.G.

### B. New Modulation Technologies

The highest order modulation of IEEE 802.11ac is 256-QAM; therefore, one modulation symbol can carry 8 bits. With the improvement of the device's processing capability and the demodulation algorithm, to further improve the peak transmission rate, IEEE 802.11ax introduces higher order modulation: 1024-QAM. In this case, a symbol carries 10 bits. Therefore, by introducing 1024-QAM, IEEE 802.11ax can achieve a 25% gain in the theoretical maximum transmission rate compared to IEEE 802.11ac in the high SINR region. In addition to adopting a more fine-grained (four times of 802.11a/g/n/ac) subcarrier division as well as the new designed guard interval (GI), IEEE 802.11ax achieves a maximum transmission rate of 9607.8 Mbps.

To enhance the SNR and transmission robustness, IEEE 802.11ax introduces DCM, enabling the information modulated in a pair of subcarriers. In DCM, to reduce the peak-to-average power ratio (PAPR), the same information needs to be rotated on a pair of subcarriers:

$$s_{k+N_{SD}/2} = s_k e^{j(k+N_{SD}/2)\pi}, k = 0,1,\cdots \frac{N_{SD}}{2} - 1, \tag{1}$$

where $N_{SD}$ indicates the number of subcarriers contained in the RU or the number of subcarriers populated in the bandwidth. It is worth noting that DCM is applicable to any type of OFDMA and OFDM transmission, while IEEE 802.11ax requires that it can only be used in MCS 0, MCS 1, MCS 3, and MCS 4, and the maximum space stream number is two. This is because DCM is designed for high reliability rather than high throughput. The SINR requirement of the receiving node is significantly reduced by using DCM. For example, when DCM is adopted with MCS 0, the bit error performance will be improved by 3.5 dB. Therefore, DCM benefits the robustness of outdoor scenario and UL transmission, and reduces the packet loss rate.

### C. New Coding Strategy

BCC is mandatory while LDPC is optional for the traditional IEEE 802.11. IEEE 802.11ax requires both BCC and LDPC to be the mandatory coding technologies, but the application scenarios of them are strictly distinguished. When the transmission bandwidth is smaller than or equal to 20 MHz, it is required to use BCC. Otherwise, LDPC is required when the transmission bandwidth is above 20 MHz. The reason is that larger bandwidth, higher modulation order, and finer granularity of subcarriers results in an increase of transmission rate. In other words, the receiver needs more time to process the received data. However, interleaver is needed in the encoding and decoding process of BCC. This means that the signal has to be interleaved with filling bits. Therefore, the receiver needs to deal with all the bits and cannot warrant obtaining the sufficient transceiver conversion time. Consequently, BCC code is not suitable for the larger bandwidth.

### D. New subcarrier division mechanism

For traditional IEEE 802.11a/g/n/ac, the 20 MHz bandwidth is divided into 64 subcarriers, and the subcarrier interval is $20,000/64=312.5KHz$. As shown in Fig. 5, taking IEEE 802.11a/g as an example, the 64 subcarriers consist of 52 populated subcarriers, one direct current (DC) subcarrier, and 11 sideband subcarriers. Furthermore, the populated subcarriers consist of 48 data subcarriers and four pilot subcarriers. Since the subcarrier interval is $312.5\text{KHz}$, the duration of one OFDM symbol is $3.2\mu s$, added to the GI $0.8\mu s$; therefore, the length of a full OFDM symbol is $4\mu s$.

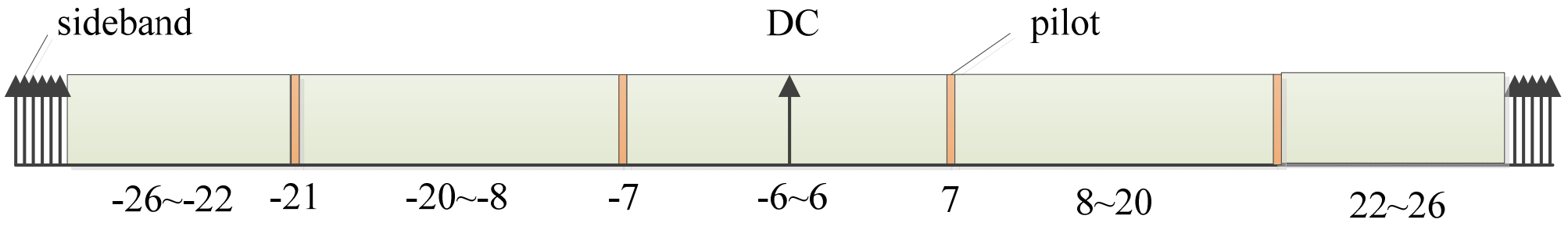

Fig. 5 Subcarrier division of the traditional IEEE 802.11.

To achieve more precise and efficient scheduling for OFDMA resources and improve the spectrum efficiency, IEEE 802.11ax proposes a much finer grained subcarrier division. As shown in Fig. 6, the 20 MHz bandwidth is divided into 256 subcarriers and consequently, the subcarrier interval is reduced to $78.125\text{KHz}$. The 256 subcarriers consist of 242 populated subcarriers, 11 sideband subcarriers, and 3 DC subcarriers. Furthermore, the populated subcarriers consist of 234 data subcarriers and eight pilot subcarriers. Since the subcarrier interval is $78.125\text{KHz}$, the OFDM symbol length of IEEE 802.11ax is 12.8 us. The GI can be selected from $0.8\mu s$, $1.6\mu s$, and $3.2\mu s$. Therefore, considering the overhead casued by GI, the spectrum utilization efficiency of IEEE 802.11ax increases from 3.2 / (3.2 + 0.8) = 0.8 to the highest 12.8 / (12.8 + 0.8) = 0.94.

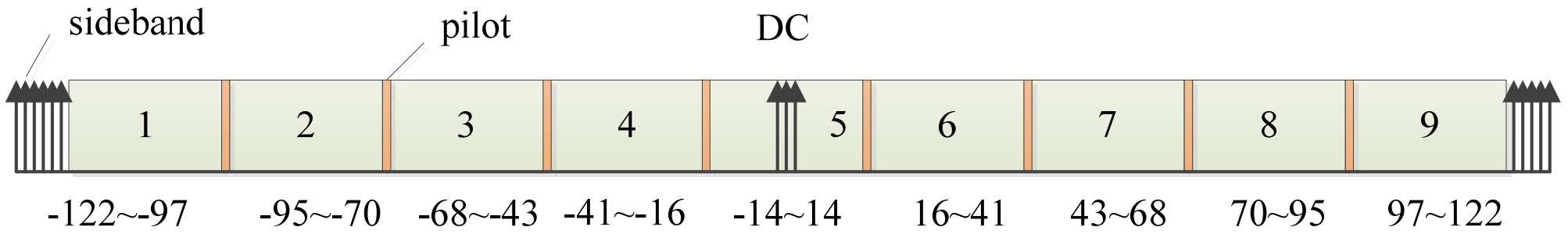

Fig. 6 Subcarrier division of IEEE 802.11ax.

### E. Enhanced Multiple Access

(1) OFDMA

OFDMA belongs to the frequency domain multiple access technology, dividing the channel(s) into multiple RUs with either the same or different bandwidth, where multiple OFDM subcarriers combine to one RU. Each RU is assigned to a specific

STA for sending or receiving frames. Thus, OFDMA possesses the advantages of both OFDM and FDMA. Multiuser diversity gain can be obtained by assigning appropriate RUs for different STAs. Therefore, to improve the efficiency of MU access in the high-dense deployment scenario, IEEE 802.11ax introduces OFDMA technology. It is worth noting that IEEE 802.11ax is the first IEEE 802.11 standard amendment to introduce OFDMA.

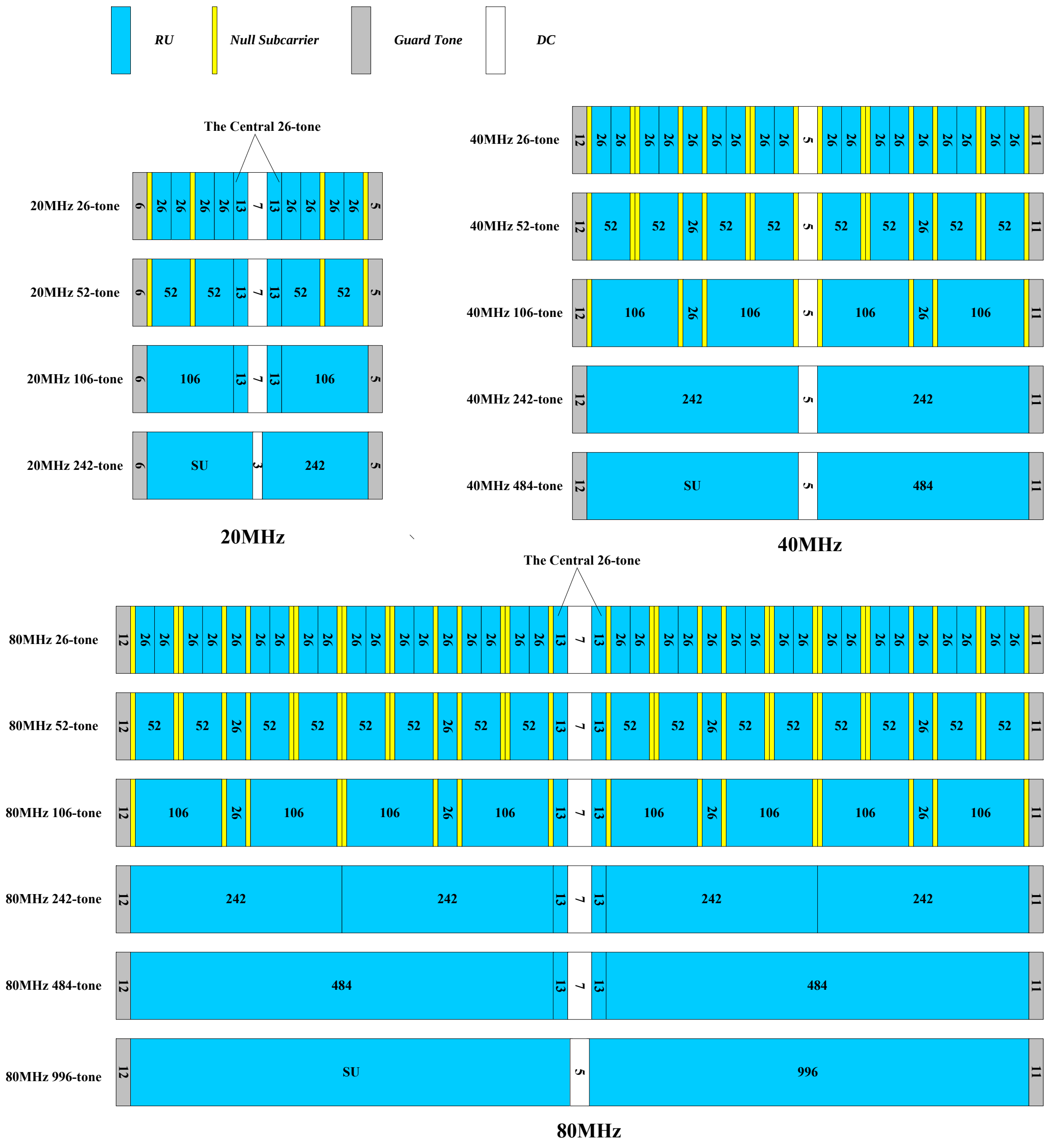

Fig. 7 RUs division for IEEE 802.11ax.

To simplify the resource scheduling for OFDMA, IEEE 802.11ax divides 20 MHz, 40 MHz, 80 MHz, and 160 MHz bandwidth into RUs with different sizes, and each STA sends or receives frames on only one RU. There are seven types of RUs specified in IEEE 802.11ax: 26-tone RU, 52-tone RU, 106-tone RU, 242-tone RU, 484-tone RU, 996-tone RU, and 2*996-tone RU. Obviously, the 484-tone RU only appears in the bandwidths of 40 MHz, 80 MHz, and 160 MHz, 996-tone RU only appears in the bandwidths of 80 MHz and 160 MHz, and 2*996-tone RU only appears in the bandwidths of 160 (80+80) MHz. Fig. 7 shows the RUs division mode for 20 MHz, 40 MHz, 80 MHz, and 160 (80+80) MHz.

(2) UL MU-MIMO

MU-MIMO belongs to the spatial domain multiple access technology. It allocates different spatial streams for different users, and the receiver needs to separate the spatial streams of different users by using signal processing technology, as shown in Fig. 8 Specifically, AP collects the channel state information (CSI) from STAs, which are further divided into several groups. The STAs in the same group can simultaneously transmit data, and the signals of different STAs can be distinguished via spatial streams. Due to the factors of cost and the implementation complexity, AP often installs mutilpe antennas, while STAs usually instrall fewer antennas. This is the reason why MU-MIMO makes full use of the advantages of multi antenna, and further improves both the transmission rate and the spectrum utilization. As IEEE 802.11ac has already adopted DL MU-MIMO, IEEE 802.11ax further improves DL MU-MIMO, and introduces UL MU-MIMO, thus enabling symmetrical high throughput of both DL and UL.

In addition, IEEE 802.11ax allows up to eight STAs to simultaneously transmit or receive through DL/UL MU-MIMO, while IEEE 802.11ac only allows four STAs to simultaneously receive through DL MU-MIMO. Moreover, it allows the MU-MIMO and OFDMA to work simultaneously; i.e., multiple STAs are allowed to send frames through MU-MIMO in the same RU, to further increase transmission efficiency.

IEEE 802.11ax designs frame structure to support UL MU-MIMO and DL MU-MIMO. For DL MU-MIMO, AP uses the HE MU PPDU format to support both OFDMA and MU-MIMO, which is more flexible than IEEE 802.11ac. Spatial stream and and RUs allocation information are embedded in the HE-SIG-B, and then AP simultaneously transmits frames to multiple STAs through allocated spatial streams. For UL MU MIMO, AP uses trigger frame (TF) to trigger multiple STAs and allocate RUs and spatial streams. After that, the STAs triggered by AP use HE TB PPDU format to simultaneously transmit frames to AP through allocated spatial streams.

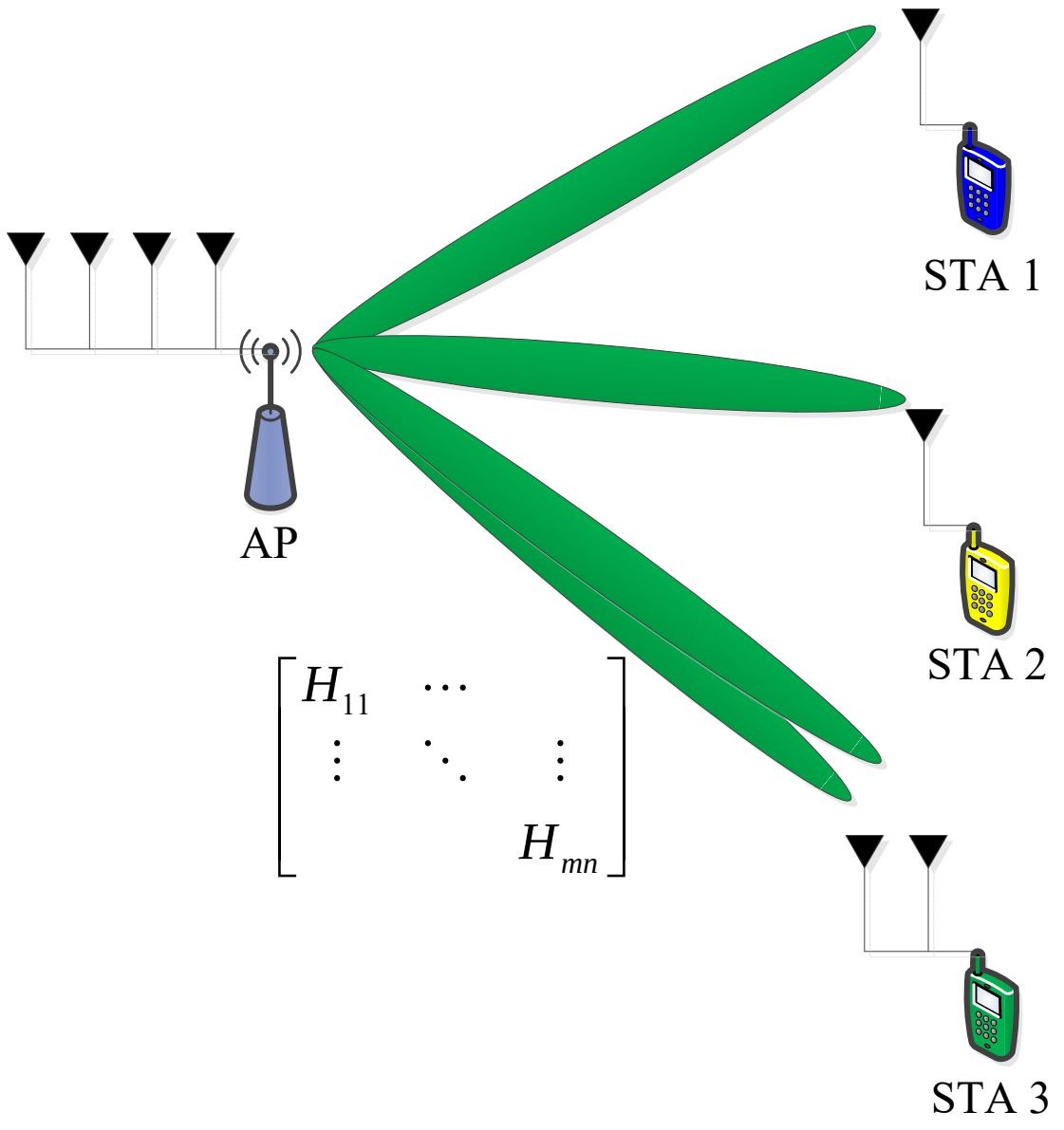

Fig. 8 Example illustration for UL MU-MIMO.

## F. Enhancement for Channel Bonding

Reviewing the development of IEEE 802.11, IEEE 802.11a/g only utilizes 20 MHz, IEEE 802.11n extends to 20/40 MHz, and IEEE 802.11ac supports 20/40/80/160(80+80) MHz. It can be observed that the maximum bonding bandwidth gradually increased with the evolution of IEEE 802.11, and the network capacity also increased accordingly. However, except for the 80+80 MHz bonding mode, all other modes require that the channel bonded must be continuous. Consequently, larger channel bandwidth cannot be bonded when multiple idle channels are separated by the busy channel. As shown in Fig. 9, given that the secondary 20 MHz is busy, even if the secondary 40 MHz is idle, AP unfortunately can only use the primary 20 MHz. This leads to a severe waste of spectrum. Predictably, this situation is more likely to occur for high-dense scenarios.

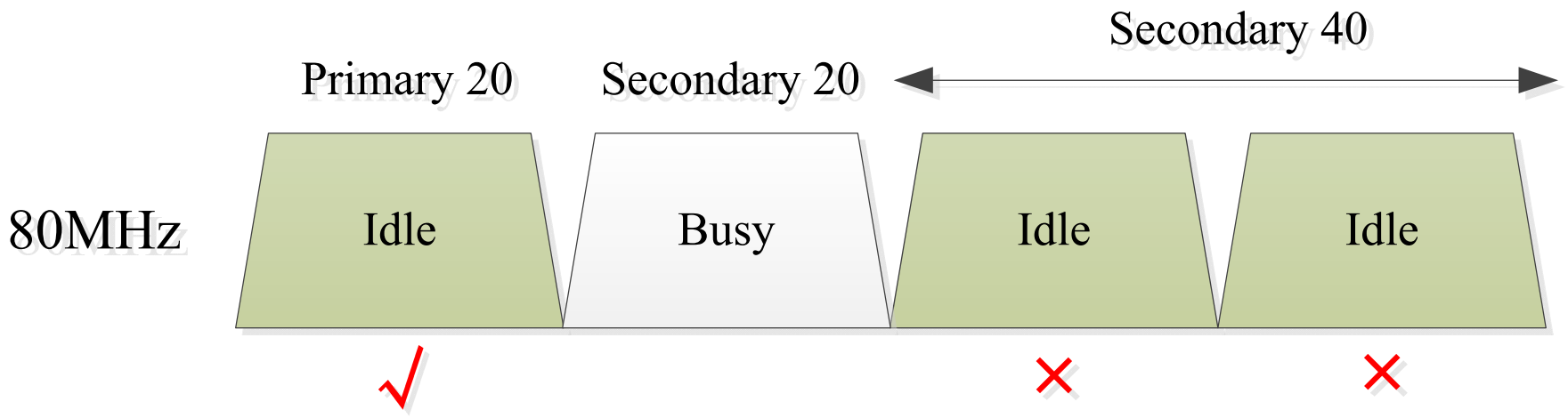

Fig. 9 Resource waste caused by the traditional CB.

To solve this problem and to further improve the spectrum efficiency, IEEE 802.11ax introduces the preamble puncturing mechanism, thus allowing the channel to be bonded in a non-continuous way. This increases the available bandwidth, makes CB more flexible, and improves the transmission rate. It is worth noting that preamble puncturing is called non-contiguous CB in previous discussion [50]. For the HE MU PPDU structure, IEEE 802.11ax extends the bandwidth field from 2 bits (mode 0 ~ 3) to 3 bits. Modes 0-3 are the same as the traditional 4 modes; therefore, we highlight mode 4-7 corresponding to the preamble puncture mechanism, as shown in Fig. 10.

- Mode 4 indicates the situation that the total bandwidth is 80 MHz, and the secondary 20 MHz is punctured. As shown in Fig. 9, if mode 4 is to be adopted, AP could bond 60 MHz to communicate with STA, and then the efficiency could be improved three-fold.
- Mode 5 indicates the situation that the total bandwidth is 80 MHz, and one 20 MHz in the secondary 40 MHz is punctured.
- Model 6 indicates the situation that the total bandwidth is 160(80+80) MHz, and the secondary 20 MHz of the primary 80 MHz is punctured. Moreover, there is no special requirement for the secondary 80 MHz.
- Model 7 indicates the situation that the total bandwidth is 160(80+80) MHz, and the primary 40 MHz of the primary 80 MHz is ilde. Actually, there are three cases covered by mode 7, as shown in Fig. 10.

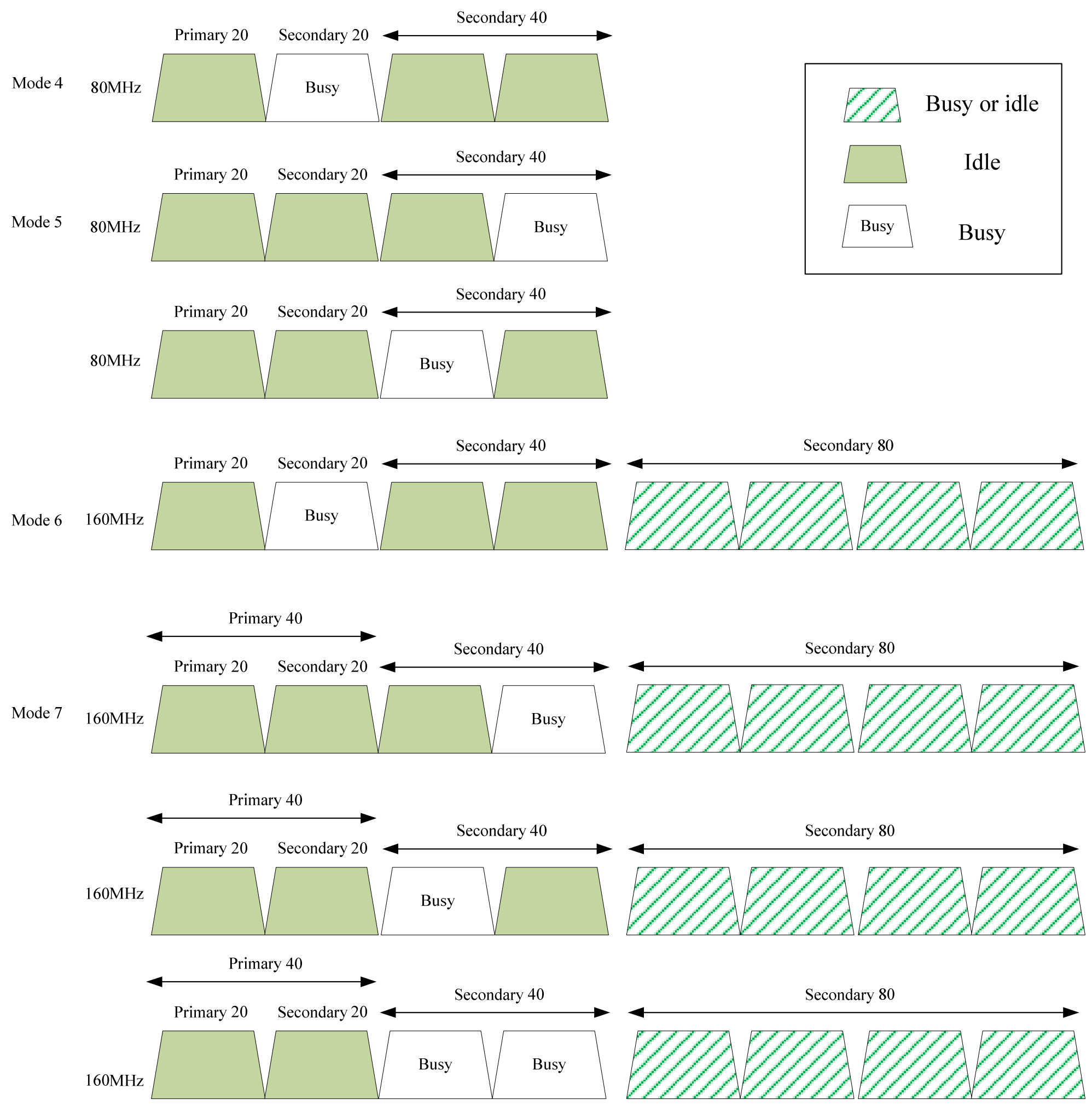

Fig. 10 Preamble puncturing mechanism in IEEE 802.11ax.

## G. PPDU

To support different technologies and scenarios, IEEE 802.11ax introduces four data packet structures.

- HE SU PPDU format

The single user format is the packet structure format between AP and one single STA, and between a single STA and another single STA. Fig. 11 shows the packet structure. Compared to IEEE 802.11ac, HE SU PPDU introduces Repeat L-SIG (RL-SIG), which is used to enhance the robustness of L-SIG, and used to confirm a PPDU is HE format through automatic detection; Packet extension (PE) extends the time for the receiver to process data.

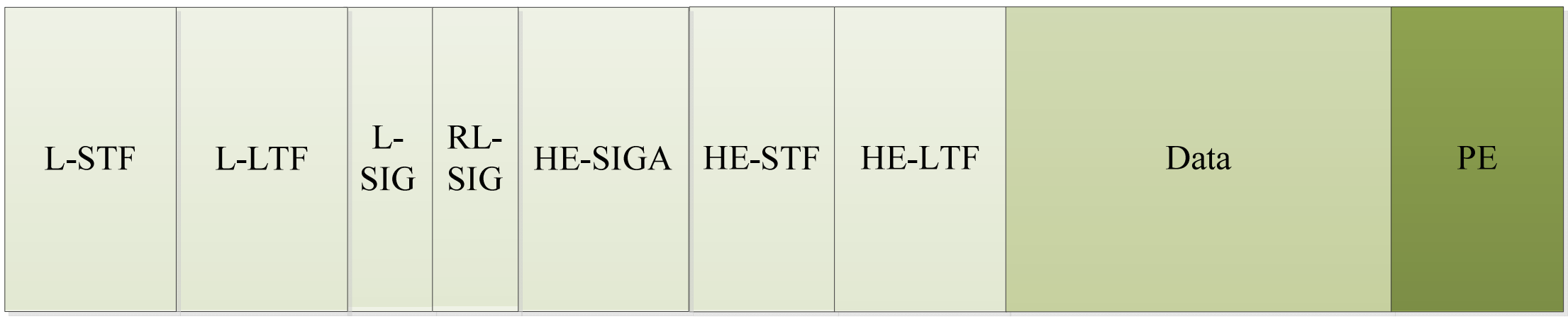

Fig. 11 HE SU PPDU format.

- HE MU PPDU format

The MU format enables simultaneous transmission among MUs via OFDMA and/or MU-MIMO. Based on the single user format, as shown in Fig. 12, the HE-SIG-B field is added to indicate the resource allocation information for multiple users.

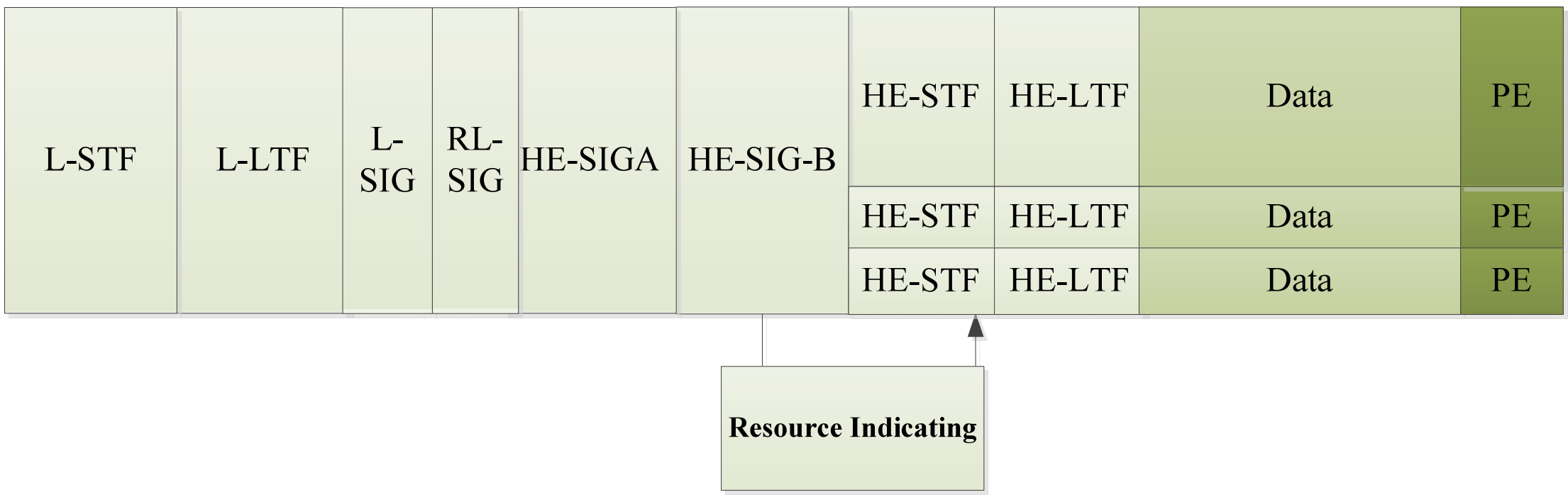

Fig. 12 HE MU PPDU format.

- HE TB PPDU format

After receiving the trigger frame, multiple STAs simultaneously transmit UL frames according to the resource allocation information in the trigger frame. Therefore, the UL MU transmission format is called a trigger-based format, as shown in Fig. 13. Compared to HE MU PPDU, this format does not have HE-SIG-B because AP indicates the resource allocation information in the trigger frame.

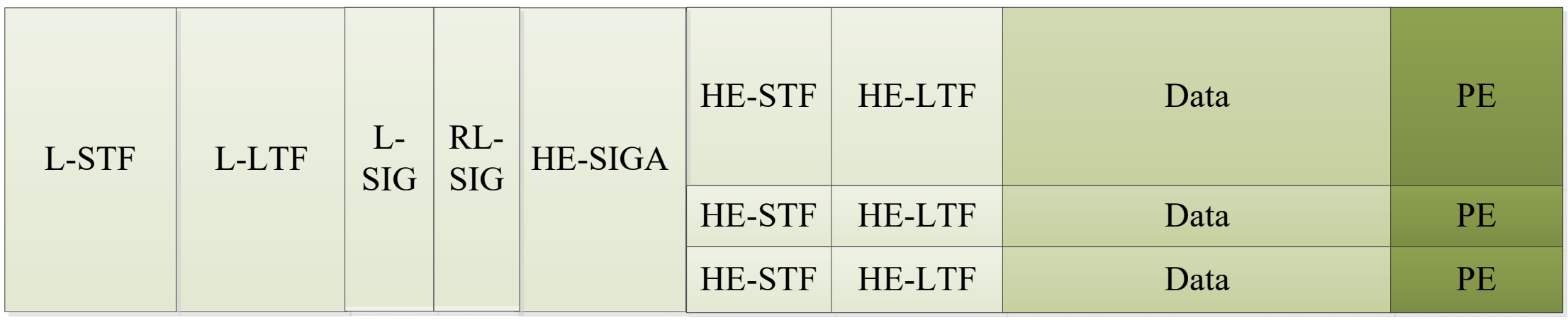

Fig. 13 HE TB PPDU format

- HE ER SU PPDU format

To improve the transmission robustness of the outdoor scenario, IEEE 802.11ax repeats the HE-SIG-A field by directly extending two symbols to four symbols, as shown in Fig. 14. Moreover, IEEE 802.11ax boosts the power of the traditional preamble, HE-STF and HE-LTF to further extend the transmission coverage. Finally, the transmission range of the data field can be extended via DCM narrowband RU transmission.

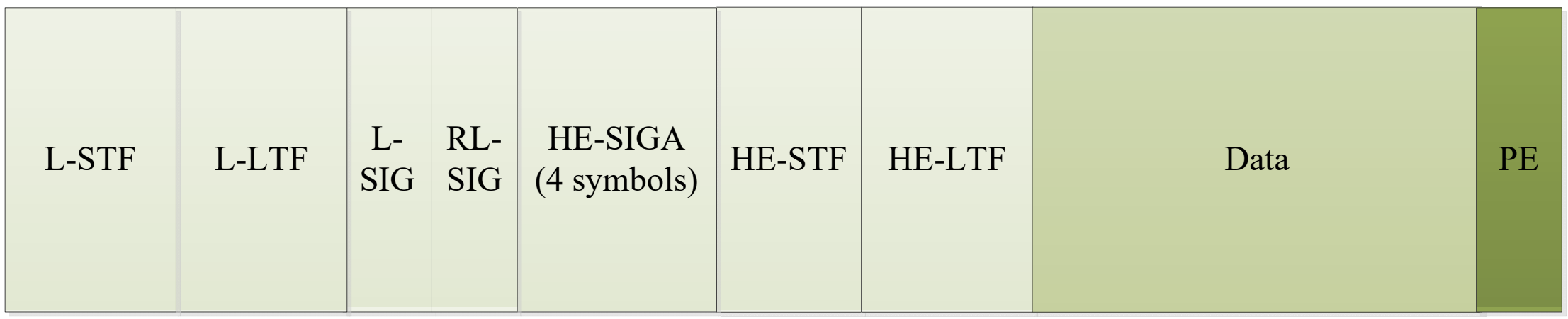

Fig. 14 HE ER SU PPDU format

# IV. MU-MAC Enhancements

## A. Introduction of MU-MAC in WLANs

MU-MAC is a type of high efficiency multiple access technologies, which enables multiple users to obey certain access rules to transmit UL data concurrently (known as UL MU-MAC), or to transmit DL data concurrently (known as DL MU-MAC) through given network access resources (i.e. space domain, time domain, and frequency domain resources). The improvement of channel access efficiency by using MU-MAC is caused by two reasons: first, through the parallel channel access of multiple users, the channel access efficiency is improved, and the utilization rate of the channel resources is also improved; second, the signaling for channel access and data transmission of multiple users can often be reasonably combined and compressed, thus reducing the signaling overhead of channel access. MU-MAC technology has been widely applied in cellular mobile communication systems [51] [52], including OFDMA [53]-[55] and MU-MIMO [56]-[58], which achieves MU channel access and data transmission in parallel from frequency domain and space domain, respectively. With regard to the high efficiency of MU-MAC, the IEEE 802.11 work group has introduced the support of DL MU-MIMO in the IEEE 802.11ac standard amendment. Nevertheless, prior to IEEE 802.11ax, the support of MU-MAC still has some limitations in IEEE 802.11ac, which makes the network equipment in most cases still work in single user MAC (SU-MAC) mode, and limits the improvement of IEEE 802.11 MAC efficiency. To meet the requirements of significantly improving the efficiency of WLANs, the comprehensive support of MU-MAC has been regarded as one of the core tasks of IEEE 802.11ax standard development. Specifically, IEEE 802.11ax retains the support of DL MU-MIMO, and introduces MU-MAC based on OFDMA and UL MU-MIMO.

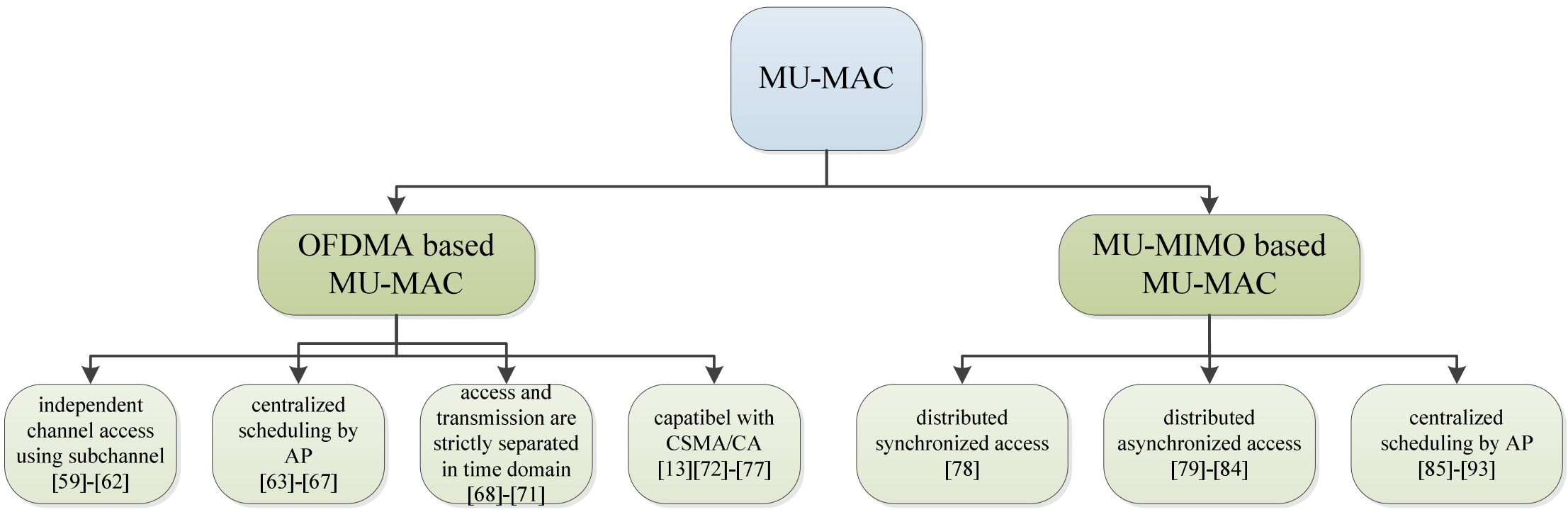

Fig. 15 A survey of MU-MAC related research results for WLANs.

OFDMA and MU-MIMO are two typical technologies for MU-MAC. Therefore, we divide the related studies of MU-MAC into two branches: OFDMA based MU-MAC and MU-MIMO based MU-MAC, as shown in Fig. 15.

**1) OFDMA based MU-MAC**

Since most of the physical layer defined by IEEE 802.11 standard uses OFDM as data transmission technology, it is very easy to extend from OFDM single user transmission to MU OFDMA transmission, to improve the frequency utilization efficiency.

i. Related work of independent channel access using subchannel

Kwon et al. [59] propose that each node needs to listen to all OFDMA subchannels, and implements a backoff procedure according to subchannels. After the backoff counter reaches 0, a subchannel is randomly selected to transmit data in the form of OFDMA. Kwon et al. [60] further propose that nodes implement independent backoff on each subchannel. The independent backoff includes the independence of backoff procedure, backoff counter, and the access procedure. The nodes which finish backoff procedure at first conduct channel access and data transmission on the subchannel. Wang et al. [61] also propose the method of independent backoff on the subchannel, but when the node finishes backoff procedure, it can access the channel and simultaneously transmit data on multiple subchannels to improve transmission efficiency. Ferdous et al. [62] propose that nodes can be grouped according to subchannels, so each subchannel corresponds to a part of the nodes, and each node only access to the corresponding subchannels.

ii. Related work of centralized scheduling in AP

Lou et al. [63] propose that the AP is able to achieve DL and UL OFDMA transmission through the centralized scheduling approach. For DL transmission, AP transmits the scheduling frame, and nodes implement DL OFDMA transmission after response on the corresponding subchannel. For UL transmission, AP transmits scheduling frame, nodes response on the corresponding subchannel, and AP schedules UL OFDMA transmission. Valentin et al. [64] design a MAC protocol oriented for DL OFDMA transmission. AP firstly sends the DL request on the entire channel, and schedules nodes to reply in turn. Then, AP allocates the subchannel for the nodes and performs DL transmission. Kamoun et al. [65] propose a MAC protocol that supports UL and DL OFDMA transmission. Mishima et al. [66] propose that AP sends RTS frame to multiple nodes, the nodes reply CTS frames sequentially, and then AP respectively schedules OFDMA DL transmission and OFDMA UL transmission. Qu et al. [67] introduce the full duplex technology into OFDMA MU access, and design a MAC protocol for MU full-duplex OFDMA, which can greatly improve throughput. Due to the simple process of centralized scheduling in AP, IEEE 802.11ax adopts OFDMA access based on centralized scheduling in AP.

iii. Related work of channel access and data transmission are strictly separated in time

Fallah et al. [68], Haile and Lim [69], and Deng et al. [70] propose that the time will be divided into two stages. AP collects the UL transmission request sent by nodes during the first stage, and AP schedules multiple nodes to achieve parallel data transmission using OFDMA during the second stage. Jung and Lim [71] propose that AP divides the nodes into several groups, and then AP schedules each group for channel access using OFDMA.

iv. Related work of OFDMA access compatible with CSMA/CA

Most of the above studies differ from the existing CSMA/CA based DCF and EDCA access mechanisms. Since backward compatibility is an important factor that must be considered in the evolution of IEEE 802.11 version, the authors conducted a series of studies on OFDMA access compatible with CSMA/CA. Qu et al. [72]

propose that to maintain compatibility and to reduce the complexity, all nodes listen to the whole channel to perform fast backoff and randomly select a subchannel to perform UL access, and AP schedules multiple user UL data transmission using OFDMA. Furthermore, Qu et al. [73] analyze and demonstrate that in the dense deployment WLANs scenarios, the STAs have good carrier sensing consistency with each other, using a previously proposed access method [72]. On the basis of [72], Zhou et al. [74], Zhou et al. [75], and Yan and Zuo [76] design single channel and multiple channel based OFDMA MAC protocols for QoS guarantee of WLANs. In addition, Li et al. [77] propose OFDMA access MAC based on spatial clustering group, which inhibits the interference diffusion of OFDMA in space, and improves the area throughput. Li et al. [13] survey the WLAN MAC based on OFDMA access.

**2) MU-MIMO based MU-MAC**

MU-MIMO is also an important MU MAC. MIMO technology is first introduced by IEEE 802.11n, and then DL MU-MIMO is introduced in IEEE 802.11ac.

i. Related work of distributed synchronous parallel access

Jin et al. [78] propose the idea of the MAC protocols based on MU-MIMO for the UL MU concurrent transmission in WLANs, which assumes that multiple nodes based on the backoff rules of IEEE 802.11 independently implement a backoff procedure, and conduct synchronous data transmissions after simultaneously completing the backoff procedure. However, the design and the details of the protocols is not described. Since the requirement of synchronous parallel transmission to the system is relatively stringent and its universality is poor, the follow-up scholars paid more attention to the MU-MIMO MAC based on asynchronous transmissions.

ii. Related work of distributed asynchronous access

Tan et al. [79] present a carrier counting multiple access (CCMA) method, where each node counts the number of nodes that are currently transmitting data by using the preamble detection of the data frame from other nodes, to determine which channel to access at this time. Babich and Comisso [80] conduct the mathematical modeling and theoretical analysis of asynchronous data parallel transmission, and it shows that compared to synchronous data transmission, a larger performance gain can be achieved via asynchronous data transmission. Ettefagh et al. [81] present a cluster-based MU-MIMO MAC, where the nodes are divided into several clusters, and channel competition occurs between the clusters. The nodes in the same cluster have the same backoff counter, so that when the backoff procedure is completed, the nodes in the same cluster simultaneously send UL data frames. Mukhopadhyay et al. [82] propose that to solve the ACK delay problem in MU-MIMO transmission, the operation mode of the waiting ACK timeout timer is modified, i.e., the waiting ACK timeout timer only starts to work after the channel is idle for DCF inter-frame space (DIFS), to effectively alleviate the system performance degradation caused by the ACK delay problem. Lin and Kung [83] propose a MU-MIMO MAC scheme combining asynchronous and synchronous data transmission, and verify the conclusion that AP has better performance when the number of antennas is comparatively large. Kuo et al. [84] present a leader-based UL MU-MIMO MAC, in which nodes use CSMA/CA rules to compete channel resources. After a node (the first node is leader) win the competition, other nodes set the backoff counter by computing the angle between them and the leader and channel parameters for channel access and asynchronous parallel transmission.

iii. Related work of AP scheduling and coordination access

Tandai et al. [85] propose a MU-MIMO MAC based on AP coordination. When the AP receives an RTS frame of any node, it sends an A-CTS frame to collect the UL transmission request to other nodes. After these other nodes reply A-RTS frames, AP sends a pR-CTS frame to measure the channel state. Finally, AP sends an N-CTS frame to schedule multiple nodes for UL transmission using MU-MIMO after receiving channel state information feedbacked by other nodes. Li et al. [86] propose that AP firstly sends an authorization frame, and the nodes feedback their UL demand and channel state using OFDMA. Then, AP schedules multiple nodes to conduct parallel transmission based on MU-MIMO according to the feedback information. Zheng et al. [87] propose that multiple nodes still obey the CSMA/CA backoff procedure, and synchronously send RTS frames after the backoff procedure. After AP receives the RTS frames from multiple nodes, it sends the CTS frame to schedule the nodes perform parallel transmission. Barghi et al. [88] introduce the concept of waiting time window, i.e., AP waits for a period of time to receive the second RTS frame from other nodes after receiving the first RTS frame, and then replies the CTS frame to schedule MU-MIMO transmission. Zhou and Niu [89], Liao et al. [90], and Zhang et al. [91] all propose that the time domain needs to be strictly divided into a request collection phase and a data transmission phase, and MU-MIMO transmission is adopted in the data transmission phase.

Jung et al. [92] propose that after the first node successfully sends RTS frames, AP sends CTS frame to trigger parallel transmission. The other nodes directly send data frames with the first node through both synchronous and asynchronous transmission modes after the other nodes received the CTS frame, and no extra requirements collection process is conducted. In addition, Liao et al. [93] summarize the MU-MIMO MAC in WLANs. Since the procedure of AP scheduling method is relatively simple, IEEE 802.11ax adopts MU-MIMO access based on AP centralized scheduling.

## B. MU-MAC Framework in IEEE 802.11ax

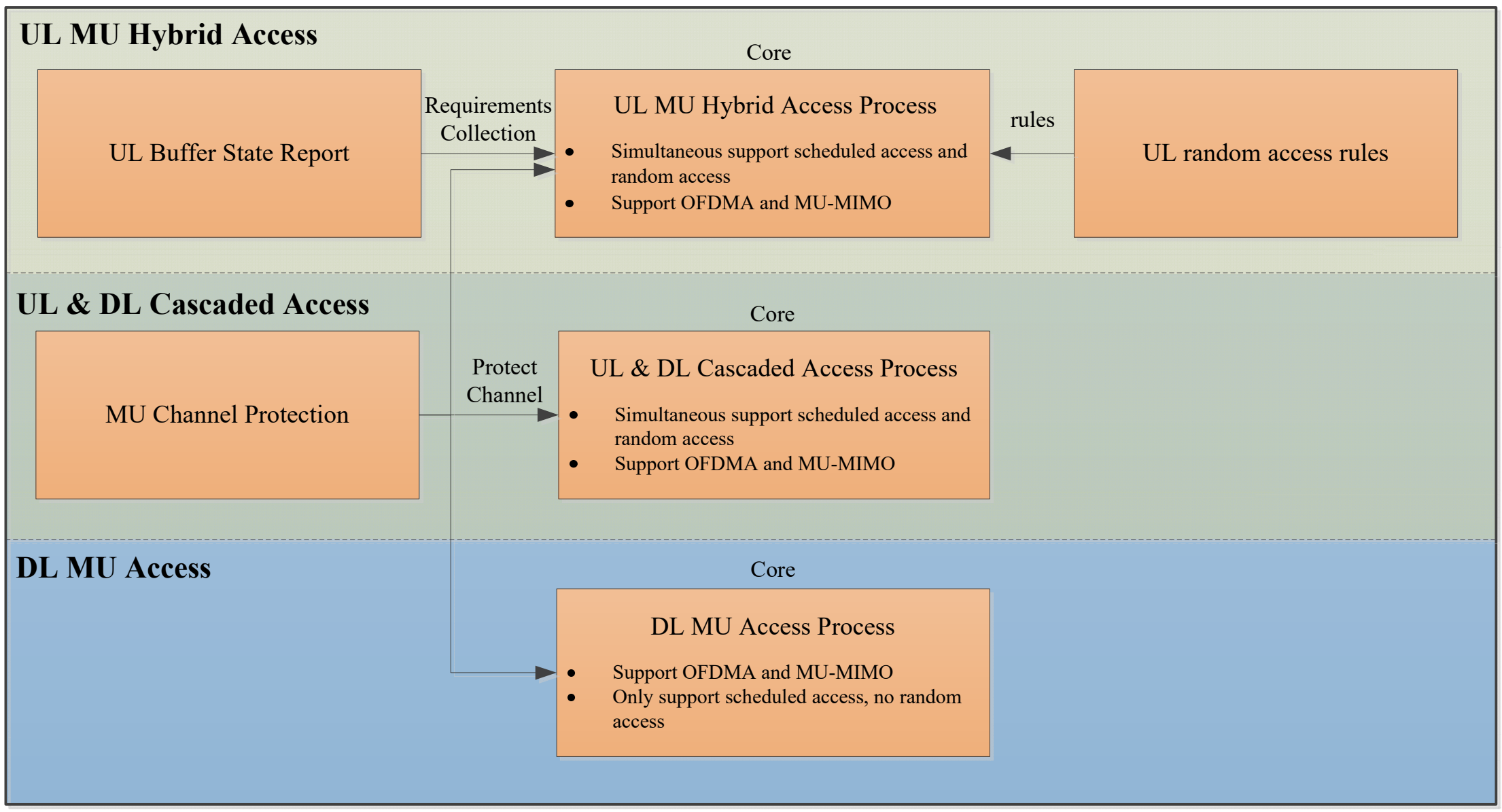

Fig. 16 MU-MAC framework in IEEE 802.11ax.

As shown in Fig. 16, the MU-MAC of IEEE 802.11ax consists of the following six technological components:

- UL MU hybrid access

On the one hand, the traditional 802.11 adopts the CSMA/CA based random access method, while on the other hand, after introducing MU-MAC, IEEE 802.11ax enhances the ability of AP scheduling based multiple access method. Therefore, IEEE 802.11ax first proposes the UL hybrid access framework by jointly bearing scheduling access and random access. Coincidentally, in the cellular network, focusing on the non-orthogonal multiple access (NOMA), a key technology of the fifth generation mobile communication system (5G), researchers introduced random access as an important complement of the traditional scheduling based access [94]-[96]; however, the integrity of the standardization process still needs to be further developed. To the best of our knowledge, IEEE 802.11ax is the first to explicitly propose a standardized framework for jointly carrying scheduling access and random access in wireless network standards. The framework supports both OFDMA and MU-MIMO. This part is discussed in detail in Sec. IV.C.

- UL OFDMA based random access rule

UL MU hybrid access in IEEE 802.11ax supports both scheduling access and random access. The resource allocation and indications for scheduling access are more explicit, whereas it is necessary to introduce more detailed rules for random access. IEEE 802.11ax specifies the detailed rules for UL OFDMA-based random access (UORA), referred to as UORA access rules, so that STAs use UORA access more efficiently. It should be pointed out that the current MU random access can only adopt the OFDMA mode, and it does not support MU-MIMO. UORA access rules are strongly related to UL MU-MAC, and will be discussed in detail in Sec. IV.C.

- UL MU buffer status report

AP is responsible for scheduling UL MU accesss. However, AP itself does not know which STAs buffered data and the buffer status of these STAs. Therefore, to support UL MU transmission, a specific UL MU buffer status report (BSR) process is required. IEEE 802.11ax supports two procedures of BSR processes: explicit BSR and implicit BSR. For the explicit BSR, AP needs to send a special trigger frame to start the process, while for the implicit BSR method, the STAs piggyback their buffer status with other UL frames. Both procedures have their own advantages: the piggyback procedure saves signaling overhead, while a special frame procedure decouples the BSR process and the data transmission. The UL MU buffer status report is strongly related to UL MU-MAC, and will be discussed in detail in Sec. IV.C.

- DL MU access

Channel access, resource scheduling, and simultaneous transmissions in DL MU transmission are controlled by AP. Specifically, AP implements the information of user allocation and resource allocation in the HE-SIG, and then sends DL frames to multiple STAs through OFDMA and/or MU-MIMO. STAs need to reply block ACK (BA) through MU transmission. This part will be discussed in detail in Sec. IV.D.

- UL& DL cascaded access

To further improve the efficiency of WLANs, IEEE 802.11ax also supports UL & DL MU cascaded accesss, enabling multiple UL and DL transmissions to alternatively cascade in one TXOP duration to further reduce the access overhead. This part will be discussed in detail in the Sec. IV.E.

- MU channel protection

In the traditional SU access process, IEEE 802.11 adopts RTS/CTS mechanism to protect the channel to address the hidden node problem. For both MU access, the hidden node problem will be much more severe without efficient channel protection means since more transmissions may be interfered simultaneously. Therefore, IEEE 802.11ax introduces the MU channel protection mechanism. This not only matches the MU access process well, but is also compatible with legacy IEEE 802.11. MU channel protection will be discussed in detail in Sec. IV.E.

### C. UL Hybrid MU-MAC

Comparing with supporting parallel UL access, we suggest that the more important contribution of the UL MU-MAC proposed by IEEE 802.11ax is to firstly explicitly propose a standardized framework to jointly carry scheduling access and random access in the wireless network standards. Scheduling access means that after receiving the requests from STAs, the base station or the AP allocates the resources and indicates access mode for users based on the predetermined resource allocation algorithms. Usually, this approach does not lead to conflicts between users. Random access means that the users do not have to report their requests, and contend for the resources in a random and opportunistic way. This approach however, may lead to access conflicts among users. Both types of access methods have advantages and disadvantages: on the one hand, from the angle of improving the resource utilization and QoS, scheduling access outperforms random access; on the other hand, from the angle of lowering the access delay and signaling overhead, random access performs better. In recent years, both cellular networks and WLANs have been paying attention to jointly support scheduling access and random access. In cellular networks, several studies on NOMA, a key technology of 5G, introduce random access as an important complement of the traditional scheduling based access. However, the standardization process still needs to be further developed. For the first time, IEEE 802.11ax proposes the hybrid access process framework based on TF for joint bearing scheduling access and random access.

1） Access process

The access process is the core of the UL hybrid MU MAC; we will illustrate it with the following four aspects: channel access, TF transmission and hybrid scheduling, UL MU data transmission, and multi-user BA (MBA).

i. Channel access

When the AP needs to trigger UL data transmissions, the carrier sensing and backoff processes are first performed to contend for the channel resources. The backoff process still follows the traditional IEEE 802.11 DCF or the EDCA process.

ii. TF transmission and hybrid scheduling

After completing the backoff process, AP needs to schedule the resources based on the UL buffer size of the STAs (the UL BSR mechanisms will be discussed in the following subsubsction) and the current state of the network, which includes four tasks: first, determine the RUs number, RUs position, and the bandwidth of each RU; then, determine each RU that is used for scheduling access or for random access; next, if multiple STAs are carried in one RU through MU-MIMO, AP needs to indicate the spatial sream information; finally, allocate the RUs used for scheduling access to the corresponding STAs.

Fig. 17 shows an example of the framework for UL hybrid access process.

After determining the scheduling results, AP transmits TF over the whole 20 MHz channel, which contains scheduling results. In the frame body of TF, AP assigns a particular RU to the corresponding STA through the AID12 field, i.e., the latter 12 bits of the STAs' AID, and the RU allocation field. Specifically, if the AID12 field corresponding to one RU is neither 0 nor 4095, the RU is assigned for the specific STA with the corresponding AID to transmit UL data. If AID12 = 0, the RU is used for random access. In particular, AP may configure all the RUs for scheduling (the AID12 corresponding to each RU does not equal to 0), or all the RUs for random access (all AID12 = 0). If AID12 = 2045, the RU is assigned for any un-associated STA to random access.

IEEE 802.11ax can simultaneously support OFDMA and MU-MIMO. In the case that the spatial streams can be differentiated, different STAs can be carried through MU-MIMO in the same RU. Specifically, if the MU-MIMO LTF Mode subfield in the Common Info field in TF equals 0, the single spatial stream is adopted; while, if the MU-MIMO LTF Mode subfield equals 1, the MU-MIMO mode is selected. Furthermore, trying to schedule multiple STAs transmitting in one RU through MU-MIMO, the AP first needs to inform multiple STAs to share the same RU through AID12 field and RU Allocation field in TF, and then allocate different spatial streams to different STAs through the space stream (SS) allocation subfield. However, it is worth noting that OFDMA is allowed to support MU random access while MU-MIMO is not allowed. In addition, IEEE 802.11 requires that MU-MIMO is allowed to only be used when the RU bandwidth is wider that or equal to 106-tone. For the RU whose bandwidth is more narrow than 106-tone, OFDMA is the only choice.

As the example shown in Fig. 17, the 20 MHz channel is divided into three RUs through TF, where the RU1 and RU3 are used for scheduling access and whose bandwidths are 106-tone. Of course, they are used to carry scheduled UL transmissions. Specifically, STA1 is scheduled to transmit on RU1, while STA2 and STA3 scheduled on RU 3 using MU-MIMO. In contrast, the RU2 is configured to be used for random access (AID12 = 0).

iii. UL MU data transmission

The STAs that successfully receive the TF determine whether they are scheduled by TF, i.e., whether the AID12 corresponding to any RU matches its own AID. If yes, the STAs need to transmit UL data on the corresponding RUs after waiting for SIFS. Furthermore, if MU-MIMO is adopted, the STAs need to transmit UL data by using the specific spatial stream in the corresponding RU. Otherwise, if there is no RU scheduled for the STAs, these STAs need to decide whether to send UL data or not. If so, the STAs buffered UL data can try to randomly access the RUs through UORA rules (see Sec. IV-C2); otherwise, i.e., there is no buffered UL data, they are required to set the NAV and keep silent during the concurrent UL MU transmission. It is necessary to point out that not every STA's UL data sent to AP has the same time length and exactly fills the entire transmission time. Those STAs whose UL data are unable to fill the entire transmission time need to add extra padding to guarantee the transmission time ends the same time.

At the same time, after sending TF, if the AP does not receive any STA's UL data in SIFS, it judges the channel access as failure. In contrast, if the AP receives at least one UL data sent by STAs, the UL transmission is considered as successful.

As shown in Fig. 17, after STAs successfully receive TF, STA1 sends the UL data in the specified RU1, STA2, and STA3 transmit UL data through MU-MIMO in RU3, and other STAs may try to randomly access the RU2 based on UORA rules. Finally, in this example, STA5 contends successfully and transmits UL data on RU2.

iv. MBA frame

If AP successfully receives at least one UL data sent by STAs, the MBA frame will be sent in the whole 20 MHz channel after waiting for SIFS. It is worth noting that although MBA is not the unique acknowledgement mode, but it is an efficient one. Compared to the traditional BA frame in IEEE 802.11, the MBA frame contains the acknowledgement information for multiple users. As shown in Fig. 17, AP replies with MBA to confirm the UL transmission sent from STA1, STA2, STA3, and STA5.

If one STA successfully receives the acknowledgement of its transmissions by the MBA frame, the transmission is successful. Otherwise, the transmission fails, and the STA will need to wait for the next TF or to select DCF/EDCA to access the channel.

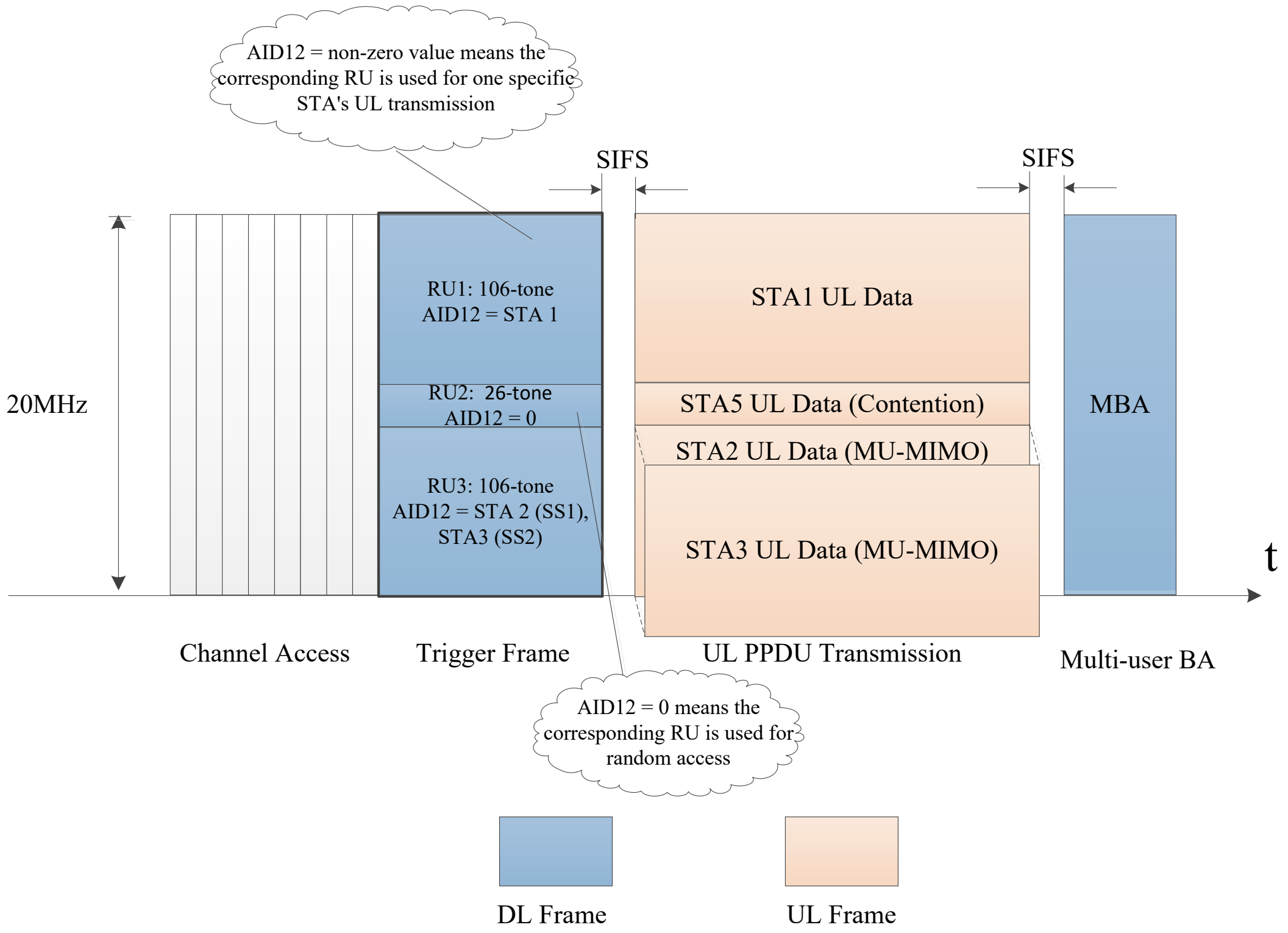

Fig. 17 Illustration for the UL hybrid access.

2） UORA Rules

The rules of UORA are introduced in detail, as shown in Fig. 18.

Each STA that wants to perform UORA needs to maintain an OFDMA backoff (OBO) value, which is randomly selected from the 0~OFDMA contention window (OCW). The related paramenters, including OCWmin and OCWmax, are carried in Beacon frame.

After the AP sends TF, any STA that wants to transmit UL data through random access needs to determine whether there is at least one RU indicated for random access. If not, this TF process does not support random access; otherwise, this STA randomly selects an OBO value in ~ OCW.

Next, STAs need to determine whether the OBO value is below the number of RUs configured for random access in TF. If yes, the OBO value is directly updated to 0; otherwise, the OBO value is updated to the current OBO value minus the number of RUs configured for random access.

After the above operation, if the updated OBO value of one STA is above 0, this STA is not allowed to perform random access in this TF round. Furthermore, then the STA maintanes the updated OBO value, and waits for the next TF opportunity; Otherwise, if the updated OBO value of one STA is equal to 0, the STA randomly selects one RU from the RUs configured for random access as candidate. As shown in Fig. 18, in the first TF round, five RUs are configured for random access and the initial OBO values of STA1-6 are 3, 5, 7, 8, 7, and 0, respectively. Therefore, based on the UORA rules, the OBO values of STA1, STA2, and STA6 are directly updated to 0, and the updated OBO values of STA3, STA4, and STA5 are 2, 3, and 2. Thus, STA1, STA2, and STA6 randomly select RU3, RU1, and RU4 for their UL transmissions, respectively.

The STAs that successful finish the OBO backoff procedure need to perform physical carrier sensing and virtual carrier intercept on the 20 MHz channel in which the candidate RU located to determine the channel state during the period SIFS after TF. If both these carrier-sensing mechanisms indicate the channel as free, the STAs will transmit UL data in the candidate RU; otherwise, the STAs abandon the access process, and randomly select an RU as acandidate RU in the next TF without re-executeing the OBO backoff procedure. As shown in Fig. 18, during the first TF round, the channel keeps idle during SIFS; therefore, the STA2 and the STA6 send UL data on the RU1 and the RU4, respectively. However, carrier sensing indicates the channel as busy; therefore, STA1 gives up the transmission opportunity and keeps OBO = 0, while randomly selecting an RU as the candidate RU in the next TF.

After that, AP sends MBA to confirm that the UL transmissions are successful. If STAs successfully receive the acknowledgement, the transmission is considered as successful, and the corresponding OCW is restored to OCWmin; otherwise, if STA does not receive acknowledgement, the transmission is considered as failure and the corresponding OCW is doubled. As shown in Fig. 18, during the first TF round, AP sends MBA to confirm the successful UL transmissions of STA2 and STA6.

In the second TF round, still shown in Fig. 18, STA2 and STA6 no longer contend to channel since they have successfully transmitted all required UL data. The OBO value of STA1 is set to 0 since STA1 has finished its backoff, while giving up the transmission opportunity since the carrier sense indicates busy in the first TF round; The OBO values of STA3, STA4, and STA5 are also updated to 0; STA7 is a new STA joining the competition, whose OBO value is also updated to 0 due to its OBO value (4) below that of the RUs number specified for random access (5). Thus, each STA selects one candidate RU and executes physical carrier sensing and virtual carrier sensing during SIFS. Since the virtual carrier sensing indicates channel busy (set NAV), STA7 gives up this opportunity; STA3 and STA5 choose the same RU and unfortunately, collision happens; STA1 and STA4 successfully transmit UL data. Finally, AP replies with MBA to confirm that the UL transmissions sent from STA1 and STA4 are successful. The OCW of STA3 and STA5 are doubled since they do not receive acknowledgement.

It is worth noting that, as shown in Fig. 18, different RUs are not required to have the same size according to the IEEE 802.11ax standard, but in practice different RUs having the same size are easy to implement for industry.

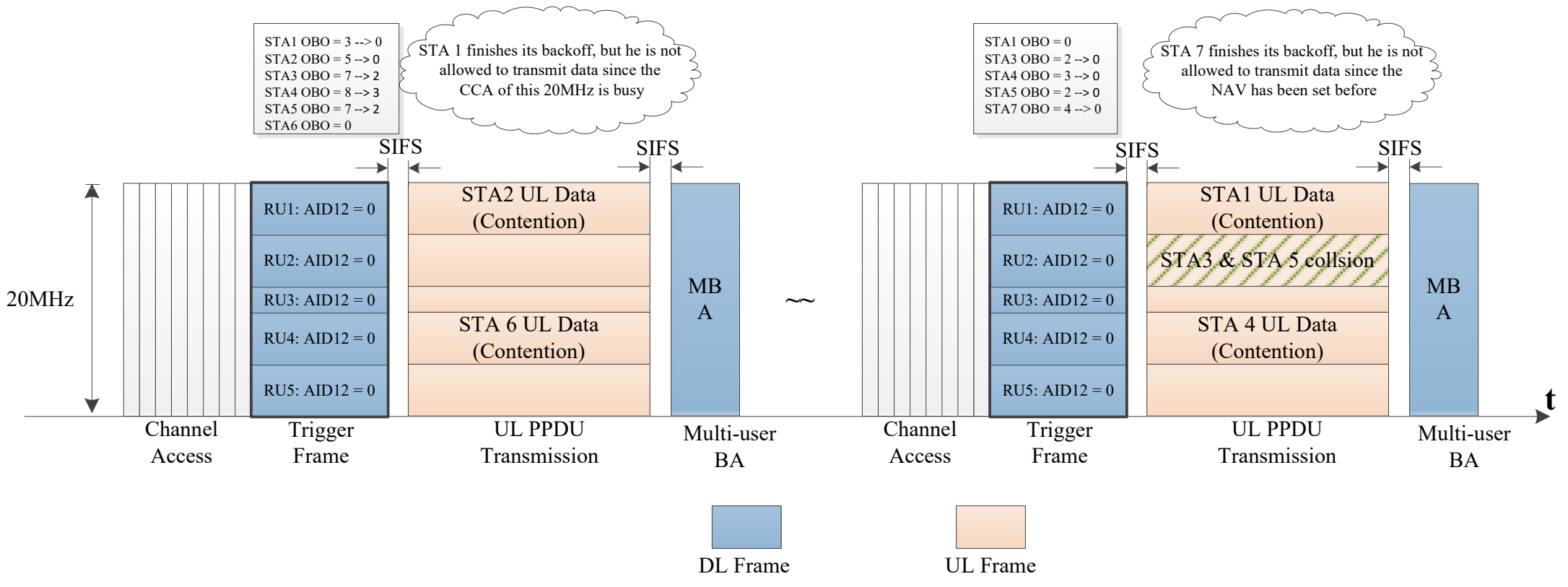

Fig. 18 Example for UORA.

3） UL MU buffer status report

AP is the sink point of all UL transmissions, but AP neither know which STAs have UL transmission demands, nor does it know the buffer status of the STAs. However, this information is very important for resource scheduling. Thus, collecting the buffer status of STAs is a very important technology. The behavior of reporting UL data requirements from STAs to AP in IEEE 802.11ax is called buffer status report (BSR). Specifically, in IEEE 802.11ax, there are two procedures to implement BSR: the piggyback procedure and the buffer status report poll (BSRP) procedure.

The piggyback procedure means that STAs can add subfields in the MAC header in other frames, or use frame aggregation to feedback buffer status. The advantage of the piggyback procedure is to save signaling overhead. However, since it always relies on STAs sending other data to AP, the buffer status cannot be reported in real-time when there is no data or no opportunity to use the channel.

The BSRP procedure refers to a special TF sent by AP. The Trigger Type field subfield in the Common Info field in TF is set to 4, which requires multiple STAs to immediately feedback BSR. The advantage of the BSRP frame is that the BSR process and the data transmission are decoupled, while the disadvantage is that it will consume extra network resources. It is quite important to note that BSPR feedback does not require AP to reply MBA.

An example of two BSR procedures is shown in Fig. 19. After finishing random backoff, AP sends DL MU data to STA1-STA3. After STA1 received the DL data and further waited for SIFS, the the BSR information is piggybacked in the BA frame. After receiving the STA1's BSR, AP sends BSRP, requiring STA4 and STA5 to feedback BSR by OFDMA. After that, STA4 and STA5 feed the BSR information back to the specific RUs. After receiving all the BSR information of STA1, STA4, and STA5, AP sends TF to ask STA1, STA4, and STA5 to transmit their UL data in the specific RUs. Finally, the AP replies with MBA to confirm the successful reception of the UL MU data.

In addition to the above two mechanisms of reporting buffer status, there is also a method for STAs to efficiently feedback some short information in high-density scenarios. This method is more suitable for cases where feedback information is very short, e.g., whether STA has power efficiency ability, whether the buffered data exceeds the buffer length threshold, and whether the channel is idle. This information can often be carried by only 1 bit. In this case, it we adopted the piggyback procedure, the real-time performance of these information would be affected, while the BSRP procedure results in a larger system overhead. To rapidly collect large amounts of STAs' information, IEEE 802.11ax proposes a mechanism that uses non data packet (NDP) to transmit a small amount of STAs information. It should be noted that the NDP is not used for channel measurement, but for information feedback. To further reduce the overhead, subcarriers are re-grouped in the short NDP access. Each group has 12 subcarriers carrying 1 bit of information. The first six subcarriers in each group having transmission energy represent 0, while the latter six subcarriers having transmission energy represent 1. Therefore, the whole 20 MHz channel can be divided into 18 groups of subcarriers, which can be used for 18 STAs to transmit 1 bit, or 9 STA to transmit 2 bits, respectively.

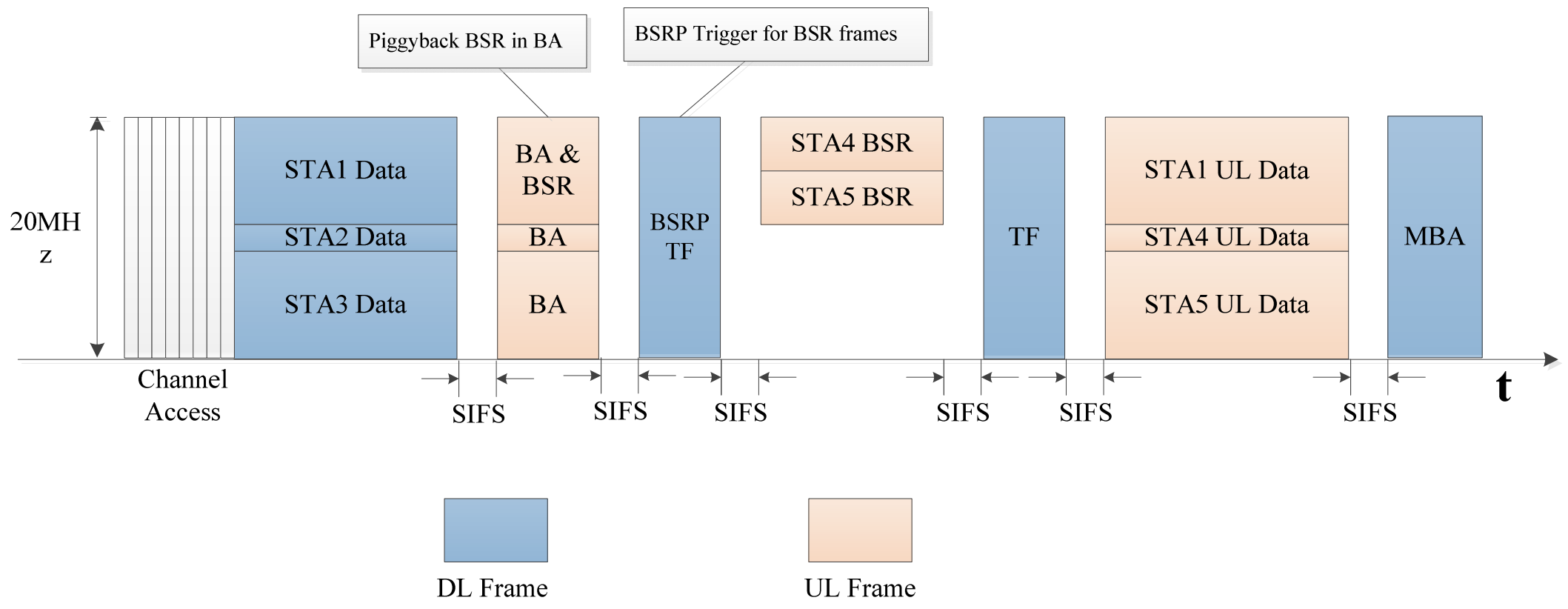

Fig. 19 Two modes of BSR.

### D. Downlink MU-MAC

Since AP knows all the DL traffic requirements, compared with the UL MU-MAC, DL MU-MAC is relatively simple. Next, we will briefly describe the DL MU-MAC access process.

1） Access process

The access process is as follows:

i. Channel access

When the AP has DL requirements and needs to transmit data to multiple users, it is still necessary to first perform both the carrier sensing and backoff process to contend for the channel. The backoff process still follows the traditional IEEE 802.11 DCF or EDCA procedure.

ii. DL MU transmission

After finishing the backoff process, AP needs to perform resource scheduling according to the DL requirements, and then transmit the DL MU data. It is worth noting that unlike UL MU transmission, DL MU-MAC does not require AP to additionally send a TF since AP fills the resource scheduling results in HE-SIG-B. Thus STAs can directly obtain the resource allocation results and receive data on the specific RUs.

Specifically, the MU PPDU sent by AP contains both the resource scheduling information and the DL data for multiple STAs. The PPDU uses the specific frame format specified by IEEE 802.11ax, which contains the resource allocation results in HE-SIG-B. First, the RUs division is conducted in the RU Allocation subfield of the Common Block field. Next, resource allocation is performed by AP for each STA in the per-user content field, indicating each STA's RU and spatial stream (if MU-MIMO is used). Multiple STAs DL data is transmitted in the payload of the PPDU. The STAs receive DL data from their corresponding RUs after receiving the resource scheduling results carried in the HE-SIG-B. Again, it is necessary to point out that not every STA's DL data sent from AP has the same time length and exactly fills the entire PPDU transmission time. For those STAs whose DL data is unable to fill the entire transmission time, the AP needs to add extra paddings to guarantee the transmission time ending the same time.

iii. OFDMA based BA

After the STAs successfully receive the DL data on the corresponding RUs, they need to reply with BA to confirm the successful transmission to the AP. Unlike UL MU transmission where AP just replies with one MBA frame, multiple STAs in DL

MU-MAC need to reply with BA via OFDMA. Specifically, two ways are supported: the first way is to indicate each specific RU in which each STA replies with BA in the HE-SIG-A sent by AP. The other way is that AP sends the MU-BAR frame to ask STAs to reply BA, where the MU-BAR frame is a special type of TF frame by setting the Trigger Type subfield in the Common Info field to 2. After that, STA replies with BA on the corresponding RUs. If AP does not receive any STA's BA until the extended interframe space (EIFS) expires, the DL MU transmission is considered as failed. Then, AP needs to double the contention window and re-select backoff values.

Fig. 20 shows an example of DL MU-MAC. AP first performs channel access on the primary 20 MHz channel. After finishing the backoff procedure, AP directly sends the DL data to multiple STAs, and indicates the scheduling results in HE-SIG-B. Specifically, AP divides the primary channel into four RUs: RU1 (52-tone), RU2 (52-tone), RU3 (26-tone), and RU4 (106-tone) and further assigns these four RUs to STA1-STA4, respectively. The PPDUs of STA2 and STA3 need padding to ensure the alignment of transmission time. After successfully receiving the DL PPDU and then waiting for SIFS, the STAs send BA to the AP based on OFDMA. When the AP successfully received the BA frames sent from STAs, the MU DL transmission succeeds.

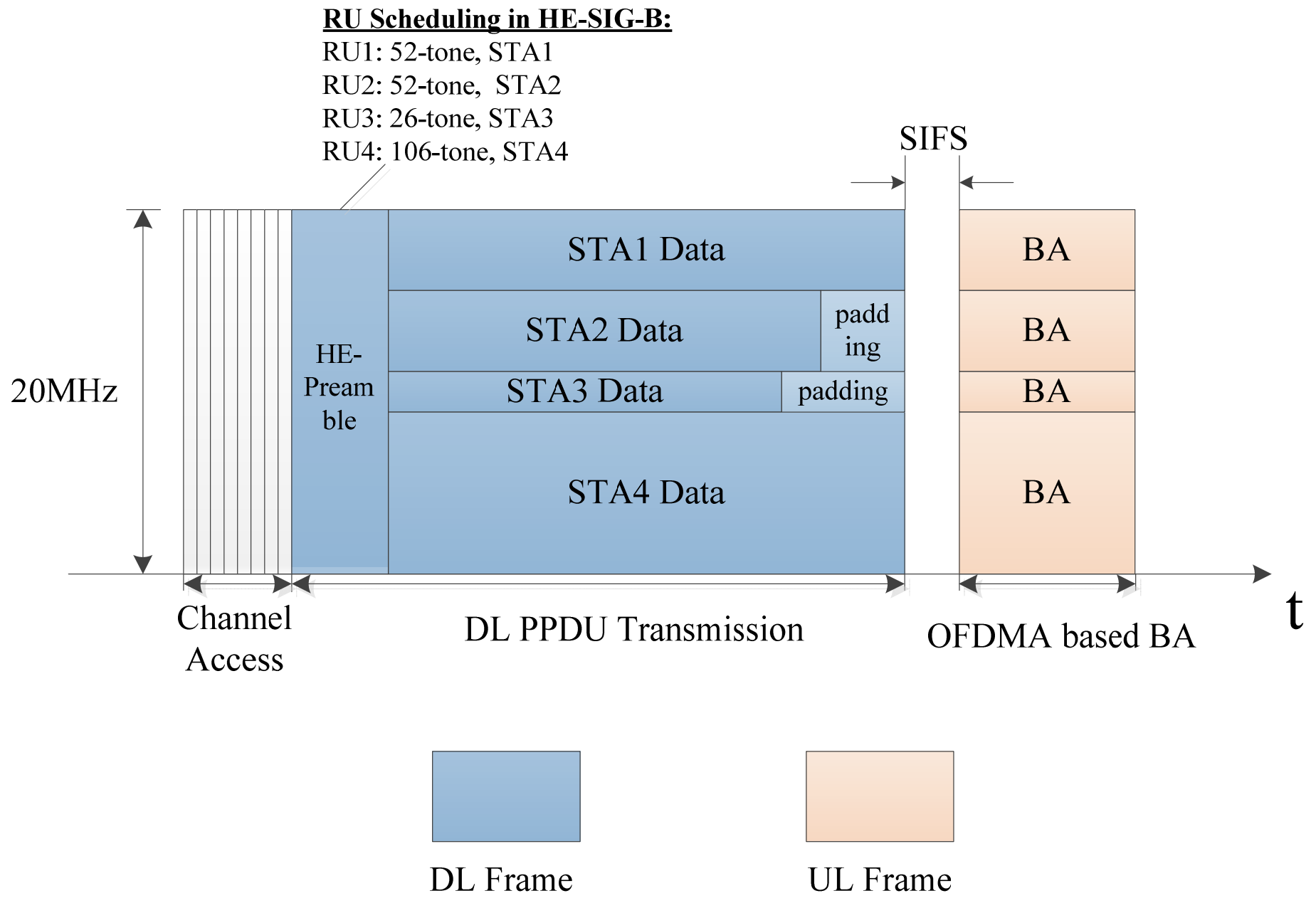

Fig. 20 Example of DL MU-MAC.

## E. Cascaded MU-MAC

Exept for the independent UL MU-MAC and DL MU-MAC, IEEE 802.11ax further introduces a UL & DL cascaded access (cascaded MU-MAC). In other words, after obtaining a TXOP, as long as not exceeding the TXOP duration, UL transmissions and DL transmissionscan alternatively cascade to fully use the TXOP opportunity. This section firstly introduces the cascaded MU-MAC process, and then introduces the MU channel protection mechanism.

1） Access processes

In the cascade transmission process, the scheduling information for UL MU transmission can be aggregated in the DL A-MPDU, which saves the overhead of TF sending and frame spaces. Tab. 5 shows that the DL A-MPDU sent by AP contains not only the DL data, but also the acknowledgement frames (confirming the previous UL frames) and TF (scheduling the next UL MU transmission); while the UL A-MPDU sent to AP contains not only the UL data, but also the acknowledgement frames (confirming the previous DL frames). Due to the great flexibility of cascade transmission, this study discusses the cascaded process with the example depicted in Fig. 21.

i. Channel access and protection

The Channel access process for the UL MU-MAC, DL MU-MAC, and the cascaded MU-MAC are identical. Firstly, the AP performs channel access process in the primary channel. After finishing the backoff procedure, AP and STAs can achieve a TXOP and protect the channel through the interaction of MU-RTS and CTS (channel protection mechanism will be discussed in the next subsubsection). Specifically, in Fig. 21, the AP schedules STA1-STA3 to reply with CTS on the secondary 20 MHz channel, while it asks STA4-STA8 to reply CTS on the primary channel. Furthermore, other STAs receiving MU-RTS or CTS and that are not involved in this transmission should set NAV and remain silent during the concurrent

TXOP.

ii. Cascaded transmission and acknowledgement

As shown in Fig. 21, during the first transmission round, AP sends DL MU data and allocates the RUs for the next UL MU transmission. For DL transmission, the primary channel is divided into five RUs, carrying the DL data of STA4-STA7 and the TF content of STA8, respectively; the secondary channel is divided into three RUs for the DL transmissions of STA1-STA3, respectively. Except for receiving their own DL A-MPDU in the corresponding RUs, the STAs also obtain the scheduling results for succeesive next UL MU transmission from the TF. Consequently, the first transmission round is completed and the second transmission round for UL is started after SIFS.

The second transmission round: The primary channel is divided into five RUs. STA4, STA6, and STA7 reply with BA for the precious DL transmission on the specified RUs respectively. STA5 aggregates the BA frame and UL data in the corresponding RU. Moreover, STA8 transmit UL data in the corresponding RU according to the TF. For the secondary channel, STA1-STA3 aggregate BA frames and UL data in their own RUs. Until now, the second transmission round is finished and the third transmission round for DL starts after SIFS.

The third transmission round: For primary channels, AP scheduling the entire 20 MHz for the DL transmission of STA7. The secondary channel is divided into five RUs to reply with BA frame for the previous UL transmission of STA1, transmit TF of STA5 and STA8, aggregate BA frames and DL data of STA2 and STA3, and transmit DL data of STA6, respectively. After receiving the DL data, the STAs obtain the RUs scheduling information for the next UL MU transmission. Then, the third transmission round is finished and the fourth transmission round for UL is started after SIFS.

The fourth transmission round: The primary channel is divided into three RUs, where two 106-tone RUs transmitted UL data by STA5 and STA8, respectively, according to the TF, and the 26-tone RU carries the BA frame of STA7. The secondary channel is divided into three RUs, where two 106-tone RUs are used by STA2 and STA3 to aggregate the BA frame and UL data, and the 26-tone RU only carries BA frame of STA6.

Finally, the AP sends MBA frames to confirm the UL transmissions, and the cascaded transmission ends.

Tab. 5 UL frame and DL frame for cascaded MU-MAC

| Frame Type | Contents | Number Constraints | Descriptions |
|---|---|---|---|
| DL A-MPDU | Acknowledgement | At most one ACK, BA, orMBA | Acknowledgements for the previous UL frames |
| | MPDU | 0 or more | DL data |
| | TF | At least one TF if extra UL transmissions follow;<br>0 TF if UL transmission | Scheduling the next UL transmission |
| UL A-MPDU | Acknowledgement | At most one ACK or BA | Acknowledgements for the previous DL frames |
| | MPDU | 0 or more | UL data |

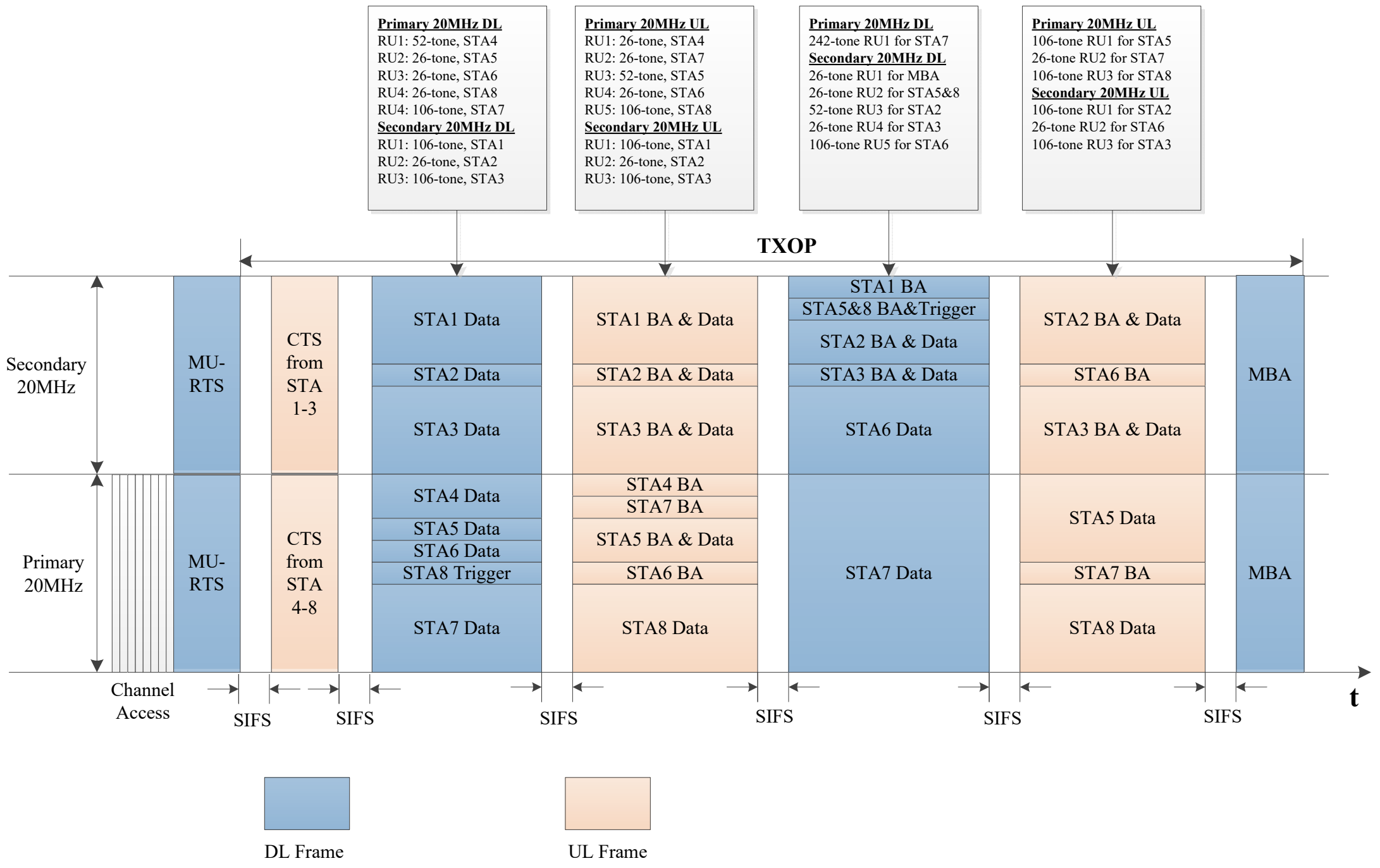

Fig. 21 Example for the cascaded UL & DL MU-MAC.

2） MU channel protection mechanism

In the traditional DCF and EDCA channel access, AP and STA protect the channel via the RTS/CTS mechanism, inhibiting the hidden node problem. IEEE 802.11ax introduces MU-MAC, indicating that more user transmissions will be interfered at the same time, thus the threat caused by the hidden nodes problem becomes severe. As shown in Fig. 22, the AP in the current BSS simultaneously sends DL data to STA1-STA4. On the one hand, because the OBSS STA is far from the AP, the OBSS STA fails to decode the frame sent by the AP, and the physical carrier sensing of the OBSS STA indicates idle. On the other hand, AP and STAs do not adopt the channel protection mechanism similar to RTS/CTS; therefore, the virtual carrier sensing of OBSS STA is also indicated as idle. This means that the OBSS STA does not notify the concurrent MU transmission. After successfully contending the channel, the OBSS STA performs single-user transmission on the entire 20 MHz channel, and then conflicts with the concurrent MU transmission, i.e., seriously interfering the reception of the STA1-STA4. Based on the above reasons, MU-RTS and CTS mechanism are introduced in IEEE 802.11ax to protect the wireless channel for MU access.

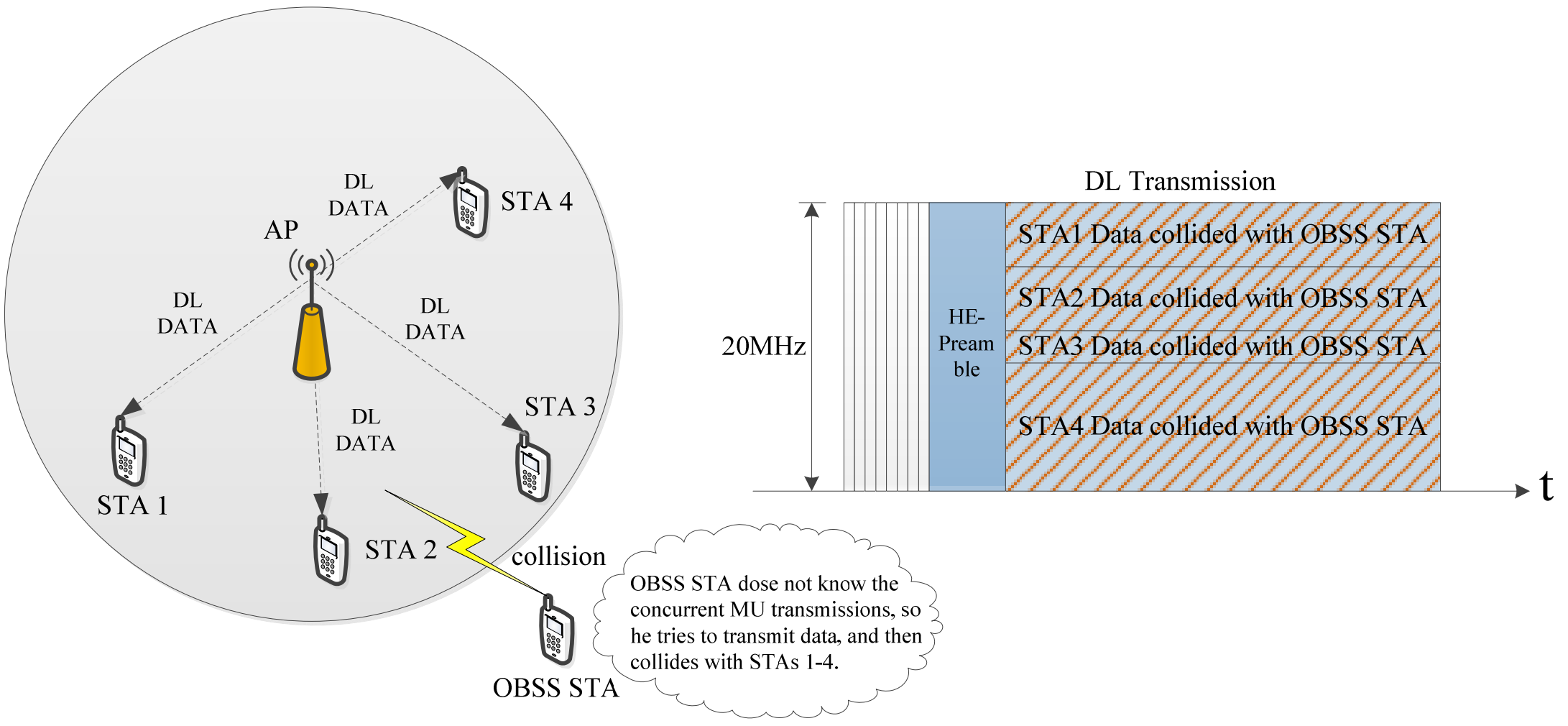

Fig. 22 Challenges for the MU-MAC caused by the hidden nodes problem.

In the MU-RTS/CTS mechanism, when AP wants to send DL MU data to multiple STAs or trigger UL MU data from multiple STAs, it sends MU-RTS frame. MU-RTS is a special TF by setting the Trigger Type subfield in the Common Info field in TF to 3. MU-RTS explicitly indicates that multiple STAs should simultaneously reply CTS in one or more 20 MHz channels. After the STAs receiving the MU-RTS frame, they send CTS in the specific 20 MHz channel(s) according to the indications in the MU-RTS. Although multiple STAs will likely send CTS frames in the same 20 MHz channel, the overlapping CTS can be decoded by other nodes since they share the same CTS frame content and adopt the same scrambling codes. Other STAs receiveing MU-RTS and/or CTS should set NAV to avoid collision. Thus, MU-RTS and CTS greatly alleviate the affection caused by the hidden nodes problem. Both MU-RTS and CTS utilize the BPSK modulation adopted by the legacy WLANs; thus, the compatibility and coexistence can be guaranteed well.

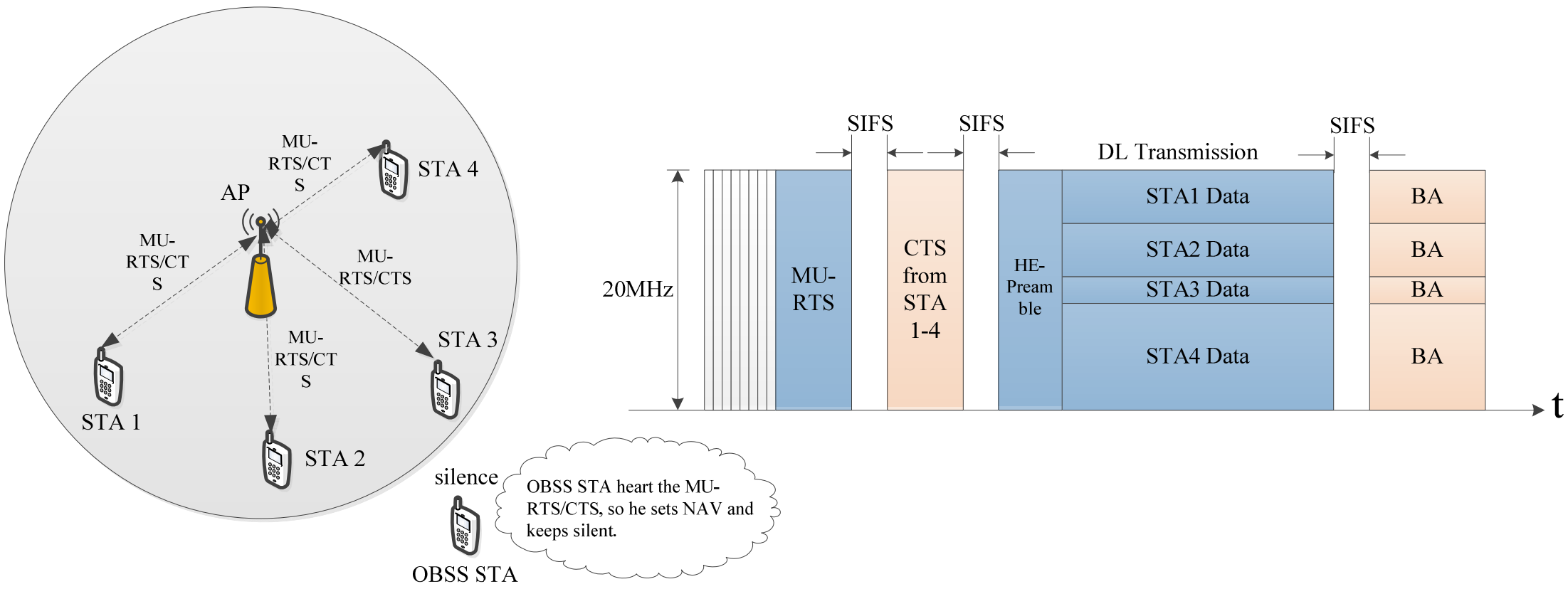

Fig. 23 MU-RTS/CTS mechanism.

As shown in Fig. 23, after completing the backoff procedure, AP sends the MU-RTS frame firstly. When STA1-4 successfully receive MU-RTS, they send the CTS frames on the same 20 MHz channel. After receiving the CTS, AP starts to transmit the DL MU data. At the same time, after the OBSS STA receiving MU-RTS and/or CTS, it sets the NAV and remains silent during concurrent transmissions. Evidently, compared to Fig. 22, the MU-RTS/CTS mechanism can protect the channels for the UL and DL access process, thus greatly reducing the impact of the hidden nodes problem. It is worth noting that the MU-RTS/CTS mechanism is optional, which means that the AP can directly perform UL and DL MU-MAC without channel protection.

## F. Overall access scenes for IEEE 802.11ax

According to Fig. 23, IEEE 802.11ax significantly enhances the ability of parallel access for multi-users, thus significantly improving the efficiency of WLANs. In this subsection, we describe and analyze the overall access process of the IEEE 802.11 after the introduction of the IEEE 802.11ax standard amendment.

First, IEEE 802.11 can use the HCF to plan and manage wireless resources. HCF runs based on the periodic super frame structure. Each super frame is initiated by a Beacon frame and lasts for several Beacon intervals. In each super frame, the network first enters CFP. The HCCA is adopted as the channel access means. For HCCA, AP completely controls access, while STAs are not allowed to spontaneously access channels. The remaining time of the super frame is called the contention period (CP). As the name suggests, CP allows both AP and STAs to spontaneously access the channels. Both HCCA and EDCA can be adopted in CP.

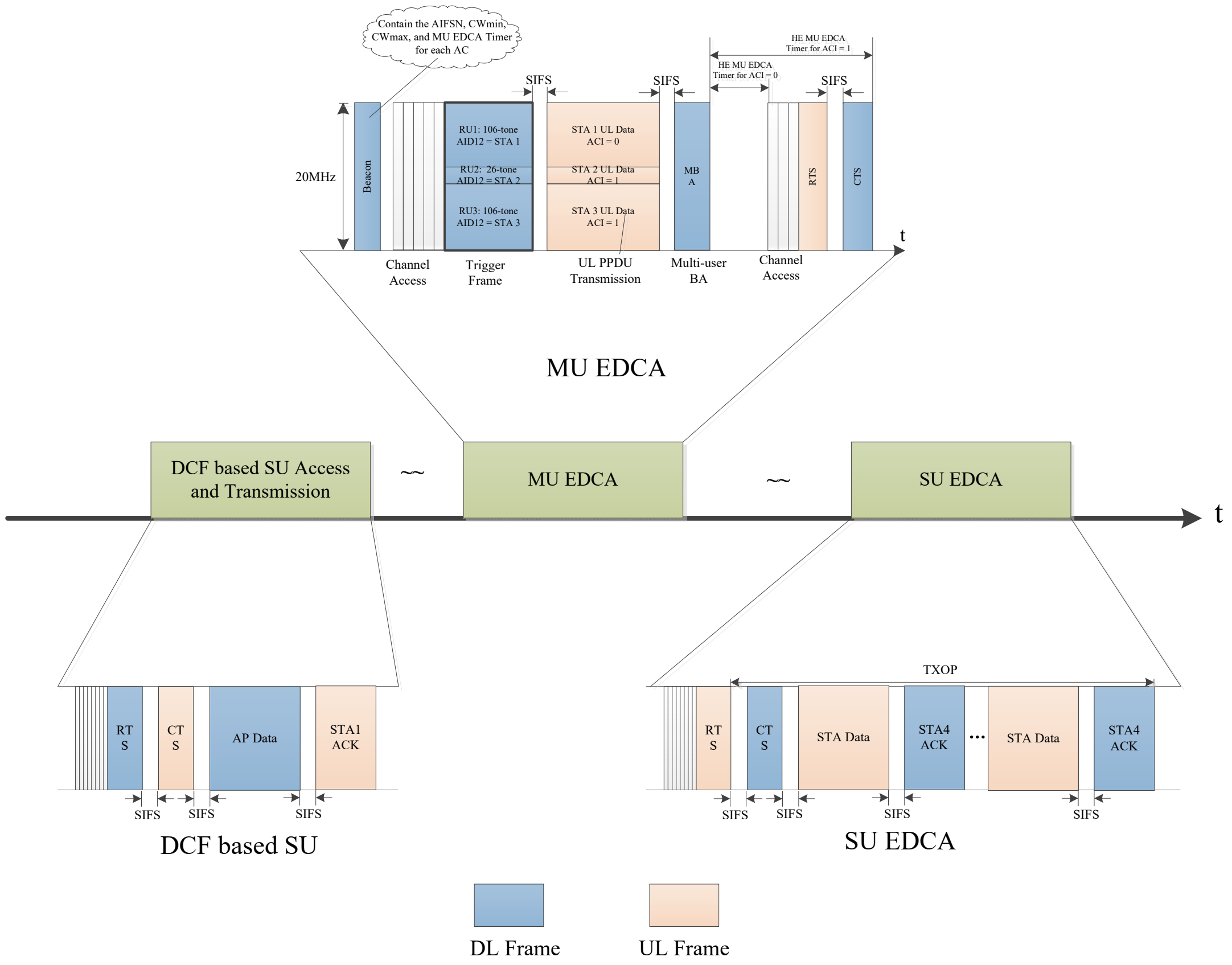

Fig. 24 Access and transmission overview for IEEE 802.11ax.

However, the compatibility between the MAC process in IEEE 802.11ax and HCCA needs to be further studied and designed. Therefore, for now, the MU-MAC in IEEE 802.11ax is only compatible with the EDCA and DCF, as shown in Fig. 24.

Both AP and STAs try to access the channel via random access based on EDCA or DCF. Furthermore, EDCA can be divided into SU EDCA as well as MU EDCA. Among these, SU access is still based on the traditional IEEE 802.11, and no longer on details. Furthermore, MU access mainly refers to the MU-MAC of IEEE 802.11ax. In conclusion, by introducting the MU-MAC, on the one hand, IEEE 802.11ax greatly enriches the access means for IEEE 802.11, and efficiently meets the requirelents of different network scenarios via flexible configurations of different access means; on the other hand, MU-MAC and the legacy IEEE 802.11 are perfectly compatible with one another.

It is worth noting that MU EDCA proposed in IEEE 802.11ax achieves a tradeoff between scheduling access and random access. As shown in Fig. 24, when a STA needs to transmit UL data, it can either use EDCA, or wait for the TF to use UL hybrid access. It is unfair that a STA can still access channel using EDCA after it transmits UL data through UL hybrid access procedure. Therefore, IEEE 802.11ax proposes to solve this problem using MU EDCA parameters. Specifically, after a STA transmits data through UL hybrid access, it cannot contend channel resource using EDCA for a period of time. As shown in Fig. 24, AP sends beacon frame to STAs, in which contains arbitration inter frame spacing number (AIFSN), ECWmax, ECWmin, and MU EDCA Timer for each AC. After AP wins the channel contention, it sends TF to allocate RU1, RU2, and RU3 to STA1, STA2, and STA3 respectively. Then, STA1 sends UL data on RU1 with AC index (ACI) equal to 0, STA2 sends UL data on RU2 with ACI equal to 1, and STA3 sends UL data on RU3 with ACI equal to 1. After receiving the MBA frame sent by AP, STA1, STA2 and STA3 all starts their MU EDCA timer according to the value of MU EDCA Timer in beacon frame. Since the value of MU EDCA Timer for ACI equal to 0 is shorter, MU EDCA timer of STA1 expires firstly. Then, STA1 carries out channel access though EDCA using MU EDCA parameters, including AIFSN, ECWmin, and ECWmax contained in beacon frames.

# V. Spatial Reuse Enhancements

## A. Introduction to Spatial Reuse Enhancements

In WLANs, due to the broadcast nature of wireless communication, the link in current BSS inevitably suffers from interference caused by simultaneous transmission of surrounding BSS using the same channel; conversely, the communication link in current BSS also interferes with the other communication link in the adjacent BSS using the same channel. The mutual interference caused by overlapping coverage between BSS can seriously affect the throughput and user experience in the network. However, on the other hand, increasing the interferences among multiple BSS at a certain extent may also increase the number of the concurrent transmission links, and further enhances the overall throughput. Based on this situation, IEEE 802.11ax focuses on high-dense deployment scenarios and emphasizes the need to significantly enhance the spectrum efficiency of dense deployments in OBSS WLAN situations.

IEEE 802.11ax introduces multiple key technologies, including enhanced physical carrier sensing mechanism, enhanced virtual carrier sensing mechanism, and transmission power control mechanism; collectively, these are termed SR technology. The basic idea of SR technology is to combine the enhanced carrier sensing mechanism with the transmission power control mechanism to increase the concurrent transmission probability of neighboring BSS links as much as possible, thereby improving the area throughput of WLANs. Prior to introducing the SR technology employed in IEEE 802.11ax, we will provide an overview of pertinent SR technology research to facilitate readers' understanding of our study. Fig. 25 outlines the key technologies of SR and the relationships between them.

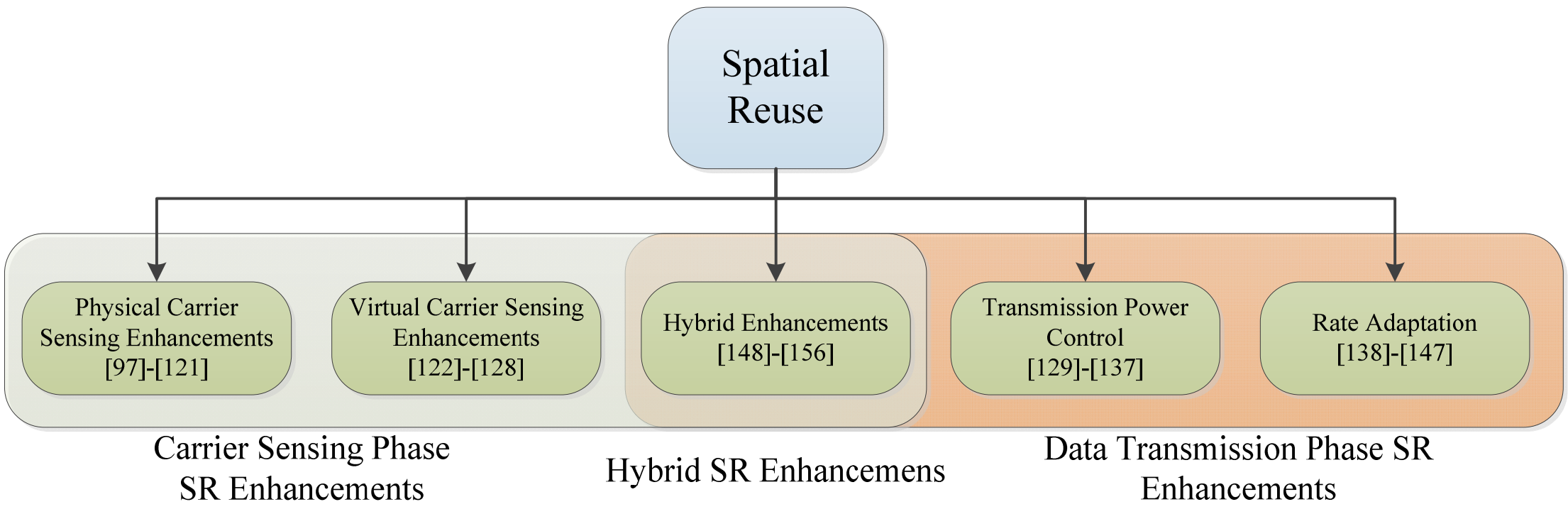

Fig. 25 Key technologies and their relationships in SR

1) Physical Carrier Sensing Enhancements

Physical carrier sensing means the node judges the channel state (i.e., busy or idle) by listening to the channel's existing wireless signal energy. The process of performing physical carrier sensing is called CCA. Specifically, based on CSMA/CA, when the transmitter detects that the channel is idle, it needs to retreat for a period of time to send data. IEEE 802.11 standard stipulates that in the backoff process, it is necessary to sense channel energy through physical carrier sensing in every time slot. If the energy is greater than a certain threshold, known as the CCA threshold, then the channel is considered busy; otherwise, the channel is idle.

i. Related work on fixed CCA

Jamil et al. [97] verify the influence of CCA threshold variations on whole-network throughput in an outdoor cellular network topology scenario using simulation. The results show an optimal CCA threshold in multiple BSS scenario (i.e., -75dBm in the simulation). The simulation results also indicate that the average throughput of each STA and of the 5% STAs behind can be maximized by adjusting the CCA threshold in [98]-[100]. However, on one hand, the optimal CCA threshold is related to the network scenarios and parameters; on the other hand, the optimal CCA threshold of the maximum average throughput of each STA and of the 5% STAs behind are not the same; thus, the two optimal CCA thresholds are difficult to satisfy simultaneously. Through analysis and simulation methods, extensive research [101]-[104] point out the challenge in achieving whole-network optimization through a fixed CCA threshold. Therefore, many scholars and enterprises have studied dynamic CCA threshold adjustment.

ii. Related work on dynamic CCA

Nakahira et al. [105] propose a CCA threshold dynamic adjustment method using centralized control wherein a central AP or controller optimizes the CCA threshold and bandwidth according to the received signal strength indicator (RSSI) between the AP and channel state information. Murakami et al. [106] also use centralized control in which the controller dynamically optimizes the CCA threshold based on the geographical location of STAs. In the authors' opinion, centralized control is better suited to pre-planned WLAN networks such as those in businesses and on campuses. It is not appropriate for hot-spot scenarios in general; the distributed dynamic CCA adjustment method is a better option. Hua et al. [107] propose a distributed dynamic CCA adjustment method in which each AP predicts the current network state by detecting its own packet loss rate and busy channel ratio and adjusts the CCA accordingly. Similarly, Kim et al. [108] suggest that in the DL scenario, the CCA threshold should be set differently according to interference between the AP and the receiving STAs. Smith [109][110] notes that STA can adjust its CCA threshold as a value by subtracting a preset margin from the RSSI of the Beacon frame transmitted by the AP to enhance the spatial multiplexing capability.

Through simulation methods, Zhong et al. [111] further verify Smith's proposed scheme, which they found to improve network throughput compared to the fixed CCA threshold adjustment. Nevertheless, improvement in network throughput is strongly related to network topology and node density. Coffey et al. [112] propose that STAs adjust the CCA threshold according to whether they received the ACK frame of the UL data frame. That is, when a node successfully receives the ACK frame several times (e.g., 16 times), the CCA threshold advances one level; when no ACK frame is received, the CCA threshold decreases one level. The range of carrier sensing and communication interference has also been examined in [113]-[115]. It is proposed that the radius of the optimal carrier sensing range is equal to the sum of the range of communication interference and the distance between the transmitter and receiver.

iii. Related work on CCA based on data source differentiation

Hedayat et al. [116] suggest that STAs need to maintain a neighbor list, and when the destination address of the receiving frame is not in a neighbor list, a higher CCA threshold is used. Son and Kwak [117], Jiang et al. [118], and Choudhury et al. [119] propose that the BSS color mechanism in IEEE 802.11ah should be introduced, so that when the OBSS frame is monitored, the CCA threshold of IEEE 802.11ax increases to a certain value. Ishihara et al. [120] also point out that BSS color and CCA threshold adjustment can be used as two methods to improve IEEE 802.11ax OBSS performance. In addition, Hedayat et al. [121] summarize and combine related work on TGax SR technology. It should be noticed that the OBSS_PD technology introduced by IEEE 802.11ax is this type.

2) Virtual Carrier Sensing Enhancements

Virtual carrier sensing refers to informing the surrounding nodes of impending data transmission through broadcast signaling. Specifically, before a link initiates data transmission, the transmitter and receiver first exchange short frames (e.g., RTS/CTS) to reserve the channel resource; then, a network allocation vector (NAV) is carried in the short frames to indicate the duration of data transmission (i.e., the duration that channel is busy) to their surrounding node data. When both the virtual carrier sensing mechanism and the physical carrier sensing mechanism indicate that the channel is idle, the STA will consider its surrounding channel to be idle.

Ye and Sikdar [122] first demonstrate that the traditional geographic scope covered by NAV in RTS/CTS is not optimal and thus influences the spatial multiplexing effect. They further explore the need to calculate optimal virtual carrier sensing coverage according to the distance between the receiver and transmitter, and propose that RTS/CTS carried channel reservation information and information about the coverage of virtual carrier sensing, which can effectively control the geographic scope of reservation and enhance the spatial multiplexing effect. Seok et al. [123] and Luo et al. [124] point out that an STA could adopt higher CCA threshold while receiving inter-BSS frame. However, if the receiving frame is RTS/CTS frame, this STA needs to set NAV according to the existing rules and cannot transmit a data frame (i.e., there is no spatial multiplexing gain). Therefore, it is recommended that the virtual carrier sensing rules must also make corresponding adjustments. When an STA receives the RTS/CTS inter-BSS frames (i.e., those sent by another BSS) and considers that the channel is idle according to a higher CCA threshold, NAV is not updated. At the same time, Luo et al. [124] also notes that if STAs receive a frame, the channel should be considered busy before determining whether the received frame is an inter-BSS frame. In addition, Fang et al. [125] put forward two NAVs ideas: SBNAV and OBNAV. SBNAV is used to update the NAV of the current BSS, and OBNAV is used to update that of inter-BSS. OBNAV is subdivided into OBNAV-CP and OBNAV-CFP, which are applied to CP and CFP, respectively; the longer NAV is selected as the OBNAV. In addition, Huang et al. [126] and Huang [127] also point out the drawbacks of legacy IEEE 802.11 are only using one NAV, suggesting that different NAVs should be used for intra-BSS and inter-BSS scenarios. This idea has been adopted by IEEE 802.11ax. Li et al. [128] propose that CTS frames carry the interference tolerance and acceptance of NAV probability p. After the neighbor node successfully receives NAV, if it determines that the interference caused by its transmission does not exceed the interference tolerance, it accepts the NAV in the probability p and ignores the NAV in probability *(1-p)*. Thus, the probability of concurrent transmission in the network improves. The idea of controlling the potential concurrent link through the current link is consistent with the SRP based SR mechanism proposed in IEEE 802.11ax.

3) Transmission Power Control

Shih et al. [129] focus on improving the number of concurrent communications pairs and propose a distributed spatial reuse (DSR) protocol. The protocol is divided into the contention stage and transmission stage, where communication pairs seize the transmission opportunity in the competition stage and increase the number of concurrent transmissions through power control in the transmission stage. On one hand, this protocol uses power control to reduce interference between the communication links; on the other, it can significantly increase the number of communication links. They also propose a maximum independent set (MIS) algorithm that maximizes the independent set. The algorithm is centralized to maximize the number of communication links. Su et al. [130] assume that AP can support the capture effect. AP divides the coverage of BSS into several regions according to the geographical location (e.g., annular regions). Hence, AP can differentially control STAs' transmit power in different regions so the received power of data frames from STAs in different regions can all reach the preset received power level. When STAs from different regions simultaneously transmit data according to the preset transmit power, AP can parse the signal produced by the STA with the strongest signal. Similarly, Patras et al. [131], Vukovic and Smavatkul [132], and Sutton [133] also use power control and the capture effect to enhance spatial reuse capability and reduce contention. Gandarillas et al. [134] demonstrate that WLANs require high-speed transmission often in DL transmission scenarios, and propose a method for AP to control the transmit power according to the channel state (e.g., channel occupancy rate, MCS, RSSI, packet retransmission times) over time. If the retransmission rate over a period of time is high, implying there are too many concurrent links and extreme collision, AP needs to increase the transmit power to reduce concurrent transmission. If the retransmission rate over a period of time is low, suggesting fewer concurrent links, then AP must reduce its transmit power to encourage concurrent transmission. This method is based on the channel state for statistics and does not require additional signaling. Li et al. [135] are the first to analyze and test the criteria for improving energy efficiency focusing on the energy efficiency problem in WLANs. Then, they put forth a method of dynamically adjusting the transmit power according to whether ACK or BA is successfully received to achieve the goal of improving energy efficiency. When the interference or collision is serious and successive, ACK or BA cannot be received successfully, and the transmit power needs to be reduced. When the channel quality is good and successive, ACK or BA are successfully received, the transmit power needs to be increased. Oteri and Yang [136] and Arjun et al. [137] design the power control rule and corresponding signaling for multi-user transmission. They also analyze the necessary power control for improving MU access performance and highlight the need to bring relevant parameters in TF, including the transmit power of AP, MCS of each scheduled STA for data transmission, and expected received power at the AP side. After receiving TF, STA can calculate the transmit power for its transmission according to the above parameters.

4) Rate Adaptation

Lower transmission rate enhances the link reliability, which enables more concurrent links; while higher transmission rate decreases the link reliability, which enables less concurrent links. Pefkianakis et al. [138] propose history-aware robust rate adaptation (HA-RRAA) with the premise that nodes predict current channel quality based on the frame error rate in a previous short-term period to optimize and adjust the sending rate. When the frame error rate exceeds the maximum tolerable threshold, the transmission rate is reduced; otherwise, the transmission rate increases when the frame error rate is below the minimum tolerable threshold. Yang et al. [139] propose that by adjusting the transmission rate, the interference range would be included in the carrier sensing range to avoid the hidden nodes problem. Acharya et al. [140] demonstrate that the traditional rate adaptation strategy has a poor effect in a congested network, and propose the wireless congestion optimized fallback (WOOF) mechanism. The basic idea of this mechanism is to dynamically adjust the transmission rate according to network congestion as measured by the channel busy time (CBT). Shen et al. [141] find that the traditional rate adaptation method does not apply to MU-MIMO and propose a rate adaptation mechanism called TurboRate. In it, the STA determines the transmission rate according to SNR and the signal receiving direction of AP. SNR and the signal receiving direction can be calculated via a passive method (i.e., continuous listening Beacon frame without additional signaling). Similarly, Makhlouf and Hamdi [142] and Pefkianakis et al. [143] extend the characteristics of MIMO based on legacy rate adaptation, which is matched with SU-MIMO. Noubir et al. [144] put forward two link adaptive algorithms for industrial application scenarios and improve the classical ARF algorithm.

The ARF algorithm [145] requires the following: if the successive K transmission fails, the transmission rate is reduced by one level; if the successive N transmission is successful, the transmission rate is increased by one level. Noubir et al. [145] also propose two improved algorithms: static retransmission rate ARF (SARF) and fast rate reduction ARF (FARF). The minimum rate must be adopted when SARF requires retransmission, but any retransmission success is not recognized as one successful ARF transmissions. FARF requires that the minimum rate be used for the next transmission after a failed transmission. Cardoso et al. [146] design an automatic link adaptation algorithm for the dense deployed IEEE 802.11 network based on the classical sample rate link adaptation algorithm. Biaz et al. [147] summarize early IEEE 802.11 rate adaptive algorithms.

5) Hybrid Enhancements Mechanism of SR

The aforementioned technologies can further enhance SR performance through a joint design approach. Wang et al. [148] suggest a dynamic CCA mechanism jointly designed with transmit power control in IEEE 802.11ax. That is, when the transmit power is reduced, inter-BSS CCA can be enhanced accordingly. This idea is eventually adopted by IEEE 802.11ax. Zhou et al. [149] model the relationship between physical carrier sensing, transmit power control, and transmission rate and proposed a joint adjustment algorithm. Smith et al. [150] simulate and analyze the joint optimization of TPC and dynamic CCA for enterprise WLAN scenarios. The results show that performance gain is obvious only when all STAs use TPC; otherwise, performance gain is limited. In [151], we jointly optimize the carrier sensing range and transmission rate control. Specifically, the carrier sensing range contain the interference range, and the area throughput of the network is improved. Similarly, Zhang et al. [151] and Chen et al. [152] also focus on joint optimization of dynamic CCA and rate adaptation to enhance network throughput. Ma et al. [153] propose that the transmission rate can be initially adjusted according to the distance between the transmitter and receiver, and then the CCA threshold can be adjusted accordingly. Our early work [154] points out that physical carrier sensing is closely related to the backoff procedure because the existing carrier sensing mechanism only has busy and free states. Thus, some nodes perform backoff almost uniformly, and the other nodes are in retreat pending state identification. This coarse-grained carrier sensing and backoff procedure suppresses the regional throughput. Therefore, we design a more fine-grained decimal backoff mechanism, the results of which demonstrate significantly improve area throughput in dense deployment scenarios.

In [155], we propose that the transmitting node predicts the channel state of the receiving node based on its historical transmission quality to perform a more accurate backoff procedure. We also suggest the idea of a receiver executing the backoff procedure. In [156], we assume that STAs can be associated with multiple neighboring BSS, thereby enhancing the probability of concurrent transmissions. It is necessary to point out that the SR enhancement technology in IEEE 802.11ax combines an enhanced physical carrier sensing mechanism, enhanced virtual carrier sensing mechanism, transmit power control mechanism, and link adaptation mechanism that have been standardized to improve throughput and area throughput in high-dense deployment scenarios.

### B. SR Framework in IEEE 802.11ax

In this section, the framework of SR in IEEE 802.11ax is discussed.

First, there is a basic technical prerequisite for various SR technologies: the nodes in WLANs must easily distinguish whether the received packet is from intra-BSS or inter-BSS. Thus, IEEE 802.11ax introduces the BSS color mechanism. Because different BSS often adopt different colors, the nodes can distinguish the source of the packet and then perform appropriate SR operations. This function is detailed in Sec. V.C.

Secondly, IEEE 802.11ax introduces two NAVs counters: intra-BSS NAV and basic NAV. When the node judges that the received frame comes from its own BSS, the intra-BSS NAV is updated; when the node determines that the received frame is from inter-BSS or cannot judge whether it is from intra-BSS, the basic NAV is updated. Thus, the rule of virtual carrier sensing is modified as follows: if and only if both NAVs are 0, the virtual carrier detection is determined to be idle; otherwise, it is busy. The two NAVs based virtual carrier sensing mechanism not only matches MU-MAC but also avoid the TXOP-ending chaos problem in high-dense deployment scenarios. This function will be described in Sec. V.D.

Thirdly, in order to encourage concurrent transmission, IEEE 802.11ax adopts the OBSS power detection (OBSS_PD) mechanism. OBSS_PD specifies that when a node receives a PPDU from inter-BSS, it can use a higher CCA level, called the OBSS_PD level, to increase the probability of concurrent transmission while improving spatial reuse capability. At the same time, the transmission power needs to be controlled when adjusting the OBSS_PD level: when the transmission power is small, the OBSS_PD level can be enhanced; when the transmission power is large, the OBSS_PD level can be reduced. Hence, a flexible tradeoff between interference and spatial reuse capability is possible, to be explained in Sec. V.E.

Finally, IEEE 802.11ax introduces the spatial reuse parameter (SRP) mechanism. This mechanism specifies that when STAs receive the TF content from inter-BSS, and the corresponding fields in the HE-SIG-A and TF indicate that SRP is allowed, the STAs obtain an SRP opportunity during the concurrent PPDU transmission. As a result, STAs can continue to perform backoff and attempt to access the channel during the SRP opportunity. Of course, the transmit power needs to meet certain conditions, which will be detailed in Sec. V.F.

Except for the BSS color mechanism as the basic technical prerequisite, the other three SR technologies introduced in IEEE 802.11ax are consistent with the virtual carrier sensing mechanism, the physical carrier sensing mechanism, and transmission power control. Collectively, they help to realize the objective of significantly increasing spectral frequency reuse and interference management in high-dense OBSS scenarios. It is worth noting that because IEEE 802.11ax continues and extends the MCS, another SR technique, rate adaptation, is inherently supported in IEEE 802.11ax.

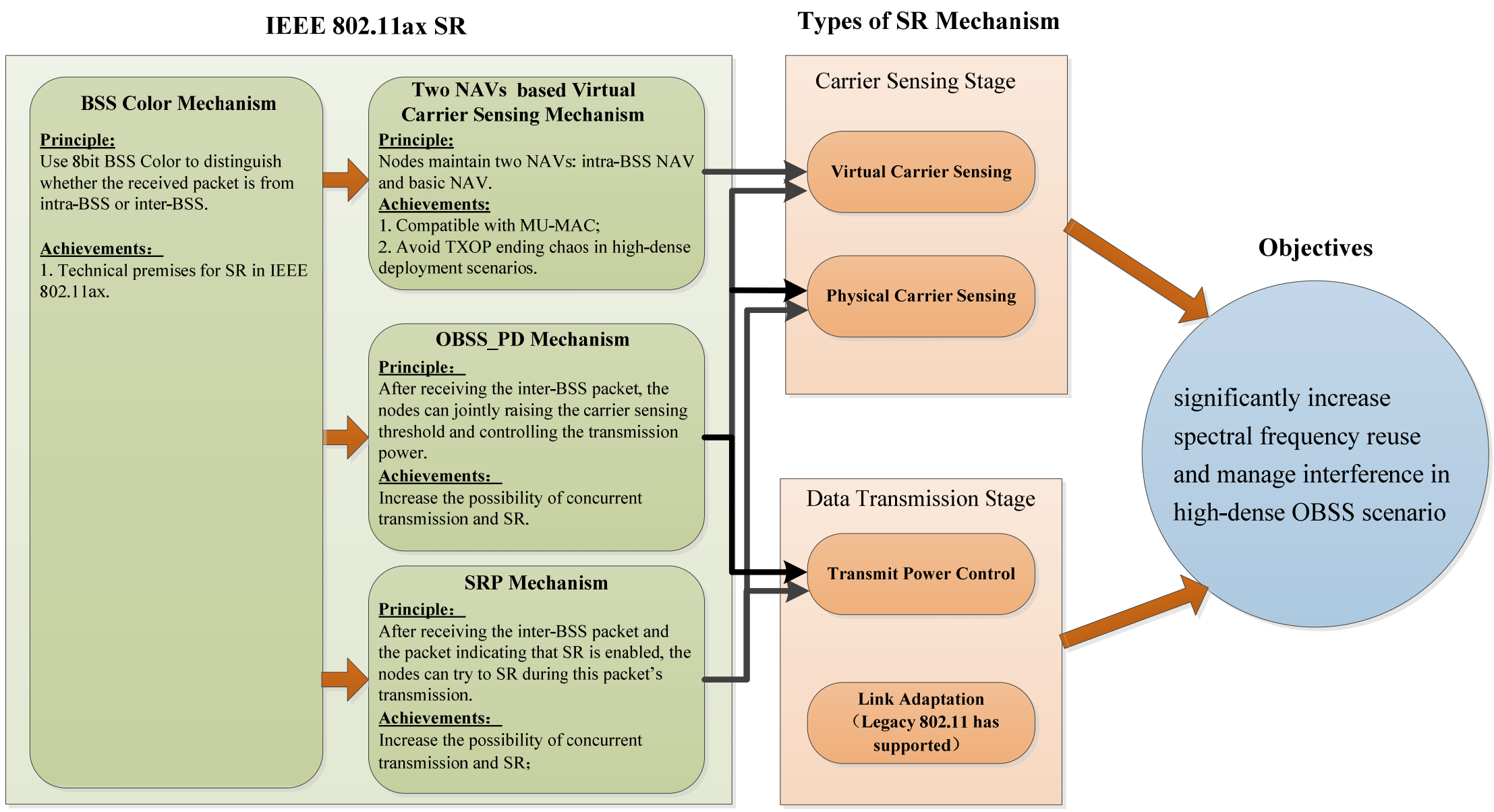

Fig. 26 IEEE 802.11ax SR framework.

## C. BSS Color Mechanism

When introducing SR, a technical prerequisite is that the nodes can accurately identify the received packets as either intra-BSS or inter-BSS; otherwise, SR would be hard landing. Therefore, IEEE 802.11ax applies the concept of BSS color to the SR mechanism.

1) The Concept of BSS Color

BSS color is an unsigned integer value of 8 bits whose range is 1~63. It is used to distinguish different BSS. The concept of BSS color is first proposed in the IEEE 802.11ah standard amendment by filling the color field in SIG to distinguish whether the receiving packet is from intra-BSS or inter-BSS. However, IEEE 802.11ah does not focus on how to detect and avoid conflict in color BSS. IEEE 802.11ax extends the BSS color into the SR mechanism. Based on BSS color, STAs can distinguish the received packets as being intra-BSS or inter-BSS. After that, additional SR technologies can be adopted to improve SR capability. Specific SR technologies are discussed in Sec. V.D, Sec. V.E, and Sec. V.F. Tab. 6 lists all the possible methods supported in IEEE 802.11ax to distinguish packets as inter-BSS or intra-BSS. After receiving the packets, if the BSS color, RA/TA/BSSSID field, or partial AID is contained in the packet, the STAs or AP can directly determine whether the packet is inter-BSS or intra-BSS. Further, if an IEEE 802.11ax AP receives a non-HE PPDU (legacy PPDU) or HE MU PPDU, it can identify the packet as being inter-BSS because its own BSS will not send MU PPDU without a trigger from the AP itself.

Tab. 6 Determination conditions for whether a received frame is inter-BSS or intra-BSS.

| **Determination Conditions** | | **Intra-BSS** | **Inter-BSS** |
|---|---|---|---|
| BSS color in the received frame | | same as the BSS color announced by the AP with which the STA is associated | not 0 and not matching the BSS color announced by the AP with which the STA is associated |
| RA field, TA field, or BSSID field of the received frame | | same as the BSSID of the AP with which the STA is associated | none of the address fields match the BSSID of the AP with which the STA is associated |
| Partial AID | Case 1: group ID = 0 | partial AID same as the BSSID[39:47] of the AP with which the STA is associated | different from the BSSID[39:47] of the AP with which the STA is associated |
| | Case 2: group ID = 63 | same as the partial BSS color announced by the AP with which the STA is associated when the partial BSS color field in the most recently received HE operation element is 1 | different from the partial BSS color announced by the AP with which the STA is associated when the partial BSS color field in the most recently received HE operation element is 1 |
| RA field of a control frame that does not have a TA field | | matches the saved TXOP holder address for the BSS with which it is associated | ———— |
| Multiple BSSID | | the RA field, TA field, or BSSID field of the received frame is same as the BSSID of any member of the multiple BSSID set<br>a control frame that does not have a TA field, and the RA matches the saved TXOP holder address for a BSS that is a member of the multiple BSSID set | the BSSID field of the received frame does not match the BSSID of any member of the multiple BSSID set<br>none of the address fields of the received frame match the BSSID of any member of the multiple BSSID set<br>partial AID is different from the BSSID[39:47] of any member of the multiple BSSID set |
| MU PPDU received by AP | | ——— | An HE AP receives either a VHT MU PPDU or an HE MU PPDU |

2) BSS Color Conflict Detection

BSS color conflict is the event in which two or more neighboring BSS have the same BSS color, which affects the judgment of packet sources, SR rules, and so on. The AP determines that BSS color conflict events occur in two ways: first, when the AP receives a packet from the inter-BSS and finds that the inter-BSS and its own BSS have the same color, color conflict occurs; second, when the STAs in its own BSS autonomously report that another BSS and its own BSS have the same color, color conflict occurs. Specifically, when STAs detect that another BSS has the same color as their own BSS, they need to send an event report frame to their own AP. The event report frame contains the color information for the other BSS. It should be emphasized that the reported color information should contain the color information of all other BSS detected by STAs, not just the BSS whose color conflicts with its own BSS. This helps its own AP to calculate and update a new and more appropriate BSS color.

When the BSS color conflict lasts for a period of time, specified by the standard, the AP shall execute the BSS color change process, which is described in the next subsection. To distinguish BSS, different BSS often set different BSS colors. But when an AP uses multiple BSSID, all the virtual APs deployed in one physical AP need to use the same BSS color to prevent SR from occurring in one physical AP.

3) BSS Color Change Process

When the AP is sure that one or more neighboring BSS color is the same as its own, it can decide to change its BSS color in one of two ways. The first way is fill the color change information in the BSS Color Change Notification Announcement element in the Beacon frame, probe response frame, or (re-) association response frame; the second way is to send a dedicated HE BSS Color Change Announcement frame. Using either method, the notification of the BSS color change must be sent several times because the unreliability of wireless communications may cause packet loss, and there may be some STAs running in power save mode that miss the notification. Therefore, the Color Switch Countdown field is introduced during the process of changing the BSS color. Specifically, the counter subtracts 1 after every Beacon frame or each HE BSS Color Change Announcement frame. Each Beacon frame needs to broadcast the BSS color change notification until the counter reaches 0. This mechanism is consistent with our previous work proposed in [157] and [158]. We propose a multi-step channel reservation mechanism (i.e., the current transmitted data broadcasting the backoff values) for the following data. Thus, the transmission time of each packet is reserved many times, further guaranteeing reliability.

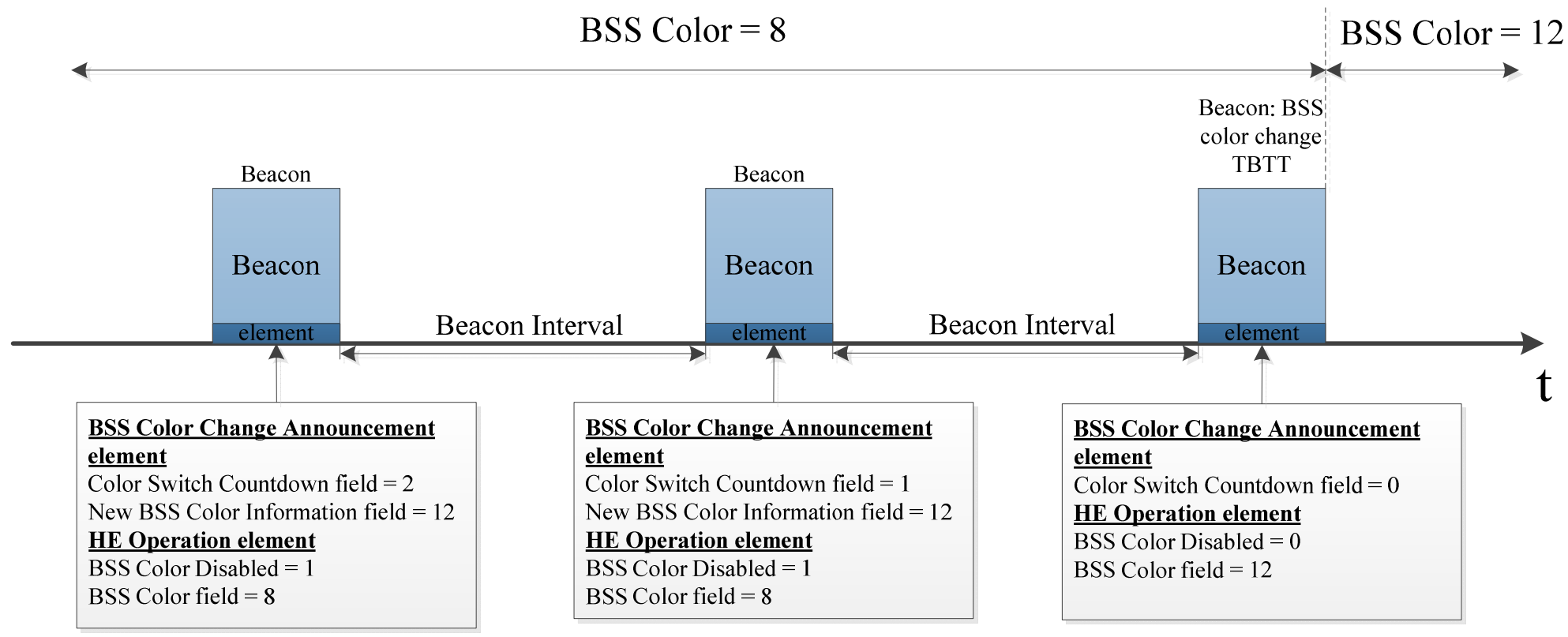

(a) Beacon-based BSS color changing process

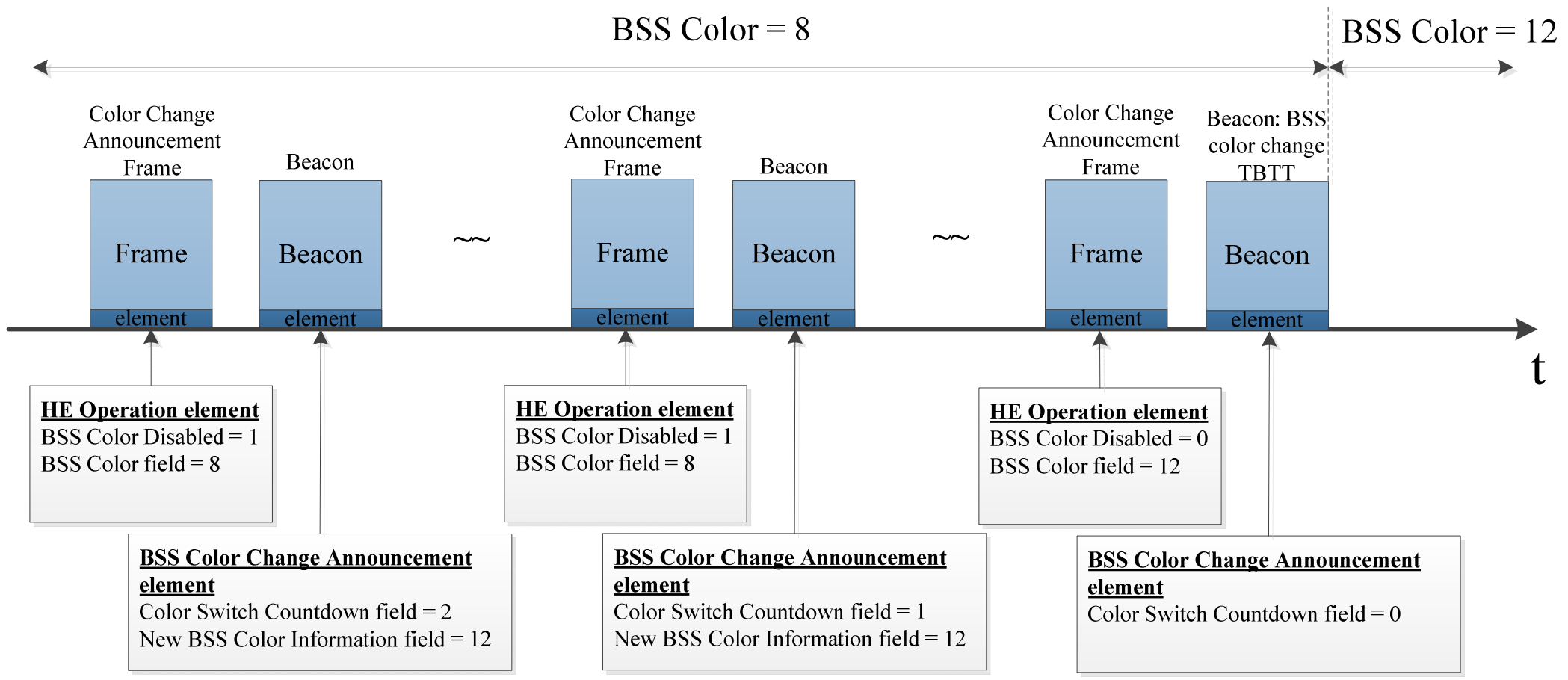

(b) Color Change Announcement frame based BSS color changing process

Fig. 27 Example for BSS color change process in IEEE 802.11ax.

i. Color change time duration validation phase

At this stage, the AP needs to determine after how many target Beacon transmission times (TBTT), denoted as $N^{Color_Change}$, the new BSS color starts to work. A more detailed method is determined by a specific algorithm, which is beyond the scope of IEEE 802.11ax. In this example, the AP decides to complete the color update process after 2 TBTTs (i.e., $N^{Color_Change} = 2$ ).

ii. Middle color change phase

The AP needs to successively send $N^{Color_Change} + 1$ Beacon frames to complete the whole process. In the previous $N^{Color_Change}$ beacon, the BSS Color Disabled field in the HE Operation element should to be set to 1, and the BSS Color field is still set as the existing BSS color, such as 8 in Fig. 27(a). The new BSS Color Information field in the BSS Color Change Announcement element needs to be set as the new color value. The Color Switch Countdown field is reduced by 1 after each Beacon frame. It should be emphasized that the new BSS color information field is not allowed to change throughout this entire process.

iii. Color change completion phase

In the last Beacon frame, the BSS Color Disabled field in the HE Operation element needs to be set to 0, and the BSS Color field must be updated to the new BSS color, such as 12 in the example. At this time, the Color Switch Countdown field in the BSS Color Change Announcement element is reduced to 0. When the STA receives this Beacon frame, it is determined that the BSS color has been updated to a new value.

Fig. 27(b) describes how the AP changes BSS color through the Color Change Announcement frame. The process is similar to the Beacon-based method; the only difference is that the AP needs to send the Color Change Announcement frame separately and carries the BSS Color Change Announcement element in the frame.

**D. Enhanced NAV mechanism**

1) Problems with the Traditional NAV Mechanism

Traditional virtual carrier sensing in IEEE 802.11 is realized using a NAV mechanism. However, because the nodes in the traditional IEEE 802.11 only record one NAV, a series of problems will occur in high-dense deployment scenarios.

First, the existing NAV mechanism is incompatible with MU-MAC. As shown in Fig. 28(a), STA1~STA4 are associated with AP1. AP1 and STA1 obtain the TXOP by exchanging RTS and CTS. After successfully receiving RTS and/or CTS, all the STA2, STA3, and STA4 set NAV. However, after AP1 and STA1 accomplish their transmissions in the TXOP, AP1 wants to schedule STA1~STA4 to perform UL MU transmission because the TXOP time has not elapsed. Therefore, AP1 sends TF to schedule STA1-STA4 to transmit UL data on RU1-RU4, respectively. Unfortunately, according to the existing NAV rules, because STA2, STA3, and STA4 have set NAV, they believe the channel is busy and then abandon this UL transmission opportunity. In this case, only STA1 is successfully transmitted on RU1, seriously wasting channel resources. Therefore, the existing NAV mechanism is incompatible with MU-MAC.

Secondly, the existing NAV mechanism causes the TXOP-ending chaos problem in high-dense deployment scenarios. As shown in Fig. 28(b), STA1~STA4 are associated with AP1, while STA5 is associated with AP2. Thus, any AP or STA may receive multiple frames with channel protection information from both intra-BSS and inter-BSS. According to the existing NAV mechanism, the node only updates the NAV whose ending time is later without identifying the source node. So, when a TXOP holder (i.e., the node that initiates the TXOP) sends the CF-End frame to end its TXOP, STAs will cancel their current NAV in any case. This will lead to TXOP-ending chaos in some circumstances. For example, at time t1, STA5 in the intra-BSS receives frames from BSS2 and sets NAV. At time t2, STA5 receives frames from inter-BSS (BSS1) and updates the NAV because the ending time is later. However, at time t3, the intra-BSS AP sends CF-End to end its TXOP, and then STA5 cancels the NAV. Next, because the physical carrier sensing indicates the channel is idle, STA5 successfully accesses the channel and sends data. After that, a serious collision occurs between the UL SU transmission of STA5 and the UL MU transmissions in BSS1, leading to failed receipt of all four RUs for AP1. This problem is becoming increasingly serious in high-dense deployment scenarios.

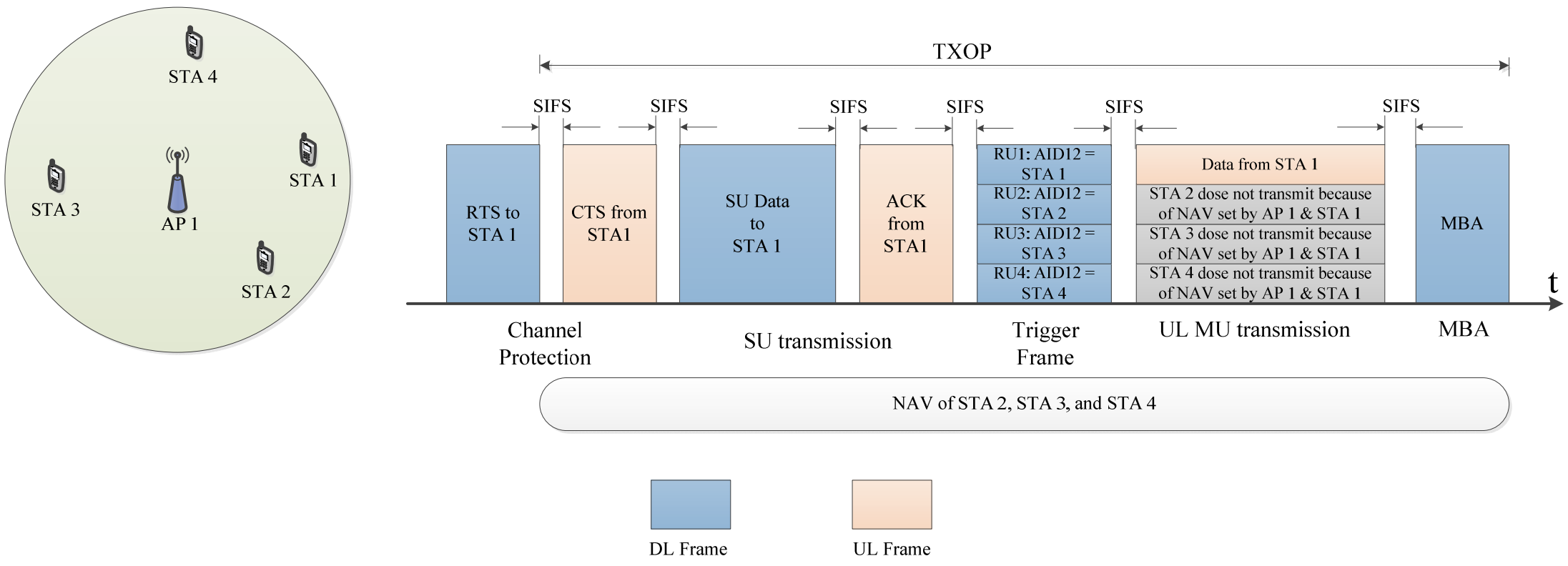

(a) Incompatible with MU-MAC

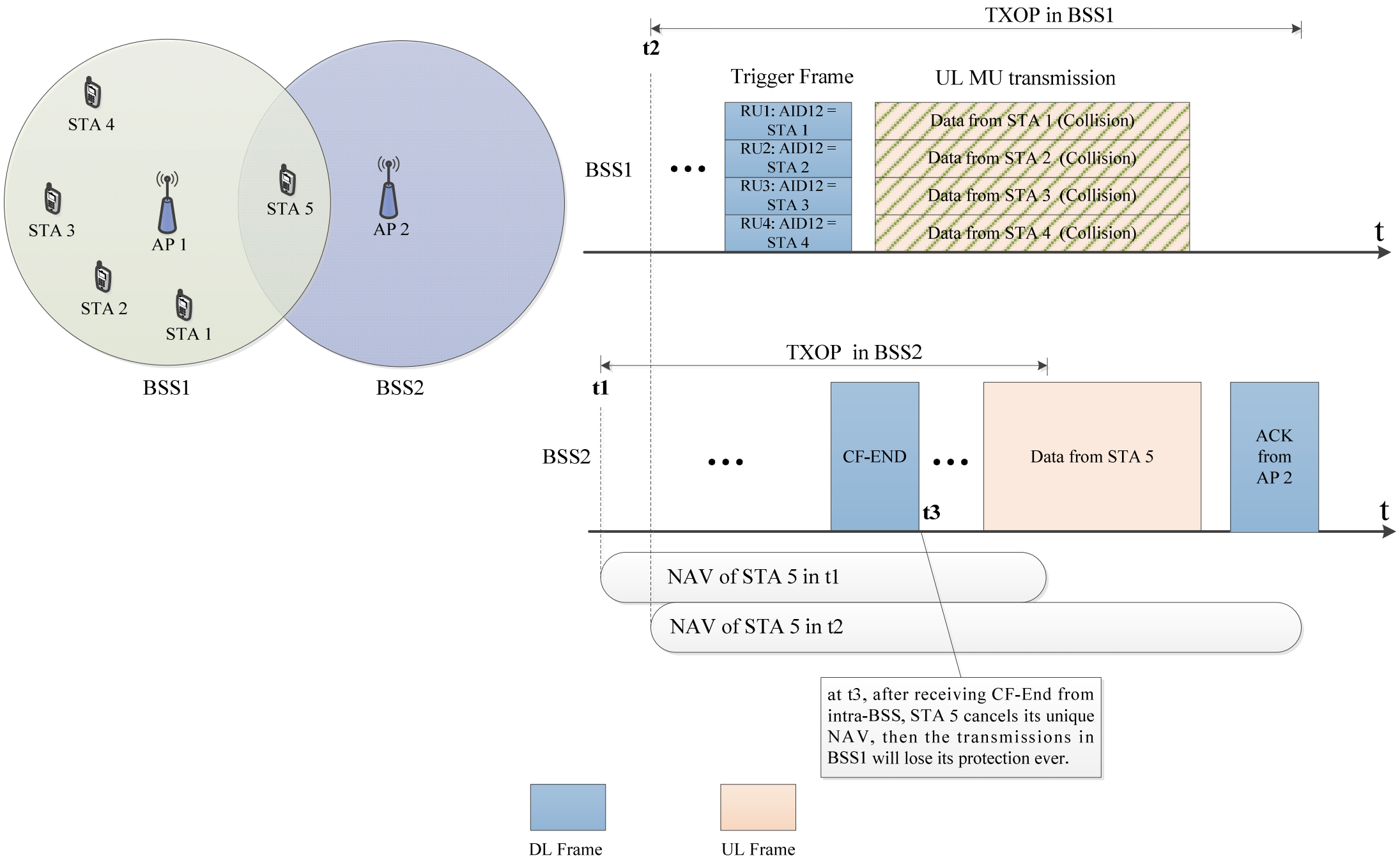

(b) TXOP-ending chaos problem

Fig. 28 Problem caused by the existing NAV mechanism.

2) Two NAVs based Virtual Carrier Sensing Mechanism in 802.11ax

To adapt to MU-MAC while simultaneously addressing the TXOP-ending chaos problem, IEEE 802.11ax introduces the two NAVs based virtual carrier sensing mechanism. In IEEE 802.11ax, the node will maintain two NAVs counters: intra-BSS NAV and basic NAV. When the node judges that the received frame is from intra-BSS, the intra-BSS NAV is updated; when the node determines that the received frame is from inter-BSS or cannot judge whether it is from the intra-BSS, the basic NAV is updated. Thus, the rule of virtual carrier sensing can be modified as follows: if and only if the two NAVs are both 0, the virtual carrier sensing result is idle; otherwise, it is busy.

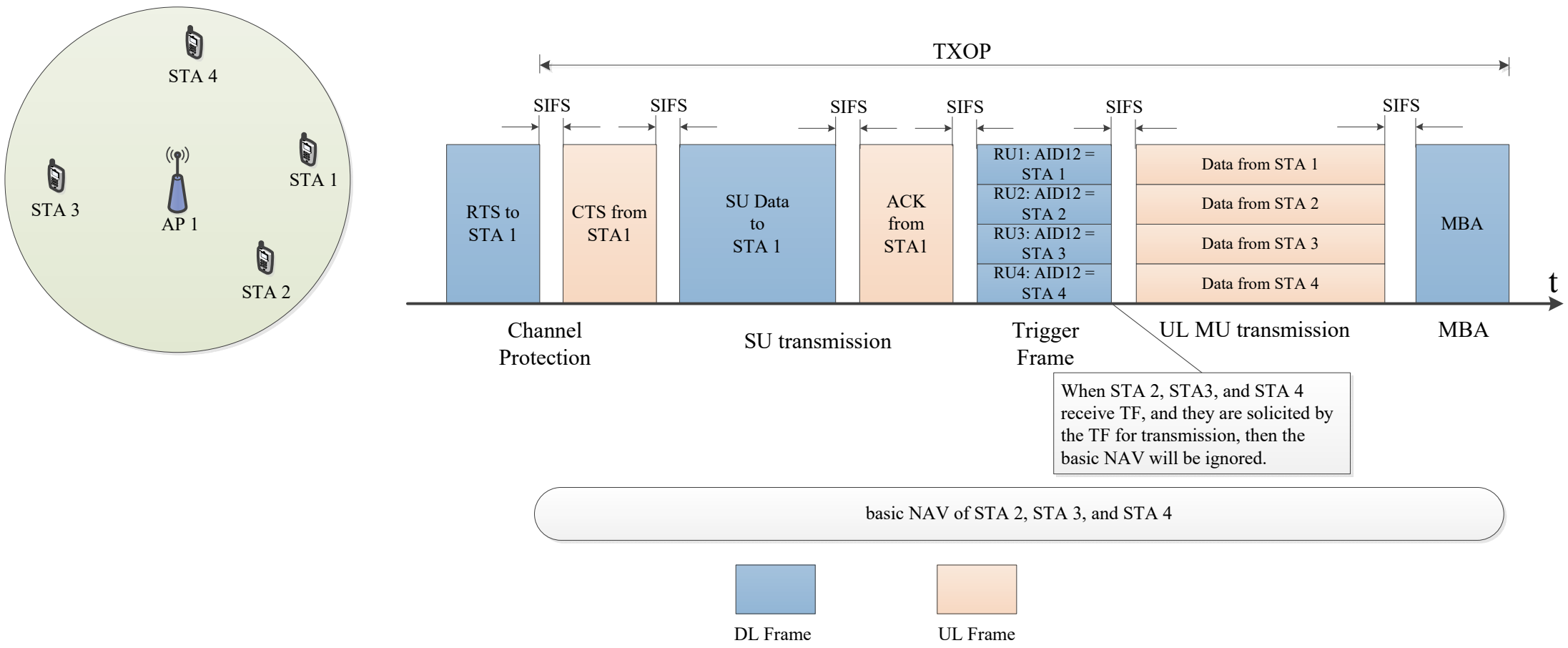

(a) Highly compatible with MU-MAC

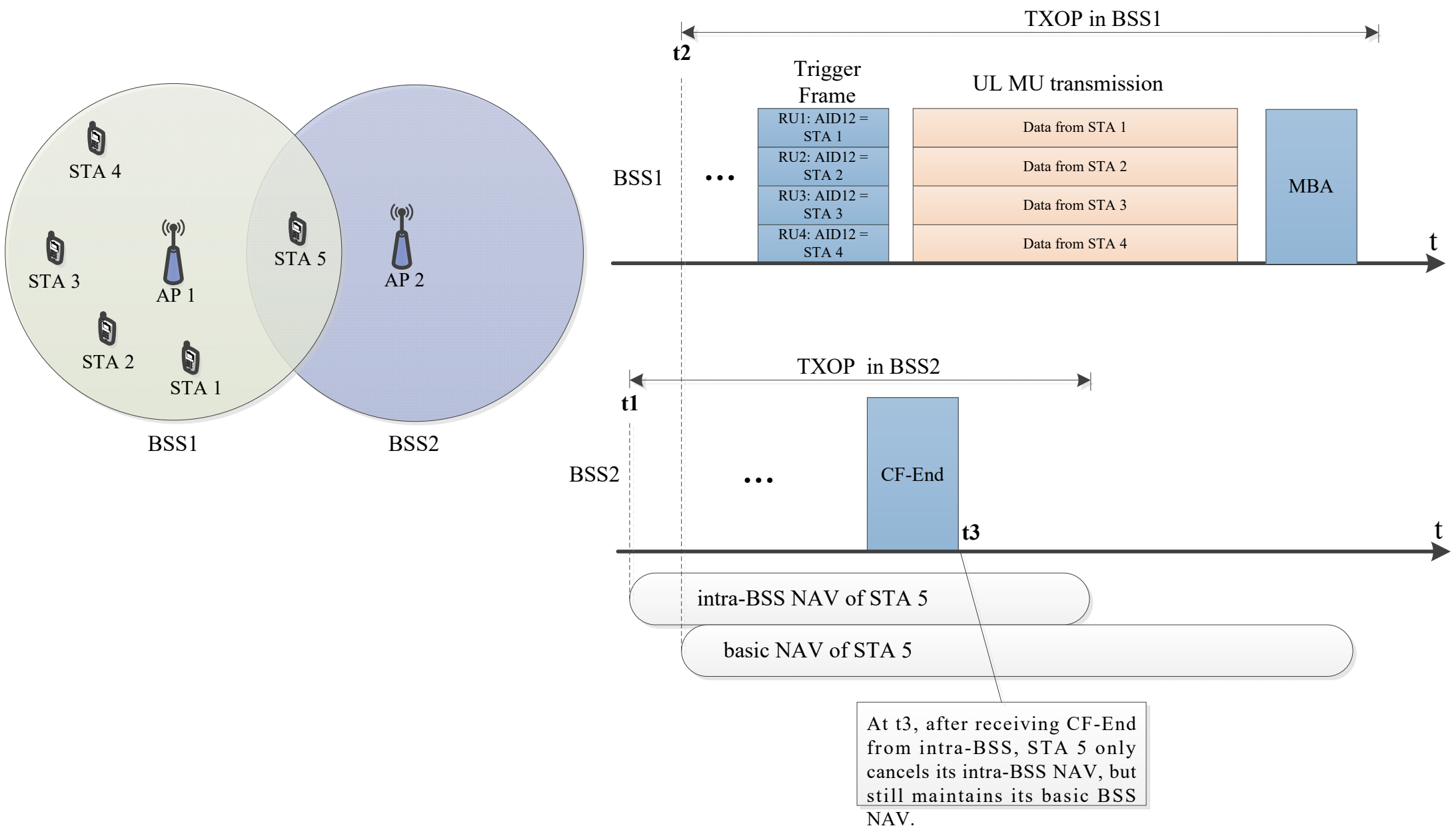

(b) Avoid the TXOP-ending chaos problem

Fig. 29 Advantages of the two NAVs based virtual carrier sensing mechanism.

IEEE 802.11ax specifies that when any STA receives the TF from intra-BSS and itself is scheduled in the TF for UL transmission, the intra-BSS NAV can be ignored. Then, as shown in Fig. 29(a), the two NAVs based virtual carrier sensing mechanism is highly compatible with MU-MAC.

In addition, AP and STAs can maintain the two NAVs counters differently according to the received frames sources. As shown in Fig. 29(b), STA5 receives the CF-End frame and then determines that this frame to be from intra-BSS. After that, STA5 cancels the intra-BSS NAV, maintains the basic NAV, and stays silent. Thus, STA5 will not collide with the concurrent UL MU transmission in BSS1, alleviating the TXOP-ending chaos problem. Frankly, because the basic NAV may still be set from many BSS, the TXOP-ending chaos problem cannot be fully addressed. However, we believe the two NAVs based virtual carrier sensing mechanism to be a big step forward.

## E. OBSS_PD based SR Mechanism

Traditional IEEE 802.11 adopts a more conservative physical carrier sensing mechanism (i.e., a lower CCA threshold). This severely suppresses concurrent transmissions and leads to poor spatial reuse. To significantly enhance the spectrum reuse in high-dense scenarios and improve the area throughput, IEEE 802.11ax introduces the OBSS_PD based SR mechanism as shown in Fig. 30.

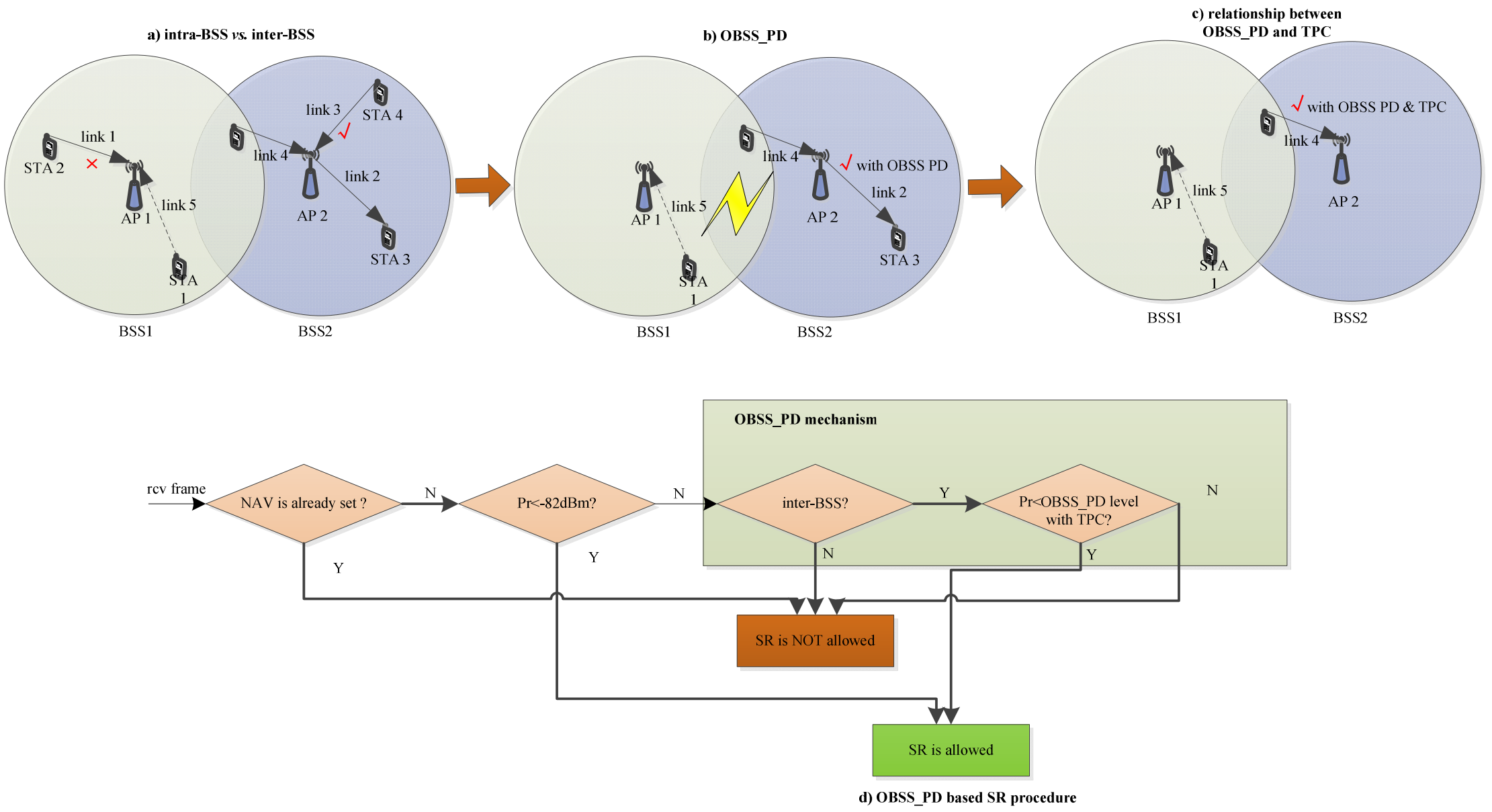

Fig. 30 Overview of OBSS-PD.

### 1) Distinguishing Intra-BSS and Inter-BSS

There are two BSSs shown in Fig. 30, where STA1 expects to send UL data to AP1 (i.e., link5). To simplify, in the following analysis, it is assumed the NAV is not yet set for STA1. It can be observed from the scenario that if an intra-BSS transmission such as link1 exists, regardless of whether the transmission direction is UL or DL or whether the channel state is busy or idle, STA1 is not allowed to perform SR because these two links (i.e., link1 and link5) share the same endpoint (AP). Thus, concurrent transmissions between intra-BSS cause collisions, hence the SR prerequisite to distinguish inter-BSS and intra-BSS.

In contrast, consider an inter-BSS transmission: if the sensed energy of STA1 is lower than -82 dBm, such as link3, the channel is determined to be idle and the SR is allowed; if the sensed energy is greater than or equal to -82 dBm, the channel is considered as busy state according to the rules of the legacy IEEE 802.11, and the SR is not allowed. However, introducing OBSS_PD in IEEE 802.11ax will introduce additional SR opportunities.

2) OBSS_PD based SR Mechanism

The OBSS_PD based SR mechanism specifies that when a node receives a PPDU from inter-BSS, a higher CCA level (OBSS_PD level) can be used. Once a PPDU is successfully received, the physical carrier sensing rule in IEEE 802.11ax is amended as follows: if the PPDU is from intra-BSS, the traditional CCA threshold (i.e., -82dBm) is adopted to perform physical carrier sensing; if the PPDU comes from inter-BSS, it can be divided into two subcases. Subcase 1: if the received power is greater than or equal to the OBSS_PD level, the physical carrier sensing is determined to be busy, and the basic NAV needs to be set. Subcase 2: if the received power is below the OBSS_PD level, the physical carrier sensing is determined to be idle, and the basic NAV is not set. This encourages concurrent transmissions among inter-BSS, such as link2 in Fig. 30(b). OBSS_PD based SR mechanism is one important SR technologies in IEEE 802.11ax.

3) OBSS_PD based SR Mechanism and TPC

The OBSS_PD mechanism needs to cooperate with TPC. Different levels of transmission power lead to different amounts of interference. For example, larger power results in relatively serious interference, while lower power leads to comparatively low interference. Therefore, IEEE 802.11ax specifies that when one node utilizes more transmission power, it should use a lower OBSS_PD level; when one node selects less transmission power, it can use a higher OBSS_PD level, such as link4 in Fig. 30(c).

IEEE 802.11ax specifies that the range of OBSS_PD is from OBSS_PDmin and OBSS_PDmax, where the default value of OBSS_PDmax is -62 dBm and that of OBSS_PDmin is -82 dBm. The range of OBSS_PD can be constrained by:

$$OBSS_PD\ Level \le \max(OBSS_PD\ Level_{\min}, \min(OBSS_PD\ Level_{\max}, OBSS_PD\ Level_{\min} + (TXPWR_{ref} - TXPWR))),$$

where TXPWR represents the current transmission power of the transmitter node, and TXPWRref represents a reference power value and is generally set to 21 dBm. The following conclusions can be drawn from the formula:

- The OBSS_PD level never exceeds the OBSS_PD Levelmax;
- The OBSS_PD level is never lower than OBSS_PD Levelmin;
- The OBSS_PD level is related to the transmit power.

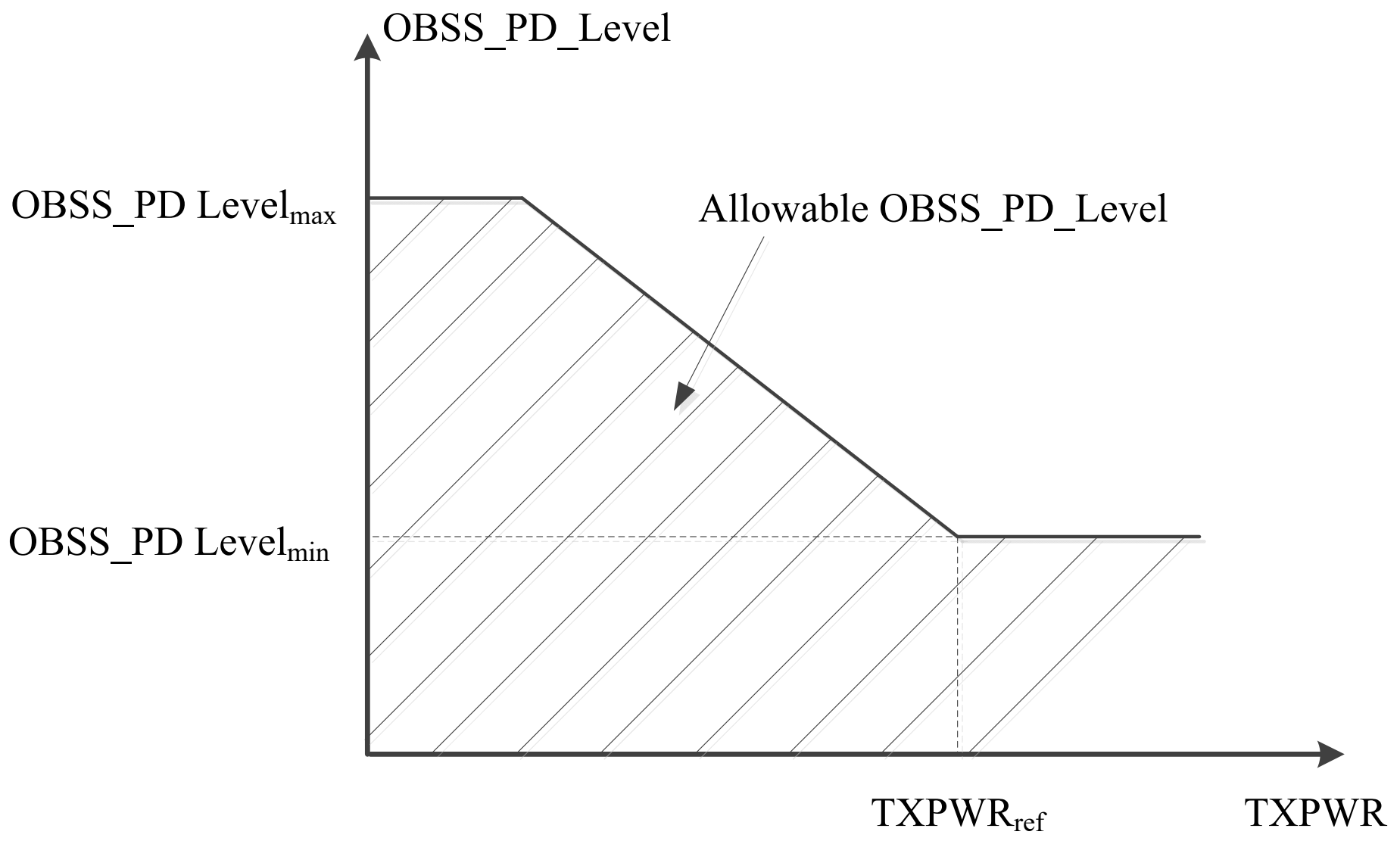

Fig. 31 Relationship between the OBSS_PD level and transmission power.

Fig. 31 shows the relationship between the OBSS_PD level and the transmission power.

Together with Fig. 30(d), we briefly introduce a possible SR process based on the OBSS_PD based SR mechanism. Once an STA receives a frame and identifies that the destination address is not its own,

Step 1: if the NAV has been set, SR cannot be executed and NAV needs to be updated (if necessary); otherwise, proceed to Step 2.

Step 2: using the traditional CCA threshold to sense the channel, if the sensed result is less than the traditional CCA threshold, the channel is determined to be idle. Proceed to Step 5; otherwise, proceed to Step 3.

Step 3: if the frame is from intra-BSS or unknown BSS, SR cannot be executed and NAV needs to be updated (if necessary); otherwise, proceed to Step 4.

Step 4: using the OBSS_PD level to sense the channel, if the detected signal power is less than the OBSS_PD level, the channel is determined to be idle and the transmission power needs to be controlled. Proceed to Step 5; otherwise, SR cannot be executed and NAV needs to be updated (if necessary).

Step 5: the STA contends for SR transmission.

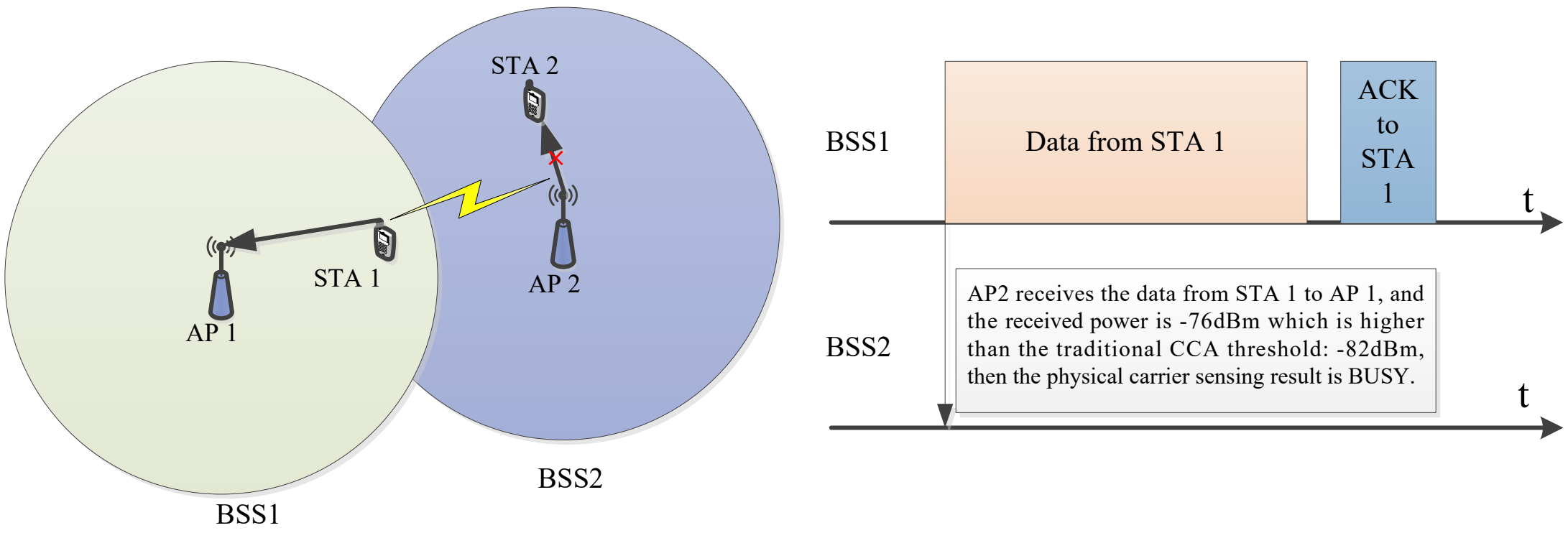

(a) Traditional CCA suppresses the SR

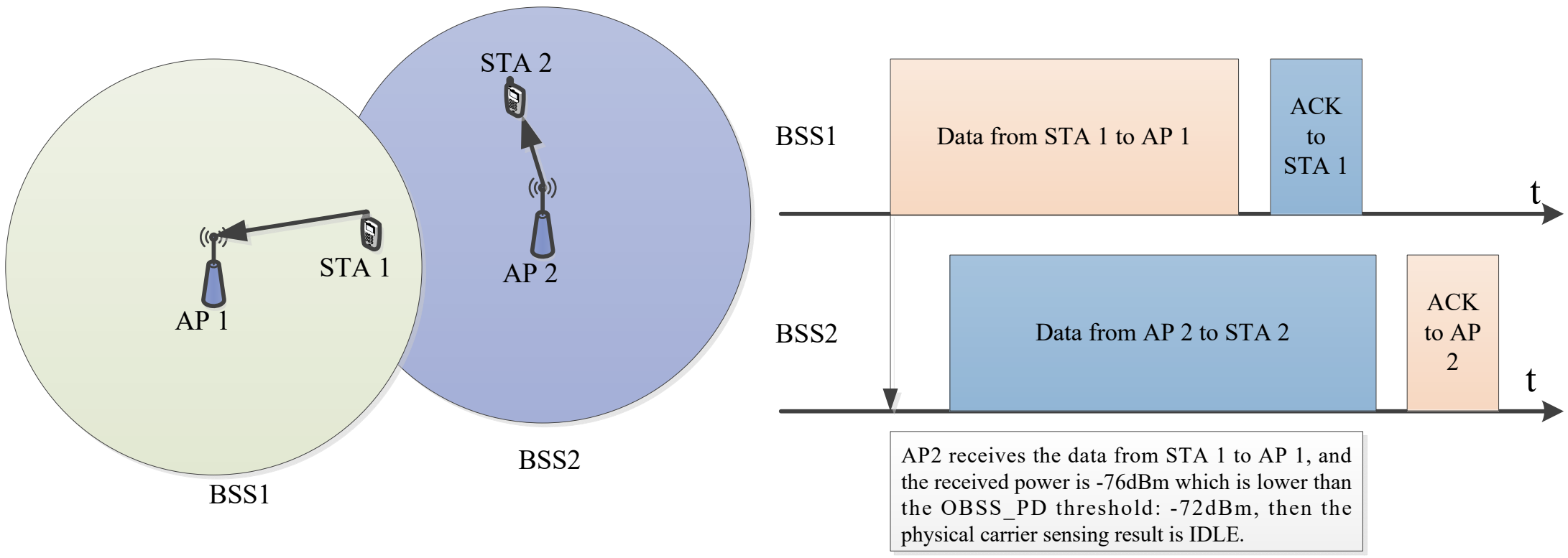

(b) OBSS_PD improves the SR

Fig. 32 Comparison between OBSS_PD and traditional CCA.

As shown in Fig. 32(a), for the traditional IEEE 802.11, the fixed CCA threshold (-82 dBm) is too conservative, so the adjacent BSS mutually suppress each other and further restrict SR capability. In Fig. 32(b), introducing the OBSS_PD based SR mechanism into IEEE 802.11ax enhances SR capability; thus, the spectrum reuse ability in high-dense deployment scenarios is significantly improved.

### F. SRP based SR Mechanism

Ongoing communication link is called the current link, while the link that wants to concurrently transmit is called the potential concurrent link. The OBSS_PD based SR mechanism mentioned above focuses on the case in which the sender of the potential concurrent link spontaneously performs SR through the OBSS_PD operation and rules. OBSS_PD based SR is not controllable and unknowable by either the sender or the receiver of the current link. In contrast, IEEE 802.11ax also proposes another important SR mechanism: SRP based SR mechanism. The biggest difference between OBSS_PD and SRP is that the current link decides whether to allow SR. Tab. 7 compares the OBSS_PD based SR mechanism and the SRP based SR mechanism.

Tab. 7 The OBSS_PD mechanism compared with the SRP mechanism

| | **OBSS_PD mechanism** | **SRP mechanism** |
|---|---|---|
| Frame source | Only works after receiving an inter-BSS frame | Only works after receiving an inter-BSS frame |
| Whether the current link joins or controls the SR | No | Yes |
| Relationship with TPC | Jointly adjusting | Jointly adjusting |
| Relationship with physical carrier sensing | Uses the OBSS_PD level to perform physical carrier sensing | Uses intra-high CCA level (e.g., positive infinite) to perform physical carrier sensing |
| Relationship with virtual carrier sensing | If the SR can initialize, ignore NAV | If the SR can initialize, ignore NAV |
| Time limits for SR | No restriction | Intra-PPDU |

Specifically, as shown in Fig. 33, the SRP based SR mechanism can be divided into three stages.

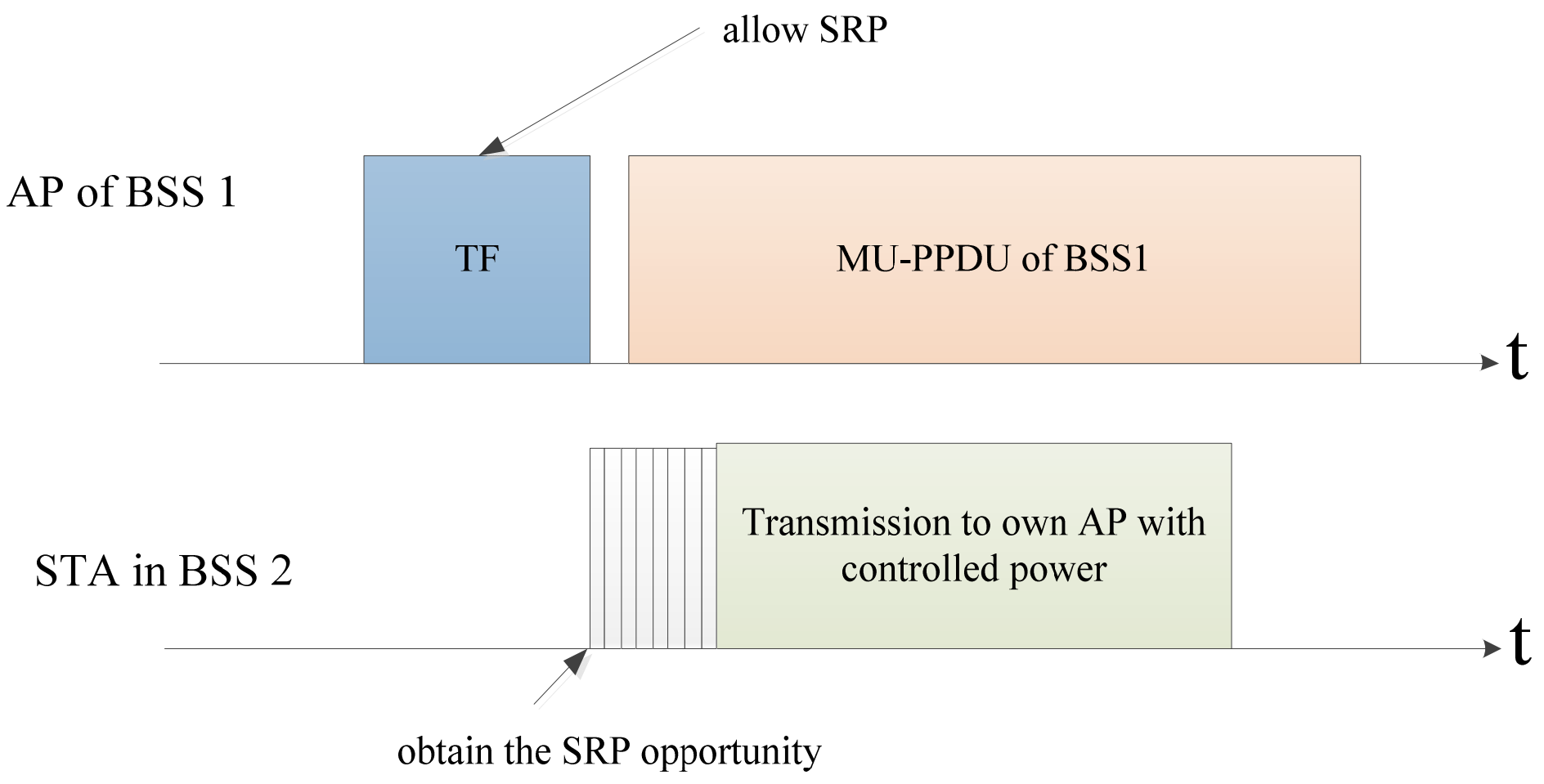

Fig. 33 SRP based SR mechanism.

**SRP allowing phase.** The AP indicates whether SRP is supported by configuring the Beacon frame. In the following discussion, we assume the SRP is supported. When an AP is ready to send a TF to schedule the UL MU transmission, and it allows the AP or STAs in the other BSS to perform SR during its own UL MU transmission process, the AP needs to allow the SRP by setting the HE-SIG-A and TF and further sets the SRP value according to the tolerated interference level.

**SRP opportunity achieving phase.** When 1) an STA or AP receives TF from inter-BSS, 2) the HE-SIG-A and TF indicate that SRP is allowed, and 3) the transmitting power to be used is not higher than the SRP value-RPL, then an SRP opportunity is achieved, and the NAV can be ignored; otherwise, the SRP is not allowed during the current transmission, and the NAV needs to be set. The SRP value is the number carried in the Spatial Reuse subfield in the HE-SIG-A. A larger SRP value means that the current link can tolerate greater interference. RPL indicates the receiving power of TF; a higher RPL indicates the initialization of the potential link should be more cautious.

**Access phase for the potential concurrent links.** Considering that there may be a number of APs or STAs obtaining the SRP opportunity, in order to avoid collisions, the APs or STAs that obtain the SRP opportunities need to continue to implement the backoff process. Only when the backoff procedure is finished before the end of the current transmission can the SP be performed. During the backoff process for the SRP opportunity, if the nodes receive a new TF and it does not allow SRP, then the backoff procedure needs to be suspended immediately. Otherwise, the backoff procedure still needs to be suspended until the nodes confirm a new SRP opportunity. In addition, if the nodes finish the backoff procedure and obtain the transmission opportunity, the end time of the concurrent link must not exceed that of the current PPDU.

In the SRP based SR mechanism, because the current sender or receiver of the link can control and predict the SR, the authors believe the SRP based SR mechanism is more suitable for enterprise networks, campus networks, and other WLAN centralized deployment scenarios.

# VI. Other Technologies

## A. TWT and Service Time Reservation

### 1) Overview of TWT and Reservation

Compared to a centralized network, one disadvantage of the distributed network is that its channel access and data transmission are spontaneous and random, which may affect the network's overall performance (e.g., collision can lead to reduced throughput). From the service point of view, in WLANs, the AP is the focal point of all UL services and the starting point of all DL services. At the same time, the AP can obtain or configure the network states through management frames and control frames. In other words, for a particular BSS, the AP has a global perspective. Therefore, the authors believe that an important breakthrough in IEEE 802.11ax is to enhance WLANs' scheduling capability through the AP's global vision and control and management capabilities. This scheduling ability is reflected in the resource allocation in the MU access procedure (see Sec. IV) as well as in SP reservation wherein certain users, services, and traffic reserve the right to use the channel at a specific time (or for a certain period). Specifically, IEEE 802.11ax implements SP reservations by adopting TWT mechanisms.

The TWT mechanism is first proposed in IEEE 802.11ah, and the TWT requester and responder need to exchange the TWT request frame and reply frame to reserve the next wake-up time and requester service time. To adapt to the relevant technology in IEEE 802.11ax, especially with MU-MAC, IEEE 802.11ax has greatly enhanced the TWT mechanism. Specifically, IEEE 802.11ax uses individual TWT mechanisms, particularly the broadcast TWT mechanism to enable STAs to sleep for a period of time, and then exchanges data with the AP at the reserved time. The AP can define a series of TWTs and the reserved service time, called TWT SP. Each TWT represents the time for STAs to wake up, and each TWT SP represents a reserved service time for STAs. Through the TWT mechanism, reservation services and planning are introduced in the time dimension to reduce access collision in WLANs. Furthermore, the STAs are made dormant when they do not transmit data frames, which reduces energy cost.

Before further discussing TWT in IEEE 802.11ax, especially in terms of reserving service times, we will provide an overview of relevant academic research related to channel reservation. In Fig. 34, we classify the existing work on channel reservation. Distributed channel reservation from the resource dimension can be divided into three categories: *time reservation* is the duration of the reserved transmission time duration, and other nodes in that time are not allowed to transmit data frames; *space reservation* indicates that a certain geographical area is reserved, and other nodes in this area are not allowed to transmit data frames simultaneously; and *frequency reservation* means a certain channel or sub-channel is reserved, and other nodes are not allowed to transmit data frames on the reserved channel or sub-channel.

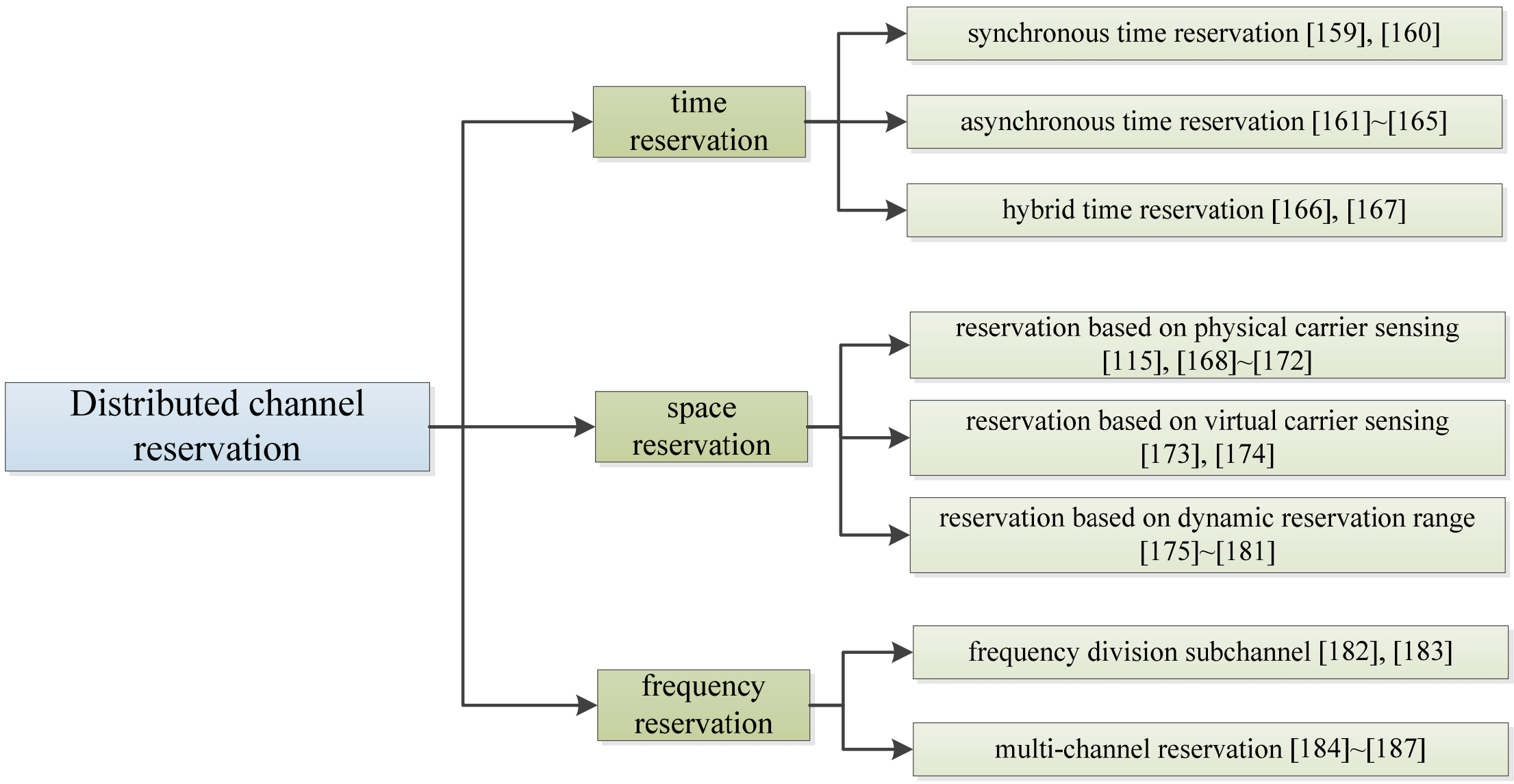

Fig. 34 Overview of existing research on channel reservation.

i. Related work on time reservation

- Synchronous time reservation

In synchronous time reservation, the nodes need to obtain the channel resources of certain time slots by using reservation. A typical protocol is soft reservation multiple access based on priority assignment (SRMA/PA) [159] and distributed packet reservation multiple access (D-PRMA) [160]. In SRMA/PA, the access procedure of a time slot includes six sub-slots: synchronization (SYNC), soft reservation (SR), reservation request (RR), reservation confirmation (RC), data sending (DS), and acknowledgement (ACK). Specifically, SYNC is used for slot synchronization; SR is used to send reservation information; a node with a data frame for transmission can request a channel reservation by sending an RTS frame in the RR sub-slot; the corresponding node replays a CTS frame in the RC sub-slot; the DS sub-slot is used for data transmission; and the ACK sub-slot is used for ACK confirmation.

Once the node succeeds in accessing the channel through RTS/CTS, the slot at the same location in the subsequent frames is reserved until its data transmission is completed. D-PRMA divides a slot into multiple sub-slots. From the first sub-slot, the nodes compete for channel reservation with RTS/CTS. Once a node wins the competition, it can use all the remaining sub-slots. For high-priority nodes, the same slot in subsequent frames can be reserved, while the low-priority nodes can only use the current slot if it wins the competition.

- Asynchronous time reservation

The asynchronous time reservation protocol can be further subdivided into slot reservation and time length reservation. EBA [161] is a typical slot reservation mechanism: the node obtains the channel resources of the corresponding slot by broadcasting the backoff counter used in the next channel contention. Specifically, EBA puts the backoff counter used in the next channel contention into the data frame and broadcasts it to neighbor nodes. Once the neighbor nodes receive it, they can avoid selecting the same backoff counter to ensure reliable channel access of the reservation node. Further, to avoid reservation information failing to be effectively transmitted due to unreliable link transmission, in [157] and [158], we propose the idea of a multi-step reservation termed m-DIBCR in which backoff counters for the contention of the next multiple frames are carried in the data frame. Thus, the transmit time of each frame is reserved for various times to ensure increased reservation reliability. This idea has been applied to the TWT mechanism in IEEE 802.11ax.

For the time length reservation, the node obtains channel control in the corresponding period by broadcasting the start and end times of the next transmission in DATA and ACK. Lin and Gerla [162] propose that the channel reservation information be carried in the current data frame initially. However, due to the frequent exchange of channel reservation tables between nodes, the throughput of the network is reduced. Manoj and Murthy [163] and Ying et al. [164] improve MACA/PR, respectively, so the transmitter and receiver can make the correct access judgment by analyzing their maintenance channel resource reservation table, which avoids frequent exchange of the reservation table between the nodes. Singh et al. [165] takes full advantage of the periodic characteristics of real-time services and proposes an implicit reservation mechanism based on Sticky CSMA/CA. The node ``insists'' on accessing a channel with a fixed cycle, and this regular access behavior can ``implicitly'' notify the surrounding nodes to reserve channel resources.

- Hybrid time reservation

R-CSMA/CA [166] and DBAS [167] are typical hybrid time reservation protocols. The hybrid time reservation protocol consists of a contention-free period (CFP) and controlled access phase (CAP). The CFP stage is divided into equal-length slots for transmission and reservation; nodes in the CAP stage compete for channel access. The nodes that are allowed to reserve channel resources can reserve the slot in CFP after successful competition, while the nodes that are not allowed to reserve channel resources can only compete for channel access for data transmission in the CAP stage.

ii. Related work on space reservation

- Space reservation based on physical carrier sensing

In the cognitive radio, Cho et al. [168] adjust secondary users' node density by adjusting the physical carrier sensing threshold, thus reducing their interference with primary users. Kaynia et al. [169] study the optimal carrier sensing threshold for minimizing outage probability. Yang and Vaidya [170] study the physical carrier sensing threshold optimization considering the MAC layer overhead, and it is concluded that the optimal physical carrier sensing threshold is related to the contention window size, frame length, and MAC layer overhead. Zhang et al. [171] propose that the surrounding areas of the current transmission are divided into three parts: hidden nodes range, exposed node range, and mixed overlapping range. The study implements local optimization of the physical carrier sensing threshold through simplified RTS/CTS interaction.

Furthermore, Zhang et al. [172] introduce a throughput penalty function to characterize the impact of each range on throughput, and design an iterative algorithm based on statistical channel state to dynamically adjust the physical carrier sensing threshold. Yang et al. [115] study the impact of transmit power and the physical carrier sensing threshold on the network capacity and points out that when the physical carrier sensing range just covers the interference range of the receiver, the network throughput can reach the optimal approximation. In addition, academics and industry have also developed a library of research on dynamic physical carrier sensing thresholds, and IEEE 802.11ax proposes the concepts of OBSS-PD and CCA thresholds for distinguishing intra-BSS and inter-BSS (see Sec. V).

- Space reservation based on virtual carrier sensing

In addition to the physical carrier sensing mechanism, IEEE 802.11 also uses a virtual carrier sensing mechanism, namely through the interaction of RTS/CTS control frames, and the subsequent transmission time is broadcast to neighbor nodes using NAV to avoid interference. Unlike physical carrier sensing, virtual carrier sensing reservation is for a period of time. Therefore, virtual carrier sensing can be regarded as the space reservation in the NAV period as well as the time reservation in the corresponding space within the transmission range. The traditional IEEE 802.11 standard defines the calculation and carry mode of NAV time and stipulates that the neighbor nodes that successfully listen to and parse the NAV information assume the virtual carrier is busy at the time as indicated by the NAV. Therefore, the effective range of virtual carrier interception is the reservation of the transmission range of the node to reduce interference.

There are mainly two types of research related to the virtual carrier sensing mechanism. In the first, the effectiveness of the mechanism is related to the distance between the transmitter and receiver. When the transmitter and receiver are close to each other, virtual carrier detection can be completely obscured by physical carrier sensing; when the transmitter and receiver are far apart, the limited transmission range means the virtual carrier sensing has no reservation function [173][174]. Secondly, IEEE 802.11ax proposed a double NAV differentiation mechanism according to the characteristics of high-dense deployment scenarios: NAV information in intra-BSS frames is fully accepted, while that in inter-BSS is selectively accepted or rejected according to the OBSS_PD mechanism (see Sec. V).

- Space reservation based on dynamic reservation range

Unlike physical carrier sensing and virtual carrier sensing, a space reservation MAC protocol based on a dynamic reservation range allows nodes to ensure reliable space reservation via reservation information forwarding. Compared with physical carrier sensing, it is more flexible in controlling the size of the reservation range. Compared with virtual carrier sensing, it can provide more reliable space reservation. Hasan and Andrews [175] propose a space reservation mechanism based on a guard zone, which takes the receiver as the center of the circle to delimit a reservation range and protect the receiver from interference of neighbor nodes. For a single link, a larger reservation range can ensure correct reception of the receiver. However, because the nodes cannot compete for channel resources within the scope of the reservation, a larger reservation range limits the number of simultaneous transmissions that can be accommodated in the network, resulting in a relatively low-space multiplexing that reduces the throughput of the whole network.

To help ensure that nodes within the range of interference and beyond the transmission range receive reservation information, we propose that the neighbor nodes forward the reservation information to ensure reliable channel access to the reservation link as described in [176] and [177]. Furthermore, when the number of neighbor nodes is small and the forwarding of reservation information is inefficient, we adopt the method of forwarding different reservation information at once to further improve reservation reliability as proposed in [178]. In [151], we jointly optimize the carrier sensing range and transmission rate control so the carrier sensing range contains the interference range, thereby improving the network's area throughput. Meanwhile, we combine channel reservations with relay cooperative technology in the [179]-[181] and prove that reservation and cooperation can be highly complementary. We also derive the optimal cooperative node location and optimal reservation radius and design an efficient MAC protocol. The results show that a cooperative reservation mechanism can greatly enhance network throughput and area throughput.

iii. Related work on frequency reservation

The MAC protocols based on frequency reservation generally divide the channel into a number of different frequency sub-channels or different time frequency resource blocks and reduce interference between wireless links when multiple nodes transmit in different sub-channels or time frequency resource blocks. Multi-channel CSMA (MCSMA) is a typical MAC protocol based on frequency reservation [182], and its principle is as follows: each node maintains a free sub-channel list, and nodes randomly select one sub-channel for the first transmission; if that transmission is successful, the last sub-channel for transmission is chosen for a subsequent transmission; and if the selected sub-channel is busy, then nodes randomly select other idle sub-channels for transmission. In a cognitive radio network, potential secondary users and new users bring interference to the current users in the network. Channel access for potential secondary users and new users is optimized by reserving a certain number of sub-channels as presented in [183]. We introduce multi-step channel reservation into a multi-channel MAC in [184]-[187]. Common control is used for channel reservation information transmission, including the data channel index and transmission time; the data channel is used for data transmission. The results show that the channel reservation mechanism can significantly improve the system throughput of a multi-channel MAC.

2) TWT and SP Reservation in 802.11ax

First of all, IEEE 802.11ax still supports the traditional individual TWT based on the request and response mechanism; that is, STAs can negotiate independently with the AP for TWT. Referring to Fig. 35, the STAs need to send a TWT request frame to the AP to submit their TWT SP request after successful channel competition. Then, the AP replies to the TWT respond frame to indicate it accepts or rejects the STAs' request. This mechanism is consistent with IEEE 802.11ah.

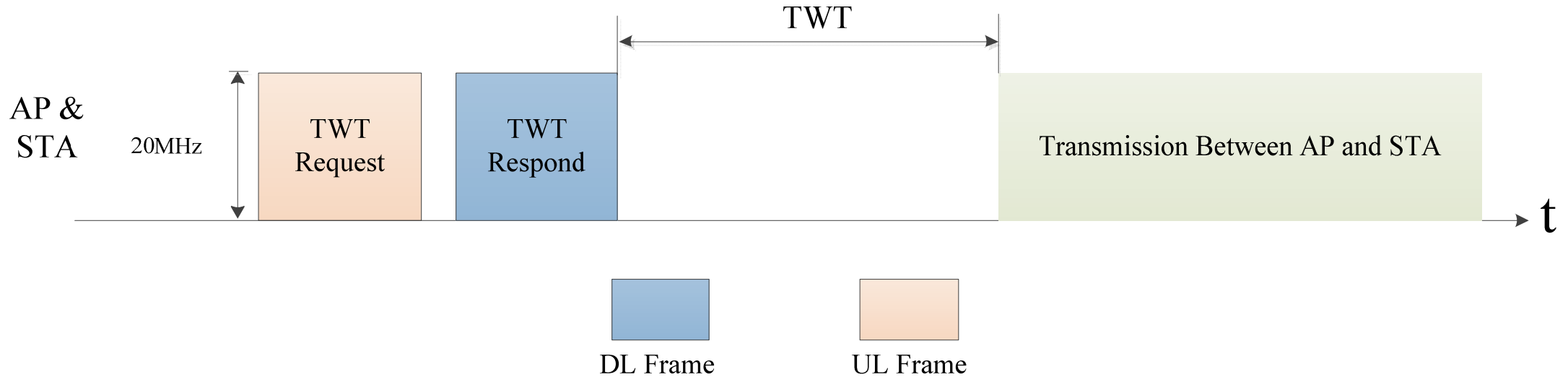

Fig. 35 Individual TWT example.

More importantly, to match MU-MAC, IEEE 802.11ax introduces broadcast TWT in addition to maintaining the traditional individual TWT mechanism. Broadcast TWT does not require STAs to negotiate with the AP alone; rather, the AP configures TWT in Beacon frames and makes reservations for different TWT SPs. Specifically, the AP adds a TWT element to the Beacon frame to broadcast TWT configuration results. In the TWT element, the starting time of each TWT SP, the shortest duration of TWT SP, and some other information (e.g., whether to support OFDMA random competition in the TWT SP) is explicitly defined. Next, STAs receive the Beacon frame and acquire TWT configuration information to select the starting time of the TWT SP associated with its waking up and accepting the service.

IEEE 802.11ax specifies that if a TWT SP contains at least one TF, then the Trigger field in the corresponding TWT Request Type field must be set to 1; otherwise, it is set to 0. In addition, the TWT Flow Identifier field specifies the constraints on the data frames that are transmitted in the TWT SP, as shown in Tab. 8. If a TWT SP has multiple TF cascade transmissions, the Cascade Indication Field in the other TF is set to 1, but the Cascade Indication field in the last TF is set to 0. Explanations of several important fields are shown in Tab. 8.

Tab. 8 function description of TWT-related fields

| **Field Name** | **Value** | **Brief description** |
|---|---|---|
| Target Wake Time | Continuous value | The wake-up time of the TWT |
| Nominal Minimum TWT Wake Duration | Continuous value | The shortest duration of the TWT SP corresponds to the shortest active time of the STAs |
| Trigger | 0 | There is no TF in the corresponding TWT SP |
| | 1 | There is at least one TF in the corresponding TWT SP |
| TWT Flow Identifier | 0 | There is no limit to the data frames in the corresponding TWT SP |
| | 1 | OFDMA random access is not supported in TF |
| | 2 | At least one TF in RU can be used for OFDMA random access |
| | 3 | AP needs to send TIM frames or FILS discovery frames containing TIM at the beginning of the TWT SP |
| | 4-7 | Reserved |

The AP has a more global vision in WLANs and a more comprehensive grasp on the demand of UL and DL users, services, and network congestion. Thus, adopting broadcast TWT in IEEE 802.11ax can more optimally divide time into several stages, namely the reservation of multiple SPs, and is helpful to planning service time from a global perspective. For example, for STAs with a feedback UL requirement through BSR, the AP could reserve the SP without OFDMA random contention (TWT Flow SP Random Competition Identifier = 1) to allow scheduling access; and for STAs requiring random access via OFDMA, the AP could reserve the SP with OFDMA random contention (TWT Flow Random Competition Identifier = 2) for them.

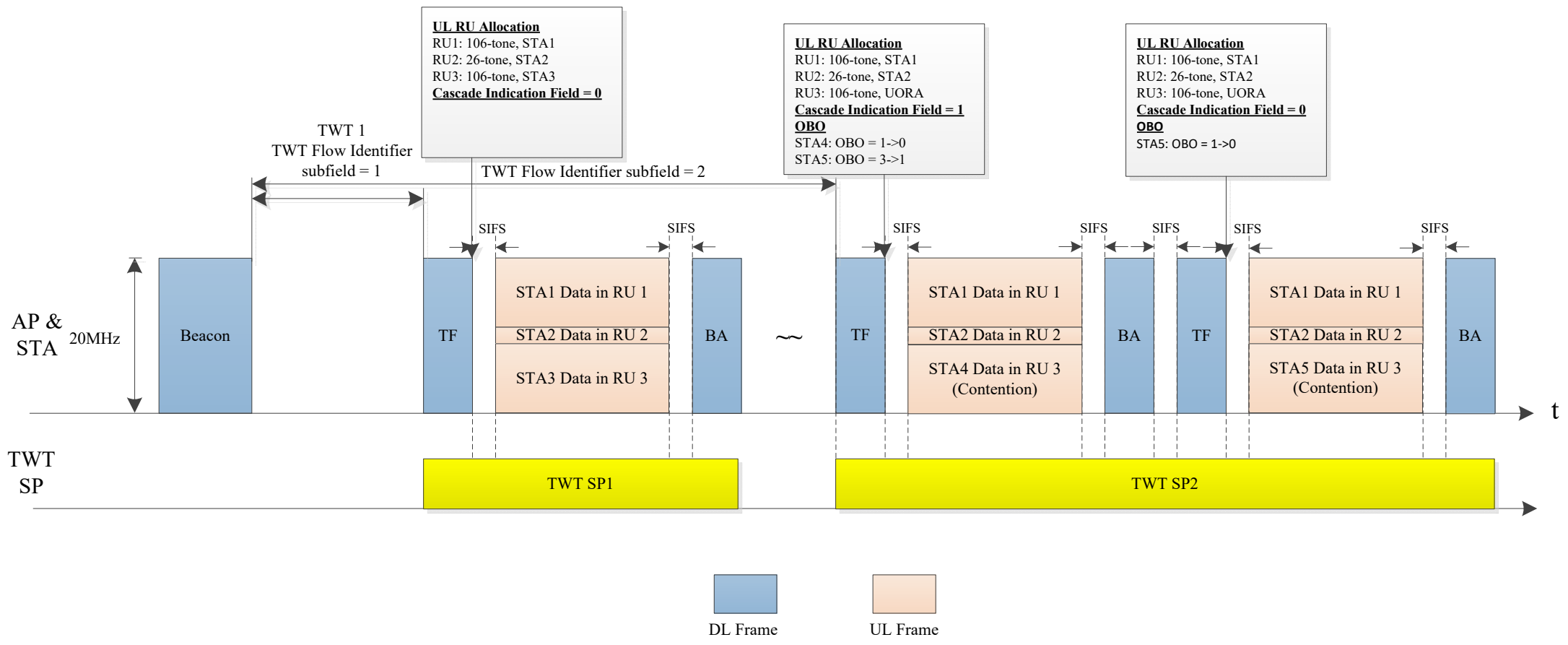

Fig. 36 Example of TWT mechanism in IEEE 801.11ax.

As shown in Fig. 36, the AP makes two TWT SP reservations through Beacon frames and explicitly indicates that TWT SP1 does not contain OFDMA random access, while TWT SP2 supports OFDMA random access through the TWT Flow Identifier subfield. STA1~STA3 have a feedback UL demand through BSR, so they must wait for the AP to schedule UL transmission while STA4 and STA5 hope to send data frames through OFDMA random access. In TWT SP1, the AP schedules STA1~STA3 for UL transmission. In TWT SP2, the AP sends the first TF to schedule STA1 and STA2 to perform UL transmission on RU1 and RU2, respectively, and to indicate that RU3 is used for random contention access. Thus, STA4 succeeds in accessing RU to transmit UL data through random contention.

However, because the OBO of STA5 is not reduced to 0 at this time, STA5 cannot access the channel. Nevertheless, the Cascade Indication Field in this TF is 1, indicating there is a cascaded TF in the future. The second TF sent by AP continues to schedule STA1 and STA2 on RU1 and RU2, respectively, while RU3 is still used for random contention access. At this point, the OBO of STA5 is reduced to 0 so the UL data can be transmitted successfully on RU3. At the same time, the current TF Cascade Indication Field is 0, indicating there is no cascaded TF in this TWT SP. The TWT mechanism is also naturally conducive to improving power efficiency. After broadcasting TWT in particular, it can be more efficient to save power, which will be explained further in Sec 6.B

## B. Power Efficiency Enhancements

### 1) Overview of Power Efficiency Enhancements in IEEE 802.11ax

Improving power efficiency is a common technical objective in the evolution of IEEE 802.11. Power efficiency issues are particularly prominent in next-generation wireless networks. The contradiction between the increasingly powerful function of smart terminals and the lagging technology of the power supply makes it necessary for the wireless network to account for power-saving problems from the standard setting; but more importantly, after IEEE 802.11ax introduced MU-MAC, it became necessary to introduce a new power-saving technology to achieve good adaptation. Thus, IEEE 802.11ax introduces two mechanisms to improve power efficiency.

First, there is a power-saving mechanism based on TWT. The AP can schedule SP through TWT in Beacon frames to allow STAs to accept services at different TWT SPs and sleep at other stages; at the same time, STAs can apply independently for TWT by negotiating with the AP. Specifically, the power-saving mechanism based on TWT includes three specific parts: the TWT-oriented power-saving mechanism for UORA, the periodic TWT opportunistic power-saving mechanism, and the individual TWT power-saving mechanism. These will be discussed in detail in Sec. VI.B.

Second, IEEE 802.11ax allows STAs to sleep actively during intra-PPDU transmission. Based on the sleep mechanism, when an STA hears that an intra-PPDU is transmitting and confirms that the transmission is independent of itself, it can actively enter a sleep state but needs to wake up at the end of the PPDU transmission. This will be discussed further in Sec. VI.B.

All power-saving mechanisms introduced later in this article provide the time duration for STAs to sleep, but this is not mandatory; that is, STAs do not necessarily have to sleep.

In recent years, academia has paid close attention to research on power-saving mechanisms in WLANs.

**Modeling and performance analysis of power-saving mechanisms.** In the power-saving mechanism of legacy IEEE 802.11, the AP periodically carries TIM information in the Beacon frame, and STA periodically wakes up according to the listen interval. The listen interval is usually one or more Beacon intervals. Each STA listens to the Beacon after waking up. If TIM indicates there is data for itself, it needs to wake up until the data transmission is completed; otherwise, it can continue to sleep. STAs wake up after listening to a Beacon. If TIM indicates it has its own data, it needs to wake up until the data transfer is completed, or it can continue to sleep. Lei and Nilsson [188] analyze the energy-saving mechanism of Legacy IEEE 802.11 by establishing the M/G/1 queuing model and D/G/1 queuing model. The M/G/1 model is more intuitive, while the D/G/1 model is less complex. Through them, the system's energy efficiency and throughput performance under the corresponding configuration can be predicted. Sangkyu and Bong [189] use an embedded Markov chain to model the power-saving mechanism of Legacy IEEE 802.11, and this model not only matches in accuracy but also deduces the packet delay variance [188]. Agrawal et al. [190] use the Markov process to model the energy consumption of the system for a TCP-based file download scenario. The simulation results show the model could accurately describe the system energy consumption in continuously active mode (CAM) and power save mode (PSM). Zhu et al. [191] use a random process to model the power-saving mode; derive the probability of STAs being in an active and idle state; and assess the STAs' energy consumption, cache data, average packet waiting time, and other indicators to put forward an algorithm to realize energy consumption optimization. Perez-Costa et al. [192] summarize and analyze the QoS mechanism and power-saving mechanism proposed in IEEE 802.11e and verify the system performance of different mechanisms and their combinations using simulation. Tauber et al. [193] measure the PSM and derive many valuable measurements.

**Related work on power-saving mechanism design.** Agrawal et al. [194] propose an opportunity-based power-saving mechanism (OPSM) for web browsing applications. The STAs that adopt OPSM need to listen or determine whether AP is currently providing web page download services to other STAs. If the AP is providing service to other STAs, the STA needs to enter a sleep state and try the operation again after waiting for a period of time; otherwise, if the AP is not providing service to other STAs, the STA can initiate the web browsing file download service. This mechanism prevents the AP from providing multiple STAs with web browsing file download services at the same time, which would require multiple STAs to be continuously awake, hence impairing power efficiency. Tabrizi et al. [195] present an intelligent power-saving mode (IPSM) that no longer uses the fixed sleep cycle mode of STAs in PSM. In IPSM, when STAs wake up and communicate with the AP, the AP optimally calculates how many Beacon intervals there are before the STAs' next wakeup. This idea is similar to the reservation wakeup time through individual TWT in IEEE 802.11ax. Anastasi et al. [196] propose a cross-layer energy management mechanism based on traffic differences in the application layer (e.g., web browsing, e-mail, file transfer, etc.) that makes STAs switch between an active state, sleep state, and off mode to enhance power-saving efficiency. Si et al. [197] are the first to model and analyze the DCF protocol based on PSM using a Markov model and find that after each AP Beacon transmission, there are more STA contention channels, resulting in excessive collision and throughput degradation. To solve this problem, Si et al. propose a strategy to control the number of users allowed to access the channel in Beacon TIM, thereby improving throughput. Similarly, in the periodic TWT opportunistic power-saving mechanism of IEEE 802.11ax, the AP also needs to send TIM frames at the beginning of each cycle. Chen et al. [198] propose an enhanced PSM mechanism (M-PSM) for mobile WiFi nodes. Based on PSM, M-PSM further increases the opportunity to save power according to mobility and business status. For example, STAs are allowed to sleep under low traffic load and delay insensitivity in the case of imperfect channel states and at a low transmission rate. Liu et al. [199] find that the current power-saving mechanism could cause significant collisions in machine-to-machine communication scenarios. Thus, an offset listen interval (OLi) algorithm is designed to stagger the listening wakeup time of different M2M communications as much as possible to reduce collisions. Tsao et al. [200] provide a comprehensive overview of the IEEE 802.11 power-saving MAC protocol.

2) TWT-Based Power Efficiency Mechanism

TWT based power efficiency mechanisms includes UORA-oriented TWT based, Periodic TWT based, and individual TWT based mechanism.

i. UORA-oriented TWT power-saving mechanism

The object of the UORA power-saving mechanism is the STAs that want to access the channel through OFDMA random contention. In the UORA power-saving mechanism, the AP reserves one or more TWT SPs in the Beacon frame, some or all of which support UORA (the TWT Flow Identifier field = 2; for those that do not, the TWT Flow Identifier field = 1). As a result, the STAs that wish to access the channel through UORA can sleep while waiting for the TWT SPs that support UORA. The STAs that wish to access the channel through UORA need to wake up and attempt to access. In other words, in this process, STAs do not need to wake up for TWT SPs that do not support UORA.

In a TWT SP, if the STAs receive a TF, and the Cascade Indication Field is 1 but the STAs' OBO is not reduced to 0 (i.e., does not transmit UL DATA), then the STAs may remain awake for a cascade TF; if the STAs receives a TF, and the Cascade Indication Field is 0 (indicating the end of this TWT SP) but the STA's OBO is not reduced to 0, then the STA may continue to sleep.

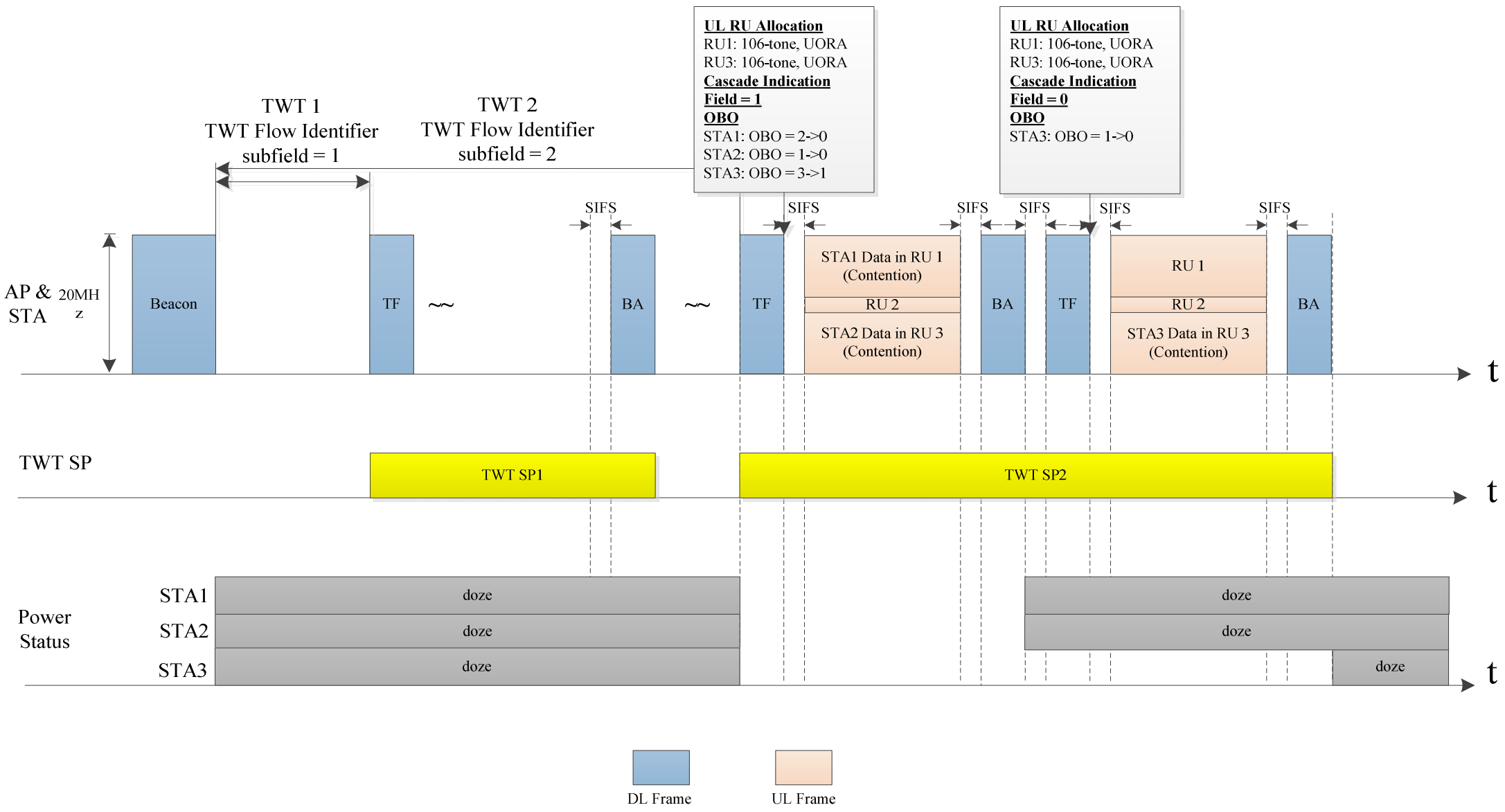

Fig. 37 Power Saving with UORA.

As shown in Fig. 37, the AP plans two TWT SP through Beacon frames and explicitly indicates that TWT SP1 does not contain OFDMA random contention access; instead, TWT SP2 supports OFDMA random contention access through the TWT Flow Identifier subfield. STA1~STA3 want to send UL data through OFDMA random contention access, so they do not need to wake up at the beginning of TWT SP1 until TWT SP2 starts. In TWT SP2, the AP sends the first TF to allocate RU1 and RU3 for random contention access, while STA1 and STA2 successfully access the corresponding RU to transmit UL data through random contention access. However, because the OBO of STA3 has not been reduced to 0, it has no right to access the current TF. After the current cycle, STA1 and STA2 go to sleep. STA3 is awared that the Cascade Indication Field is 1 in TF, indicating there is a cascade of TF. Thus, STA3 decides not to go to sleep and continues to wait. The AP sends the second TF to allocate RU1 and RU3 for random contention access. At this point, the OBO of STA3 is reduced to 0 so that STA3 can successfully transmit UL using RU3. Meanwhile, the Cascade Indication Field of the current TF is 0, indicating that this TWT SP no longer has a cascade TF; thus, STA3 can enter a sleep state.

ii. Periodic TWT opportunistic power-saving mechanism

IEEE 802.11ax supports an opportunistic power-saving mechanism based on TWT. The AP reserves multiple periodic TWT SPs in time and broadcasts scheduling information at the beginning of each TWT SP. After each STA receives the scheduling information, it enters sleep state selectively. The mechanism provides a sleep opportunity for STAs with periodic TWT SP. The following describes this specific process.

First of all, the AP needs to complete the periodic TWT SP configuration in Beacon: it needs to configure the corresponding field of TWT Beacon in order to support periodic TWT based on the description in Tab. 9.

Tab. 9 The configuration of the periodic TWT opportunistic power-saving mechanism

| **Field** | **Value and brief description** |
|---|---|
| TWT Flow Identifier | TWT Flow Identifier |
| Wake Interval Exponent | These two fields are used to jointly set the cycle lengths between the two adjacent TWT SPs, and the length is equal to $(TWT\ Wake\ Interval\ Mantissa) * 2^{(TWT\ \text{Wake Interval Exponent})}$。 |
| Wake Interval Mantissa | |
| Target Wake Time | The start time of the first TWT SP after Beacon |
| Nominal Minimum TWT Wake Duration | The shortest duration of each TWT SP corresponds to the shortest active time of STAs |

Next, the AP needs to send the TIM or FILS frame containing the TIM information discovery information (FILS is proposed IEEE 802.11ai) at every start of TWT SP, which is used to broadcast scheduling information and to indicate to each STA whether DL service is cached.

After an STA receives the resource scheduling information, if the TIM in its corresponding bit is 0 (i.e., the AP will not send DL data to this STA or schedule UL data transmission), this STA could enter sleep until the start of the next TWT SP cycle; conversely, if the TIM in the corresponding bit is 1, this STA should be awake to wait for DL data transmission or the scheduling of UL data transmission. Thus, this mechanism requires STAs to wake up at every TWT cycle.

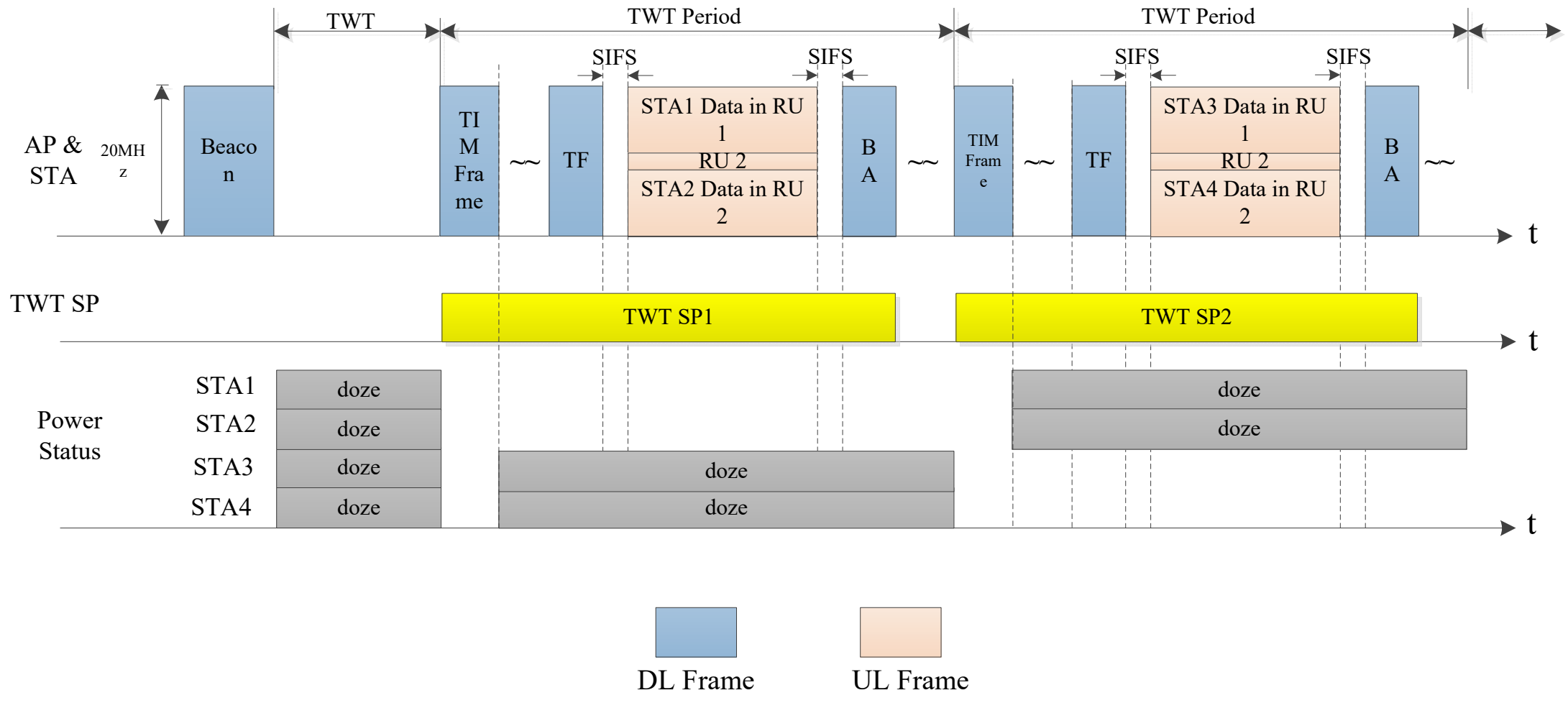

Fig. 38 Example of the periodic TWT opportunistic power saving mechanism.

As shown in Fig. 38, the AP configures the periodic TWT in Beacon. STA1~STA4 needs to wake up at the beginning of every TWT SP. The AP sends the TIM frame at the beginning of the first TWT SP to indicate that STA1 and STA2 have UL/DL transmission planning. STA3 and STA4 in the subsequent time of the first TWT cycle can enter sleep until the start time of the second TWT SP. Similarly, the AP sends TIM frame at the beginning of the second TWT SP to indicate the STA3's and STA4's UL/DL transmission planning in the second TWT SP. STA1 and STA2 in the subsequent time of the second TWT cycle can enter sleep until the start time of the third TWT SP.

iii. Individual TWT power-saving mechanism

STAs can also negotiate SP time with the AP by using individual TWT and can enter sleep state before SP arrives. In particular, STAs can negotiate the interval of listening Beacon frames (i.e., the listen interval) with the AP via the TWT request frame. As shown in Fig. 39, STA first sends the TWT request frame in which the Wake TBTT Negotiation field is set to 1 to request its expected listen interval to the AP through TWT Wake Interval Mantissa and Wake Interval Exponent fields to implement configuration. Next, the AP needs to reply the TWT respond frame to indicate that it accepts or rejects the listen interval, the next time for listening for the Beacon frame in the Target Wake Time field, and its ultimate listen interval (again, through the TWT Mantissa and Wake Interval Wake Interval Exponent field to implement configuration).

Different power save mechanisms are compared in Tab. 10.

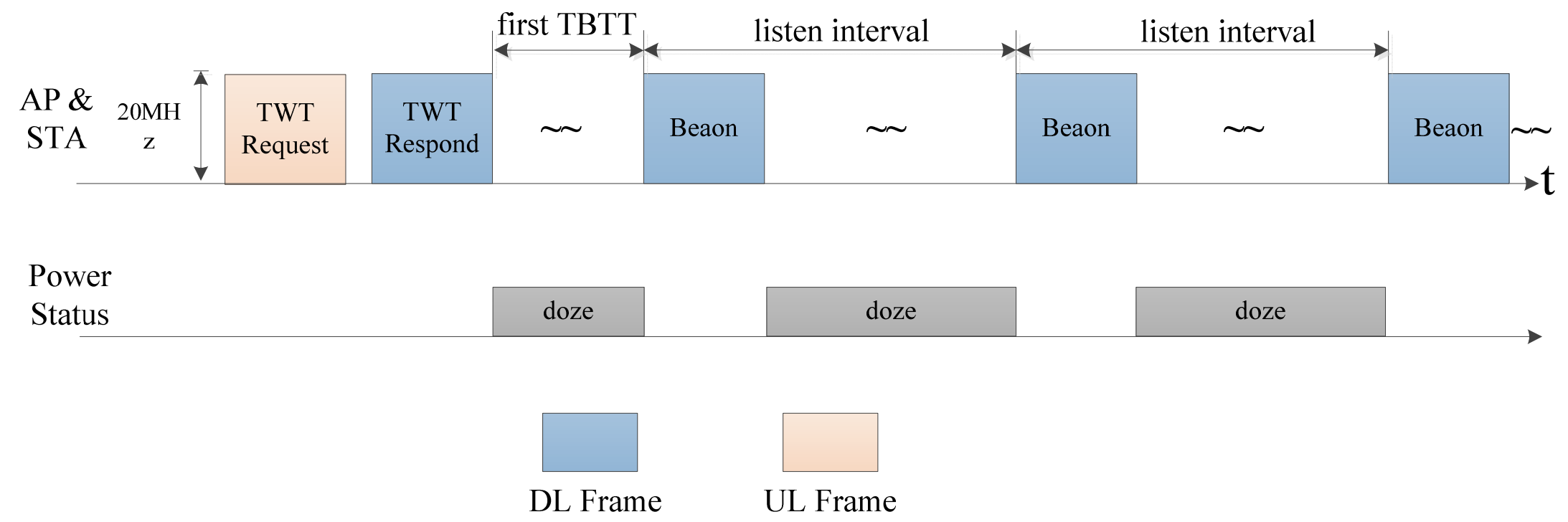

Fig. 39 An example of booking Beacon listening time through an individual TWT.

Tab. 10 Comparison of energy mechanisms in IEEE 802.11ax

| **Mechanisms** | **Sub mechanisms** | **Correlation with MU** | **Dependence of Beacon configuration** | **Sponsor** | **Power Saving Duration** |
|---|---|---|---|---|---|
| Power saving mechanisms based on TWT | TWT-oriented power saving mechanism for UORA | Relevant | Dependent | STA makes decision based on AP's planning | Relatively long |
| | Periodic TWT opportunistic power saving mechanism | Relevant | Dependent | STA makes decision based on AP's planning | Moderate |
| | Individual | Irrelevant | Not dependent | STA negotiates | Long |
| | TWT power saving mechanism | | | with AP individually | |
| Intra-PPDU power saving mechanism | —— | Relevant | Not dependent | STA decides independently | Short |

IEEE 802.11ax also introduces the mechanism by which STAs can implement the sleep operation autonomously (i.e., the intra-PPDU power-saving mechanism). The intra-PPDU power-saving mechanism refers to the duration that the STA receives a PPDU. If the PPDU is from an intra-BSS PPDU and is not involved in the transmission, the STAs can enter sleep state until the end of the PPDU and then wake up. The method to determine the origin of data packets belonging to the intra-BSS or inter-BSS is referenced to Sec. V.C. The standard specifies that if the STA is in a sleep state during the intra-BSS PPDU transmission, the NAV timer should continue running during dormancy. The sleep time of STAs using this mechanism is relatively short.

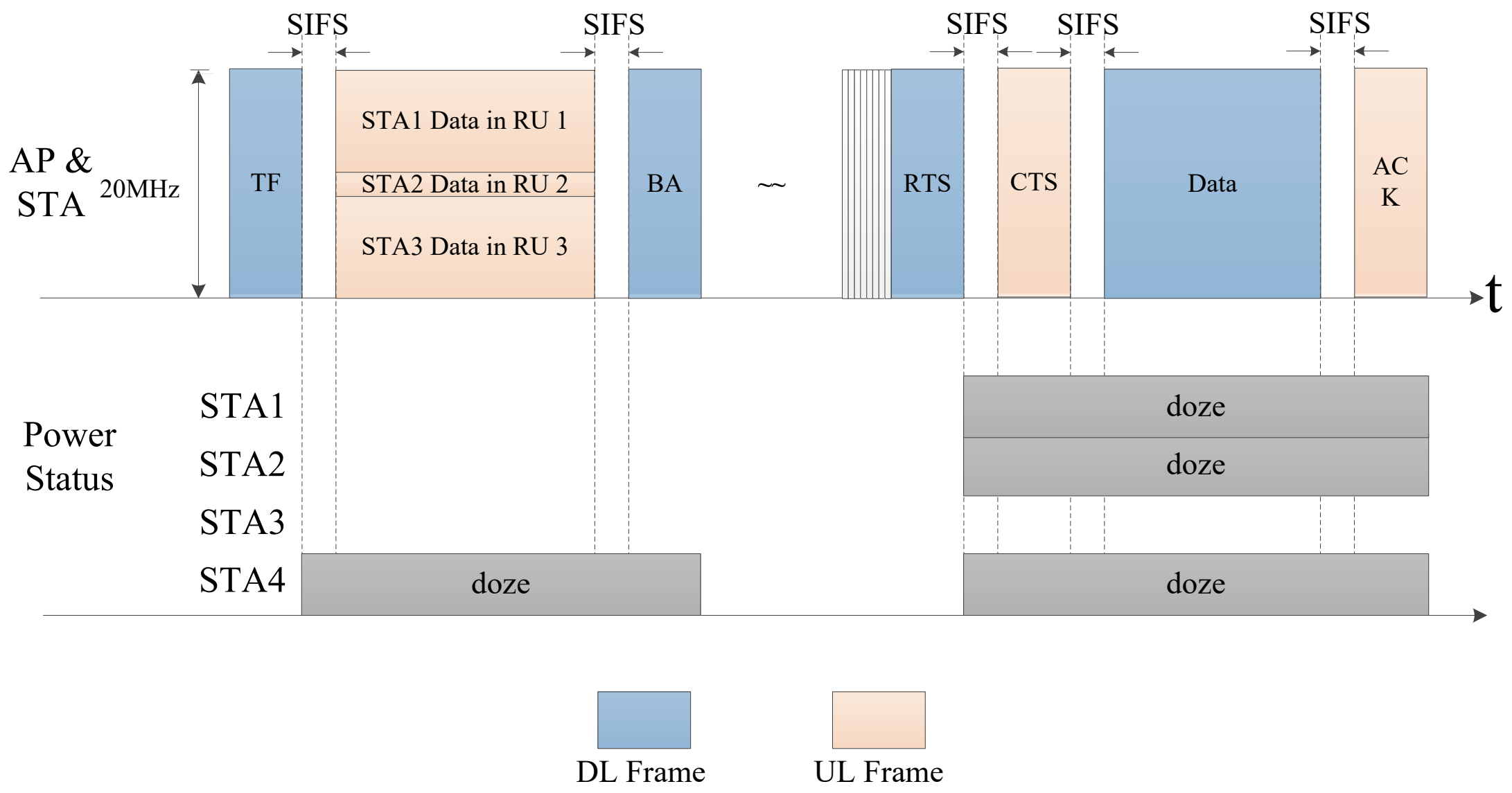

Fig. 40 Example of intra-PPDU power-saving.

As shown in Fig. 40, the AP first sends TF and schedules STA1~STA3 to send

UL data on RU1~RU3, respectively. At this point, STA4 notices that the transmission is intra-PPDU and confirms that the transmission is independent of itself. Thus, STA4 enters sleep state after TF and wakes up at the end of the transmission (i.e., at the end of the BA transmission). After a period of time, the AP completes the backoff process and sends the RTS frame to STA3, hoping for single-user data transmission, and STA3 replies with the CTS frame. Next, the AP transmits the DL data to STA3. At this point, STA1, STA2, and STA4 notice that this is an intra-PPDU that has nothing to do themselves, respectively, so they all enter sleep until the end of the ACK transmission.

# VII. Simulation Platform and Performance Evaluation

## A. Overview of Simulation Platform for WLANs

For any new or revised wireless network standard, standardization is necessary to fully demonstrate and validate the key technologies and potential technical solutions and algorithms. In addition to the necessary theoretical analysis, computer simulation has become indispensable to key technology analysis, verification, and optimization. The simulation platform for wireless networks is generally divided into two categories: system level simulation and link level simulation.

System level simulation platforms are mainly focused on the network behavior of the MAC layer and the higher-layer protocol. The advantages of system simulation platforms are that they can accurately depict the network behavior of a higher layer protocol. The disadvantage is that the link simulation, namely PHY and wireless channel, is too simple and abstract (e.g., the channel model, interference calculation, calculation of PER and SINR, and authenticity and objectivity of the simulation results are affected).

Link level simulation platforms focus largely on the link description. They have the advantage of being able to accurately portray the characteristics of point-to-point communication links, but the disadvantage is that their simulation and equivalence of higher layer behaviors is too simple and cannot even characterize higher-layer protocols. The authenticity and objectivity of simulation results are also affected.

Before we introduce the requirements of simulation platforms defined by IEEE 802.11ax, more details about our proposed system and link level platform SLISP, and the overall analysis and evaluation of IEEE 802.11ax based on SLISP, we will review valuable research based on WLAN simulation platforms in recent years. Fig. 41 classifies the existing work on simulation platforms.

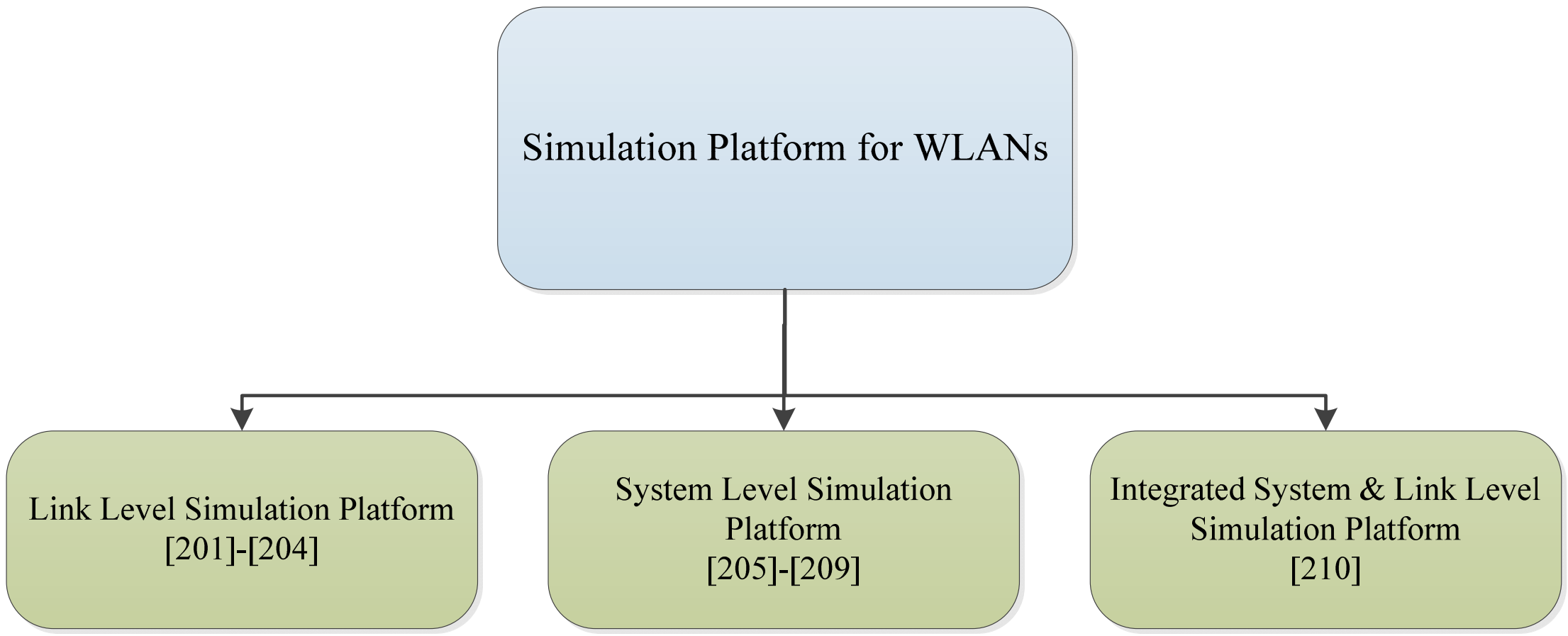

Fig. 41 Overview of existing research on WLAN simulation platforms.

In terms of link level simulation platforms, Khosroshahy [201] builds the legacy IEEE 802.11 and channel model based on NS3 [202]. The main work is the integration of the large-scale fading, small-scale fading, and computational model of bit error rate (BER) and PER for legacy IEEE 802.11 in the NS3 network simulation tool. Jacob et al. [203] model the channel dynamic characteristics of IEEE 802.11ad and analyz the barrier effect of the human body on wireless transmission. Brueninghaus et al. [204] summarize various channel models and PER computing methods and extracted a general PER computing method that adopts the instantaneous SINR of each symbol for PER prediction.

In system level simulation, Jonsson et al. [205][206] build an IEEE 802.11ac system simulation platform for enterprise networks based on NS3. The channel model in this platform uses the log-distance path loss model in NS3 only contained large-scale fading. Chen et al. [207] build an IEEE 802.11a/b/g system level simulation platform based on NS2 [208], and the authors focus on the implementation of the DCF MAC protocol based on CSMA/CA. Assasa and Widmer [209] build an IEEE 802.11ad simulation platform based on NS3. The simulation platform uses a free space fading model to calculate the received signal energy. In addition, NS2, NS3, OPNET, and other network simulation tools have also built their own WLAN simulation platforms.

In terms of system and link level integration simulation platforms, this article designs and implements a simulation platform [210] that integrates the system and link level and aims to fulfill IEEE 802.11ax standardization requirements. The simulation platform integrates a PHY and channel model into the system level simulation platform to ensure the authenticity and objectivity of system simulation. The key technologies of IEEE 802.11ax (e.g., OFDMA, UL MU-MIMO, and non-continuous channel binding) are designed and implemented as well. To the best of our knowledge, this platform is the first simulation platform for IEEE 802.11ax, and it is the first integrated simulation platform for IEEE 802.11. In order to fully verify the performance of IEEE 802.11ax, the simulation platform is further enhanced, and the specific content is outlined in Sec. VII.B. In addition, academia and industry have designed and implemented a series of simulation platforms for cellular networks [211]-[214].

## B. Introduction of Proposed SLISP

1) Requirements for the IEEE 802.11ax Simulation Platform

To more objectively verify the performance of key technologies in IEEE 802.11ax to effectively promote the standardization process, the TGax working group presents requirements for the simulation platform in the Evaluation Methodology document [29]. The document demonstrates the advantages and disadvantages of system level simulation, link level simulation, and integrated simulation as described in Tab. 11. The advantage of simply using a system or link level simulation is to simplify the design and expedite the simulation. Of course, this advantage is not obvious in the fast growth of computing. But, its shortcomings are more obvious: it is different from a real-world scenario, which is the advantage of an integrated simulation platform. Because the authenticity and objectivity of performance verification are more important targets, the document pointed out that it is necessary to build an integrated simulation platform for IEEE 802.11ax. Therefore, the authors build the SLISP simulation platform.

Tab. 11 Demonstration of the advantages and disadvantages of TGax on an IEEE 802.11ax simulation platform

| | **Advantages** | **Disadvantages** |
|---|---|---|
| Link Level Simulation | 1）Simplify the details of the MAC layer for faster simulation speed;<br>2）Isolate the MAC layer and focus on the performance gains or losses brought by the PHY layer technology itself. | 1）The MAC layer is too simplified, the simulation performance is biased by the real-world scenario, and the objectivity and authenticity are poor;<br>2）Only focusing on PHY is not conducive to a global view of PHY and MAC to jointly analyze the performance of IEEE 802.11ax. |
| System Level Simulation | 1）Simplify the details of the PHY layer for faster simulation speed;<br>2）Isolate the PHY layer and focus on the performance gains or losses brought by the MAC layer technology itself. | 1）PHY is too simplified, the simulation performance is different from the real-world scenario, and the manageability and authenticity are poor;<br>2）Only focusing on MAC is not conducive to the global view of PHY and MAC to jointly analyze the performance of IEEE 802.11ax. |
| Integrated Simulation | 1）Simulation performance is close to the real-world scenario;<br>2）Stand on the global perspective to analyze the influence of interoperability, cooperation, and mutual restriction between MAC and PHY on IEEE 802.11ax performance. | 1）Simulation speed is relatively slow;<br>2）Increasing complexity. |

2) SLISP Architecture

The system structure of SLISP is illustrated in Fig. 42, and the integration of link and system level simulation is implemented. SLISP mainly includes the system level simulation unit, link level simulation unit, and integrated entity unit, each of which is discussed briefly below.

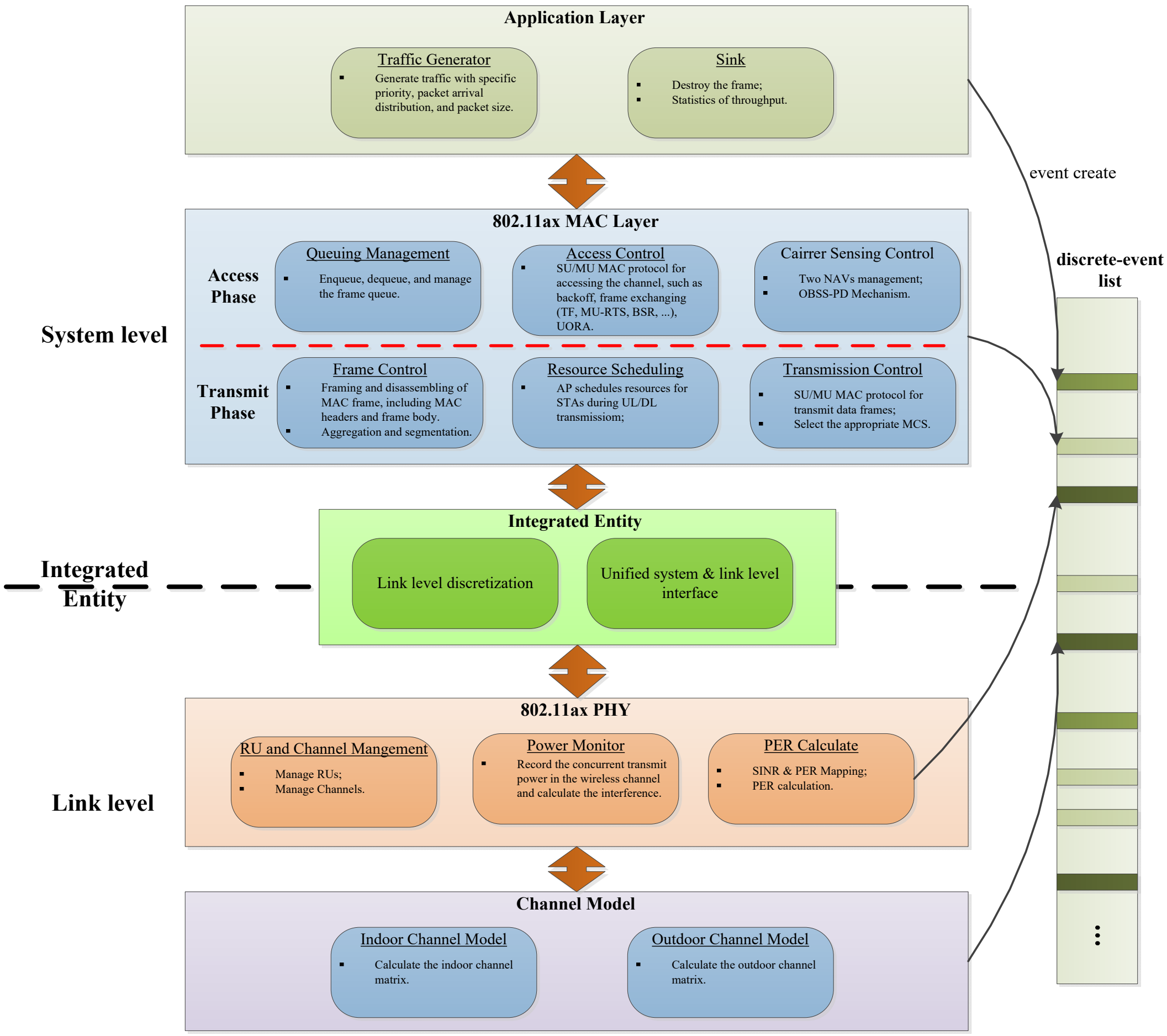

Fig. 42 Framework of integrated simulation platform.

The system level simulation unit mainly contains the application layer and IEEE 802.11ax MAC layer function, focusing on nodes and their behavior in the network. The application layer is mainly responsible for the generation and destruction of service data, which consists of the traffic generator module and sink module. IEEE 802.11ax MAC is primarily responsible for the implementation of the IEEE 802.11ax layer MAC layer protocol. The access phase includes a queuing management module, access module, and control carrier sensing control module. The transmission phase includes a frame control module, resource scheduling module, and transmission control module.

The link level simulation unit includes the IEEE 802.11ax PHY and channel model, which focuses on the communication link. The IEEE 802.11ax PHY is mainly responsible for the implementation of IEEE 802.11ax PHY-related functions, composed of RU and channel management module, power monitor module, and PER calculate module. RU and channel management are largely responsible for the management of RUs and channels; the power monitor module is mainly responsible for energy calculation and statistics; and the PER calculate module is mainly used for the node to calculate the PER of the received data frame according to the channel matrix obtained from the channel model, the process also called physical layer abstraction. The channel model needs to generate the H matrix of the channel according to the spatial correlation coefficient of the transmitting antenna and the receiving antenna and other parameters for the IEEE 802.11ax PHY.

The integrated entity unit is responsible for the integration of system and link level simulation, consisting of two modules: link level discretization and a unified system and link level interface. The link level discretization module is responsible for incorporating link level simulation behavior into the discrete event simulation mechanism, which is an event-driven simulation mechanism. Specific to the network simulation, it considers all network behavior as events, and the simulation system only changes when a new event occurs (i.e., events trigger the background processing program). Thus, the simulation time is not uniform but instead triggered by events. Traditional link level simulation usually adopts a process-oriented or continuous time-oriented simulation mechanism, which is not consistent with the discrete event simulation mechanism. The biggest challenge in integrated simulation is to use the discrete event simulation mechanism in both link and system level simulation. As shown in Fig. 42, the link level discretization module is responsible for inserting the behavior of the IEEE 802.11ax PHY and the channel model as events into the discrete-event list. Therefore, both the system level (i.e., application layer and IEEE 802.11ax MAC layer) and link level simulation (i.e., IEEE 802.11ax PHY and channel model) will be incorporated into the discrete simulation event mechanism, which achieves the integrated design principle.

The unified system and link level interface is responsible for the implementation of a unified interface for the link and system levels, and the main interface is a MAC layer and PHY data transmission interface. Specifically, the IEEE 802.11ax MAC needs to send the encapsulated data frames to the IEEE 802.11ax PHY when triggered by the transmission event; in turn, the IEEE 802.11ax PHY needs to send the received data frames to the IEEE 802.11ax MAC layer when triggered by the transmission complete event. In addition, some state settings and query interfaces also need to be implemented, such as the PHY carrier sensing state query, SINR query, and so on.

3) Satisfaction of SLISP to the requirements

The design of the integrated simulation platform is restrained and suggested by IEEE 802.11ax [29][48]. As shown in Tab. 12, √ represents SLISP support this feature, and * represents that this feature is not required by IEEE 802.11ax documents but is implemented by SLISP additionally. The integrated simulation platform (SLISP) designed and built by the authors not only meets all basic requirements specified in the document but those of its enhanced feathers, which are key to core technologies in IEEE 802.11ax (marked with * in Tab. 12), such as MU control frame, MU operation, SR, 1024-QAM, and so on.

Tab. 12 The satisfaction of SLISP to IEEE 802.11ax simulation platform

| | **Basic Features** | | **Enhancement Features** | |
|---|---|---|---|---|
| | **Features** | **SLISP** | **Features** | **SLISP** |
| Architecture | Discrete event-driven based | √ | —— | —— |
| Scenario | High-dense indoor | √ | —— | —— |
| | High-dense outdoor | √ | —— | —— |
| MAC | CCA | √ | Multiple channels | √ |
| | Control frame (RTS/CTS/ACK/block ACK) | √ | MU control frame (TF, MBA, OFDMA BA) | * |
| | EDCA | √ | Management frame | √ |
| | Aggregation (A-MPDU in 11ac) | √ | UL MU operation | * |
| | Link adaption | √ | DL MU operation | * |
| | Transmission mode (SU-OL, Beamforming,…) selection | √ | OBSS_PD mechanism | * |
| | Power save | √ | Two NAVs | * |
| | mechanism | | | |
| PHY | Beamforming vector | √ | MU-MIMO | √ |
| | MMSE | √ | RU allocation | * |
| | Effective SINR mapping and PER prediction | √ | 1024-QAM | * |
| | Energy detection | √ | | |

## C. Performance Evaluation for Single BSS

### 1) Simulation Scenarios and Settings

In order to make the verification results more convincing, single BSS scenarios in the simulations follow the TGax Simulation Scenarios document presented by the TGax workgroup [48].

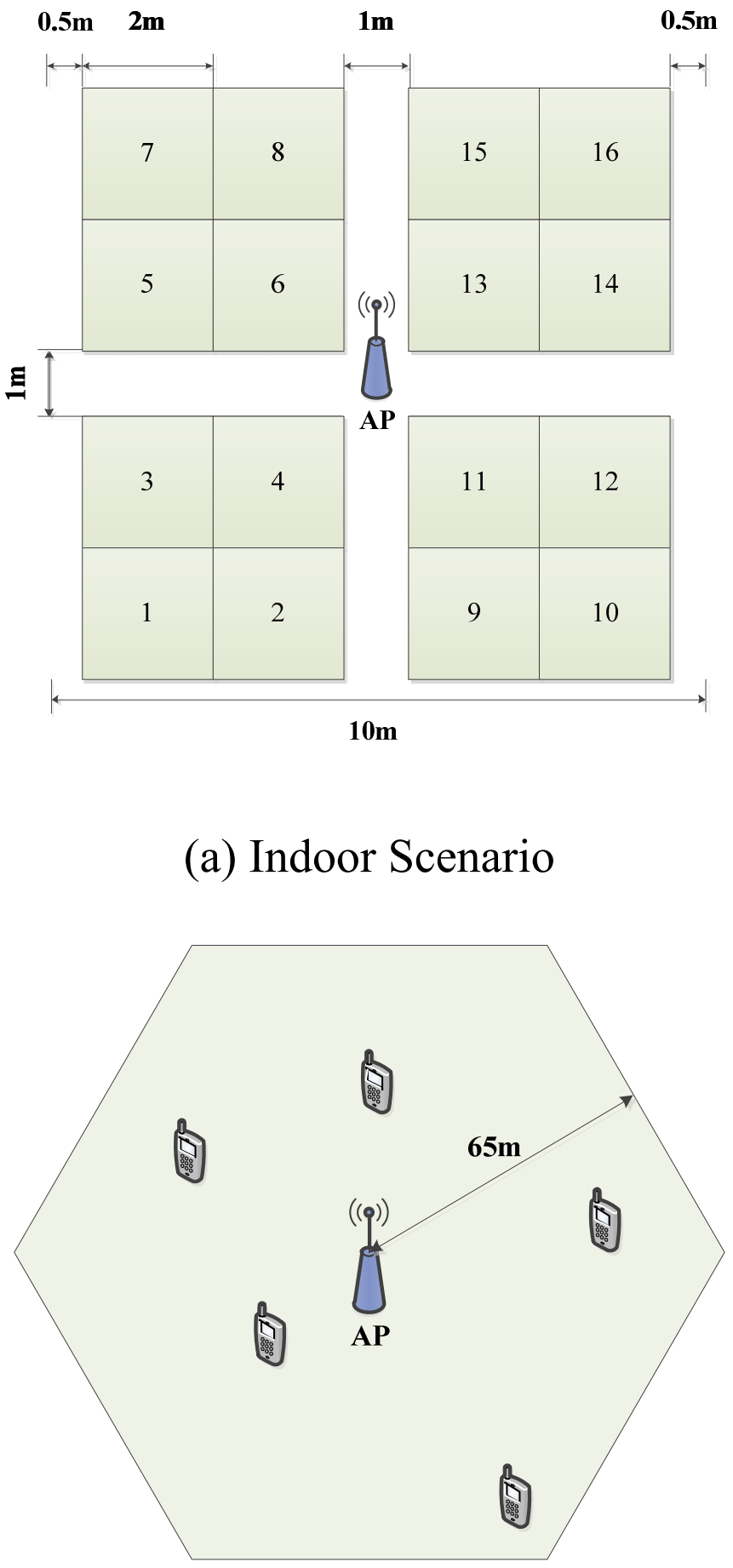

(a) Indoor Scenario

(b) Outdoor Scenario

Fig. 43 Single BSS scenarios in the simulations.

The indoor single BSS scenario of IEEE 802.11ax is shown in Fig. 43(a). Each square represents a room with an area of $4m^2$. The AP is located in the center of the district and serves the 16 rooms symmetrically. There is an aisle between the middle and the four sides, and the distance between the rooms is 1m. There are four STAs randomly distributed in each room and 64 total STAs in BSS. The detailed simulation parameter configuration of the indoor single BSS scenario is shown in Tab. 13.

Tab. 13 Parameter configuration of indoor single BSS scenario in the simulations

| Parameters | Description |
|---|---|
| Service type | CBR (constant bit rate) |
| Per-STA service rate | 1M~13Mbps (20MHz), 4M~52Mbps (20MHz), 8M~104Mbps (160MHz) |
| Number of STAs | 64 |
| STAs' position | Randomly distributed in each room |
| MCS index | 0~11; MCS 10 and MCS 11 are only employed when RU is wider than or equal to 242-tone |
| AP transmit power | 18dBm |
| STA transmit power | 18dBm |
| AP antenna height | 1.5m |
| STA antenna height | 1.5m |
| Frequency | 5.57G |
| CCA threshold | -82dBm |
| SIFS | 16 μs |
| DIFS | 34 μs |
| CWmin | 15 |
| CWmax | 1023 |
| TXOP duration | 3.008ms |
| Number of AP antennas | 8 |
| Number of STA antennas | 4 |

The outdoor single BSS scenario of IEEE 802.11ax is shown in Fig. 43(b). Each BSS has a hexagonal shape, and the AP is located at the center of the hexagon. The radius of the inner circle of the cell is 65m, and each cell contains 50~100 STAs. The detailed simulation parameter configuration of the outdoor single BSS scenario is shown in Tab. 14.

Tab. 14 The parameter configuration of outdoor single BSS scenario in the simulations.

| Parameters | Description |
|---|---|
| Service type | CBR |
| Per-STA service rate | 1M~13Mbps (20MHz), 4M~52Mbps (20MHz), 8M~104Mbps (160MHz) |
| The number of STAs | 64 |
| STAs Position | Randomly distributed in each room |
| MCS index | 0~11; MCS 10 and MCS 11 are only employed when RU is wider than or equal to 242-tone |
| AP transmit power | 18dBm |
| STA transmit power | 18dBm |
| AP antenna height | 1.5m |
| STA antenna height | 1.5m |
| Frequency | 5.57G |
| CCA threshold | -82dBm |
| SIFS | 16 μs |
| DIFS | 34 μs |
| CWmin | 15 |
| CWmax | 1023 |
| TXOP duration | 3.008ms |
| Number of AP antennas | 8 |
| Number of STA antennas | 4 |

In this simulation, we compare the performance of three schemes as follows:

**Scheme 1: 802.11ac**

Scheme 1 implements a channel access and data transmission procedure for IEEE 802.11ac (i.e., EDCA channel access). According to the priority of frames in the node, this scheme sets different AIFS, CW, and TXOP. The AP and STA contend for channel access randomly through the exchange of RTS and CTS frames. The node requires MAC to be able to aggregate frames, transmit frames, split frames, and reply BA frames in a TXOP duration (higher priority) and releases the channel after data transmission to contend for the next channel access opportunity.

**Scheme 2: 802.11ax with OFDMA**

In the UL channel access and data transmission, the AP performs RU allocation according to the buffer status of STAs and sends TF to broadcast the RU allocation information. MU-MIMO is not supported in this scheme. After receiving the TF, each scheduled STA sends the data frame on the corresponding RU in the way of A-MPDU according to the RU allocation information in the TF frame combined with the current channel state and individual queue situation. After receiving the UL packets in RU, the AP delivers the frames to the upper layer and sends MBA to UL STAs. In a TXOP duration, there may be multiple UL transmissions. When an STA does not have UL data need to transmit or the remaining time is insufficient to carry out a complete data transmission, AP sends CF-End frame to finish the UL transmission.

In DL channel access and data transmission, the AP allocates RUs according to its own service request after acquiring the channel resources. Similar to UL transmission, the AP sends DL frames for multiple STAs on different RUs by the way of OFDMA. STAs receive data frames on different RUs, respectively, according to the RU allocation information in HE-SIG field and reply BA frames on corresponding RUs. In a TXOP duration, there may also be multiple DL data transmissions. When AP has no data frame for DL transmission or the remaining time is insufficient for a complete data transmission (i.e., the end of the TXOP), the AP sends CF-End frame to finish the DL transmission.

Because the scheduling algorithms are beyond the scope of this article, the scheduling algorithm we adopt is the method of randomly selecting an STA for each RU, which can reveal the essential characteristics of IEEE 802.11ax data transmission.

**Scheme 3: IEEE 802.11ax with OFDMA and MU-MIMO**

Based on Scheme 2, Scheme 3 adds UL MU-MIMO capability to the data transmission (i.e., supports OFDMA and MU-MIMO) as illustrated in Fig. 44. Therefore, the AP needs to allocate the resources on RUs and spatial streams according to STAs' buffer status. Multiple STAs transmit data frames in the same RU using different spatial streams simultaneously. According to the IEEE 802.11ax standard, we only allow STAs to use MU-MIMO on RUs wider than or equal to 106-tone. Because IEEE 802.11ac has already supported DL MU-MIMO and this article focuses on performance gains in IEEE 802.11ax, Scheme 3 only contained UL

MU-MIMO, not DL MU-MIMO.

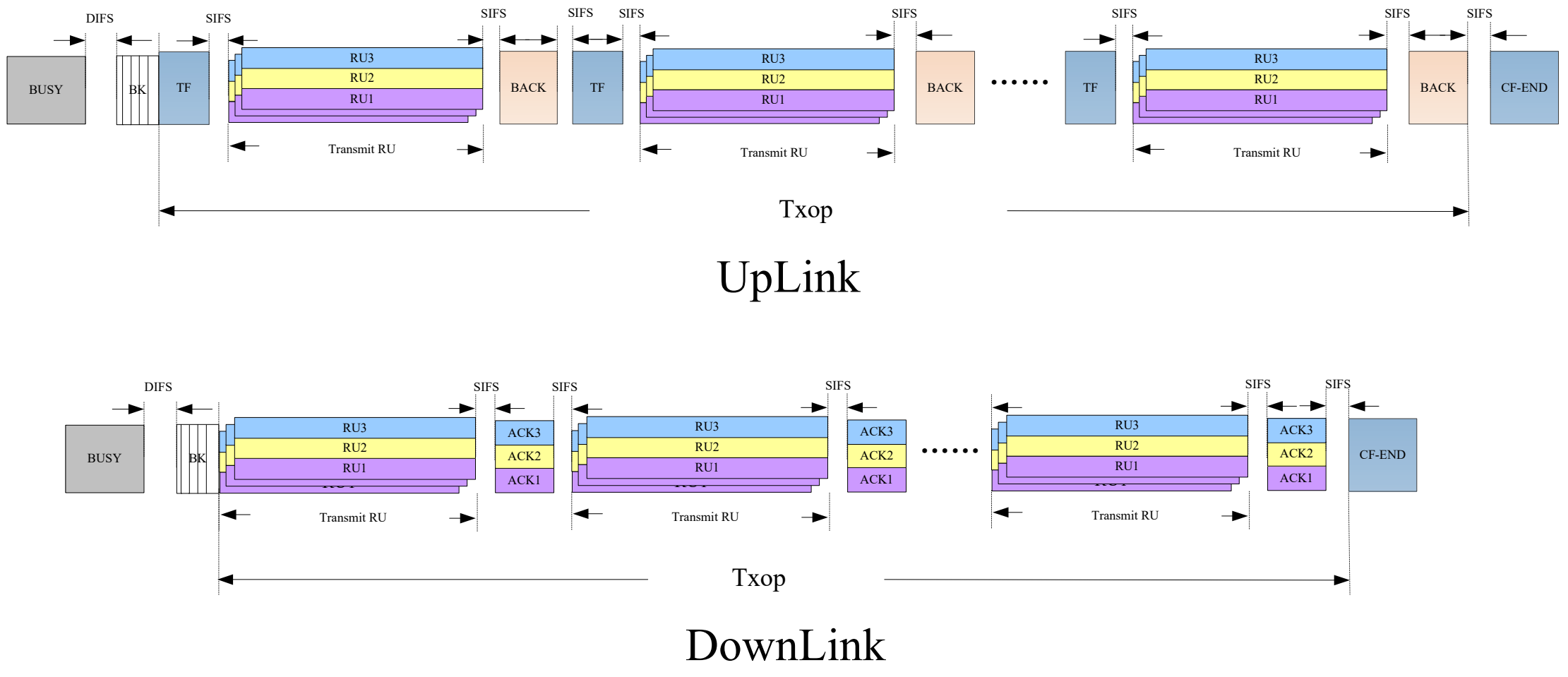

Fig. 44 IEEE 802.11ax with OFDMA and MU-MIMO.

The functional comparison of the three schemes is shown in Tab. 15.

Tab. 15 Comparison of 3 schemes for single BSS simulations.

| | **802.11ac** | **802.11ax with OFDMA** | **802.11ax with OFDMA and MU-MIMO** |
|---|---|---|---|
| EDCA-based Channel access | √ | √ | √ |
| TXOP | √ | √ | √ |
| Aggregation and segmentation | √ | √ | √ |
| OFDMA | × | √ | √ |
| UL MU-MIMO | × | × | √ |
| Multi-channel | √ | √ | √ |

2) Simulation results

Several simulations are deployed and analyzed as follow.

i. Performance analysis of indoor single BSS

Fig. 45 shows that the network performance of different schemes varied with the service rate in the indoor single BSS scenario, corresponding to the simulation results of bandwidth of 20MHz, 80MHz, and 160MHz, respectively.

Take UL simulation results as an example (DL simulation results is similar): the saturation throughput of IEEE 802.11ax with OFDMA reaches 131%, 206%, and 273% performance gain in 20MHz, 80MHz and 160MHz, respectively, compared to IEEE 802.11ac. Obviously, the performance gain of IEEE 802.11ax increases with the increase of bandwidth. This is because when the bandwidth increases, the MAC efficiency of IEEE 802.11ac and the stable MAC efficiency of IEEE 802.11ax with OFDMA will decrease. In IEEE 802.11ax with OFDMA, the overhead of channel access and data transmission is shared with multiple STAs. Thus, the MAC efficiency of IEEE 802.11ax with OFDMA could maintain a relatively high level when the bandwidth is wide. Moreover, in the >20MHz scenario, additional gains can be obtained from 1024-QAM.

The performance of IEEE 802.11ax with OFDMA and MU-MIMO scheme is theoretically about two times that of IEEE 802.11ax with OFDMA scheme. Every time the former supports twice the number of STAs for channel access and data transmission. However, in IEEE 802.11ax, MU-MIMO can only be used when the number of sub-carriers is greater or equal to 106 tone, so some RUs (such as 26-tone) could not use MU-MIMO. In addition, when the number of STAs increases, the overhead of corresponding scheduling signaling and BA frames will increase accordingly. The actual simulation results show that the gain is approximately 1.8 times. In particular, the throughput of IEEE 802.11ax with OFDMA and MU-MIMO reaches 5.67Gbps at 160MHz, and the UL throughput of IEEE 802.11ac is 1.2Gbps. The throughput of IEEE 802.11ax with OFDMA and MU-MIMO achieves 4.7 times as much as IEEE 802.11ac. As the numbers of STAs in the three schemes are the same in the simulations, the per-STA average throughput also reached 4.7 times. In summary, IEEE 802.11ax can meet the requirements of four times throughput gain in PAR compared to legacy IEEE 802.11.

Notably, regardless of the IEEE 802.11ac or IEEE 802.11ax scheme, network throughput in DL transmission is always greater than that of UL transmission. For DL transmission in IEEE 802.11ac, only the AP needs to contend for channel resources, whereas for UL transmission in IEEE 802.11ac, multiple STAs contend for channel access. Therefore, signaling overhead for channel access and data transmission in DL is less than in UL in IEEE 802.11ac. For IEEE 802.11ax, UL and DL data transmission are scheduled by the AP, and the overhead for both are the same. However, UL transmission in IEEE 802.11ax brings additional TF transmission and SIFS interval overhead, so the DL overhead is slightly lower.

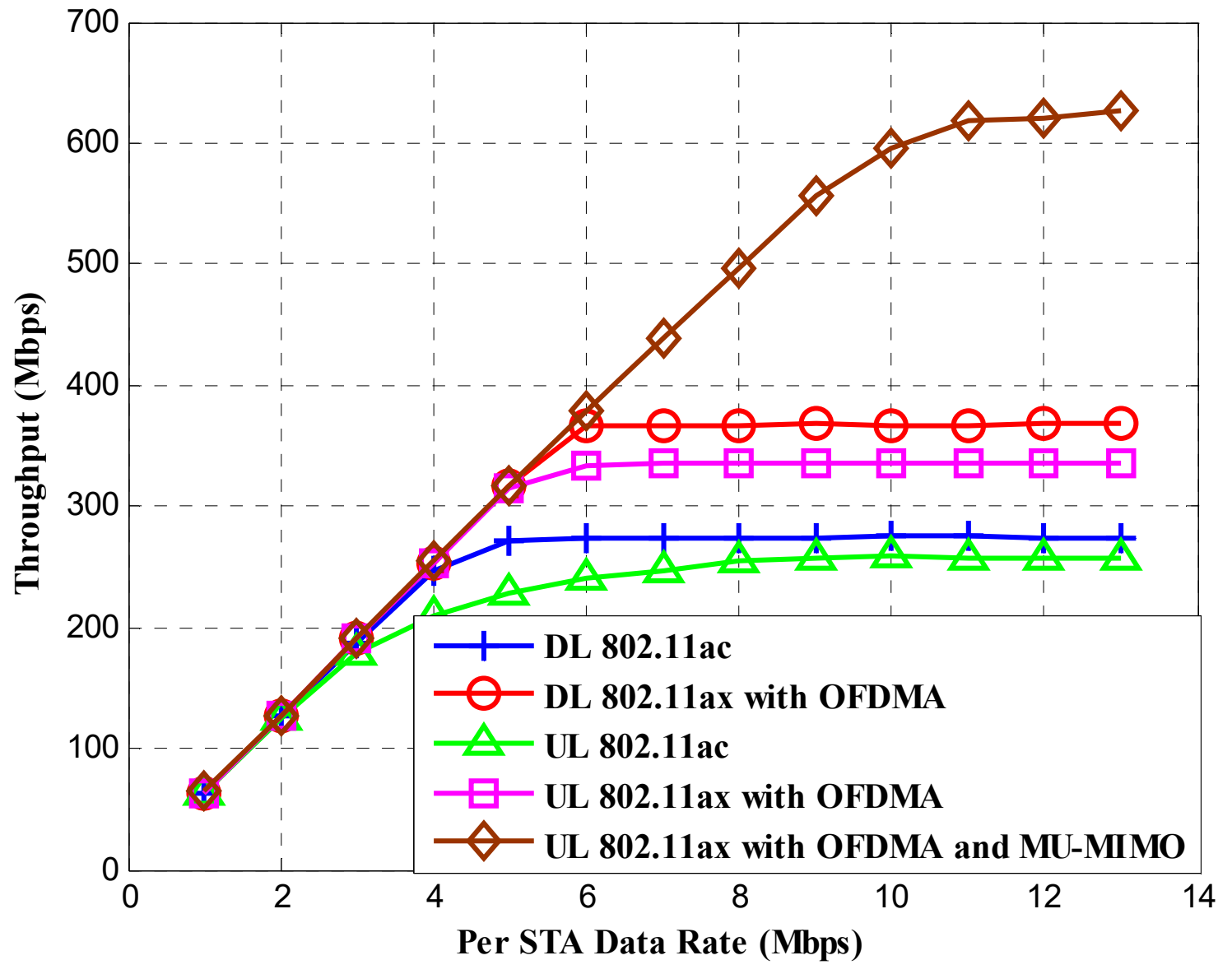

(a) Performance comparison of 20MHz in indoor scenarios

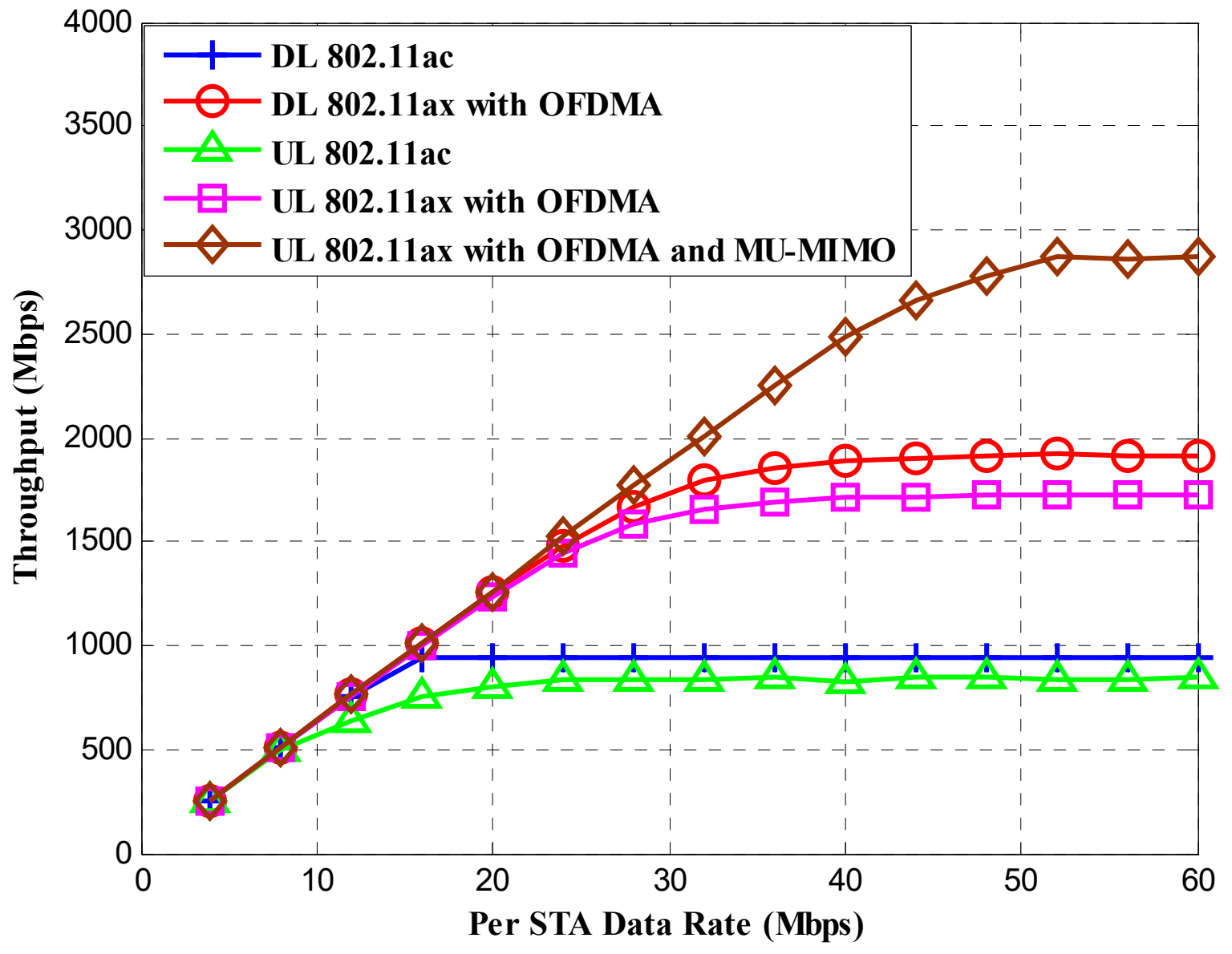

(b) Performance comparison of 80MHz in indoor scenarios

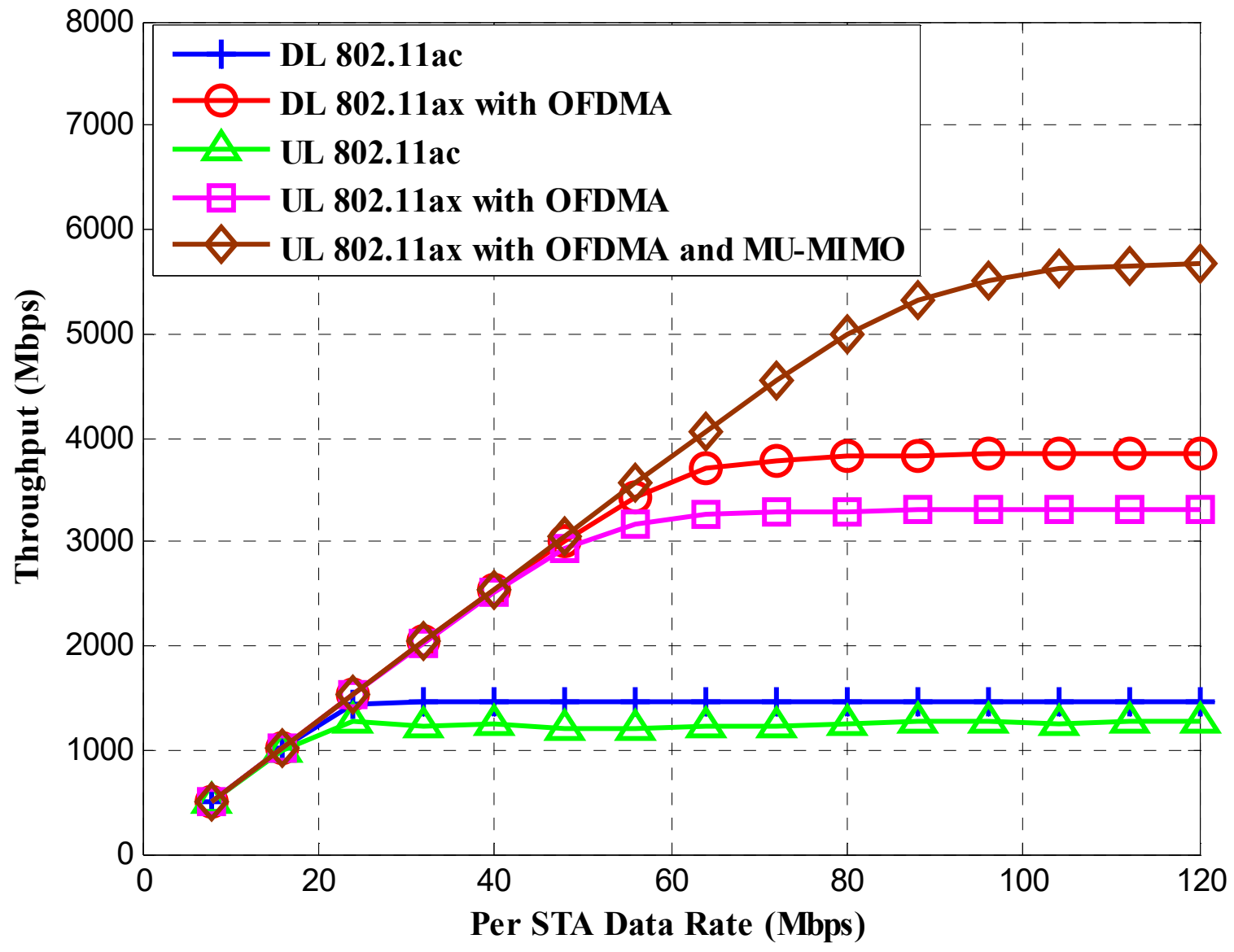

(c) Performance comparison of 160MHz in indoor scenarios

Fig. 45 The influence of service rate variation on network performance in indoor single BSS scenarios.

ii. Performance analysis of outdoor single BSS

Fig. 46 shows that the network performance of different schemes varies with the service rate in the outdoor single BSS scenario, corresponding to the simulation results of bandwidth of 20MHz, 80MHz, and 160MHz, respectively. Compared to the indoor scenarios, the performance fluctuation in outdoor scenarios is larger because the distance between the AP and STAs is uneven in the outdoor scenarios, and channel fading has a great influence on MCS selection. We average the simulation results of the outdoor single BSS scenarios by multiple simulations. In the indoor single BSS scenarios, AP and all STAs are very close, so the influence of channel fading is limited to the simulation results and higher MCS is probably selected.

Consider the UL simulation results for example (DL simulation results were similar): the saturation throughput of IEEE 802.11ax with OFDMA achieves 145%, 221%, and 292% performance gain in 20MHz, 80MHz, and 160MHz, respectively, compared to IEEE 802.11ac. With an increase in bandwidth, the performance gain of IEEE 802.11ax with OFDMA also increases due to the stable MAC efficiency of IEEE 802.11ax with OFDMA.

One concern is that in the outdoor single BSS scenario, the performance of IEEE 802.11ax with OFDMA and MU-MIMO is roughly the same as that of 802.11ax with OFDMA, and throughput only increased slightly. Because of the poor quality of the wireless channel in the outdoor scenarios, STAs using MU-MIMO have to use the lower MCS in the data transmission procedure, so the throughput of IEEE 802.11ax with OFDMA and MU-MIMO is not significantly increased. In addition, SU-MIMO is used in IEEE 802.11ax with OFDMA, where throughput is improved by the antenna diversity gain because the AP has eight antennas and each STA have four. In summary, these findings indicate that data transmission with MU-MIMO is not suitable for outdoor scenarios or poor channel states.

Another interesting phenomenon is that, unlike in the indoor scenario, the UL saturation throughput of all the three schemes in the outdoor scenario is greater than the DL saturation throughput. For IEEE 802.11ac, the AP sends RTSs successively, and the chance for STAs to receive an RTS is nearly equal. However, when UL STAs sent RTS frames, the AP is more likely to receive the RTS sent by the STAs closer to it, and the channel quality of STAs near the AP is usually better than those far away from the AP. Thus, the UL throughput is larger than the DL throughput in IEEE 802.11ac. For IEEE 802.11ax, because the transmission power of the AP and STAs is the same, the AP needs to allocate power in more than one RU in the DL transmission while the STAs use the maximum transmit power for UL transmission. Thus, the overall transmit power increases, and STAs have more opportunity to select a higher-order MCS.

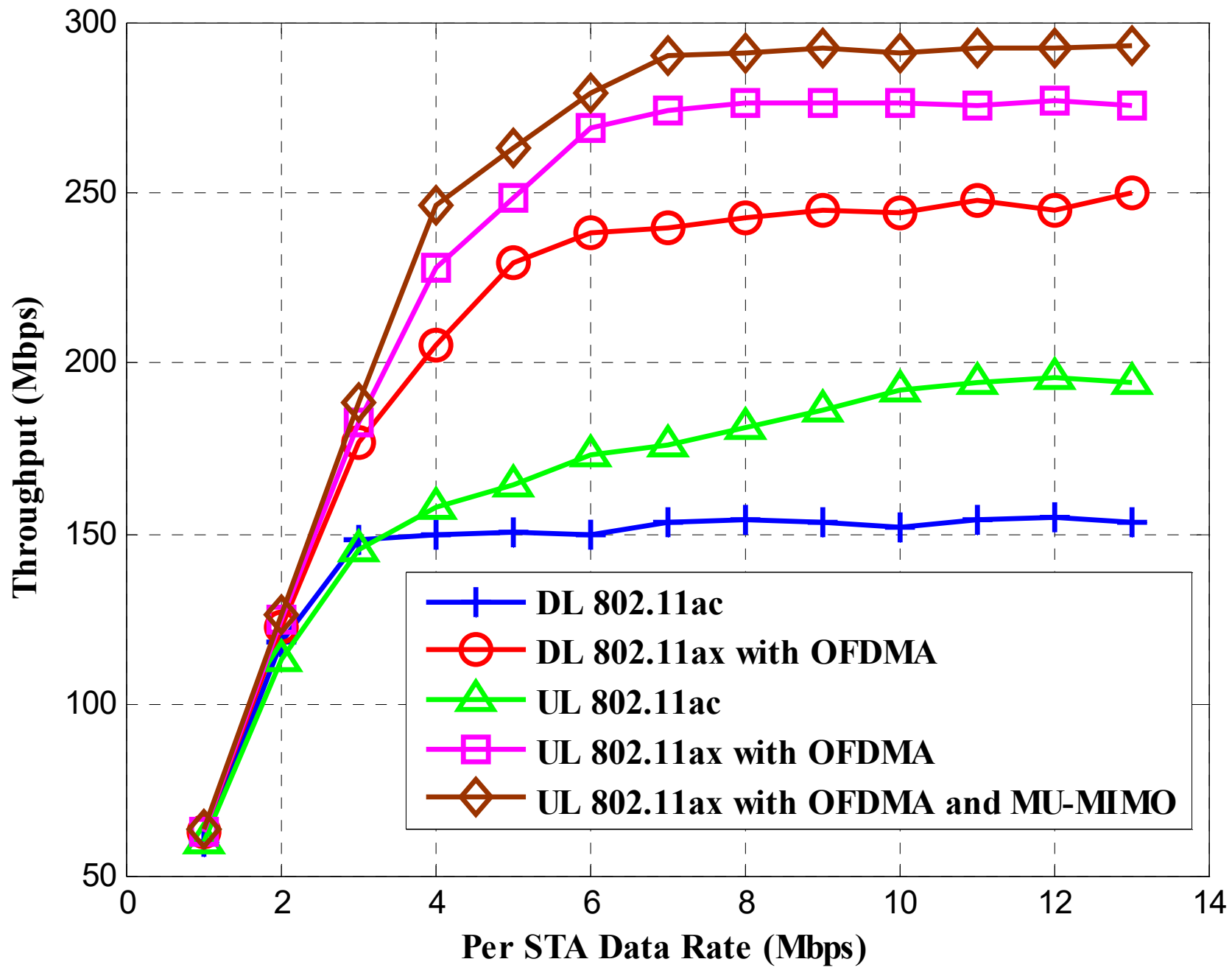

(a) Performance comparison of 20MHz in outdoor scenarios

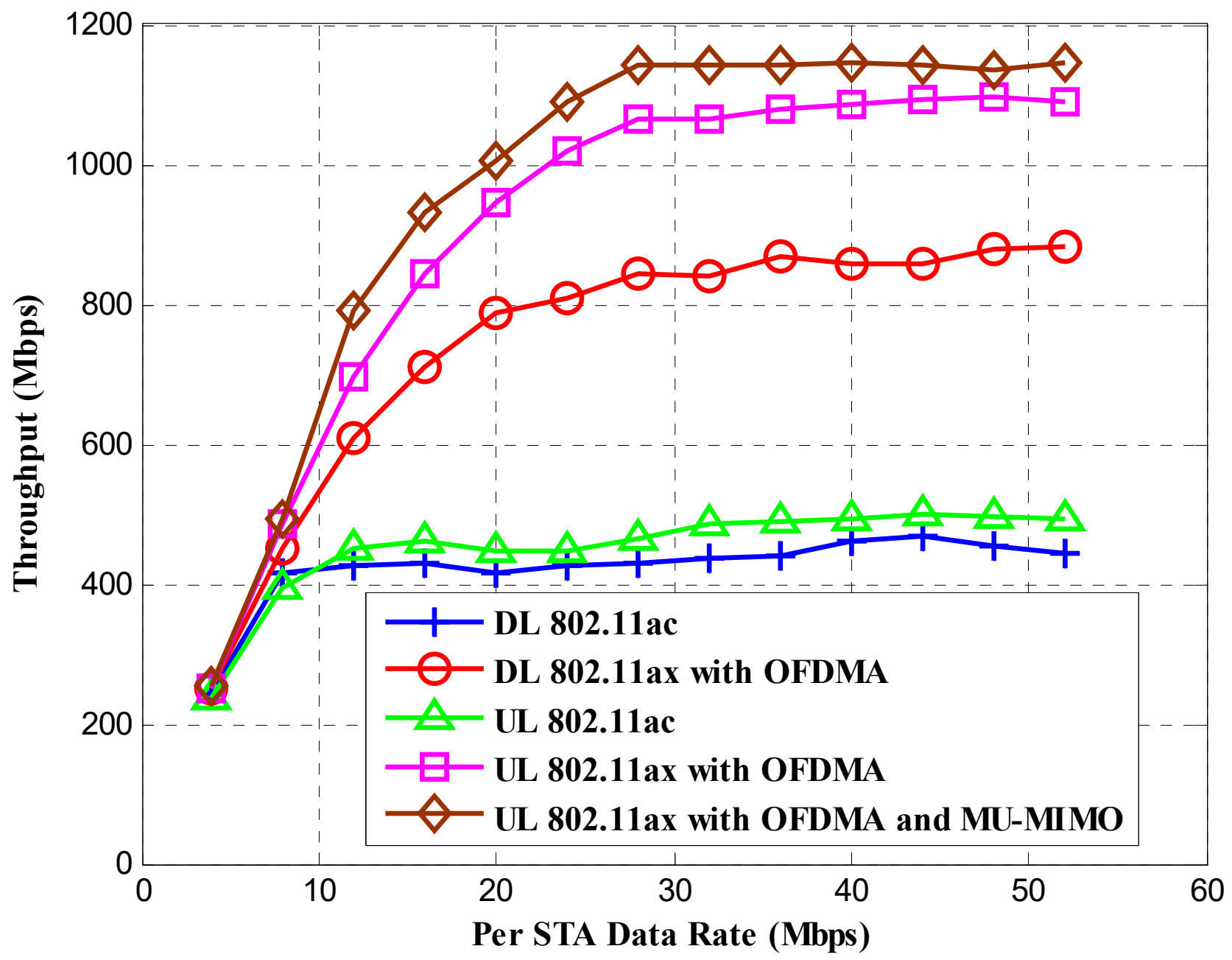

(b) Performance comparison of 80MHz in outdoor scenarios

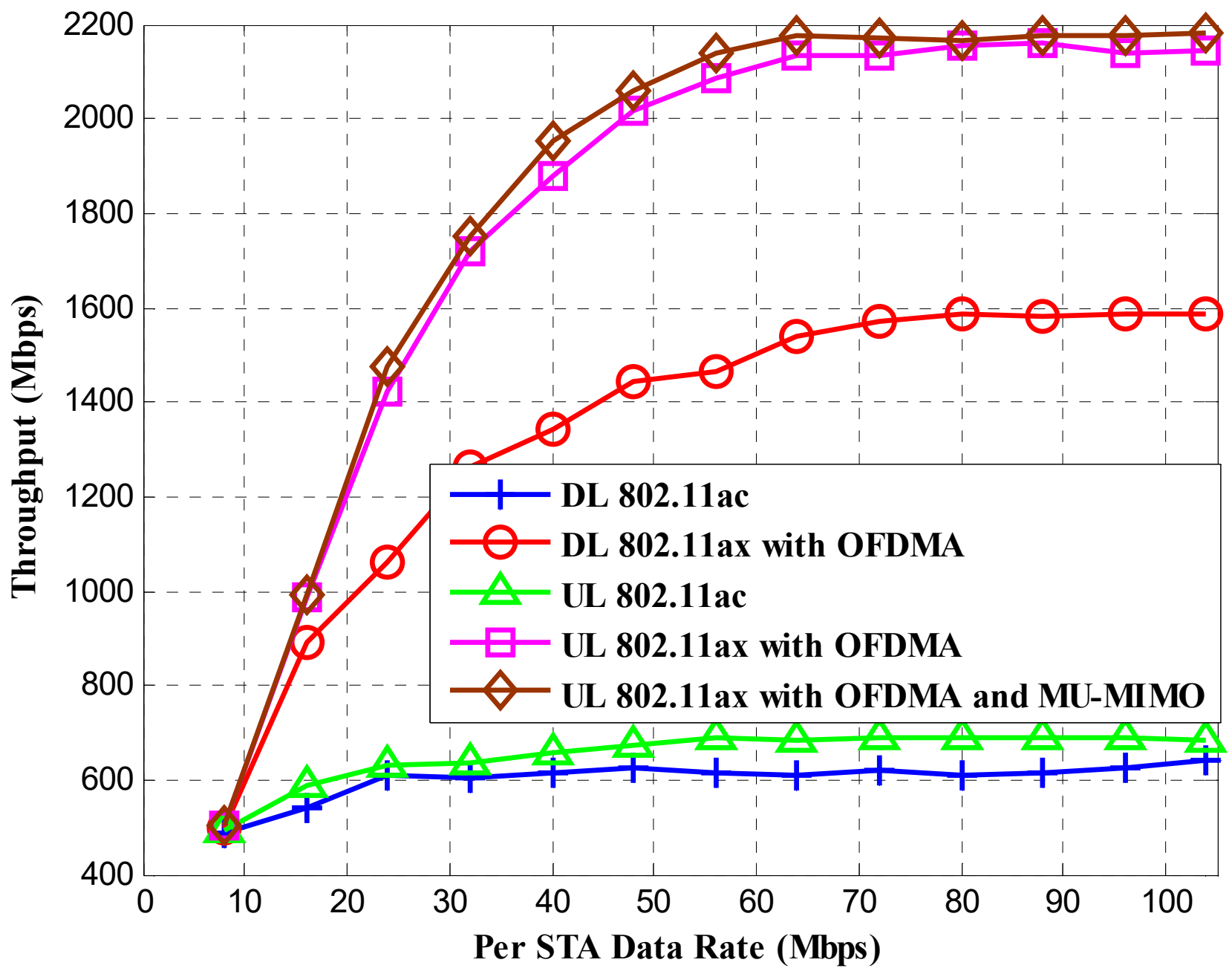

(c) Performance comparison of 160MHz in outdoor scenarios

Fig. 46 The influence of service rate variation for network performance in outdoor single BSS scenarios.

## D. Performance Evaluation for High-Dense Deployed Multiple BSS

### 1) Simulation Scenarios and Settings

To make the verification results more convincing, the scenario setting of multiple BSS still follows the TGax Simulation Scenarios document presented by the TGax [48].

The indoor multiple BSS scenario in IEEE 802.11ax is shown in Fig. 47(a). The whole network topology is composed of 32 indoor single BSS scenarios depicted as 4×8 matrices. The detailed network parameters of the indoor multiple BSS scenarios in our simulations are shown in Tab. 16.

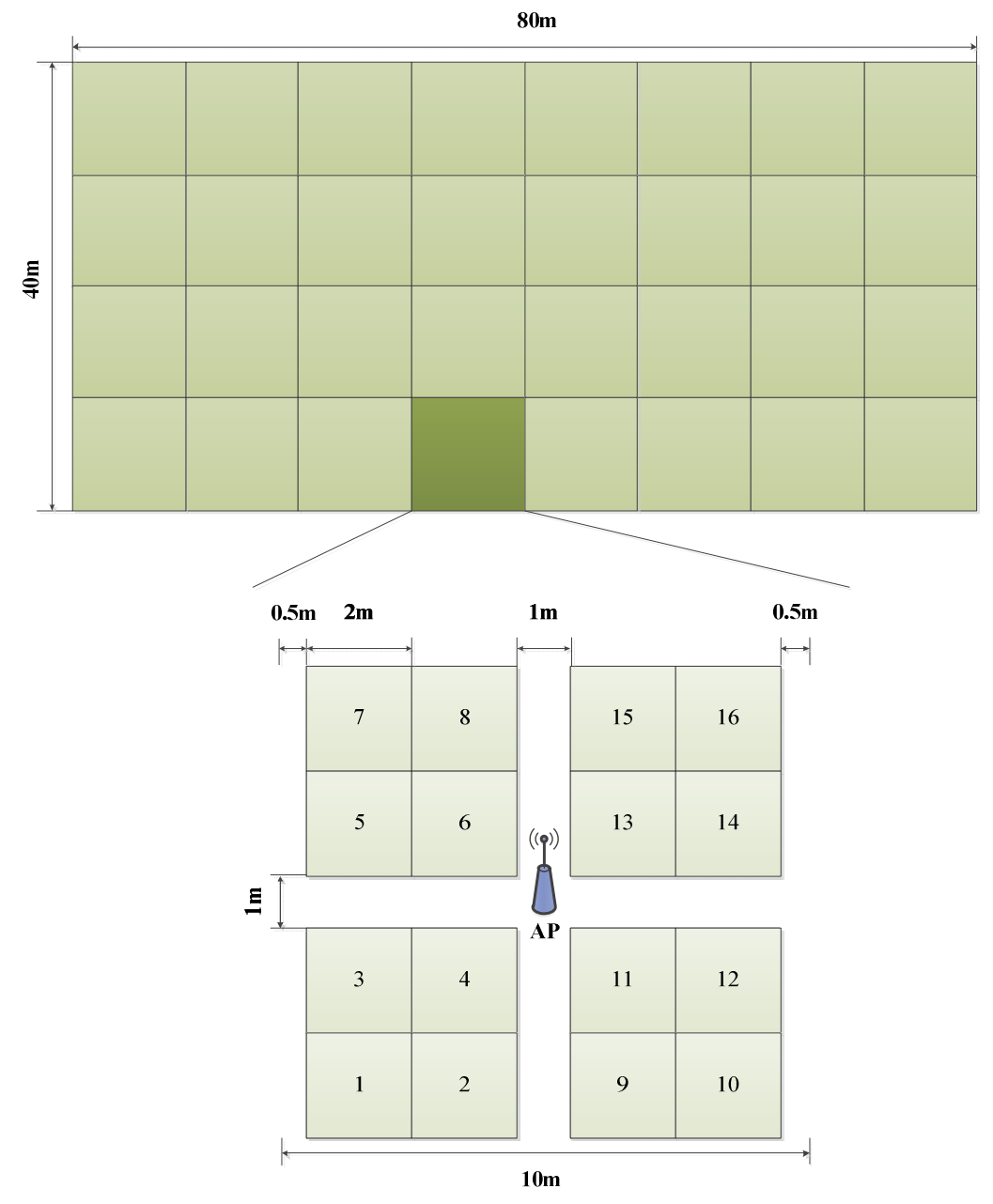

(a) Indoor scenario

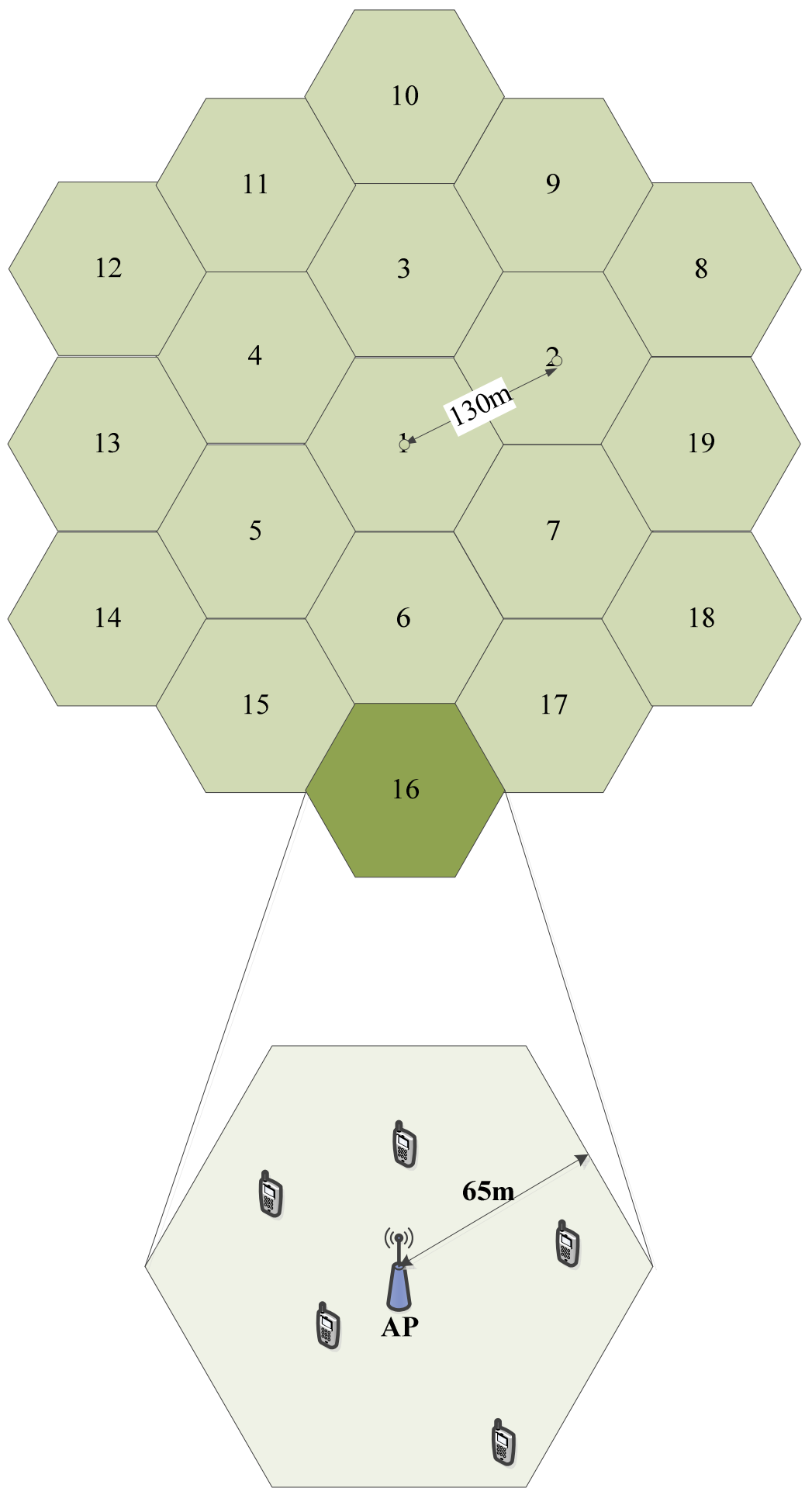

(b) Outdoor scenario

Fig. 47 Multiple BSS scenarios in the simulations.

Tab. 16 The parameter configuration of indoor multiple BSS scenario in the simulations.

| **Parameters** | **Description** |
|---|---|
| Service type | CBR |
| Per STA service rate | 0.05Mbps~3Mbps |
| Bandwidth | 20MHz |
| Number of BSS | 32 |
| The number of STAs in a BSS | 64 |
| STAs position | Randomly distributed in each room |
| MCS index | 0~11, MCS 10 and MCS 11 are only employed when RU is wider than or equal to 242-tone |
| AP transmit power | 18dBm |
| STA transmit power | 18dBm |
| AP antenna height | 1.5m |
| STA antenna height | 1.5m |
| Frequency | 5.57G |
| CCA threshold | OBSS-PD level: -62dBm<br>Traditional CCA level: -82dBm |
| SIFS | 16 μ s |
| DIFS | 34 μ s |
| CWmin | 15 |
| CWmax | 1023 |
| TXOP duration | 3.008ms |
| Number of AP antennas | 2 |
| Number of STA antennas | 2 |

The outdoor multiple BSS scenario in IEEE 802.11ax is shown in Fig. 47(b). The whole network topology is divided into three layers of BSS, and each BSS follows the rules of the outdoor single BSS scenario described previously. The distance between each adjacent AP is 130m. The detailed simulation parameter configuration of the outdoor multiple BSS scenario is shown in Tab. 17.

Tab. 17 Parameter configuration of outdoor multiple BSS scenario in the simulations.

| **Parameters** | **Description** |
|---|---|
| Service type | CBR |
| Per STA service rate | 0.05Mbps~3Mbps |
| Bandwidth | 20MHz |
| Number of BSS | 19 |
| Number of STAs in a BSS | 64 |
| STAs' position | Random Distributed in Each Room |
| MCS index | 0~11, MCS 10 and MCS 11 are only employed when RU is wider than or equal to 242-tone |
| AP transmit power | 18dBm |
| STA transmit power | 18dBm |
| AP antenna height | 1.5m |
| STA antenna height | 1.5m |
| Frequency | 5.57G |
| CCA threshold | OBSS-PD level：-62dBm<br>Traditional CCA level：-82dBm |
| SIFS | 16 μ s |
| DIFS | 34 μ s |
| CWmin | 15 |
| CWmax | 1023 |
| TXOP duration | 3.008ms |
| Number of AP antennas | 2 |
| Number of STA antennas | 2 |

The following three schemes are compared in our simulations:

**Scheme 1: IEEE 802.11ac**

The AP and STA of each BSS randomly contend for channel access using EDCA, and the carrier sensing control module used the traditional carrier monitoring threshold (-82dBm) to evaluate the physical carrier states. Similar to IEEE 802.11ax without SR, the virtual carrier sensing mechanism maintained a single NAV to implement the virtual carrier monitoring function.

**Scheme 2: IEEE 802.11ax without SR**

The AP and STAs have physical carrier sensing and virtual carrier sensing abilities with a physical carrier sensing threshold of -82dBm. Whether nodes receive the intra-BSS or inter-BSS data frame, the carrier sensing control module compared the received energy with -82dBm to evaluate the physical carrier state. Additionally, only OFDMA is used in this scheme; MU-MIMO is not adopted.

**Scheme 3: IEEE 802.11ax with SR**

In addition to the implementation of traditional physical carrier monitoring and virtual carrier monitoring, this scheme also adds the following mechanism: the BSS color field, two NAVs counters, and OBSS_PD. By reading the BSS color field in each frame, a node could identify the intra-BSS frame and inter-BSS frame. If the received frame is intra-BSS, the received energy is compared to the traditional CCA threshold; if the received frame was inter-BSS, the received energy is compared to the OBSS_PD level. After identifying a frame as either intra-BSS or inter-BSS according to BSS color, a node needs to establish and update intra-BSS NAV and basic NAV. Only when the values of the two NAVs counters are both equal to 0, the virtual carrier sensing is considered as idle; otherwise, the virtual carrier sensing is considered busy. As in Scheme 2, only OFDMA is used in this scheme; MU-MIMO is not adopted.

The functional comparison of the three schemes is shown in Tab. 18.

Tab. 18 The comparison of three schemes for multiple BSS simulations.

| | **IEEE 802.11ac** | **IEEE 802.11ax without SR** | **IEEE 802.11ax with SR** |
|---|---|---|---|
| EDCA-based channel access | √ | √ | √ |
| TXOP | √ | √ | √ |
| Aggregation and segmentation | √ | √ | √ |
| OFDMA | × | √ | √ |
| Spatial reuse | × | × | √ |

2) Simulation results

Several simulations are deployed and analyzed as follows.

i. Performance analysis of indoor multiple BSS

● Throughput performance

Fig. 48 shows the network performance of the DL transmissions in the high-dense indoor scenarios. The throughput of IEEE 802.11ax without SR is significantly larger than that of IEEE 802.11ac. Specifically, IEEE 802.11ax with SR achieves 52.7% performance gain compared with IEEE 802.11ax without SR. Similarly, Fig. 49 shows the network performance of UL transmission, and its trend is basically consistent with that of DL transmission. When compared to IEEE 802.11ax without SR, IEEE 802.11ax with SR achieves 34.3% performance gain. Thus, it is clear that the SR technology introduced by IEEE 802.11ax can significantly improve the network throughput.

The performance of IEEE 802.11ac is slightly higher than that of IEEE 802.11ax without SR when the service rate is extremely low. The reason for this phenomenon is that the number of STAs for MU transmission is less than three (i.e., the RU number in the simulations is three) when the service rate is low, resulting in wasted resources. This problem can be addressed via a more optimized scheduling algorithm, which is out of the scope of this article.

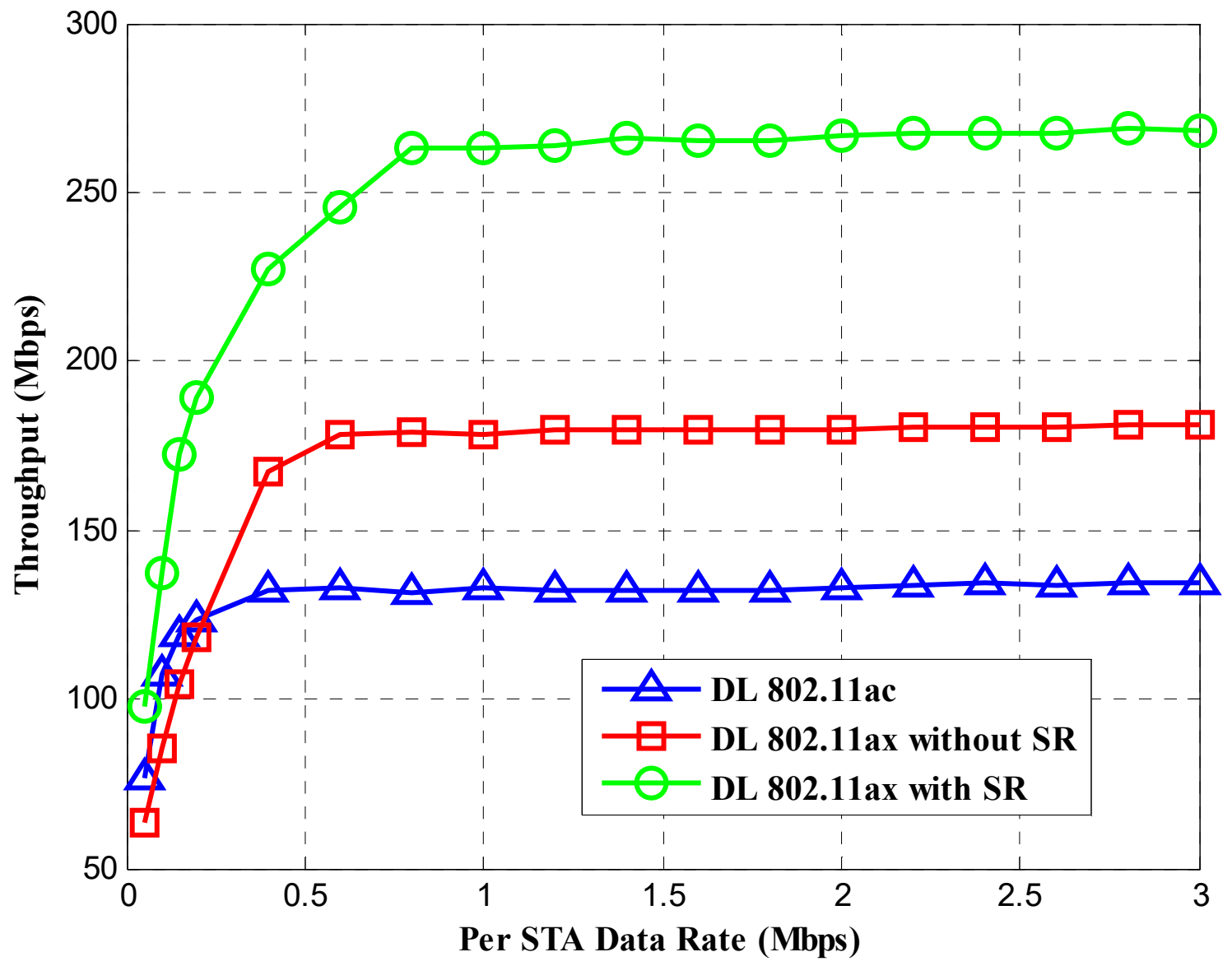

Fig. 48 Performance of DL transmission in outdoor scenarios with multiple BSS.

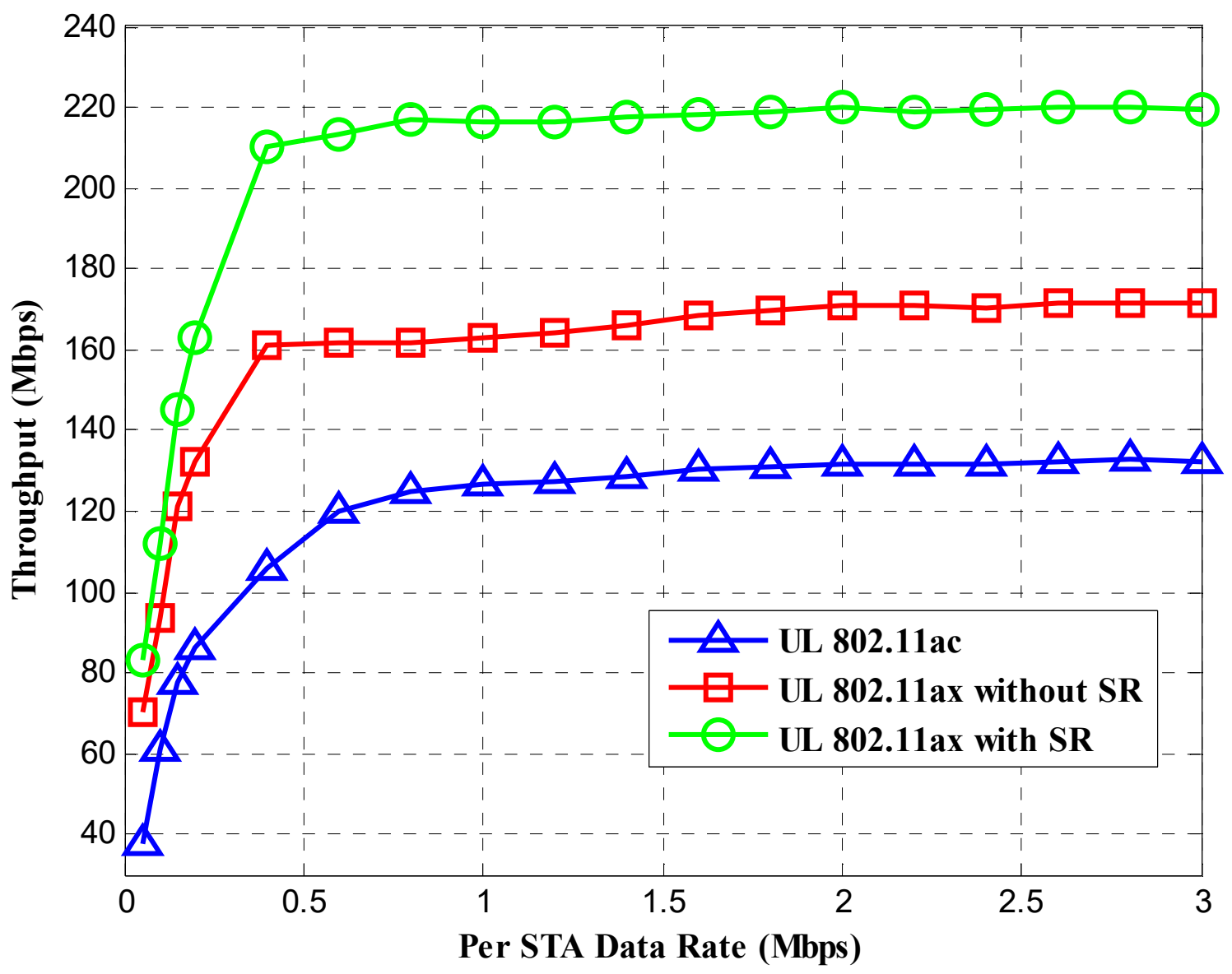

Fig. 49 Performance of UL transmission in outdoor scenarios with multiple BSS.

- The influence of BSS position on throughput

In multiple BSS scenarios, a valuable and interesting notion is the performance between different BSS varies, which is important for network planning, especially for enterprise and campus networks. Fig. 50 shows the throughput of 32 BSS adopting IEEE 802.11ax with SR. There are 64 STAs equipped with two antennas in each BSS, and the traffic rate of each STA is 3Mbps. If the BSS is closer to the corner of the network, it will probably achieves higher average BSS throughput; otherwise, if the BSS is closer to the center of the network, it will probably achieves lower average BSS throughput. In other words, interference in the corner is slight, whereas the interference in the middle position is relatively large.

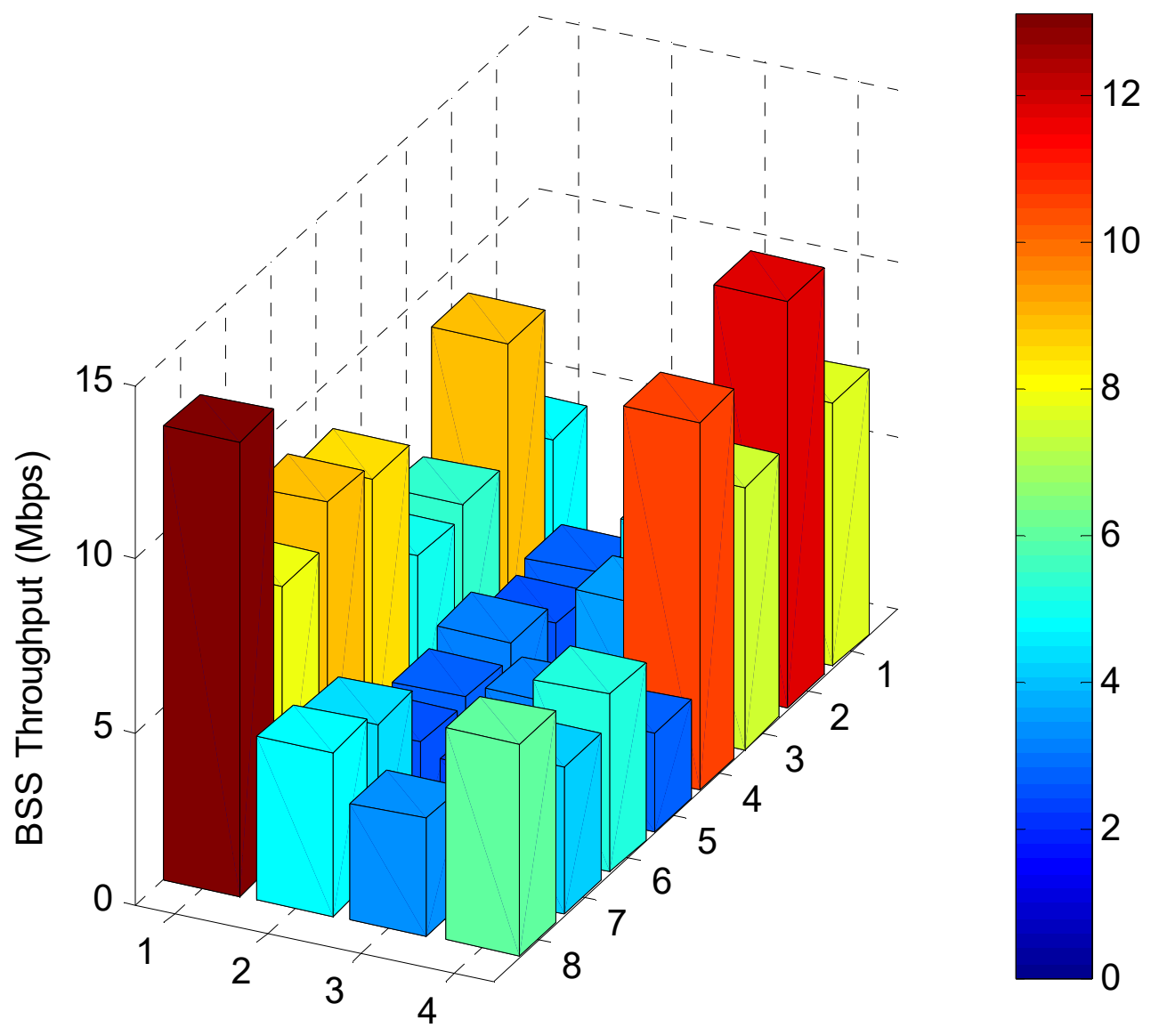

Fig. 50 Distribution of multiple BSS throughput in indoor scenarios.

- The distribution of per-STA throughput

We perform a statistical analysis of per-STA throughput using the cumulative distribution function (CDF) curve in Fig. 51. There are 64 STAs equipped with two antennas in each BSS, and the traffic rate of each STA is 3Mbps. First, we observe that per-STA throughput of IEEE 802.11ax with SR is larger than that of IEEE 802.11ax without SR and significantly larger than that of IEEE 802.11ac, consistent with previous simulation results. Next, we observe that IEEE 802.11ax improves the throughput of 5% of STAs, which meets the requirements of PAR. Finally, we also observe that the throughput of some STAs in the IEEE 802.11ac scheme is very large, suggesting that EDCA does not guarantee STA fairness. In contrast to IEEE 802.11ax, although there are differences in throughput among STAs and multiple BSS, fairness can be guaranteed to some extent in IEEE 802.11ax. We believe that the fairness of IEEE 802.11ax among STAs and multiple BSS can be further improved provided that more reasonable scheduling algorithms are designed.

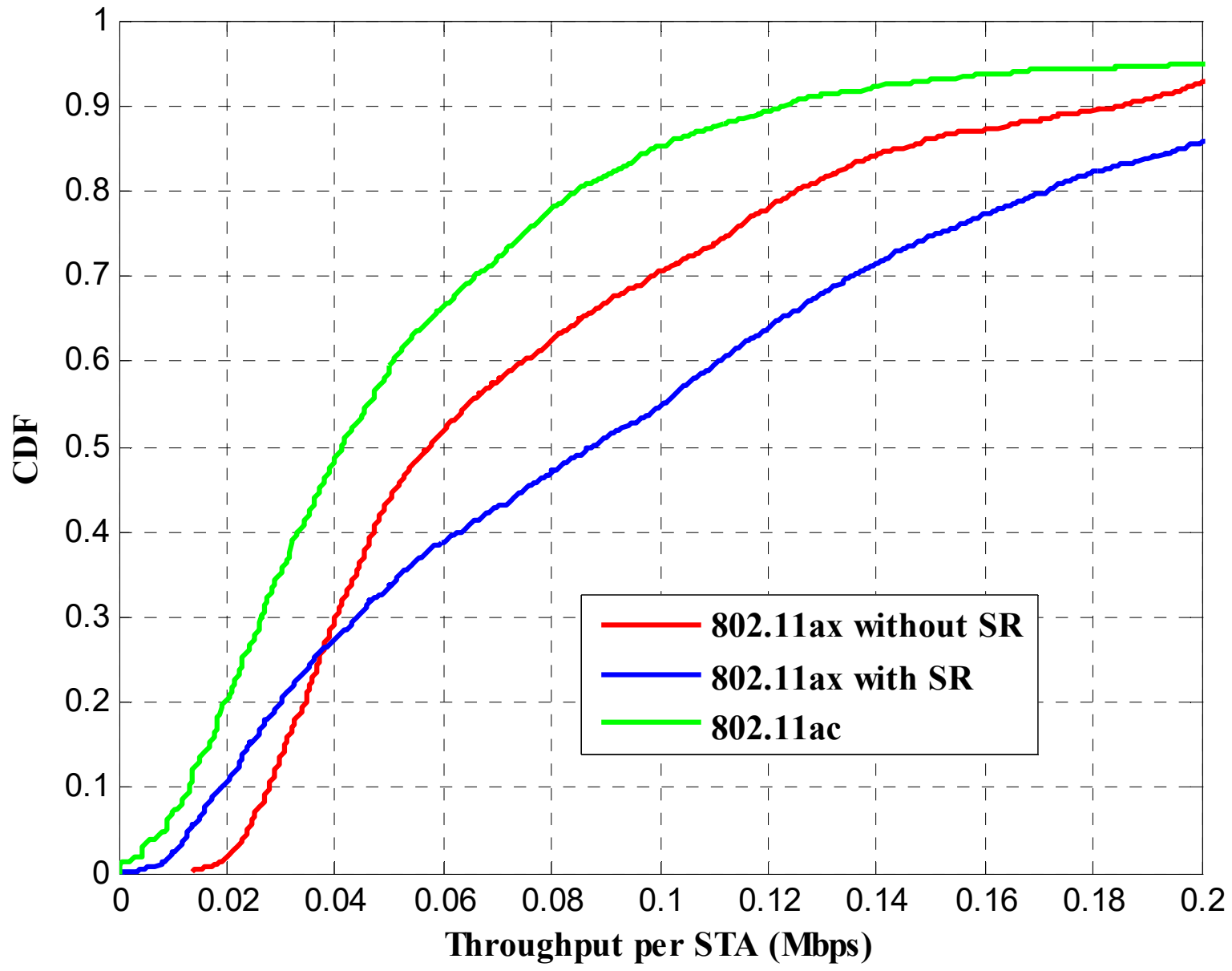

Fig. 51 Per-STA throughput CDF

ii. Performance analysis of multiple BSS in outdoor scenarios

Fig. 52 and Fig. 53 show the respective performance of DL and UL throughput of multiple BSS in outdoor scenarios. There is a negligible difference between the performance of IEEE 802.11ax with SR and IEEE 802.11ax without SR; hence, adopting SR in outdoor scenarios does not bring performance gain. The distance between two adjacent BSS is too long, and the data frames received in adjacent BSS usually experience severe attenuation, leading to small inter-cell interference and difficulty in receiving data frames. Thus, it is challenging to use the OBSS_PD based SR mechanism in outdoor scenarios. In addition, IEEE 802.11ax can still bring performance gain in outdoor scenarios, as the UL and DL throughput of IEEE 802.11ax increases to 127% and 117% compared to IEEE 802.11ac.

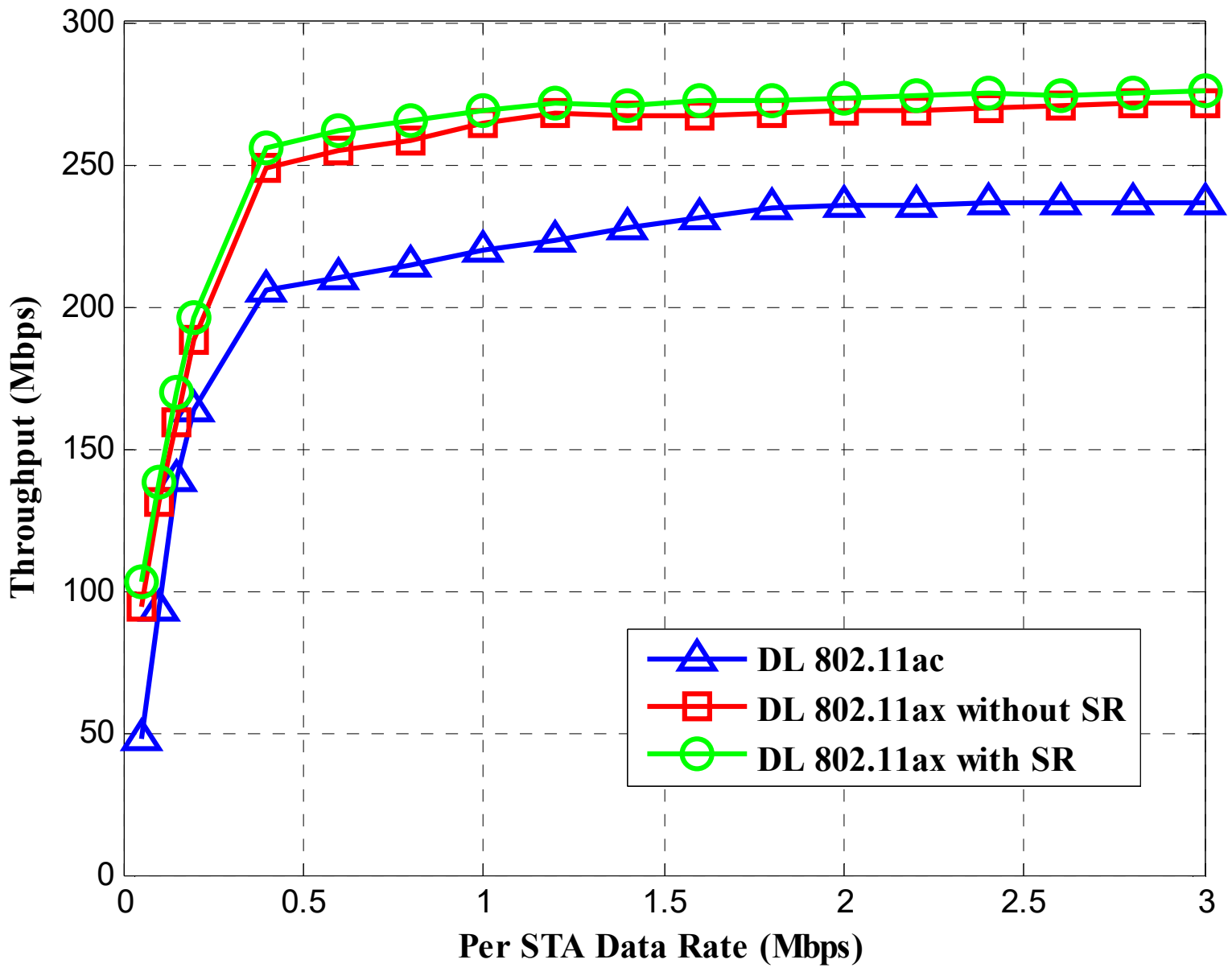

Fig. 52 Performance of DL transmission in outdoor scenarios with multiple BSS.

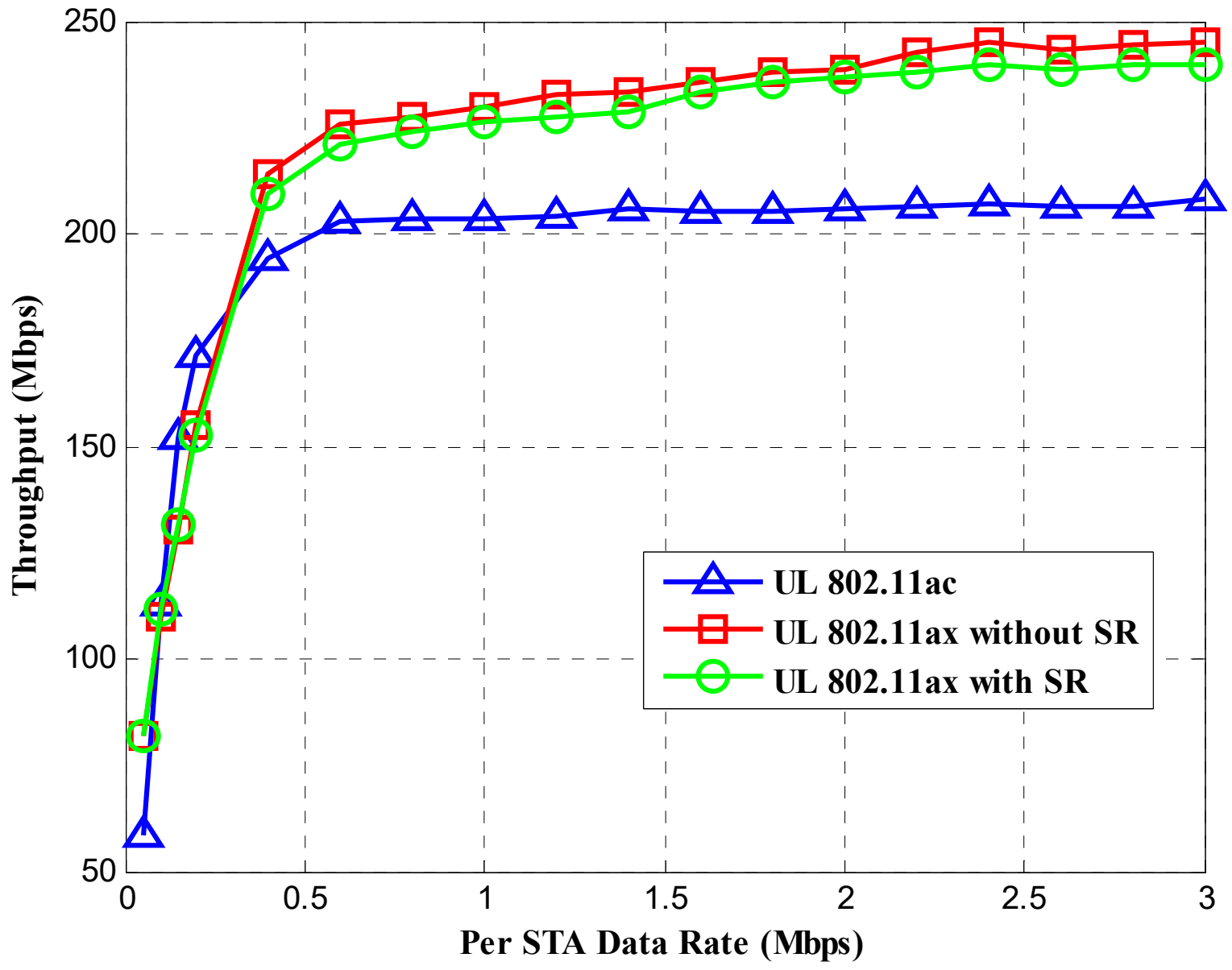

Fig. 53 Performance of UL transmission in outdoor scenarios with multiple BSS.

## E. Conclusion and Discussion

Tab. 19 summarizes and compares some important simulation results; we set the performance of the IEEE 802.11ac to 100% in this table.

- Compared to legacy IEEE 802.11, IEEE 802.11ax shows a significant improvement in throughput and per-STA throughput. In a specific scenario, it exceeds the overall throughput by 4 times that required by PAR, up to 4.74 times.
- The throughput of IEEE 802.11ax in single BSS and multiple BSS has been significantly improved, and the outdoor scenarios have been well supported while meeting the technical requirements of improving WLAN efficiency in PAR.
- In office scenarios, including single BSS and multiple BSS, DL throughput performance is better than UL throughput performance, mainly due to the small overhead of DL MAC. In the outdoor scenarios, including single BSS and multiple BSS, UL throughput performance is better than DL throughput performance. The main reason is that AP allocates transmit power to multiple RUs in DL MU transmission, while each of the UL STAs uses the maximum transmit power on its allocated RUs, leading to transmit power concentration on each RU and a greater chance of selecting the higher-order MCS.
- SR has obvious performance gains in office scenarios, but it needs a large number of BSS (i.e., a large area) to work. However, the performance gain of SR in outdoor scenarios is less obvious because the distance between adjacent BSS is too far, which leads to signal attenuation.
- The scheduling algorithm should be further studied for specific technical performance goals so the performance can be further improved. The scheduling algorithm is out of the scope of this article.
- The throughput of 5% of users has been improved to meet the requirements of PAR.
- Compared to Legacy IEEE 802.11, IEEE 802.11ax shows a significant improvement in throughput and per-STA throughput. In a specific scenario, it exceeds the overall throughput by 4 times that required by PAR, up to 4.74 times.

Tab. 19 Summary of performance evaluation results.

| | | | **IEEE 802.11ac** | **IEEE 802.11ax with OFDMA** | **IEEE 802.11ax with OFDMA and MU-MIMO** |
|---|---|---|---|---|---|
| **Single BSS scenario** | Indoor（UL） | 20MHz saturation throughput | 100% | 131% | 243% |
| | | 80MHz saturation throughput | 100% | 206% | 333% |
| | | 160MHz saturation throughput | 100% | 273% | **474%（Exceeds the performance goal in 802.11ax PAR: 4 times）** |
| | Outdoor（UL） | 20MHz saturation throughput | 100% | 145 % | 152% |
| | | 80MHz saturation throughput | 100% | 221% | 236% |
| | | 160MHz saturation throughput | 100% | 292% | 299% |
| **Multiple BSS scenario** | | | **IEEE 802.11ac** | **IEEE 802.11ax without SR** | **IEEE 802.11ax with SR** |
| | Indoor | UL | 100% | 130% | 171% |
| | | DL | 100% | 140% | 206% |
| | Outdoor | UL | 100% | 127% | 127% |
| | | DL | 100% | 117% | 117% |

# VIII. Conclusions and Perspectives

## A. Conclusions

Due to its unique advantages, WLAN has become one of the most important service providers for wireless networks. In this article, the latest generation of WLAN standards (IEEE 802.11ax) is introduced, including scenarios, technical requirements, standardization procedure, key technologies, and performance evaluation. First of all, from the view of the core application scenarios of IEEE 802.11ax, this article comprehensively summarizes and introduces its target requirements and standardization process. Next, the key technologies of IEEE 802.11ax are discussed in detail, including enhanced PHY technologies (discussed from aspects of new modulation and encoding technology, sub-carrier division, new multiple access technology based on OFDMA and UL MU-MIMO, and enhanced CB technology), enhanced MU-MAC technologies (discussed from aspects of UL MU-MAC, DL MU-MAC, and cascaded MU-MAC), SR technologies (discussed from aspects of BSS Color, two NAVs based virtual carrier sensing mechanism, OBSS_PD based SR mechanism, and SPR mechanism), and power efficiency enhancement technology. To the best of our knowledge, this article is the first to directly investigate and analyze the IEEE 802.11ax standard stable version (Draft 2.0). This article focuses on the standardization process of IEEE 802.11ax, and not only covers the standard technology, but also introduces the new progress of the related academic research. Thus, this article achieves a better fit with the latest developments of standardization process. We then comprehensively evaluate the performance of IEEE 802.11ax via the SLISP simulation platform, which is built by the authors and integrates both link level simulation and system simulation. The results indicate that IEEE 802.11ax satisfies the overall goal of significantly improving the user experience in high dense deployment scenarios, and achieves the single user throughput requirements of PAR, i.e., single user throughput needs to increase to four times compared to legacy IEEE 802.11. Furthermore, we also found some meaningful conclusions via the simulation results. Through the research, demonstration and simulation in this article, Tab. 20 shows that all the requirements of IEEE 802.11ax PAR have been met in the revision of IEEE 802.11ax standard. As far as we know, this article is the first to comprehensively and deeply evaluate the performance of IEEE 802.11ax with regard to the performance requirements of IEEE 802.11ax.

Tab. 20 Summary of the satisfaction of IEEE 802.11ax Draft2.0 to PAR

| **Requirement classification** | **Specific demand** | **The satisfaction of IEEE 802.11ax Draft2.0 to PAR** |
| --- | --- | --- |
| Overall goal: a significant increase in user experience for high-dense deployment scenarios. | In the same specific scenario, IEEE 802.11ax needs to achieve four times the average throughput of single STA compared to Legacy IEEE 802.11 | Satisfied (exceed expected goal) |
| | In addition, 5% single STA throughput, packet delay, packet error rate, and other performance also needs to be guaranteed | Satisfied |
| Improve WLAN efficiency | Higher spectral efficiency of single BSS | Satisfied |
| | Significantly enhance spectrum reuse and interference management capabilities in high-dense deployment multiple BSS scenarios | Satisfied |
| | Improve the robustness of outdoor scenarios and UL transmission | Satisfied |
| Support parallel | —— | Satisfied |
| transmission in frequency domain and spatial domain | | |
| Enhance power efficiency | —— | Satisfied |

### B. Future Prospectives of Beyond IEEE 802.11ax

WLANs and even wireless networks are carriers of the vision for highly interconnected human information. Therefore, the technological evolution of WLANs is consistent with the development of human demands for information interconnection. In the future, the ever-increasing diversity of network services and the rigor of performance requirements will doubtlessly become a development trend of wireless networks; e.g., the network service of virtual reality (VR), the IoT, ultra high-speed content delivery, social networks, and other networks presents challenging requirements for WLANs. Therefore, the IEEE 802.11 standard will continue to conduct the research on the key technology and the revision of the standard. The following forms the end of this article, and the authors would like to to share some of the key technology areas, which are expected to promote the development of WLANs in our opinion.

- Throughput enhancement technology in ultra dense scenes

This article has pointed out that dense deployment is the development trend of wireless networks, and this trend will likely continue to develop into ultra dense deployment in the future. Therefore, the improvement of network throughput will become the fundamental key technical target in the ultra dense deployment scenario. In the physical layer, the peak point-to-point transmission rate can be enhanced by using the higher order modulation (e.g., 2048-QAM and 4096-QAM), more efficient encoding technology, massive MIMO, full-duplex communication technology, and non-orthogonal multiple access technology. In the MAC layer, it is necessary to further enhance WLANs efficiency and reduce signaling overhead through protocol design. Additionally, cross-band carrier aggregation or channel bonding can achieve wider bandwidth and is therefore worthy of future research.

- Multiple BSS enhancement technologies

In ultra dense deployment wireless networks, the interference between links will become increasingly severe. Thus, how to improve the wireless network performance in multiple BSS scenarios will also be the promising technology for consideration. In the physical layer, the physical interference between links could be further reduced using interference elimination, interference coordination, and interference alignment; in the MAC layer, it would be helpful to improve the area throughput from the perspective of protocol using enhanced SR mechanism, TPC technology, channel reservation technology, inter BSS coordination, and scheduling mechanisms.

- Support for IoT

IoT is an important form of future networks. As one of the most significant wireless network bearer methods, the authors suggest that WLANs based on IEEE 802.11 also require supporting IoT. Since for the most part, IoT needs to operate in the narrow band mode, the challenge of supporting IoT in WLANs is how IEEE 802.11 supports the efficient operation of narrowband IoT in the same system (if necessary) without the degradation of wideband network performance to ensure that both types of networks have better compatibility and coexistence with each other.

- QoS guarantee technologies

The diversification of the network service leads to the diversity of QoS requirements of different users, such as bandwidth, delay, delay jitter, packet loss rate, and delay deterministic. On the one hand, IEEE 802.11 needs to consider how to support the diversity of QoS requirements in the standard; on the other hand, a more efficient scheduling mechanism is required to allocate wireless resources for users with different QoS requirements.

- Coexistence and convergence with cellular networks

The long-term coexistence of WLANs and cellular networks is the development trend of wireless networks. At present, the industry has proposed a number of solutions for cellular networks operating in the unlicensed frequency band, such as LTE unlicensed (LTE-U), Licensed Assisted Access (LAA), and MulteFire [215]. Thus, how WLANs coexist and coordinate with these networks more efficiently is the issue faced by both sides. Furthermore, the convergence of WLANs and cellular networks is also an interesting topic.

## Acknowledgment

This work was supported in part by the National Natural Science Foundations of CHINA (Grant No. 61771390, No. 61501373, No. 61771392, and No. 61271279), the National Science and Technology Major Project (Grant No. 2016ZX03001018-004), and the Fundamental Research Funds for the Central Universities (Grant No. 3102017ZY018).

## References

[1] Ericsson, “Ericsson mobility report: On the pulse of the networked society,” Ericsson, Stockholm, Sweden, Tech. Rep. EAB-16:018498, 2016.

[2] Cisco, “Cisco visual networking index: Global mobile data traffic forecast update, 2016c2021 white paper,” Cisco, Jialefuniya, America, Tech. Rep. Cisco white paper, 2017.

[3] M. S. Afaqui, E. Garcia-Villegas, and E. Lopez-Aguilera, “IEEE 802.11ax: Challenges and requirements for future high efficiency wifi,” IEEE Wireless Communications, vol. 24, no. 3, pp. 130 – 137, 2017.

[4] “Wireless lan medium access control (mac) and physical layer (phy) specifications amendment 6: Enhancements for high efficiency WLAN,” IEEE Draft 802.11ax/D2.0, Oct 2017.

[5] A. Al-Fuqaha, M. Guizani, M. Mohammadi, M. Aledhari, and M. Ayyash, “Internet of things: A survey on enabling technologies, protocols, and applications,” IEEE Communications Surveys Tutorials, vol. 17, no. 4, pp. 2347 – 2376, Fourthquarter 2015.

[6] M. Kamel, W. Hamouda, and A. Youssef, “Ultra-dense networks: A survey,” IEEE Communications Surveys Tutorials, vol. 18, no. 4, pp. 2522 – 2545, Fourthquarter 2016.

[7] IEEE, “802.11 hew sg proposed par,” IEEE, doc. IEEE 802.11-14/0165r1, 2014.

[8] “Wireless lan medium access control (mac) and physical layer (phy) specifications amendment 5: Television white spaces (tvws) operation,” IEEE Std 802.11af-2013, Feb 2014.

[9] “Wireless lan medium access control (mac) and physical layer (phy) specifications amendment 2: Sub 1 ghz license exempt operation,” IEEE Std 802.11ah-2016, May 2017.

[10] “Wireless lan medium access control (mac) and physical layer (phy) specifications amendment 3: Enhancements for very high throughput in the 60 ghz band,” IEEE Std 802.11ad-2012, Dec 2012.

[11] “Wireless lan medium access control (mac) and physical layer (phy) specifications amendment 7: Enhanced throughput for operation in license-exempt bands above 45 ghz,” IEEE Draft 802.11ay/D1.0, Nov 2017.

[12] D. J. Deng, K. C. Chen, and R. S. Cheng, “IEEE 802.11ax: Next generation wireless local area networks,” in 10th International Conference on Heterogeneous Networking for Quality, Reliability, Security and Robustness, Aug 2014, pp. 77 – 82.

[13] B. Li, Q. Qu, Z. Yan, and M. Yang, “Survey on ofdma based mac protocols for the next generation WLAN,” in 2015 IEEE Wireless Communications and Networking Conference Workshops (WCNCW), March 2015, pp. 131 – 135.

[14] W. Lin, B. Li, M. Yang, Q. Qu, Z. Yan, X. Zuo, and B. Yang, “Integrated link-system level simulation platform for the next generation WLAN - IEEE 802.11ax,” in 2016 IEEE Global Communications Conference (GLOBECOM), Dec 2016, pp. 1 – 7.

[15] W. Sun, O. Lee, Y. Shin, S. Kim, C. Yang, H. Kim, and S. Choi, “Wifi could be much more,” IEEE Communications Magazine, vol. 52, no. 11, pp. 22 – 29, Nov 2014.

[16] D. J. Deng, S. Y. Lien, J. Lee, and K. C. Chen, “On quality-of-service provisioning in IEEE 802.11ax WLANs,” IEEE Access, vol. 4, pp. 6086 – 6104, 2016.

[17] “Wireless lan medium access control (mac) and physical layer (phy) specifications,” IEEE

Std 802.11-1997, June 1997.

[18] “Wireless lan medium access control (mac) and physical layer (phy) specifications,” IEEE Std 802.11-2016 (Revision of IEEE Std 802.11-2012), December 2007.

[19] (2017) Official IEEE 802.11 working group project timelines. [Online]. Available: http://www.ieee802.org/11/Reports/802.11 Timelines.htm

[20] “Wireless lan medium access control (mac) and physical layer (phy) specifications,” IEEE Std 802.11-2007 (Revision of IEEE Std 802.11-1999), June 2007.

[21] “Wireless lan medium access control (mac) and physical layer (phy) specifications,” IEEE Std 802.11-2012 (Revision of IEEE Std 802.11-2007), March 2012.

[22] E. Khorov, A. Kiryanov, and A. Lyakhov, “IEEE 802.11ax: How to build high efficiency WLANs,” in 2015 International Conference on Engineering and Telecommunication (EnT), Nov 2015, pp. 14 – 19.

[23] B. Bellalta, L. Bononi, R. Bruno, and A. Kassler, “Next generation IEEE 802.11 wireless local area networks: Current status, future directions and open challenges,” Computer Communications, vol. 75, no. Supplement C, pp. 1 – 25, 2016. [Online]. Available: http://www.sciencedirect.com/science/article/pii/S0140366415003874

[24] N. Cheng and X. S. Shen, Next-Generation High-Efficiency WLAN. Cham: Springer International Publishing, 2017, pp. 651 – 675. [Online]. Available: https://doi.org/10.1007/978-3-319-34208-5 24

[25] M. X. Gong, B. Hart, and S. Mao, “Advanced wireless lan technologies: IEEE 802.11ac and beyond,” GetMobile: Mobile Comp. and Comm., vol. 18, no. 4, pp. 48 – 52, Jan. 2015. [Online]. Available: http://doi.acm.org/10.1145/2721914.2721933

[26] H. A. Omar, K. Abboud, N. Cheng, K. R. Malekshan, A. T. Gamage, and W. Zhuang, “A survey on high efficiency wireless local area networks: Next generation wifi,” IEEE Communications Surveys Tutorials, vol. 18, no. 4, pp. 2315 – 2344, Fourthquarter 2016.

[27] B. Bellalta, “IEEE 802.11ax: High-efficiency WLANs,” IEEE Wireless Communications, vol. 23, no. 1, pp. 38 – 46, February 2016.

[28] S. Merlin, G. Barriac, H. Sampath, and et al, “TGax simulation scenarios,” IEEE, doc. IEEE 802.11-14/0980r16, 2015.

[29] R. Porat, M Fischer, S. Merlin, and et al, “11ax evaluation methodology,” IEEE, doc. IEEE 802.11-14/0571r12, 2016.

[30] J. Liu, R. Porat, N. Jindal, and et al, “IEEE 802.11ax channel model document,” doc. IEEE 802.11-14/0882r4, 2014.

[31] M. Kamel, W. Hamouda, and A. Youssef, “Ultra-dense networks: A survey,” IEEE Communications Surveys Tutorials, vol. 18, no. 4, pp. 2522 – 2545, Fourthquarter 2016.

[32] R. Kudo, Y. Takatori, B. A. H. S. Abeysekera, Y. Inoue, A. Murase, A. Yamada, H. Yasuda, and Y. Okumura, “An advanced wi-fi data service platform coupled with a cellular network for future wireless access,” IEEE Communications Magazine, vol. 52, no. 11, pp. 46 – 53, Nov 2014.

[33] V. Jones and H. Sampath, “Emerging technologies for WLAN,” IEEE Communications Magazine, vol. 53, no. 3, pp. 141 – 149, March 2015.

[34] Y. Ma, J. Li, H. Li, H. Zhang, and R. Hou, “Multi-hop multi-ap multi-channel cooperation for high efficiency WLAN,” in 2016 IEEE 27th Annual International Symposium on Personal, Indoor, and Mobile Radio Communications (PIMRC), Sept 2016, pp. 1 – 7.

[35] Y. Kim, M. S. Kim, S. Lee, D. Griffith, and N. Golmie, “Ap selection algorithm with adaptive ccat for dense wireless networks,” in 2017 IEEE Wireless Communications and Networking Conference (WCNC), March 2017, pp. 1 – 6.
[36] (2012) The future of your office is wireless. [Online]. Available: https://www.inc.com/christina-desmarais/the-future-of-youroffice-is-wireless.html
[37] E. Perahia, R. Abiri, H. Yin, “Hew usage scenarios categorization,” doc. IEEE 802.11-13/0795r0, 2013.
[38] (2012) Study: 25 percent of all households use wifi.[Online]. Available: https://www.neowin.net/news/study-25-percentof-all-households-use-wifi
[39] (2012) Study: 61% of u.s. households now have wifi. [Online]. Available: https://techcrunch.com/2012/04/05/study-61-ofu-s-households-now-have-wifi/
[40] J. Soder, F. Mestanov, E. Sakai, K. Sakoda, and K. Agardh, “Stadium scenario for hew,” IEEE, doc. IEEE 802.11-14/0381r0, 2014.
[41] H. Persson, J. Soder, F. Mestanov, and et al, “Text proposal of a stadium scenario to ax,” doc. IEEE 802.11-14/0860r2, 2014.
[42] (2014) Status of IEEE 802.11 hew study group high efficiency WLAN (hew). [Online]. Available: http://www.ieee802.org/11/Reports/hew update.htm
[43] O. Aboul-Magd, “IEEE 802.11 hew sg proposed csd,” doc. IEEE 802.11-14/0169r1, 2014.
[44] (2017) Status of project IEEE 802.11ax high efficiency (he) wireless lan task group. [Online]. Available: http://www.ieee802.org/11/Reports/tgax update.htm
[45] R. Stacey, “Specification framework for tgax,” IEEE, doc. IEEE 802.11-15/0132r15, 2016.
[46] “Wireless lan medium access control (mac) and physical layer (phy) specifications amendment 6: Enhancements for high efficiency WLAN,” IEEE Draft 802.11ax/D1.0, Nov 2016.
[47] IEEE, “802.11ax selection procedure,” IEEE, doc. IEEE 802.11-014/0938-03-00ax, 2014.
[48] S. Merlin, G. Barriac, H. Sampath, and et al, “TGax simulation scenarios,” doc. IEEE 802.11-14/0980r16, 2015.
[49] L. Wang, H. Lou, H. Zhang, and et al, “Proposed 802.11ax functional requirements,” IEEE, doc. IEEE 802.11-14/1009r2, 2014.
[50] Y. Li, Y. Li, L. Liu, and et al, “Non-contiguous channel bonding in 11ax,” IEEE, doc. IEEE 802.11-16/0059r1, Jan. 2016.
[51] A. Ghosh, R. Ratasuk, B. Mondal, N. Mangalvedhe, and T. Thomas, “Lte-advanced: next-generation wireless broadband technology [invited paper],” IEEE Wireless Communications, vol. 17, no. 3, pp. 10 – 22, June 2010.
[52] M. Dehghani, K. Arshad, and R. MacKenzie, “Lte-advanced radio access enhancements: A survey,” Wireless Personal Communications, vol. 80, no. 3, pp. 891 – 921, Feb 2015. [Online]. Available: https://doi.org/10.1007/s11277-014-2062-y
[53] S. Srikanth, P. A. M. Pandian, and X. Fernando, “Orthogonal frequency division multiple access in wimax and lte: a comparison,” IEEE Communications Magazine, vol. 50, no. 9, pp. 153 – 161, September 2012.
[54] H. D. H. and T. D. A., LTE for UMTS: OFDMA and SC-FDMA Based Radio Access. John Wiley & Sons, 2009.
[55] E. Yaacoub and Z. Dawy, “A survey on uplink resource allocation in ofdma wireless networks,” IEEE Communications Surveys Tutorials, vol. 14, no. 2, pp. 322 – 337, Second

2012.

[56] L. Liu, R. Chen, S. Geirhofer, K. Sayana, Z. Shi, and Y. Zhou, “Downlink mimo in lte-advanced: Su-mimo vs. mu-mimo,” IEEE Communications Magazine, vol. 50, no. 2, pp. 140–147, February 2012.

[57] C. Lim, T. Yoo, B. Clerckx, B. Lee, and B. Shim, “Recent trend of multiuser mimo in lte-advanced,” IEEE Communications Magazine, vol. 51, no. 3, pp. 127–135, March 2013.

[58] L. Nagel, S. Pratschner, S. Schwarz, and M. Rupp, “Efficient multiuser mimo transmissions in the lte-a uplink,” in 2016 1st International Workshop on Link- and System Level Simulations (IWSLS), July 2016, pp. 1–6.

[59] H. Kwon, H. Seo, S. Kim, and B. G. Lee, “Generalized csma/ca for ofdma systems: protocol design, throughput analysis, and implementation issues,” IEEE Transactions on Wireless Communications, vol. 8, no. 8, pp. 4176–4187, August 2009.

[60] H. Kwon, S. Kim, and B. G. Lee, “Opportunistic multi-channel csma protocol for ofdma systems,” IEEE Transactions on Wireless Communications, vol. 9, no. 5, pp. 1552–1557, May 2010.

[61] X. Wang and H. Wang, “A novel random access mechanism for ofdma wireless networks,” in 2010 IEEE Global Telecommunications Conference GLOBECOM 2010, Dec 2010, pp. 1–5.

[62] H. S. Ferdous and M. Murshed, “Enhanced IEEE 802.11 by integrating multiuser dynamic ofdma,” in 2010 Wireless Telecommunications Symposium (WTS), April 2010, pp. 1–6.

[63] H. Lou, X. Wang, J. Fang, M. Ghosh, G. Zhang, and R. Olesen, “Multiuser parallel channel access for high efficiency carrier grade wireless lans,” in 2014 IEEE International Conference on Communications (ICC), June 2014, pp. 3868–3870.

[64] S. Valentin, T. Freitag, and H. Karl, “Integrating multiuser dynamic ofdma into IEEE 802.11 WLANs - llc/mac extensions and system performance,” in 2008 IEEE International Conference on Communications, May 2008, pp. 3328–3334.

[65] M. Kamoun, L. Mazet, and S. Gault, “Efficient backward compatible allocation mechanism for multi-user csma/ca schemes,” in 2009 First International Conference on Communications and Networking, Nov 2009, pp. 1–6.

[66] T. Mishima, S. Miyamoto, S. Sampei, and W. Jiang, “Novel dcf-based multi-user mac protocol and dynamic resource allocation for ofdma WLAN systems,” in 2013 International Conference on Computing, Networking and Communications (ICNC), Jan 2013, pp. 616–620.

[67] Q. Qu, B. Li, M. Yang, Z. Yan, and X. Zuo, “Mu-fuplex: A multiuser full-duplex mac protocol for the next generation wireless networks,” in 2017 IEEE Wireless Communications and Networking Conference (WCNC), March 2017, pp. 1–6.

[68] Y. P. Fallah, S. Khan, P. Nasiopoulos, and H. Alnuweiri, “Hybrid ofdma/csma based medium access control for next-generation wireless lans,” in 2008 IEEE International Conference on Communications, May 2008, pp. 2762–2768.

[69] G. Haile and J. Lim, “C-ofdma: Improved throughput for next generation WLAN systems based on ofdma and csma/ca,” in 2013 4th International Conference on Intelligent Systems, Modelling and Simulation, Jan 2013, pp. 497–502.

[70] D. J. Deng, K. C. Chen, and R. S. Cheng, “IEEE 802.11ax: Next generation wireless local area networks,” in 10th International Conference on Heterogeneous Networking for Quality,

Reliability, Security and Robustness, Aug 2014, pp. 77 – 82.

[71] J. Jung and J. Lim, “Group contention-based ofdma mac protocol for multiple access interference-free in WLAN systems,” IEEE Transactions on Wireless Communications, vol. 11, no. 2, pp. 648 – 658, February 2012.

[72] Q. Qu, B. Li, M. Yang, and Z. Yan, “An ofdma based concurrent multiuser mac for upcoming IEEE 802.11ax,” in 2015 IEEE Wireless Communications and Networking Conference Workshops (WCNCW), March 2015, pp. 136 – 141.

[73] Q. Qu, B. Li, M. Yang, Z. Yan, X. Zuo, and Y. Zhang, “The neighbor channel sensing capability for wireless networks,” in 2016 IEEE International Conference on Signal Processing, Communications and Computing (ICSPCC), Aug 2016, pp. 1 – 6.

[74] H. Zhou, B. Li, Z. Yan, M. Yang, and Q. Qu, “An ofdma based multiple access protocol with qos guarantee for next generation WLAN,” in 2015 IEEE International Conference on Signal Processing, Communications and Computing (ICSPCC), Sept 2015, pp. 1 – 6.

[75] H. Zhou, B. Li, Z. Yan, and M. Yang, “A channel bonding based qos-aware ofdma mac protocol for the next generation WLAN,” Mobile Networks and Applications, vol. 22, no. 1, pp. 19 – 29, Feb 2017. [Online]. Available: https://doi.org/10.1007/s11036-015-0670-8

[76] R. Zhou, B. Li, M. Yang, Z. Yan, and X. Zuo, “Qos-oriented ofdma mac protocol for the next generation WLAN (accepted),” Journal of Northwestern Polytechnical University, 2017.

[77] Y. Li, B. Li, M. Yang, and Z. Yan, “Spatial clustering group based ofdma multiple access scheme for the next generation WLAN (accepted),” in 2017 EAI International Conference on IoT as a Service (IoTaaS), 2017.

[78] H. Jin, B. C. Jung, H. Y. Hwang, and D. K. Sung, “Performance comparison of uplink WLANs with single-user and multi-user mimo schemes,” in 2008 IEEE Wireless Communications and Networking Conference, March 2008, pp. 1854 – 1859.

[79] K. Tan, H. Liu, J. Fang, W. Wang, J. Zhang, M. Chen, and G. M. Voelker, “Sam: Enabling practical spatial multiple access in wireless lan,” in Proceedings of the 15th Annual International Conference on Mobile Computing and Networking, ser. MobiCom ’09. New York, NY, USA: ACM, 2009, pp. 49 – 60. [Online]. Available: http://doi.acm.org/10.1145/1614320.1614327

[80] F. Babich and M. Comisso, “Theoretical analysis of asynchronous multi-packet reception in 802.11 networks,” IEEE Transactions on Communications, vol. 58, no. 6, pp. 1782 – 1794, June 2010.

[81] A. Ettefagh, M. Kuhn, C. Es, li, and A. Wittneben, “Performance analysis of distributed cluster-based mac protocol for multiuser mimo wireless networks,” EURASIP Journal on Wireless Communications and Networking, vol. 2011, no. 1, p. 34, Jul 2011. [Online]. Available: https://doi.org/10.1186/1687-1499-2011-34

[82] A. Mukhopadhyay, N. B. Mehta, and V. Srinivasan, “Acknowledgement-aware mpr mac protocol for distributed WLANs: Design and analysis,” in 2012 IEEE Global Communications Conference (GLOBECOM), Dec 2012, pp. 5087 – 5092.

[83] T. H. Lin and H. T. Kung, “Concurrent channel access and estimation for scalable multiuser mimo networking,” in 2013 Proceedings IEEE INFOCOM, April 2013, pp. 140 – 144.

[84] T. W. Kuo, K. C. Lee, K. C. J. Lin, and M. J. Tsai, “Leader-contentionbased user matching for 802.11 multiuser mimo networks,” IEEE Transactions on Wireless Communications, vol. 13, no. 8, pp. 4389 – 4400, Aug 2014.

[85] T. Tandai, H. Mori, K. Toshimitsu, and T. Kobayashi, “An efficient uplink multiuser mimo protocol in IEEE 802.11 WLANs,” in 2009 IEEE 20th International Symposium on Personal, Indoor and Mobile Radio Communications, Sept 2009, pp. 1153–1157.

[86] H. Li, K. Wu, Q. Zhang, and L. M. Ni, “Cuts: Improving channel utilization in both time and spatial domain in WLANs,” IEEE Transactions on Parallel and Distributed Systems, vol. 25, no. 6, pp. 1413–1423, June 2014.

[87] P. X. Zheng, Y. J. Zhang, and S. C. Liew, “Multipacket reception in wireless local area networks,” in 2006 IEEE International Conference on Communications, vol. 8, June 2006, pp. 3670–3675.

[88] S. Barghi, H. Jafarkhani, and H. Yousefi’zadeh, “Mimo-assisted mpraware mac design for asynchronous WLANs,” IEEE/ACM Transactions on Networking, vol. 19, no. 6, pp. 1652–1665, Dec 2011.

[89] S. Zhou and Z. Niu, “Distributed medium access control with sdma support for WLANs,” IEICE transactions on communications, vol. 93, no. 4, pp. 961–970, 2010.

[90] R. Liao, B. Bellalta, and M. Oliver, “Dcf/usdma: Enhanced dcf for uplink sdma transmissions in WLANs,” in 2012 8th International Wireless Communications and Mobile Computing Conference (IWCMC), Aug 2012, pp. 263–268.

[91] Y. J. Zhang, “Multi-round contention in wireless lans with multipacket reception,” IEEE Transactions on Wireless Communications, vol. 9, no. 4, pp. 1503–1513, April 2010.

[92] D. Jung, R. Kim, and H. Lim, “Asynchronous medium access protocol for multi-user mimo based uplink WLANs,” IEEE Transactions on Communications, vol. 60, no. 12, pp. 3745–3754, December 2012.

[93] R. Liao, B. Bellalta, M. Oliver, and Z. Niu, “Mu-mimo mac protocols for wireless local area networks: A survey,” IEEE Communications Surveys Tutorials, vol. 18, no. 1, pp. 162–183, Firstquarter 2016.

[94] K. Au, L. Zhang, H. Nikopour, E. Yi, A. Bayesteh, U. Vilaipornsawai, J. Ma, and P. Zhu, “Uplink contention based scma for 5g radio access,” in 2014 IEEE Globecom Workshops (GC Wkshps), Dec 2014, pp. 900–905.

[95] A. Bayesteh, E. Yi, H. Nikopour, and H. Baligh, “Blind detection of scma for uplink grant-free multiple-access, ” in 2014 11th International Symposium on Wireless Communications Systems (ISWCS), Aug 2014, pp. 853–857.

[96] L. Dai, B. Wang, Y. Yuan, S. Han, C. l. I, and Z. Wang, “Nonorthogonal multiple access for 5g: solutions, challenges, opportunities, and future research trends,” IEEE Communications Magazine, vol. 53, no. 9, pp. 74–81, September 2015.

[97] I. Jamil, L. Cariou, and J. F. Helard, “Improving the capacity of future IEEE 802.11 high efficiency WLANs,” in 2014 21st International Conference on Telecommunications (ICT), May 2014, pp. 303–307.

[98] J. Luo, J. Zhang, P. Loc, and et al, “Considerations on cca for obss opearation in 802.11ax,” IEEE, doc. IEEE 802.11-14/1225r1, Sep. 2014.

[99] N. Jindal, and R. Porat, “Performance gains from cca optimization,” IEEE, doc. IEEE 802.11-14/0889r3, Jun. 2014.

[100] G. Barriac, S. Merlin, and G. Cherian, “Changing cca in the residential environment,” IEEE, doc. IEEE 802.11-14/0846r0, Jul. 2014.

[101] H. Safavi-Naeini, E. Tuomaala, O. Alanen, S. Choudhury, E. Rantala, and J. Kneckt,

“Adapting cca and receiver sensitivity,” IEEE, doc. IEEE 802.11-14/1443r0, Nov. 2014.
[102] J. Pang, J. Zhang, L. Liu, and et al, “Performance evaluation of obss densification,” IEEE, doc. IEEE 802.11-14/0832r0, Jul. 2014.
[103] I. Jamil, L. Cariou, and T. Derham, “Obss reuse mechanism which preserves fairness,” IEEE, doc. IEEE 802.11-14/1207r1, Sep. 2014.
[104] F. L. Sita, P. Xia, J. Levy, and R. Murias, “Residential scenario sensitivity and transmit power control simulation results,” IEEE, doc. IEEE 802.11-14/0833r0, Jul. 2014.
[105] T. Nakahira, K. Ishihara, Y. Asai, Y. Takatori, R. Kudo, and M. Mizoguchi, “Centralized control of carrier sense threshold and channel bandwidth in high-density WLANs,” in 2014 Asia-Pacific Microwave Conference, Nov 2014, pp. 570 – 572.
[106] K. Murakami, T. Ito, and S. Ishihara, “Improving the spatial reuse of IEEE 802.11 WLAN by adaptive carrier sense threshold of access points based on node positions,” in 2015 Eighth International Conference on Mobile Computing and Ubiquitous Networking (ICMU), Jan 2015, pp. 132 – 137.
[107] Y. Hua, Q. Zhang, and Z. Niu, “Distributed physical carrier sensing adaptation scheme in cooperative map WLAN,” in GLOBECOM 2009 - 2009 IEEE Global Telecommunications Conference, Nov 2009, pp. 1 – 6.
[108] S. Kim, S. Yoo, J. Yi, Y. Son, and S. Choi, “Fact: Fine-grained adaptation of carrier sense threshold in IEEE 802.11 WLANs,” IEEE Transactions on Vehicular Technology, vol. 66, no. 2, pp. 1886 – 1891, Feb 2017.
[109] G. Smith, “Dynamic sensitivity control v2,” IEEE, doc. IEEE 802.11- 13/1012r4, Nov. 2013.
[110] G. Smith, “Dynamic sensitivity control practical usage,” IEEE, doc. IEEE 802.11-14/0779r2, Jun. 2014.
[111] Z. Zhong, F. Cao, P. Kulkarni, and Z. Fan, “Promise and perils of dynamic sensitivity control in IEEE 802.11ax WLANs,” in 2016 International Symposium on Wireless Communication Systems (ISWCS), Sept 2016, pp. 439 – 444.
[112] S. Coffey, D. Liu, B. Hart, and G. Hiertz, “A protocol framework for dynamic cca,” IEEE, doc. IEEE 802.11-14/0872r0, Jul. 2014.
[113] X. Zhang, H. Zhu, and G. Qiu, “Optimal physical carrier sensing to defend against exposed terminal problem in wireless ad hoc networks,” in 2014 23rd International Conference on Computer Communication and Networks (ICCCN), Aug 2014, pp. 1 – 6.
[114] T.-S. Kim, Y. Yang, and J. C. Hou, “Modeling of IEEE 802.11 based multi-rate wireless ad hoc networks and its application towards optimal network operation,” University of Illinois at Urbana-Champaign, Stockholm, Sweden, Tech. Rep. Urbana, Venezuela, 2007.
[115] Y. Yang, J. C. Hou, and L. C. Kung, “Modeling the effect of transmit power and physical carrier sense in multi-hop wireless networks,” in IEEE INFOCOM 2007 - 26th IEEE International Conference on Computer Communications, May 2007, pp. 2331 – 2335.
[116] R. Hedayat, Y. H. Kwon, Y. Seok, H. Kwon, V. Ferdowsi, and A. Jafarian, “Considerations for adaptive cca,” IEEE, doc. IEEE 802.11-14/1448r2, Nov. 2014.
[117] J. Son, and J. S. Kwak, “Further considerations on enhanced cca for 11ax,” IEEE, doc. IEEE 802.11-14/0847r1, Jul. 2014.
[118] J. Jiang, L. Chu, H. Zhang, and et al, “System level simulations on increased spatial reuse,” IEEE, doc. IEEE 802.11-14/0372r2, Mar. 2014.

[119] S. Choudhury, A. Cavalcante, F. Chaves, and et al, “Impact of cca adaptation on spatial reuse in dense residential scenario,” IEEE, doc. IEEE 802.11-14/0861r0, Jul. 2014.

[120] K. Ishihara, Y. Inoue, Y. Asai, and et al, “Consideration of asynchronous interference in obss environment,” IEEE, doc. IEEE 802.11-14/1148r1, Sep. 2014.

[121] R. Hedayat, Y. H. Kwon, Y. Seok, H. Kwon, V. Ferdowsi, and A. Jafarian, “Perspectives on spatial reuse in 11ax,” IEEE, doc. IEEE 802.11-14/1580r0, Dec. 2014.

[122] F. Ye and B. Sikdar, “Distance-aware virtual carrier sensing for improved spatial reuse in wireless networks,” in Global Telecommunications Conference, 2004. GLOBECOM ’04. IEEE, vol. 6, Nov 2004, pp. 3793 – 3797 Vol.6.

[123] IEEE, “Nav operation for spatial reuse,” IEEE, doc. IEEE 802.11-15/0797r0, Jul. 2015.

[124] J. Luo, P. Loc, L. Liu, and et al, “Obss nav and pd threshold rule for spatial reuse,” IEEE, doc. IEEE 802.11-15/1109r1, Sep. 2015.

[125] Y. Fang, D. Gu, A. B. McDonald, and J. Zhang, “A two-level carrier sensing mechanism for overlapping bss problem in WLAN,” in 2005 14th IEEE Workshop on Local Metropolitan Area Networks, Sept 2005, pp. 6 pp. – 6.

[126] P. Huang, R. Stacey, C. Ghosh, L. Cariou, and Q. Li, “Determining two network allocation vector settings,” Jan. 2017, uS Patent App. 14/973,528.

[127] IEEE, “11ax d0.3comment resolution for two navs- part ii,” IEEE, doc. IEEE 802.11-16/1173r2, Sep. 2016.

[128] B. Li, Y. Zhang, M. Yang, and Z. Yan, “A carrier detection method based on dynamic free channel estimation threshold,” 2015.

[129] K.-P. Shih, C.-M. Chou, M.-Y. Lu, and S.-M. Chen, “A distributed spatial reuse (dsr) mac protocol for IEEE 802.11 ad-hoc wireless lans,” in 10th IEEE Symposium on Computers and Communications (ISCC’05), June 2005, pp. 658 – 663.

[130] S. L. Su, Y. C. Tsai, and H. C. Liao, “Transmit power control exploiting capture effect for WLANs,” in 2015 Seventh International Conference on Ubiquitous and Future Networks, July 2015, pp. 634 – 638.

[131] P. Patras, H. Qi, and D. Malone, “Exploiting the capture effect to improve WLAN throughput,” in 2012 IEEE International Symposium on a World of Wireless, Mobile and Multimedia Networks (WoWMoM), June 2012, pp. 1 – 9.

[132] I. N. Vukovic and N. Smavatkul, “Impact of capture and varying transmit power levels on saturation throughput in IEEE 802.11,” in 2005 IEEE 16th International Symposium on Personal, Indoor and Mobile Radio Communications, vol. 3, Sept 2005, pp. 1435 – 1440 Vol. 3.

[133] G. J. Sutton, R. P. Liu, X. Yang, and I. B. Collings, “Modelling capture effect for 802.11 dcf under rayleigh fading,” in 2010 IEEE International Conference on Communications, May 2010, pp. 1 – 6.

[134] C. Gandarillas, C. Martın-Engenos, H. L. Pombo, and A. G. Marques, “Dynamic transmit-power control for wifi access points based on wireless link occupancy,” in 2014 IEEE Wireless Communications and Networking Conference (WCNC), April 2014, pp. 1093 – 1098.

[135] Y. Li, K. Li, W. Li, Y. Zhang, M. Sheng, and J. Chu, “An energyefficient power control approach for IEEE 802.11n wireless lans,” in 2014 IEEE International Conference on Computer and Information Technology, Sept 2014, pp. 49 – 53.

[136] K. Oteri, and R. Yang, “Power control for multi-user transmission in 802.11ax,” IEEE, doc. IEEE 802.11-16/0331r1, Mar. 2016.

[137] A. Bharadwaj, B. Tian, Y. Kim, and S. Vermani, “Power control for ul mu,” IEEE, doc. IEEE 802.11-16/0413r0, Mar. 2016.

[138] I. Pefkianakis, S. H. Y. Wong, H. Yang, S. B. Lee, and S. Lu, “Toward history-aware robust 802.11 rate adaptation,” IEEE Transactions on Mobile Computing, vol. 12, no. 3, pp. 502 – 515, March 2013.

[139] E. Yang, J. Choi, C. K. Kim, and J. Lee, “An enhanced link adaptation strategy for IEEE 802.11 wireless ad hoc networks,” in 2007 International Conference on Wireless Communications, Networking and Mobile Computing, Sept 2007, pp. 1672 – 1676.

[140] P. A. K. Acharya, A. Sharma, E. M. Belding, K. C. Almeroth, and K. Papagiannaki, “Congestion-aware rate adaptation in wireless networks: A measurement-driven approach,” in 2008 5th Annual IEEE Communications Society Conference on Sensor, Mesh and Ad Hoc Communications and Networks, June 2008, pp. 1 – 9.

[141] W. L. Shen, K. C. J. Lin, S. Gollakota, and M. S. Chen, “Rate adaptation for 802.11 multiuser mimo networks,” IEEE Transactions on Mobile Computing, vol. 13, no. 1, pp. 35 – 47, Jan 2014.

[142] A. B. Makhlouf and M. Hamdi, “Practical rate adaptation for very high throughput WLANs,” IEEE Transactions on Wireless Communications, vol. 12, no. 2, pp. 908 – 916, February 2013.

[143] I. Pefkianakis, Y. Hu, S. H. Wong, H. Yang, and S. Lu, “Mimo rate adaptation in 802.11n wireless networks,” in Proceedings of the Sixteenth Annual International Conference on Mobile Computing and Networking, ser. MobiCom ’10. New York, NY, USA: ACM, 2010, pp. 257 – 268. [Online]. Available: http://doi.acm.org/10.1145/1859995.1860025

[144] G. Noubir, R. Rajaraman, B. Sheng, and B. Thapa, “On the robustness of IEEE 802.11 rate adaptation algorithms against smart jamming,” in Proceedings of the Fourth ACM Conference on Wireless Network Security, ser. WiSec ’11. New York, NY, USA: ACM, 2011, pp. 97 – 108. [Online]. Available: http://doi.acm.org/10.1145/1998412.1998430

[145] A. Kamerman and L. Monteban, “Wavelan-ii: A high-performance wireless lan for the unlicensed band,” Bell Labs Technical Journal, vol. 2, no. 3, pp. 118 – 133, Summer 1997.

[146] K. V. Cardoso and J. F. de Rezende, “Increasing throughput in dense 802.11 networks by automatic rate adaptation improvement,” Wireless Networks, vol. 18, no. 1, pp. 95 – 112, Jan 2012. [Online]. Available: https://doi.org/10.1007/s11276-011-0389-9

[147] S. Biaz and S. Wu, “Rate adaptation algorithms for IEEE 802.11 networks: A survey and comparison,” in 2008 IEEE Symposium on Computers and Communications, July 2008, pp. 130 – 136.

[148] J. Wang, T. Pare, C. Wang, and et al, “Adaptive cca and tpc,” IEEE, doc. IEEE 802.11-15/1069r3, Sep. 2015.

[149] Z. Zhou, Y. Zhu, Z. Niu, and J. Zhu, “Joint tuning of physical carrier sensing, power and rate in high-density WLAN,” in 2007 Asia-Pacific Conference on Communications, Oct 2007, pp. 131 – 134.

[150] G. Smith, “Tg ax enterprise scenario, tpc and dsc,” IEEE, doc. IEEE 802.11-16/0350r0, Mar. 2016.

[151] Y. Zhang, B. Li, M. Yang, Z. Yan, and X. Zuo, “Joint optimization of carrier sensing

threshold and transmission rate in wireless ad hoc networks," in 2015 11th International Conference on Heterogeneous Networking for Quality, Reliability, Security and Robustness (QSHINE), Aug 2015, pp. 210 – 215.

[152] Z. Chen, X. Yang, and N. H. Vaidya, "Dynamic spatial backoff in fading environments," in 2008 5th IEEE International Conference on Mobile Ad Hoc and Sensor Systems, Sept 2008, pp. 255 – 264.

[153] H. Ma and S. Roy, "Simple and effective carrier sensing adaptation for multi rate ad-hoc mesh networks," in 2006 IEEE International Conference on Mobile Ad Hoc and Sensor Systems, Oct 2006, pp. 795 – 800.

[154] B. Li, M. Yang, Z. Yan, X. Zuo, and B. Yang, "Fractional-backoff procedure and dynamic cca," IEEE, doc. IEEE 802.11-16/0589r0, May 2016.

[155] B. Li, M. Yang, Z. Yan, X. Zuo, and R. Zhou, "Channel state estimation based bidirectional initialized random access," IEEE, doc. IEEE 802.11-16/0588r0, May 2016.

[156] M. Yang, B. Li, Z. Yan, X. Zuo, and L. Ji, "Multi-bss association for edge users throughput improvements," IEEE, doc. IEEE 802.11-16/0590r0, May 2016.

[157] L. Bo, T. Wenzhao, Z. Hu, and Z. Hui, "m-dibcr: Mac protocol with multiple-step distributed in-band channel reservation," IEEE Communications Letters, vol. 12, no. 1, pp. 23 – 25, January 2008.

[158] B. Li, W. Li, F. Valois, S. Ubeda, H. Zhou, and Y. Chen, "Performance analysis of an efficient mac protocol with multiple-step distributed inband channel reservation," IEEE Transactions on Vehicular Technology, vol. 59, no. 1, pp. 368 – 382, Jan 2010.

[159] C. W. Ahn, C. G. Kang, and Y. Z. Cho, "Soft reservation multiple access with priority assignment (srma/pa): a novel mac protocol for qos-guaranteed integrated services in mobile ad-hoc networks," in Vehicular Technology Conference Fall 2000. IEEE VTS Fall VTC2000. 52nd Vehicular Technology Conference (Cat. No.00CH37152), vol. 2, 2000, pp. 942 – 947 vol.2.

[160] S. Jiang, J. Rao, D. He, X. Ling, and C. C. Ko, "A simple distributed prma for manets," IEEE Transactions on Vehicular Technology, vol. 51, no. 2, pp. 293 – 305, Mar 2002.

[161] J. Choi, J. Yoo, S. Choi, and C. Kim, "Eba: an enhancement of the IEEE 802.11 dcf via distributed reservation," IEEE Transactions on Mobile Computing, vol. 4, no. 4, pp. 378 – 390, July 2005.

[162] G. Mario and C. R. Lin, "Maca/pr: An asynchronous multimedia multihop wireless network," in INFOCOM 97. Sixteenth Annual Joint Conference of the IEEE Computer and Communications Societies. Driving the Information Revolution., Proceedings IEEE, vol. 97, 1997.

[163] B. S. Manoj and C. S. R. Murthy, "Real-time traffic support for ad hoc wireless networks," in Proceedings 10th IEEE International Conference on Networks (ICON 2002). Towards Network Superiority (Cat. No.02EX588), 2002, pp. 335 – 340.

[164] Z. Ying, A. L. Ananda, and L. Jacob, "A qos enabled mac protocol for multi-hop ad hoc wireless networks," in Conference Proceedings of the 2003 IEEE International Performance, Computing, and Communications Conference, 2003., April 2003, pp. 149 – 156.

[165] S. Singh, P. A. K. Acharya, U. Madhow, and E. M. Belding-Royer, "Sticky csma/ca: Implicit synchronization and real-time qos in mesh networks," Ad Hoc Networks, vol. 5, no. 6, pp. 744 – 768, 2007, (1) Wireless Mesh Networks (2) Wireless Sensor Networks. [Online].

Available: http://www.sciencedirect.com/science/article/pii/S1570870507000121
[166] I. Joe, “Qos-aware mac with reservation for mobile ad-hoc networks,” in IEEE 60th Vehicular Technology Conference, 2004. VTC2004-Fall. 2004, vol. 2, Sept 2004, pp. 1108 – 1112 Vol. 2.
[167] S.-T. Sheu and T.-F. Sheu, “A bandwidth allocation/sharing/extension protocol for multimedia over IEEE 802.11 ad hoc wireless lans,” IEEE Journal on Selected Areas in Communications, vol. 19, no. 10, pp. 2065 – 2080, Oct 2001.
[168] B. Cho, K. Koufos, and R. Jantti, “Interference control in cognitive wireless networks by tuning the carrier sensing threshold,” in 8th International Conference on Cognitive Radio Oriented Wireless Networks, July 2013, pp. 282 – 287.
[169] M. Kaynia, N. Jindal, and G. E. Oien, “Improving the performance of wireless ad hoc networks through mac layer design,” IEEE Transactions on Wireless Communications, vol. 10, no. 1, pp. 240 – 252, January 2011.
[170] X. Yang and N. Vaidya, “On physical carrier sensing in wireless ad hoc networks,” in Proceedings IEEE 24th Annual Joint Conference of the IEEE Computer and Communications Societies., vol. 4, March 2005, pp. 2525 – 2535 vol. 4.
[171] X. Zhang, G. Qiu, Z. Dai, and D. K. Sung, “Coordinated dynamic physical carrier sensing based on local optimization in wireless ad hoc networks,” in 2013 IEEE Wireless Communications and Networking Conference (WCNC), April 2013, pp. 398 – 403.
[172] X. Zhang, H. Zhu, and G. Qiu, “Optimal physical carrier sensing to defend against exposed terminal problem in wireless ad hoc networks,” in 2014 23rd International Conference on Computer Communication and Networks (ICCCN), Aug 2014, pp. 1 – 6.
[173] K. Xu, M. Gerla, and S. Bae, “How effective is the IEEE 802.11 rts/cts handshake in ad hoc networks,” in Global Telecommunications Conference, 2002. GLOBECOM ’02. IEEE, vol. 1, Nov 2002, pp. 72 – 76 vol.1.
[174] F. Ye, S. Yi, and B. Sikdar, “Improving spatial reuse of IEEE 802.11 based ad hoc networks,” in Global Telecommunications Conference, 2003. GLOBECOM ’03. IEEE, vol. 2, Dec 2003, pp. 1013 – 1017 Vol.2.
[175] A. Hasan and J. G. Andrews, “The guard zone in wireless ad hoc networks,” IEEE Transactions on Wireless Communications, vol. 6, no. 3, pp. 897 – 906, March 2007.
[176] Y. Yunjie and L. Bo, “Performance evaluation of an efficient cooperative channel reservation mac protocol in wireless ad hoc networks,” in 2011 IEEE International Conference on Signal Processing, Communications and Computing (ICSPCC), Sept 2011, pp. 1 – 5.
[177] Y. Yuan, B. Li, Y. Chen, and H. Zhou, “Ccrm: A mac protocol with cooperative channel reservation for wireless ad hoc networks,” in 2011 7th International Conference on Wireless Communications, Networking and Mobile Computing, Sept 2011, pp. 1 – 4.
[178] Y. Yuan and B. Li, “Cooperative channel reservation multiple access protocol with multiple reservation information forwarding,” Journal of Xi’an Jiaotong University, vol. 46, no. 8, pp. 59 – 64, 2012.
[179] Y. Zhang, B. Li, M. Yang, and Z. Yan, “Capacity analysis of dense wireless networks with joint optimization of reservation and cooperation,” in 2016 IEEE Wireless Communications and Networking Conference, April 2016, pp. 1 – 6.
[180] Y. Zhang, B. Li, M. Yan, and Z. Yan, “Capacity analysis of wireless ad hoc networks with improved channel reservation,” in 2015 IEEE Wireless Communications and Networking

Conference (WCNC), March 2015, pp. 1189 – 1194.

[181] Y. Zhang, B. Li, M. Yang, Z. Yan, X. Zuo, and Q. Qu, “Ajrc-mac: An aloha-based joint reservation and cooperation mac for dense wireless networks,” in 2017 IEEE Wireless Communications and Networking Conference (WCNC), March 2017, pp. 1 – 6.

[182] S. Agarwal, R. H. Katz, S. V. Krishnamurthy, and S. K. Dao, “Distributed power control in ad-hoc wireless networks,” in 12th IEEE International Symposium on Personal, Indoor and Mobile Radio Communications. PIMRC 2001. Proceedings (Cat. No.01TH8598), vol. 2, Sep 2001, pp. F – 59 – F – 66 vol.2.

[183] T. Chakraborty and I. S. Misra, “Design and analysis of channel reservation scheme in cognitive radio networks,” Computers & Electrical Engineering, vol. 42, no. Supplement C, pp. 148 – 167, 2015. [Online]. Available: http://www.sciencedirect.com/science/article/pii/S0045790614002833

[184] B. Yang, B. Li, Z. Yan, and M. Yang, “A distributed multi-channel mac protocol with parallel cooperation for the next generation WLAN,” in 2016 IEEE Wireless Communications and Networking Conference, April 2016, pp. 1 – 6.

[185] B. Yang, B. Li, Z. Yan, M. Yang, “A channel reservation based cooperative multi-channel mac protocol for the next generation WLAN,” Wireless Networks, Aug 2016. [Online]. Available: https://doi.org/10.1007/s11276-016-1355-3

[186] B. Yang, B. Li, M. Yang, Z. Yan, and X. Zuo, “Mi-mmac: Mimo-based multi-channel mac protocol for WLAN,” in 2015 11th International Conference on Heterogeneous Networking for Quality, Reliability, Security and Robustness (QSHINE), Aug 2015, pp. 223 – 226.

[187] B. Yang, B. Li, Z. Yan, M. Yang, and X. Zuo, “A reliable channel reservation based multi-channel mac protocol with a single transceiver,” in 2015 11th International Conference on Heterogeneous Networking for Quality, Reliability, Security and Robustness (QSHINE), Aug 2015, pp. 265 – 271.

[188] H. Lei and A. A. Nilsson, “Queuing analysis of power management in the IEEE 802.11 based wireless lans,” IEEE Transactions on Wireless Communications, vol. 6, no. 4, pp. 1286 – 1294, April 2007.

[189] S. Baek and B. D. Choi, “Performance analysis of power save mode in IEEE 802.11 infrastructure WLAN,” in 2008 International Conference on Telecommunications, June 2008, pp. 1 – 4.

[190] P. Agrawal, A. Kumar, J. Kuri, M. K. Panda, V. Navda, R. Ramjee, and V. N. Padmanabhani, “Analytical models for energy consumption in infrastructure WLAN stas carrying tcp traffic,” in 2010 Second International Conference on COMmunication Systems and NETworks (COMSNETS 2010), Jan 2010, pp. 1 – 10.

[191] Y. h. Zhu and V. C. M. Leung, “Efficient power management for infrastructure IEEE 802.11 WLANs,” IEEE Transactions on Wireless Communications, vol. 9, no. 7, pp. 2196 – 2205, July 2010.

[192] X. Perez-Costa and D. Camps-Mur, “IEEE 802.11e qos and power saving features overview and analysis of combined performance [accepted from open call],” IEEE Wireless Communications, vol. 17, no. 4, pp. 88 – 96, August 2010.

[193] M. Tauber and S. N. Bhatti, “The effect of the 802.11 power save mechanism (psm) on energy efficiency and performance during system activity,” in 2012 IEEE International Conference on Green Computing and Communications, Nov 2012, pp. 573 – 580.

[194] P. Agrawal, A. Kumar, J. Kuri, M. K. Panda, V. Navda, and R. Ramjee, "Opsm - opportunistic power save mode for infrastructure IEEE 802.11 WLAN," in 2010 IEEE International Conference on Communications Workshops, May 2010, pp. 1 – 6.

[195] H. Tabrizi, G. Farhadi, and J. Cioffi, "An intelligent power save mode mechanism for IEEE 802.11 WLAN," in 2012 IEEE Global Communications Conference (GLOBECOM), Dec 2012, pp. 3460 – 3464.

[196] G. Anastasi, M. Conti, E. Gregori, and A. Passarella, "802.11 powersaving mode for mobile computing in wi-fi hotspots: Limitations, enhancements and open issues," Wirel. Netw., vol. 14, no. 6, pp. 745 – 768, Dec. 2008. [Online]. Available: http://dx.doi.org/10.1007/s11276-006-0010-9

[197] P. Si, H. Ji, F. R. Yu, and G. Yue, "IEEE 802.11 dcf psm model and a novel downlink access scheme," in 2008 IEEE Wireless Communications and Networking Conference, March 2008, pp. 1397 – 1401.

[198] X. Chen, S. Jin, and D. Qiao, "M-psm: Mobility-aware power save mode for IEEE 802.11 WLANs," in 2011 31st International Conference on Distributed Computing Systems, June 2011, pp. 77 – 86.

[199] R. P. Liu, G. J. Sutton, and I. B. Collings, "Wlan power save with offset listen interval for machine-to-machine communications," IEEE Transactions on Wireless Communications, vol. 13, no. 5, pp. 2552 – 2562, May 2014.

[200] S.-L. Tsao and C.-H. Huang, "A survey of energy efficient mac protocols for IEEE 802.11 WLAN," Computer Communications, vol. 34, no. 1, pp. 54 – 67, 2011. [Online]. Available: http://www.sciencedirect.com/science/article/pii/S014036641000424X

[201] M. Khosroshahy, "Study and implementation of IEEE 802.11 physical channel model in yans (ns3 prototype) network simulator," INRIASophia Antipolis-Plante Group, Tech. Rep., Nov. 2006.

[202] (2017) Network simulator 3. [Online]. Available: https://www.nsnam.org/

[203] M. Jacob, C. Mbianke, and T. Krner, "A dynamic 60 ghz radio channel model for system level simulations with mac protocols for IEEE 802.11ad," in IEEE International Symposium on Consumer Electronics (ISCE 2010), June 2010, pp. 1 – 5.

[204] K. Brueninghaus, D. Astely, T. Salzer, S. Visuri, A. Alexiou, S. Karger, and G. A. Seraji, "Link performance models for system level simulations of broadband radio access systems," in 2005 IEEE 16th International Symposium on Personal, Indoor and Mobile Radio Communications, vol. 4, Sept 2005, pp. 2306 – 2311 Vol. 4.

[205] A. Jonsson, D. Akerman, E. Fitzgerald, C. Nyberg, B. E. Priyanto, and K. Agardh, "Modeling, implementation and evaluation of IEEE 802.11ac in ns-3 for enterprise networks," in 2016 Wireless Days (WD), March 2016, pp. 1 – 6.

[206] D. Akerman and A. Jonsson, "Modeling, implementation and evaluation of IEEE 802.11ac in enterprise networks," 2016, student Paper.

[207] Q. Chen, F. Schmidt-Eisenlohr, D. Jiang, M. Torrent-Moreno, L. Delgrossi, and H. Hartenstein, "Overhaul of IEEE 802.11 modeling and simulation in ns-2," in Proceedings of the 10th ACM Symposium on Modeling, Analysis, and Simulation of Wireless and Mobile Systems, ser. MSWiM '07. New York, NY, USA: ACM, 2007, pp. 159 – 168. [Online]. Available: http://doi.acm.org/10.1145/1298126.1298155

[208] (2011) Network simulator 2. [Online]. Available: https:// www.isi.edu/nsnam/ns/

[209] H. Assasa and J. Widmer, "Implementation and evaluation of a WLAN IEEE 802.11ad model in ns-3," in Proceedings of the Workshop on Ns-3, ser. WNS3 '16. New York, NY, USA: ACM, 2016, pp. 57 – 64. [Online]. Available: http://doi.acm.org/10.1145/2915371.2915377

[210] W. Lin, B. Li, M. Yang, Q. Qu, Z. Yan, X. Zuo, and B. Yang, "Integrated link-system level simulation platform for the next generation WLAN - IEEE 802.11ax," in 2016 IEEE Global Communications Conference (GLOBECOM), Dec 2016, pp. 1 – 7.

[211] J. C. Ikuno, M. Wrulich, and M. Rupp, "System level simulation of lte networks," in 2010 IEEE 71st Vehicular Technology Conference, May 2010, pp. 1 – 5.

[212] L. Chenand, W. Chen, B. Wang, X. Zhang, H. Chen, and D. Yang, "System-level simulation methodology and platform for mobile cellular systems," IEEE Communications Magazine, vol. 49, no. 7, pp. 148 – 155, July 2011.

[213] C. Mehlf¨uhrer, J. Colom Ikuno, M. ˇSimko, S. Schwarz, M. Wrulich, and M. Rupp, "The vienna lte simulators - enabling reproducibility in wireless communications research," EURASIP Journal on Advances in Signal Processing, vol. 2011, no. 1, p. 29, Jul 2011. [Online]. Available: https://doi.org/10.1186/1687-6180-2011-29

[214] Y. Gao, X. Zhang, D. Yang, and Y. Jiang, "Unified simulation evaluation for mobile broadband technologies," IEEE Communications Magazine, vol. 47, no. 3, pp. 142 – 149, March 2009.

[215] Multefire: Lte-like performance with wifi-like simplicity. (2016) [Online]. Available: http://www.lightreading.com/mobile/4g-lte/multefirelte-like-performance-with-wifi-like-simplicity/v/d-id/722261